\documentclass{ptephy_v1}

\usepackage{amsmath,amssymb,amsfonts,bm,braket}
\usepackage{color}
\usepackage{graphicx}

\usepackage{threeparttable}

\makeatletter
\let\MYcaption\@makecaption
\makeatother

\usepackage[subrefformat=parens]{subcaption}

\makeatletter
\let\@makecaption\MYcaption
\makeatother

\newcommand{\slashb}[1]{\not\!\!{#1}}
\SetSymbolFont{largesymbols}{bold}{OMX}{txex}{b}{n}

\DeclareMathOperator{\km}{km}
\DeclareMathOperator{\s}{s}
\DeclareMathOperator{\MeV}{MeV}

\makeatother

\begin{document}

\title{Leptonic and semi-leptonic neutrino interactions with muons in the proto-neutron star cooling}


\author{Ken'ichi Sugiura}
\affil{Graduate School of Advanced Science and Engineering, Waseda University, 3-4-1 Okubo, Shinjuku, Tokyo 169-8555, Japan. \email{sugiura@heap.phys.waseda.ac.jp}}

\author{Shun Furusawa}
\affil{College of Science and Engineering, Kanto Gakuin University, 1-50-1 Mutsuurahigashi, Kanazawa-ku, Yokohama, Kanagawa 236-8501,
Japan.}
\affil{Interdisciplinary Theoretical and Mathematical Sciences Program (iTHEMS), RIKEN, Wako, Saitama 351-0198, Japan.}

\author{Kohsuke Sumiyoshi}
\affil{National Institute of Technology, Numazu College, Ooka 3600, Numazu, Shizuoka 410-8501, Japan.}

\author{Shoichi Yamada}
\affil{Graduate School of Advanced Science and Engineering, Waseda University, 3-4-1 Okubo, Shinjuku, Tokyo 169-8555, Japan.}


\begin{abstract}%
  It is known that muons are scarce just after the birth of a proto-neutron star via a supernova explosion but get more abundant as the proto-neutron star cools via neutrino emissions on the Kelvin-Helmholtz timescale.
  In this paper we evaluate all the relevant rates of the neutrino interactions with muons at different times in the proto-neutron star cooling.
  We are particularly interested in the late phase ($ t \gtrsim 10 \operatorname{s}$), which will be accessible in the next Galactic supernova but has not been studied well so far.
  We calculate both leptonic and semi-leptonic processes, for the latter of which we pay attention also to the form factors with their dependence on the transferred momentum as well as to the modification of the dispersion relations for nucleons on the mean field level.
  We find that the flavor-exchange reactions $\nu_e + \mu^- \rightarrow \nu_{\mu} + e^-$ and $\bar{\nu}_{\mu} + \mu^- \rightarrow \bar{\nu}_e + e^-$ can be dominant, particularly at low energies, over the capture of $\nu_e$ on neutron and the scatterings of $\bar{\nu}_{\mu}$ on nucleons as the opacity sources for these species and that the inverse muon decay $ \bar{\nu}_e + \nu_{\mu} + e^-  \leftrightarrows \mu^- $ can overwhelm the scatterings of $\bar{\nu}_e$ and $\nu_{\mu}$ on nucleons again at low energies.
  At high energies, on the other hand, the corrections in the semi-leptonic processes mentioned above are more important.
  We also show the non-trivial energy- and angular dependences of the flavor-exchange reactions and the inverse muon decay.
  In the study of the diffusion coefficients from these reactions, we find that $\bar{\nu}_{\mu}$ is most affected.
  These pieces of information are indispensable for numerical computations and the interpretation of results thereof for the proto-neutron star cooling particularly at the very late phase.
\end{abstract}

\subjectindex{E26, E32, E45}

\maketitle


\section{Introduction}
Understanding neutrino interactions in hot dense matter is indispensable not only to the investigation of the explosion mechanism of core-collapse supernovae (CCSNe), in which shock revival by neutrino heating is believed to be crucial, but also to the quantitative prediction of neutrino signals from newly born proto-neutron stars (PNSs) in CCSNe.
As a matter of fact, most of the gravitational binding energy of PNS is released in the form of neutrinos, with roughly a half of them being emitted in the explosion phase of CCSN and the remaining half being radiated in the cooling phase of PNS.

The latter phase, which lasts much longer than the former, will be more important observationally.
In fact, the PNS is formed from an unshocked inner core, which is shrouded by shocked hot matter, and is settled to a quasi-hydrostatic configuration at a few hundred milliseconds after core bounce, and its subsequent evolution is driven by diffusive neutrino emissions \cite{Burrows1986,Pons1999, Fischer2010, Hudepohl2010, Roberts2012ApJ, Nakazato2013, Roberts2017a}.
This phase is called the deleptonization phase or the cooling phase, which occurs on the Kelvin-Helmholtz (KH) timescale given as
\begin{equation}
  \tau_{\text{KH}} = \dfrac{E_b}{4 \pi R_{\nu}^2 F_{\nu}}, \label{eq:KH_timescale}
\end{equation}
where $E_b$, $R_{\nu}$ and $F_{\nu}$ are the binding energy of PNS, the radius of neutrino sphere and the neutrino energy flux, respectively.
If a Galactic CCSN occurs in the near future, we will observe neutrinos from the PNS cooling over a minute at facilities that are currently or will be soon operational such as Super-Kamiokande \cite{SK2016,Suwa2019} and Hyper-Kamiokande \cite{protocollaboration2018hyperkamiokande}, IceCube \cite{IceCube2011}, NO$\nu$a \cite{Nova2020}, DUNE \cite{DUNE2020} and JUNO \cite{JUNO2016} (for other detectors, see \cite{Scholberg2012}).

Observations of a long-term neutrino signal will provide us with invaluable information on the property of dense and hot nuclear matter (see \cite{Muller2019Review} for a recent review).
For example, the nuclear response to the weak current is modified at high densities via strong interactions among nucleons \cite{Sawyer1995, Keil1995, Martinez-Pinedo2012}.
This is true not only for scatterings through the neutral current (NC) but also for emissions and absorptions of neutrinos via the charged-current (CC).
As a result, the cooling timescale is affected.
The equation of state (EoS) of nuclear matter at very high densities is certainly another target for the observation of neutrinos into the late phase of the PNS cooling \cite{Sumiyoshi1995, Camelio2017, Kenichiro2019, Nakazato2020}.
In general, the softer the EoS is, the longer the cooling timescale becomes \cite{Kenichiro2019}.

In the past, only electrons and positrons were incorporated as the charged-lepton constituents in most of the quantitative simulations of both CCSN and PNS cooling. That is because the muon and tauon have rest masses much larger than the typical temperature in the CCSN interior and their existence is supposed to be suppressed in the early explosion phase.
Recently, the possible effects of the tiny population of muons on the supernova explosion as well as on the PNS cooling in the very early phase were investigated in supernova simulations under axisymmetry \cite{Bollig2017, Fischer2020, BolligPHDs}.
They showed that muons start to emerge soon after core bounce through the thermal pair production of mu-type neutrinos and anti-neutrinos and their subsequent conversion into muons by the following CC reactions, $\nu_{\mu} + n \rightarrow \mu^- + p $ and $\bar{\nu}_{\mu} + p \rightarrow \mu^+ + n$ .
They also demonstrated that the formation of muon softens the EoS through the conversion of thermal energy to the rest mass energy, leading to a bit more rapid contraction of PNS and higher neutrino luminosities and hence the enhanced heating of matter behind the stalled shock wave; as a result, the muon formation facilitates the neutrino-driven explosion in their model \cite{Bollig2017, BolligPHDs}.
Moreover, it was argued in \cite{Horowitz1998} that faster escapes from PNS of $\bar{\nu}_{\mu}$ than $\nu_{\mu}$ because of its smaller cross section for the NC scattering with nucleon due to the weak magnetism correction \cite{Horowitz2002} will pile up the muon lepton number later in the PNS, which is essentially vanishing at the beginning.
As a matter of fact, the non-negligible population of muon in the neutron star has been known from the studies of nuclear EoS over the years \cite{Douchin2001, Shen2002, Zhang2020a}.
It is hence obvious that the muon existence is more important at later times in the PNS cooling, affecting the neutrino transport and, as a result, the cooling.

Detailed investigations of the individual muon-related neutrino reactions by numerical evaluations are restricted to the early phase of explosion, however, in the context of supernova simulations.
In \cite{Fischer2020}, for example, the authors explore how and to what extent the muon is produced up to $ 30 \operatorname{ms}$ after core bounce.
They paid attention, in particular, to $\nu_{\mu}$ capture on neutron and $\bar{\nu}_{\mu}$ capture on proton as the main muon production reactions, and various scatterings of these $\mu$-type neutrinos and anti-neutrinos as well as flavor-exchange reactions, and evaluated their rates for two thermodynamic states appropriate at this early phase.
In \cite{Guo2020}, on the other hand, the authors evaluated these rates at a later post bounce time of $0.4 \operatorname{s}$, picking up again two representative thermodynamic conditions.
They also studied their effects on the opacities for $\bar{\nu}_e$ and $\nu_{\mu}$ and found that the inverse muon decay: $\bar{\nu}_e + \nu_{\mu} + e^- \rightarrow \mu^-$ is the dominant opacity sources for those neutrino at low energies, $\lesssim 10 \MeV$.
In addition, the flavor-exchange reactions involving $\mu^+$ were evaluated in \cite{Fischer2020} only at the very early time.

At later times into the PNS cooling, the corresponding basic information is absent in the literature.
It is not that there is no investigation at all.
To the quite contrary, PNS cooling calculations up to $10 \s$ postbounce with muon-related reactions incorporated have been published \cite{Fischer2021Axion, BolligPHDs}.
In these papers, the authors are focused only on the luminosities and average energies, two most observationally important quantities, as well as on the thermodynamic states in the PNS.
It is obvious, however, that in order to understand these results and extend them one way or another, the detailed information on the individual reactions for the thermodynamic conditions suitable for the late phase is indispensable.
It should be also pointed out that even later times up to $\sim$ a minute should be investigated, since we have a fair chance to probe that phase for the next Galactic supernova \cite{Suwa2019, Li2021}.
Not to mention, there has been no such work so far.
The purpose of this paper is hence to fill this gap.


In this paper, we evaluate the rates of various muon-related neutrino reactions for some thermodynamical conditions including the very late phase up to $50 \s$ that correspond to different times in the PNS cooling and consider their possible implications for the PNS cooling.
In so doing, it is useful to compare the rates of the muon-related reactions with those of semi-leptonic processes, which are more familiar in the conventional PNS cooling.
In particular, we evaluated the corrections from the recoil and the form factors in the weak currents of nucleons, another elaboration considered by some authors \cite{Roberts2017, Fischer2020, Guo2020} rather recently.

This paper is organized as follows.
We summarize the formulae of the individual reaction rates in the next section.
The details are presented in Appendices.
In section \ref{sec:Results}, we exhibit the results, picking up a couple of specific thermodynamical conditions of relevance until $50 \s$ and discuss possible effects of the muon existence on the PNS cooling.
We summarize our investigations in Section \ref{sec:Summary}.

\section{Weak interaction rates}
We list the neutrino reactions considered in this paper in Table \ref{tab:mu_reaction}, which are essentially the same as those in \cite{Bollig2017, Fischer2020}.
They are divided into two groups, i.e. the leptonic and semi-leptonic processes;
in the latter nucleons are involved and the form factors are fully taken into account in their weak currents;
small recoils of nucleons are also completely accounted for with in-medium modifications of the dispersion relations of nucleons at the mean field level (see subsection \ref{subsec:NNinteraction} for more details).
We also list in Table \ref{tab:mu_reaction} for readers' convenience the equation numbers that correspond to the final expressions of the reaction kernels and the inverse mean free paths for the leptonic reactions as well as to those of the structure functions and the inverse mean free paths for the semi-leptonic reactions.

Having in mind the application to the collision term of the Boltzmann equation, which may be written as
\begin{equation}
  \dfrac{1}{c}\left( \dfrac{d f_\nu}{dt} \right)_{\textrm{coll}} = - \dfrac{1}{\lambda_{\nu}} f_\nu+ j_{\nu}  (1 - f_{\nu}),
\end{equation}
where $\lambda_{\nu}$ and $j_{\nu}$ are the mean free path and emissivity of neutrino, respectively, we will give their expressions for each reaction listed in Table \ref{tab:mu_reaction} in turn.
Details of the derivations will be presented in Appendices.

\begin{table}[tbp]
  \centering
  \caption{Weak reactions considered in this paper. $l$ denotes electron or muon. Leptonic reactions are divided in two groups based on the formulae of the reaction rates. The final expressions of the corresponding reaction kernels, the structure functions and the inverse mean free paths are also listed. \label{tab:mu_reaction}}

  \hspace*{-2.5cm}
  \begin{tabular}{rlcc|lcc}
  \hline \hline
  \multicolumn{2}{c}{Leptonic reactions} &
  \begin{tabular}{c}
    Reaction \\
    kernel
  \end{tabular} &
  \begin{tabular}{c}
    Inverse \\
    mean free path
  \end{tabular} &
  \begin{tabular}{c}
    Semi-leptonic \\
    reactions
  \end{tabular} &
  \begin{tabular}{c}
    Structure \\
    function
    \end{tabular} &
  \begin{tabular}{c}
    Inverse \\
    mean free path
  \end{tabular} \\
  \hline
  &$\nu + l \to \nu + l$ & Eq. (\ref{eq:rk_lepsca}) & Eq. (\ref{eq:mfp_GroupA}) & $\nu_l + n \to l^- + p$ & Eq. (\ref{eq:sf_appendix}) & Eq. (\ref{eq:mfp_CC})\\
   &$\nu_e + \mu^- \leftrightarrows \nu_{\mu} + e^-$ & Eq. (\ref{eq:kernel_flex_out}) & Eq. (\ref{eq:mfp_GroupA}) & $\bar{\nu}_l + p \to l^+ + n$ & Eq. (\ref{eq:sf_appendix}) & Eq. (\ref{eq:mfp_CC}) \\
   Group A &$\bar{\nu}_e + \mu^+ \leftrightarrows \bar{\nu}_{\mu} + e^+$ & Eq. (\ref{eq:kernel_flex_out}) & Eq. (\ref{eq:mfp_GroupA}) & $\nu + N \to \nu + N$ & Eq. (\ref{eq:sf_appendix}) & Eq. (\ref{eq:mfp_NC})\\
   &$\bar{\nu}_{\mu} + \mu^- \leftrightarrows \bar{\nu}_{e} + e^-$ & Eq. (\ref{eq:kernel_flex_out}) & Eq. (\ref{eq:mfp_GroupA}) & $\bar{\nu}_l + p + l^- \to n$ & Eq. (\ref{eq:sf_appendix}) & Eq. (\ref{eq:mfp_beta})\\
   &$\nu_{\mu} + \mu^+ \leftrightarrows \nu_e + e^+$ & Eq. (\ref{eq:kernel_flex_out}) & Eq. (\ref{eq:mfp_GroupA}) & & &  \\ \cline{1-4}
   &$\mu^- \leftrightarrows e^- + \bar{\nu}_e + \nu_{\mu}$ & Eq. (\ref{eq:rk_mudecay}) & Eq. (\ref{eq:kernel_GroupB}) & & & \\
   &$\mu^+ \leftrightarrows e^+ + \nu_e + \bar\nu_{\mu}$ & Eq. (\ref{eq:rk_mudecay}) & Eq. (\ref{eq:kernel_GroupB}) & & & \\
   Group B&$l^- + l^+ \leftrightarrows \nu + \bar{\nu}$ & Eq. (\ref{eq:rk_pair}) & Eq. (\ref{eq:kernel_GroupB}) & & & \\
   &$e^- + \mu^+ \leftrightarrows \nu_e + \bar\nu_{\mu}$ & Eq. (\ref{eq:rk_lepani}) & Eq. (\ref{eq:kernel_GroupB}) & & & \\
   &$e^+ + \mu^- \leftrightarrows \bar\nu_e + \nu_{\mu}$ & Eq. (\ref{eq:rk_lepani}) & Eq. (\ref{eq:kernel_GroupB}) & & & \\
   \hline
  \end{tabular}

\end{table}

\subsection{Leptonic reactions}
We first summarize the reaction rates of the leptonic reactions.
We follow \cite{Guo2020, Yueh1976} for the derivation.
Although they are not original, we put them here and in appendices for self-containedness of the paper and convenience for readers.
The formulae of the reaction rates of leptonic reactions are divided into two groups further for notational convenience as shown in Table \ref{tab:mu_reaction}, which will be described in turn in the following.

\subsubsection{Group A}
This group includes the scatterings of a neutrino off a lepton, \mbox{$\nu + l \leftrightarrows \nu + l$}, and the lepton-flavor exchange/conversion reactions,
\begin{align}
  &\nu_e + \mu^- \leftrightarrows \nu_{\mu} + e^-, \\
  &\bar{\nu}_e + \mu^+ \leftrightarrows \bar{\nu}_{\mu} + e^+, \\
  &\bar{\nu}_{\mu} + \mu^- \leftrightarrows \bar{\nu}_e + e^-, \\
  &\nu_{\mu} + \mu^+\leftrightarrows \nu_e + e^+,
\end{align}
which are collectively expressed as
\begin{equation}
  \nu_1 + l_1 \leftrightarrows \nu_2 + l_2,
\end{equation}
and the 4-momenta of the incoming and outgoing neutrinos are denoted by $q_1^{\alpha}$ and $q_2^{\alpha}$, respectively, while those of the incoming and outgoing leptons are expressed as $p_1^{\alpha}$ and $p_2^{\alpha}$, respectively.
The inverse mean free path for $\nu_1$ can be written as
\begin{align}
  \dfrac{1}{\lambda_{\nu_1}} &= \dfrac{1}{2 E_1} \int
  \dfrac{d^3 \bm{q}_2}{(2 \pi)^3 2 E_2}
  \dfrac{d^3 \bm{p}_1}{(2 \pi)^3 2 p_1^0}
  \dfrac{d^3 \bm{p}_2}{(2 \pi)^3 2 p_2^0} 2 f_{l_1}\left(p_1^0\right) \left[ 1 -  f_{\nu_2}\left(E_2\right) \right] \left[ 1 -  f_{l_2}\left(p_2^0\right) \right] \notag \\
  & \ \ \ \ \ \ \ \ \ \ \ \ \ \times \left( 2 \pi \right)^4 \delta^{(4)} \left(q_1^{\alpha}+p_1^{\alpha} - q_2^{\alpha}-p_2^{\alpha}\right) \langle |\mathcal{M}|^2 \rangle \label{eq:lepsca_1} \\
  & = \int d E_2 d \cos \theta \, \dfrac{q_2 E_2 }{(2 \pi)^2} \left[ 1 -  f_{\nu_2}\left(E_2\right) \right] R^{\text{in}}_{\nu_1} \left( E_1, E_2, \cos \theta \right). \label{eq:mfp_GroupA}
\end{align}
In the above equations, $f$'s are the distribution functions in the phase space of individual particles involved and $\theta$ is the angle between $\bm{q}_1$ and $\bm{q}_2$; $E_1 = q_1^0$ and $E_2 = q_2^0$ are the energies of the incoming and outgoing neutrinos, respectively, and $q_2 = |\bm{q}_2|$. The spin-averaged matrix elements squared are expressed generally as \cite{Bruenn1985}
\begin{align}
  \langle |\mathcal{M}|^2 \rangle_{\text{lsc}} =& 16 G_F^2 \left[ \beta_1 (q_1 \cdot p_1)(q_2 \cdot p_2) + \beta_2 (q_1 \cdot p_2)(q_2 \cdot p_1) + \beta_3 m_l^2 (q_1 \cdot q_2) \right], \label{eq:M_lepsca}
\end{align}
and
\begin{equation}
  \langle |\mathcal{M}|^2 \rangle_{\text{flex}} = 64 G_F^2 \left[ \alpha_1 (q_1 \cdot p_e)(q_2 \cdot p_{\mu}) + \alpha_2 (q_1 \cdot p_{\mu})(q_2 \cdot p_{e}) \right], \label{eq:M_flex}
\end{equation}
for the leptonic scattering and the lepton-flavor exchange/conversion reactions, respectively,
where $G_F$ is the Fermi coupling constant and the coefficients $\beta_i$, $\alpha_1$ and $\alpha_2$ are given in Tables \ref{tab:beta_lepsca} and \ref{tab:alpha_flex}.
The detailed calculations of the reaction kernels $R^{\text{in}}_{\nu_1}$ for these reactions are summarized in Appendices \ref{appendix:lepsca} and \ref{appendix:flex}.

If the Boltzmann equation is employed as it is for the neutrino transport in PNS, the reaction kernel $R^{\text{in}}_{\nu_1}$ is sufficient.
In some approximate treatment, however, one may need to evaluate the
integration over the outgoing neutrino momentum in Eq. (\ref{eq:mfp_GroupA}).
We will do it in fact in the next section to compare the importance of the individual reactions quantitatively.
In so doing, we assume that the neutrino distribution is given by the Fermi-Dirac distribution for beta-equilibrium and the double integrations with respect to $E_2$ and $\cos \theta$ are done numerically.

The emissivity of neutrino is then given as
\begin{align}
  j_{\nu_1}
  & = \int d E_2 \, d \cos \theta \, \dfrac{q_2 E_2 }{(2 \pi)^2} f_{\nu_2}\left(E_2\right) R^{\text{out}}_{\nu_1} \left( E_1, E_2, \cos \theta \right), \label{eq:emi_GroupA}
\end{align}
where the notations are the same as in Eq. (\ref{eq:mfp_GroupA}) and the reaction kernel $R^{\text{out}}_{\nu_1}$ is obtained from the detailed balance condition as follows:
\begin{equation}
  R^{\text{in}}_{\nu_1} \left( E_1, E_2, \cos \theta \right) = e^{\beta \left( E_1 - E_2 + \mu_{l_1} - \mu_{l_2} \right)} R^{\text{out}}_{\nu_1} \left( E_1, E_2, \cos \theta \right). \label{eq:DetailedBalance_lsc}
\end{equation}

\begin{table*}[tbp]
  \centering
  \hspace*{-2.5cm}
  \begin{threeparttable}
  \caption{Coefficients in the matrix elements of lepton scatterings \label{tab:beta_lepsca}}

  \begin{tabular}{lccc}
  \hline \hline
  Lepton scattering \tnote{1} & $\beta_1$ \tnote{2}  & $\beta_2$ & $\beta_3$ \\
  \hline
  $\nu_l + l^- \leftrightarrows \nu_l + l^-$ or $\bar{\nu}_l + l^+ \leftrightarrows \bar{\nu}_l + l^+$ &
  $\left[ \left( C_V + 1 \right) + \left( C_A + 1 \right) \right]^2$ &
  $\left[ \left( C_V + 1 \right) - \left( C_A + 1 \right) \right]^2$ & $\left( C_A + 1 \right)^2 - \left( C_V + 1 \right)^2 $ \\
  $\bar{\nu}_l + l^- \leftrightarrows \bar{\nu}_l + l^-$ or $\nu_l + l^+ \leftrightarrows \nu_l + l^+$ &
  $\left[ \left( C_V + 1 \right) - \left( C_A + 1 \right) \right]^2$ &
  $\left[ \left( C_V + 1 \right) + \left( C_A + 1 \right) \right]^2$ & $\left( C_A + 1 \right)^2 - \left( C_V + 1 \right)^2 $ \\
  $\nu_{l_1} + l_2^- \leftrightarrows \nu_{l_1} + l_2^-$ or $\bar{\nu}_{l_1} + l_2^+ \leftrightarrows \bar{\nu}_{l_1} + l_2^+$ &
  $\left( C_V + C_A \right)^2$ &
  $\left( C_V - C_A \right)^2$ & $ C_A^2 - C_V^2 $ \\
  $\bar{\nu}_{l_1} + l_2^- \leftrightarrows \bar{\nu}_{l_1} + l_2^-$ or $\nu_{l_1} + l_2^+ \leftrightarrows \nu_{l_1} + l_2^+$ &
  $\left( C_V - C_A \right)^2$ &
  $\left( C_V + C_A \right)^2$ & $ C_A^2 - C_V^2 $ \\
  \hline
  \end{tabular}
  \begin{tablenotes}
    \item[1]{$l \in \{e, \mu\}$, $l_1 \in \{e, \mu, \tau\}$, $l_2 \in \{e, \mu\}$ and $l_1 \neq l_2$}
    \item[2]{$C_V = 2 \sin^2 \theta_W - 1/2$, $C_A = -1/2$ where $\theta_W$ is the Weinberg angle.}
  \end{tablenotes}
\end{threeparttable}
\end{table*}

\begin{table}[tbp]
  \centering
  \caption{Coefficients in the matrix elements for the lepton flavor exchange/conversion reactions \label{tab:alpha_flex}}

  \begin{tabular}{lcc}
  \hline \hline
  Reactions & $\alpha_1$ & $\alpha_2$ \\
  \hline
  $\nu_e + \mu^- \leftrightarrows \nu_{\mu} + e^-$ or $\bar{\nu}_e + \mu^+ \leftrightarrows \bar{\nu}_{\mu} + e^+$ & 1 & 0 \\
  $\bar{\nu}_{\mu} + \mu^- \leftrightarrows \bar{\nu}_e + e^-$ or $\nu_{\mu} + \mu^+\leftrightarrows \nu_e + e^+$ & 0 & 1 \\
  \hline
  \end{tabular}
\end{table}


\subsubsection{Group B}

This group includes the muon decays: $\mu^- \leftrightarrows e^- + \bar{\nu}_e + \nu_{\mu}$ and \mbox{$\mu^+ \leftrightarrows e^+ + \nu_e + \bar{\nu}_{\mu} $}, the pair creations/annihilations: $l^- + l^+ \leftrightarrows \nu + \bar{\nu}$ and the leptonic annihilations: \mbox{$e^- + \mu^+ \leftrightarrows \nu_{e} + \bar{\nu}_\mu$} and \mbox{$e^+ + \mu^-  \leftrightarrows \bar{\nu}_e + \nu_{\mu} $}.
These reactions are collectively expressed as
\begin{equation}
  l_1 \rightleftarrows l_2 + \nu_1 + \nu_2 \ \ \text{or} \ \ l_1 + l_2 \rightleftarrows \nu_1 + \nu_2,
\end{equation}
and the absorptivity and emissivity for $\nu_1$ are expressed, respectively, as
\begin{align}
  \dfrac{1}{\lambda_{\nu_1}} &= \int d E_2 d \cos \theta \, \dfrac{q_2 E_2 }{(2 \pi)^2} f_{\nu_{2}}\left(E_{2}\right) R^{\text{in}}_{\nu_1} \left( E_{1}, E_{2}, \cos \theta \right), \label{eq:kernel_GroupB}\\
  j_{\nu_1} &= \int d E_2 d \cos \theta \, \dfrac{q_2 E_2 }{(2 \pi)^2} \left[ 1 -  f_{\nu_{2}}\left(E_{2}\right) \right] R^{\text{out}}_{\nu_1} \left( E_{1}, E_{2}, \cos \theta \right).\label{eq:emi_GroupB}
\end{align}
The spin-averaged matrix elements squared of each reaction are
\begin{align}
  &\langle |\mathcal{M}|^2 \rangle_{\mu \text{decay}} = 64 G_F^2 (q_{\nu_e} \cdot p_e)(q_{\nu_{\mu}} \cdot p_{\mu}) , \label{eq:M_mudecay}\\
  &\langle |\mathcal{M}|^2 \rangle_{\text{pair}} = 32 \alpha G_F^2 \left[ \beta_1 (q_1 \cdot p_1)(q_2 \cdot p_2) + \beta_2 (q_1 \cdot p_2)(q_2 \cdot p_1) + \beta_3 m_l^2 (q_1 \cdot q_2) \right], \label{eq:M_pair}\\
  &\langle |\mathcal{M}|^2 \rangle_{\text{lep.ann.}} = 64 \gamma G_F^2 (q_{\nu_e} \cdot p_e)(q_{\nu_{\mu}} \cdot p_{\mu}), \label{eq:M_lepani}
\end{align}
for the muon decay, pair annihilation/creation and leptonic annihilation, respectively.
The coefficients $\alpha$ and $\beta_i$ in Eq. (\ref{eq:M_pair}) are given in Table \ref{tab:pair}, and $\gamma = 1/2$ for $e^{\mp} + \mu^{\pm} \rightarrow \nu_e (\bar{\nu}_e) + \bar{\nu}_{\mu} (\nu_{\mu})$ and $\gamma = 2$ for $\nu_e (\bar{\nu}_e) + \bar{\nu}_{\mu} (\nu_{\mu}) \rightarrow e^{\mp} + \mu^{\pm}$.
The detailed calculations of the reaction kernels for the these reactions are summarized in Appendices \ref{appendix:mu_decay}, \ref{appendix:pair} and \ref{appendix:lep_ani}.
The detailed balance conditions are written as
\begin{equation}
  R^{\text{in}}_{\nu_1} \left( E_1, E_2, \cos \theta \right) = e^{\beta \left( E_1 + E_2 - \mu_{l_1} + \mu_{l_2} \right)} R^{\text{out}}_{\nu_1} \left( E_1, E_2, \cos \theta \right) \label{eq:DetailedBalance_B1}
\end{equation}
for $l_1 \rightleftarrows l_2 + \nu_1 + \nu_2$ and
\begin{equation}
  R^{\text{in}}_{\nu_1} \left( E_1, E_2, \cos \theta \right) = e^{\beta \left( E_1 + E_2 - \mu_{l_1} - \mu_{l_2} \right)} R^{\text{out}}_{\nu_1} \left( E_1, E_2, \cos \theta \right) \label{eq:DetailedBalance_B2}
\end{equation}
for $l_1 + l_2 \rightleftarrows \nu_1 + \nu_2$.

\begin{table*}[htbp]
  \hspace*{-1.4cm}
  \centering
  \begin{threeparttable}
  \caption{Coefficients in matrix elements of pair annihilation \label{tab:pair}}

  \begin{tabular}{lcccc}
  \hline \hline
  Pair process \tnote{1} & $\alpha$ & $\beta_1$ \tnote{2} & $\beta_2$ & $\beta_3$ \\
  \hline
  $ l^- + l^+ \rightarrow \nu_l + \bar{\nu}_l$ & 1/4 &
  $\left[ \left( C_V + 1 \right) - \left( C_A + 1 \right) \right]^2$ &
  $\left[ \left( C_V + 1 \right) + \left( C_A + 1 \right) \right]^2$ & $\left( C_A + 1 \right)^2 - \left( C_V + 1 \right)^2 $ \\
  $ l^- + l^+ \leftarrow \nu_l + \bar{\nu}_l$ & 1 &
  $\left[ \left( C_V + 1 \right) - \left( C_A + 1 \right) \right]^2$ &
  $\left[ \left( C_V + 1 \right) + \left( C_A + 1 \right) \right]^2$ & $\left( C_A + 1 \right)^2 - \left( C_V + 1 \right)^2 $ \\
  $ l_1^- + l_1^+ \rightarrow \nu_{l_2} + \bar{\nu}_{l_2}$ & 1/4 &
  $\left( C_V - C_A  \right)^2$ & $\left( C_V + C_A\right)^2$ & $ C_A^2 - C_V^2 $ \\
  $ l_1^- + l_1^+ \leftarrow \nu_{l_2} + \bar{\nu}_{l_2}$ & 1 &
  $\left( C_V - C_A  \right)^2$ & $\left( C_V + C_A\right)^2$ & $ C_A^2 - C_V^2 $ \\
  \hline
  \end{tabular}
  \begin{tablenotes}
    \item[1]{$l \in \{e, \mu\}$, $l_1 \in \{e, \mu\}$, $l_2 \in \{e, \mu, \tau\}$ and $l_1 \neq l_2$}
    \item[2]{$C_V = 2 \sin^2 \theta_W - 1/2$, $C_A = -1/2$ where $\theta_W$ is the Weinberg angle.}
  \end{tablenotes}
  \end{threeparttable}
\end{table*}

\subsection{Semi-leptonic reactions} \label{subsec:NNinteraction}
In this subsection, we summarize the interactions that involve nucleons, that is,
the captures of electron- and muon-type neutrinos on neutron and those of the electron- and muon-type anti-neutrinos on proton,
the scatterings of all flavors of neutrinos on nucleons and the beta decay of neutron and its inverse.
These reactions, especially those via the CC are sensitive to the modifications of the dispersion relations of nucleons in the hot dense matter because they shift the thresholds in the reactions.

In this work, we take them into account at the mean-field level.
Although it is well known that the vertex corrections need to be considered simultaneously at the level of the random phase approximation (RPA) \cite{Burrows1998, Reddy1999, Yamada&Toki1999, Oertel2020}, we will defer it to a later paper as it is a major undertaking and we think it is still meaningful to make comparisons with other works that also neglected the corrections \cite{Guo2020}.
We note that our formulation is based on the structure functions of nucleons, which is suited for the incorporation of RPA later.

It should be also stressed that we employ in this paper the most generic form of the weak currents of nucleons with the weak magnetism and other form factors accounted for and fully consider the recoil of nucleons.
Although these effects have been already studied by some authors and reported sporadically in the literature \cite{Horowitz2002, Martinez-Pinedo2012, Roberts2017, Fischer2020PRC, Guo2020, Fischer2020},
we think it is useful to evaluate them for the thermodynamic conditions in the current context and gauge the importance of different muon-related reactions with respect to these corrections.

\subsubsection{CC reactions: $\nu_l + n \leftrightarrows l^- + p$ and $\bar{\nu}_l + p \leftrightarrows l^+ + n$}
We write these processes in general as
\begin{equation}
  \nu_{l,1}+N_2 \leftrightarrows l_3+N_4, \label{eq:semi_reaction}
\end{equation}
where $\nu_{l,1}$ denotes the neutrino (anti-neutrino), $N_2$ the neutron (proton), $l_3$ the charged lepton (antilepton), i.e., $e$ or $\mu$, and $N_4$ the proton (neutron).
The interaction Lagrangian at low energies may be given by the Fermi theory as
\begin{equation}
  \label{eq:lagrangian_CC}
  \mathcal{L} = \frac{G_F \cos \theta_C}{\sqrt{2}} l_{\alpha} \, j^{\alpha}_{\text{CC}},
\end{equation}
where $\theta_C$ is the Cabibbo angle; the leptonic current is given as
\begin{align}
  l_{\alpha} = \bar{l}_3 \gamma_{\alpha} \left( 1 - \gamma^5 \right) \nu_1, \label{eq:LeptonCurrent}
\end{align}
and the nucleonic charged current is expressed as
\begin{align}
  j^{\alpha}_{\text{CC}} &= \bar{\Psi}_4 \left\{ \gamma^{\alpha} \left[ G_V \left( q^2\right) - G_A \left( q^2 \right) \gamma^5 \right] + F_2\left(q^2 \right) \dfrac{i \sigma^{\alpha \beta } q_{\beta}}{M} - G_P \left( q^2\right)\gamma^5 \dfrac{q^{\alpha}}{M} \right\} \Psi_2. \label{eq:NCC_current}
\end{align}
In the above expression, $\Psi_2$ and $\Psi_4$ are the wave functions of $N_2$ and $N_4$, respectively, and $q^{\alpha} = p_1^{\alpha} - p_3^{\alpha}$ is the momentum transfer to nucleon; $M = \left(m_n + m_p \right)/2 $ is the bare average mass of nucleons; the vector, axial vector, tensor and pseudoscalar form factors are given by \cite{leitner2005neutrino}, respectively, as
\begin{align}
  G_V(q^2) &= \dfrac{  g_V \left[1-\dfrac{q^2(\gamma_p-\gamma_n)}{4M^2}\right] }{  \left(1-\dfrac{q^2}{4M^2}\right) \left(1-\dfrac{q^2}{M_V^2}\right)^2 }, \\
  G_A(q^2) &= \dfrac{ g_A }{ \left(1-\dfrac{q^2}{M_A^2}\right)^2 }, \label{eq:GA} \\
  F_2(q^2) &=
  \dfrac{\gamma_p - \gamma_n -1}{ \left(1-\dfrac{q^2}{4M^2}\right) \left(1-\dfrac{q^2}{M_V^2}\right)^2 }, \\
  G_P(q^2) &= \dfrac{2 M^2 G_A(q^2)}{ m_\pi^2-q^2},
\end{align}
where $q^2 = q_{\alpha} q^{\alpha}$;
$g_V = 1$ and $g_A = 1.27$ are the vector and axial vector coupling constants, respectively; $\gamma_p = 2.793$ and $\gamma_n = -1.913$ are the magnetic moments of proton and neutron, respectively;
$M_V = 840 \MeV$, $M_A = 1 \operatorname{GeV}$ and $m_{\pi} = 139.57 \MeV$ are the vector, axial and pion mass, respectively.
Note that the tensor contribution to Eq. (\ref{eq:NCC_current}) is nothing but the weak magnetism; we do not consider the possible reduction of $g_A$ in Eq. (\ref{eq:GA}) at high densities \cite{Carter2002} although its incorporation is trivial.
The spin-averaged matrix element squared can be expressed as
\begin{equation}
  \langle |\mathcal{M}|^2 \rangle = \beta \dfrac{G_F^2 \cos^2 \theta_C}{2} L_{\mu \nu} \Lambda^{\mu \nu},
\end{equation}
with the leptonic tensor $L_{\mu \nu}$ and the hadronic counterpart $\Lambda^{\mu \nu}$; the coefficient $\beta$ originates from the spin average; $\beta = 1/2$ for $\nu_{l,1}+N_2 \rightarrow l_3+N_4$ and $\beta = 1/4$ for $\l_3+N_4 \rightarrow \nu_{l,1}+N_2$.
The leptonic tensor is written as
\begin{equation}
  L_{\mu \nu} =8 \left( p_{3\mu} p_{1\nu} + p_{3\nu} p_{1\mu} - \eta_{\mu \nu} (p_1 \cdot p_3) \pm i p_3^\rho p_1^\sigma \epsilon_{\rho \mu \sigma \nu}\right), \label{eq:LeptonTensor}
\end{equation}
where the sign is $+$ for neutrino and $-$ for anti-neutrino and all neutrinos are assumed to be massless.
The hadronic tensor is modified in medium and, as we mentioned earlier, we take into account the modifications of propagators at the mean field level as follows \cite{Roberts2017_muon}:
\begin{align}
  \Lambda^{\mu \nu} =  & \operatorname{Tr} \left\{
    (\slashb{\tilde{p}}_4 + m_4^*)
    \left[ \gamma^{\mu} (G_V - G_A \gamma^5) + F_2 \dfrac{i \sigma^{\mu \alpha} q_\alpha}{2M} - G_P \dfrac{q^{\mu}}{M} \gamma^5 \right] \right. \notag \\
  &\ \ \ \left. \times (\slashb{\tilde{p}}_2 + m_2^*)
    \left[ \gamma^{\nu} (G_V - G_A \gamma^5) - F_2 \dfrac{i \sigma^{\nu \beta} q_\beta}{2M} + G_P \dfrac{q^{\nu}}{M} \gamma^5 \right]
    \right\},
\end{align}
where we introduce the following notation: \mbox{$\tilde{p}_2^{\alpha} = \left( E_2^*, \bm{p}_2 \right)$, $\tilde{p}_4^{\alpha} = \left( E_4^*, \bm{p}_4 \right)$} and
\begin{equation}
  E_{2,4}^* = E_{2,4} - U_{2,4}
\end{equation}
with $U_{2,4}$ being the mean field potentials of particles $2$ and $4$; $E_{2,4}^*$ is given by the on-shell condition in medium as
\begin{equation}
  E_{2,4}^* = \sqrt{ |\bm{p}_{2,4}|^2 + {m_{2,4}^*}^2 },
\end{equation}
where $m_{2,4}^{*}$ are the effective masses of particles $2$ and $4$, respectively.

The inverse mean free path of $\nu_1$ is expressed as
\begin{align}
  \dfrac{1}{\lambda(E_1)} & = \int
  \dfrac{d^3 \bm{p}_2}{(2 \pi)^3}
  \dfrac{d^3 \bm{p}_3}{(2 \pi)^3}
  \dfrac{d^3 \bm{p}_4}{(2 \pi)^3}
  2f_2(E_2^*) \left[ (1 - f_3(E_3) \right] \left[ 1 - f_4(E_4^*) \right] \notag\\
  & \ \ \ \ \  \ \  \times \dfrac{1}{2^4 E_1 E_2^* E_3 E_4^*}
  (2\pi)^4 \delta^{(4)}(p_1^\mu + p_2^\mu - p_3^\mu - p_4^\mu) | \mathcal{M}  |^2 \notag \\
  &= \dfrac{G_F^2 \cos^2 \theta_C}{2} \dfrac{1}{E_1} \int
  \dfrac{d^3 \bm{p}_3}{(2 \pi)^3 2 E_3}
  \left[ 1 - f_3(E_3) \right] L_{\mu \nu} \, \mathcal{S}^{\mu \nu}\left(q^0, q\right), \label{eq:mfp_CC}
\end{align}
where $\mathcal{S}^{\mu \nu}$ is the so-called dynamical structure function of nucleon.
It can be decomposed as follows due to the isotropy of the system:
\begin{align}
  \mathcal{S}^{\mu \nu}\left(q^0, q\right) = \bar{A} P_1^{\mu \nu} + \bar{B} P_2^{\mu \nu} + \bar{C} P_3^{\mu \nu} + \bar{D} P_4^{\mu \nu} + \bar{E} P_5^{\mu \nu}, \label{eq:structure_function}
\end{align}
where the coefficients $\bar{A}, \bar{B}, \bar{C}, \bar{D}$ and $\bar{E}$ are functions of $q^0$ and $q = |\bm{q}|$, and $P_1^{\mu \nu}, P_2^{\mu \nu}, P_3^{\mu \nu}, P_4^{\mu \nu}$ and $P_5^{\mu \nu}$ are projectors relative to momentum transfer $q^{\mu}$.
Their detailed expressions are given in Appendix \ref{appendix:CC}.
The Fermi integrals included in the coefficients from $\bar{A}$ to $\bar{E}$ and other remaining integrals in Eq. (\ref{eq:mfp_CC}) are evaluated numerically.


The emissivity can be obtained from the absorptivity by using the detailed balance condition as
\begin{align}
  j(E_1)
  &= \dfrac{G_F^2 \cos^2 \theta_C}{2} \dfrac{1}{E_1} \int
  \dfrac{d^3 \bm{p}_3}{(2 \pi)^3 2 E_3}
  f_3(E_3) \,\exp\left[\beta (- q^0 - \mu_2 + \mu_4) \right] L_{\mu \nu} \, \mathcal{S}^{\mu \nu}\left(q^0, q\right). \label{eq:emi_CC}
\end{align}

\subsubsection{NC reaction: $\nu + N \leftrightarrows \nu + N$}
This is a neutrino scattering on a nucleon, which is denoted by
\begin{equation}
  \label{eq:NC_reaction}
  \nu_{1}+N_2 \leftrightarrows \nu_3+N_4,
\end{equation}
and its rate is calculated in a similar way to the CC reactions presented above.
The interaction Lagrangian is written as
\begin{equation}
  \label{eq:lagrangian_CC}
  \mathcal{L} = \frac{G_F}{\sqrt{2}} l_{\alpha} \, j^{\alpha}_{\text{NC}},
\end{equation}
where the leptonic neutral current is expressed as
\begin{equation}
  l_{\alpha} = \bar{\nu}_3 \gamma_{\alpha} \left( 1 - \gamma^5 \right) \nu_1 \label{eq:LeptonNeutralCurrent},
\end{equation}
whereas the nucleonic neutral current is given as
\begin{align}
  j^{\alpha}_{\text{NC}} &= \bar{\Psi}_4 \left\{ \gamma^{\alpha} \left[ G_1^N \left( q^2\right) - G_A^N \left( q^2 \right) \gamma^5 \right] + G_2^N \left( q^2 \right) \dfrac{i \sigma^{\alpha \beta } q_{\beta}}{M} \right\} \Psi_2.
\end{align}
The form factors in this expression are written as
\begin{align}
  G_{1,2}^p \left( q^2\right) &= \dfrac{1}{2} \left[ \left( 1 - 4 \sin^2 \theta_W \right) F_{1,2}^p - F_{1,2}^n \right], \\
  G_{1,2}^n \left( q^2\right) &= \dfrac{1}{2} \left[ \left( 1 - 4 \sin^2 \theta_W \right) F_{1,2}^n - F_{1,2}^p \right], \\
  G_A^p \left( q^2\right) &= \dfrac{1}{2} G_A\left(q^2\right), \label{eq:GAp}\\
  G_A^n \left( q^2\right) &= - \dfrac{1}{2} G_A\left(q^2\right), \label{eq:GAn}
\end{align}
with
\begin{align}
  F_{1}^p \left( q^2\right) &= \dfrac{  1-\dfrac{q^2 \gamma_p }{4M^2} }{  \left(1-\dfrac{q^2}{4M^2}\right) \left(1-\dfrac{q^2}{M_V^2}\right)^2 }, \\
  F_{2}^p \left( q^2\right) &= \dfrac{ \gamma_p -1 }{ \left(1-\dfrac{q^2}{4M^2}\right) \left(1-\dfrac{q^2}{M_V^2}\right)^2 }, \\
  F_{1}^n \left( q^2\right) &= \dfrac{  -\dfrac{q^2 \gamma_n }{4M^2} }{  \left(1-\dfrac{q^2}{4M^2}\right) \left(1-\dfrac{q^2}{M_V^2}\right)^2 }, \\
  F_{2}^n \left( q^2\right) &= \dfrac{ \gamma_n }{ \left(1-\dfrac{q^2}{4M^2}\right) \left(1-\dfrac{q^2}{M_V^2}\right)^2 },
\end{align}
and $G_A(q^2)$ is given in Eq.(\ref{eq:GA}).
Note that the axial vector coupling $G_A^{p/n}(q^2)$ in Eqs. (\ref{eq:GAp}) and (\ref{eq:GAn}) is assumed to be isovector and the possible isoscalar component, sometimes called the strangeness contribution, is ignored here although its inclusion is trivial.
The inverse mean free path is written in general as
\begin{align}
  &\dfrac{1}{\lambda(E_1)} = \dfrac{G_F^2}{2} \dfrac{1}{E_1} \int
  \dfrac{d^3 \bm{p}_3}{(2 \pi)^3 2 E_3}
  \left[ 1 - f_3(E_3) \right] L_{\mu \nu} \, \mathcal{S}^{\mu \nu}\left(q^0, q\right), \label{eq:mfp_NC}
\end{align}
where $\mathcal{S}^{\mu \nu}$ is the structure function calculated in the same way as for the CC reaction just by replacing the form factors. The emissivity is also expressed as
\begin{align}
  j(E_1)
  &= \dfrac{G_F^2}{2} \dfrac{1}{E_1} \int
  \dfrac{d^3 \bm{p}_3}{(2 \pi)^3 2 E_3}
  f_3(E_3) \exp \left[ -\beta q^0  \right] L_{\mu \nu} \, \mathcal{S}^{\mu \nu}\left(q^0, q\right) \label{eq:emi_NC}
\end{align}
from the detailed balance condition.

\subsubsection{Beta decay and its inverse: $\bar{\nu}_l + p + l^- \leftrightarrows n$}
The interaction Lagrangian, matrix element and form factors are the same as those for the CC reactions except for the momentum transfer now given as $q^{\alpha} = p_1^{\alpha} + p_3^{\alpha}$. The inverse mean free path and emissivity are then expressed, respectively, as
\begin{align}
  \dfrac{1}{\lambda(E_1)} &= \dfrac{G_F^2}{2} \dfrac{1}{E_1} \int
  \dfrac{d^3 \bm{p}_3}{(2 \pi)^3 2 E_3}
  f_3(E_3)
  L_{\mu \nu} \, \mathcal{S}^{\mu \nu}\left(q^0, q\right), \label{eq:mfp_beta}\\
  j(E_1)
  &= \dfrac{G_F^2}{2} \dfrac{1}{E_1} \int
  \dfrac{d^3 \bm{p}_3}{(2 \pi)^3 2 E_3}
  \left[ 1 - f_3(E_3) \right] \exp \left[\beta (- q^0 - \mu_2 + \mu_4) \right] L_{\mu \nu} \, \mathcal{S}^{\mu \nu}\left(q^0, q\right). \label{eq:emi_beta}
\end{align}

\begin{table*}[tbp]
  \hspace*{-2.4cm}
  \begin{threeparttable}
   \caption{Thermodynamical conditions considered in this study. Units of temperature $T$, chemical potentials, $\mu_n$, $\mu_p$, $\mu_e$ and $\mu_{\mu}$, nucleon potential difference $U_n - U_p$ and effective mass $m^*$ are all $\MeV$. \label{tab:cond}}
 \centering
 \small
 \begin{tabular}{lcccccccccccccc}
  \hline \hline
     Model & $t_{\textrm{pb}} (\operatorname{s})$ & $r (\operatorname{km})$ & $\rho (\operatorname{g/cm^3}) $ & $T$ & $Y_e$ & $Y_{\mu}$ & $Y_n$ & $Y_p$ \tnote{1} & $\mu_n $ & $\mu_p $ & $\mu_e $ & $\mu_\mu $ & $U_n - U_p $ & $m^* $ \\
 \hline
 t1S  & 1  & 20.0 & $5.0 \times 10^{12}$ & 5.5  & 0.036 & 0.0025 & 0.95 & $2.3 \times 10^{-2}$ & 933.8  & 909.0  & 25.6  & 102.8  & 3.5  & 938.0  \\
 t3S  & 3  & 16.5 & $1.0 \times 10^{13}$ & 5.0  & 0.026 & 0.0025 & 0.94 & $1.1 \times 10^{-2}$ & 938.6  & 906.8  & 30.2  & 104.2  & 7.1  & 938.0  \\
 t10S & 10 & 14.5 & $2.0 \times 10^{13}$ & 4.0  & 0.025 & 0.0025 & 0.86 & $4.5 \times 10^{-4}$ & 943.3  & 896.2  & 42.0  & 105.7  & 13.9 & 938.0  \\
 t30S & 30 & 13.6 & $4.5 \times 10^{13}$ & 2.7  & 0.018 & 0.0025 & 0.84 & $5.0 \times 10^{-11}$& 948.9  & 878.7  & 47.3  & 107.2  & 30.2 & 938.0  \\
 t50S & 50 & 13.4 & $6.1 \times 10^{13}$ & 1.5  & 0.017 & 0.0025 & 0.79 & $2.0 \times 10^{-20}$ & 951.0  & 870.4  & 52.2  & 107.8  & 38.8 & 938.0  \\
 t1D  & 1  & 10.8 & $1.2 \times 10^{14}$ & 35.6 & 0.22  & 0.05   & 0.73 & 0.27 & 904.0  & 836.3  & 126.4 & 89.8   & 27.0 & 757.8 \\
 t3D  & 3  & 9.4  & $2.1 \times 10^{14}$ & 35.0 & 0.20  & 0.05   & 0.75 & 0.25 & 930.4  & 849.2  & 156.5 & 114.5  & 30.9 & 645.7 \\
 t10D & 10 & 6.0  & $3.5 \times 10^{14}$ & 30.2 & 0.17  & 0.05   & 0.78 & 0.22 & 981.3  & 883.2  & 186.2 & 146.3  & 30.2 & 502.5 \\
 t30D & 30 & 1.7  & $4.7 \times 10^{14}$ & 17.7 & 0.16  & 0.05   & 0.80 & 0.20 & 1037.0 & 924.8  & 210.2 & 174.1  & 28.5 & 405.2 \\
 t50D & 50 & 1.69 & $4.9 \times 10^{14}$ & 2.8  & 0.15  & 0.05   & 0.80 & 0.20 & 1046.1 & 932.3  & 216.3 & 182.8  & 28.5 & 389.0 \\
 \hline
 \end{tabular}
 \begin{tablenotes}
 \item[1]{The free proton fraction. At the neutrino sphere, protons are mainly consumed by heavy nuclei.}
 \end{tablenotes}
\end{threeparttable}
\end{table*}

\section{Results and discussions \label{sec:Results}}
Employing the formulae given in the previous section, we now evaluate the rates of all reactions listed in Table \ref{tab:mu_reaction} numerically for the thermodynamic conditions that we find typically in the PNS cooling.
This is the original part of this paper.
We compare the inverse mean free paths for these reactions to see their relative importance quantitatively.
We also discuss their possible implications for the PNS cooling.

For this purpose, we first extract the thermodynamical data at different times from a one-dimensional PNS cooling calculation conducted under spherical symmetry in \cite{Nakazato2018_muon}.
The simulation was done as follows:
core-collapse of a 15 $M_{\odot}$ progenitor \cite{WoosleyWeaver1995} was first computed with the general relativistic neutrino-radiation hydrodynamics code \cite{Sumiyoshi2005} until $t = 0.3 \operatorname{s}$ after core bounce when the shock wave is stalled.
Knowing that this model does not explode in 1D but expecting that it will explode in multi-dimensions, the authors in \cite{Nakazato2018_muon} extracted the region inside the shock wave (up to the mass coordinate of $\sim 1.47 M_{\odot} $) from the result of the first simulation and use it as an initial condition for the second simulation of PNS cooling.
In this second simulation, the quasi-static evolution of the PNS is computed by solving the Tolman-Oppenheimer-Volkov equation together with the neutrino transfer equation, the latter of which was solved with the multi-group flux limited diffusion scheme \cite{Suzuki1994};
the Shen EoS \cite{Shen2011} was adopted.

We then picked up the snapshots at $t = 1, 3, 10, 30, 50 \operatorname{s}$ post bounce.
At each time, we extracted the trio of the thermodynamical quantities: density, temperature and electron fraction $(\rho, T, Y_e)$ at the neutrino sphere for $\nu_e$ with the average energy as well as at the radius where the temperature reaches the maximum.
Note that the peak temperature occurs off center particularly at early times.
The choice of the latter position is motivated by the fact that the muon fraction is expected to be largest there, since it is known to be correlated with temperature \cite{Bollig2017}.
It is a well-known fact that the radius of the neutrino sphere depends on the species.
The difference is at most $0.3 \operatorname{km}$, however, and the thermodynamical condition is not much different among the species.
The choice of $\nu_e$ here as the representative case is hence well justified.

It is noted here that the second simulation (and the first one as well) neglected muon entirely.
We hence added it by hand as a parameter in this paper: $Y_{\mu} = 0.0025$ at the neutrino sphere and $Y_{\mu} = 0.05$ at the maximum temperature, with $Y_{\mu}$ being the muon fraction.
As already explained in detail, the rates of the semi-leptonic reactions depend also on their dispersion relations of nucleons in medium.
We re-evaluated them at the mean field level, just in the same way to produce the EoS, for the given quartet $(\rho, T, Y_e, Y_{\mu})$.
Note that the effective masses of neutron and proton are the same as those in the Shen EoS.
All the relevant quantities are listed in Table \ref{tab:cond}.

Although we treat $Y_{\mu}$ parametrically in this paper, it is actually determined by the cooling history of PNS and should be derived by detailed simulations.
As such the muon fraction depends on the EoS, particularly the symmetry energy, of nuclear matter.
The previous authors of \cite{BolligPHDs} and \cite{Fischer2021Axion} employed SFHo EoS (\cite{Steiner2013}) and HS(DD2) EoS (\cite{Typel2010, Hempel2010}), respectively, the symmetry energy of which are different from Shen EoS.
In their simulations, \mbox{$Y_{\mu} \sim 0.03 - 0.05$} were observed around the point where the temperature becomes maximum at \mbox{$t \sim 5-10 \s$} (see Figures 11.36 and 11.38 in \cite{BolligPHDs} and Figure 5 in \cite{Fischer2021Axion}):
for example, $(\rho, T, Y_e, Y_{\mu}) = (9 \times 10^{14} \operatorname{g/cm^3}, 52.3 \MeV, 0.095, 0.037)$ at $t = 5 \s$ and $r = 2 \km $ in \cite{BolligPHDs}.
In order to gauge the possible uncertainty from EoS, we calculate $Y_{\mu}$ with Shen EoS for the same $\rho$ and $T$ under the following assumptions:
beta-equilibrium, that is, $\mu_e = \mu_n - \mu_p$ and $\mu_{\mu} = \mu_n - \mu_p + \mu_{\nu_{\mu}}$, where the non-vanishing chemical potentials of $\mu_{\nu_{\mu}} = - \mu_{\bar{\nu}_{\mu}} = -52 \MeV$ is taken into account and adopted also from \cite{BolligPHDs}.
We then obtain $Y_{\mu} = 0.042$ which is indeed close to the original value of 0.037.
Although this is not a proof of anything but we think that our choice of $Y_{\mu}$ is not widely off the mark and the information obtained here will be useful for interpreting simulation results.



\subsection{Inverse mean free paths at $t_{\textnormal{pb}} = 10 \operatorname{s}$ }
We first present the inverse mean free paths for all neutrino reactions at $t_{\mathrm{pb}} = 10 \operatorname{s}$ as the fiducial model.
Note again that we are more interested in late times up to $\sim$ a minute in this paper.
For comparison, however, we include earlier times. We are also interested in the corrections from the recoil, form factors in the weak currents and in-medium modifications in the dispersion relations of nucleons, other subtle effects considered rather recently \cite{Roberts2017, Fischer2020, Guo2020}.
We will see which one, the muon-related reactions or these corrections, is more important for what neutrino energy at which time.

\begin{figure*}[htbp]
  \vspace{-0.5cm}
  \begin{minipage}[b]{0.49\linewidth}
   \centering
   \includegraphics[keepaspectratio, scale=0.51]{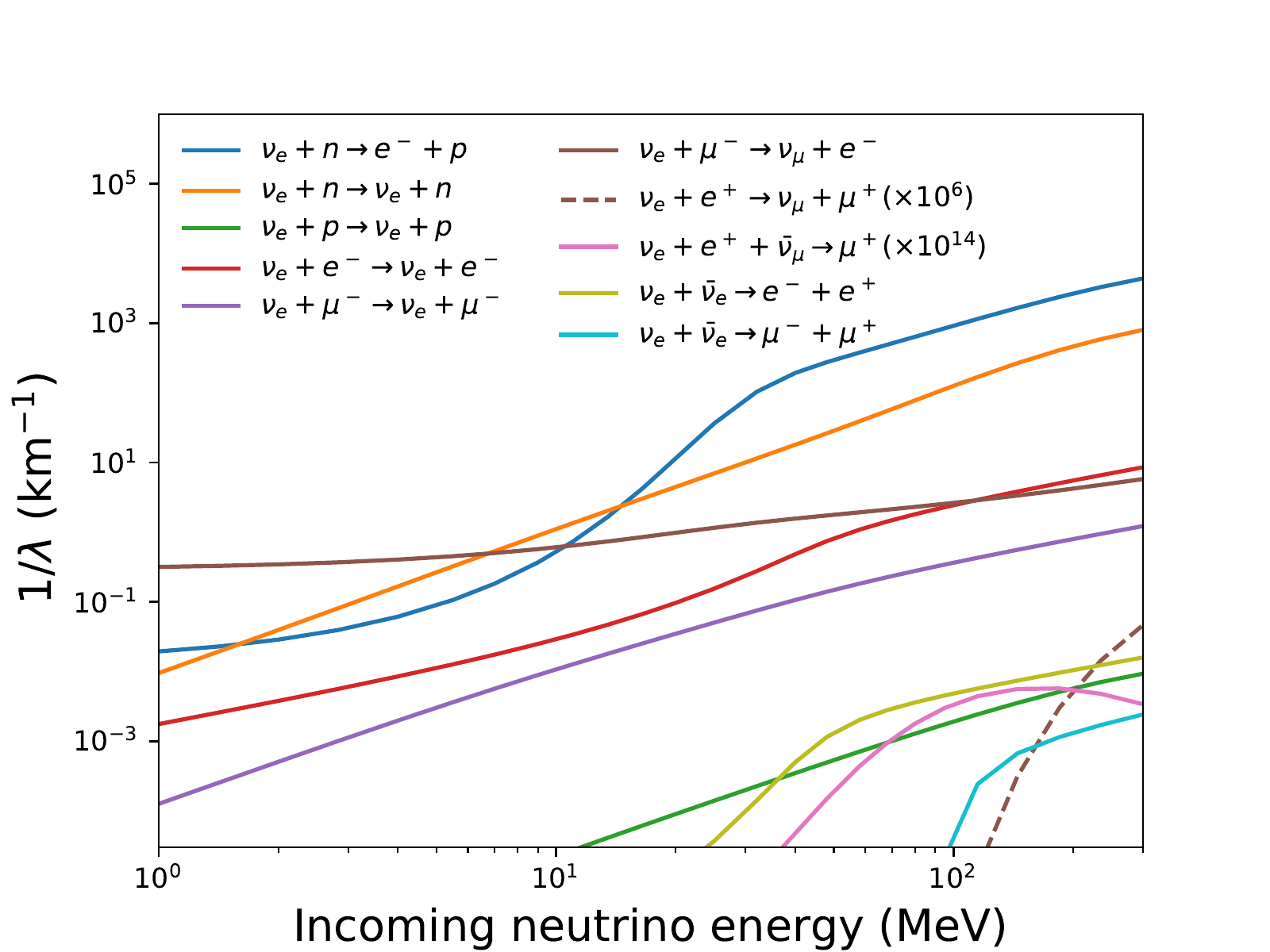}
   \subcaption{}\label{fig:t10s_NuSphere_nue}
  \end{minipage}
  \begin{minipage}[b]{0.49\linewidth}
   \centering
   \includegraphics[keepaspectratio, scale=0.51]{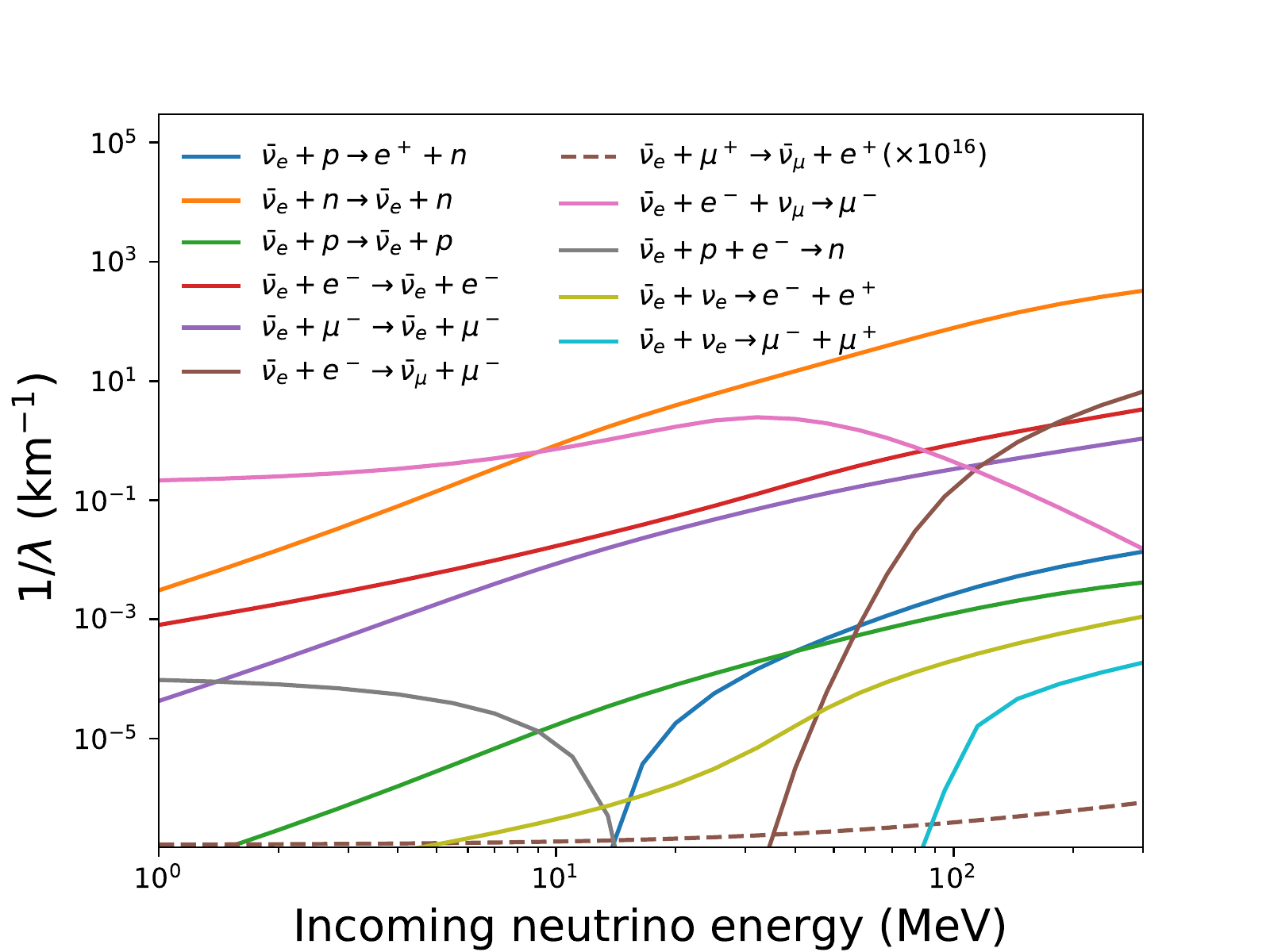}
   \subcaption{}\label{fig:t10s_NuSphere_nueb}
  \end{minipage}\\
  \begin{minipage}[b]{0.49\linewidth}
   \centering
   \includegraphics[keepaspectratio, scale=0.51]{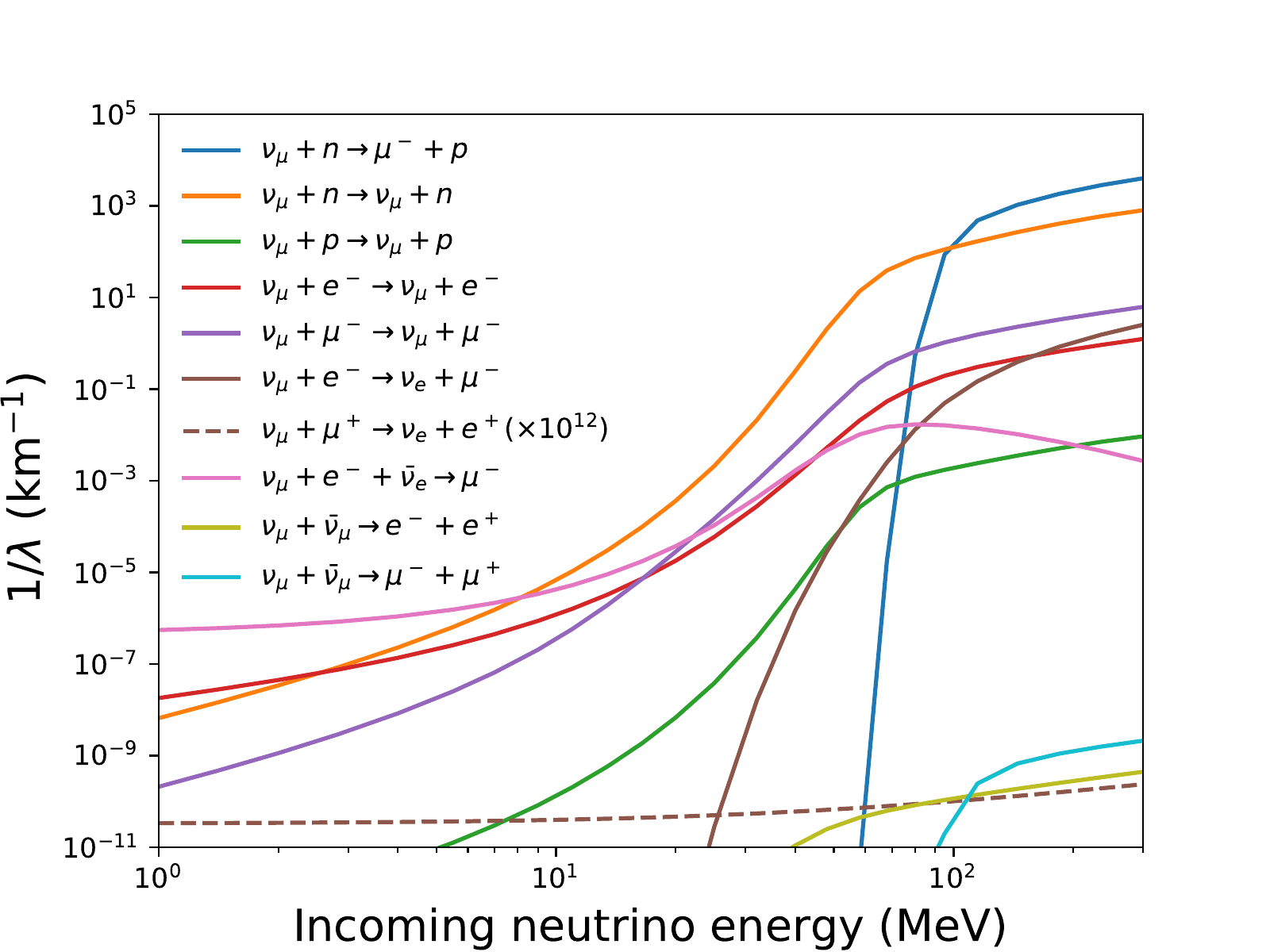}
   \subcaption{}\label{fig:t10s_NuSphere_numu}
  \end{minipage}
  \begin{minipage}[b]{0.49\linewidth}
   \centering
   \includegraphics[keepaspectratio, scale=0.51]{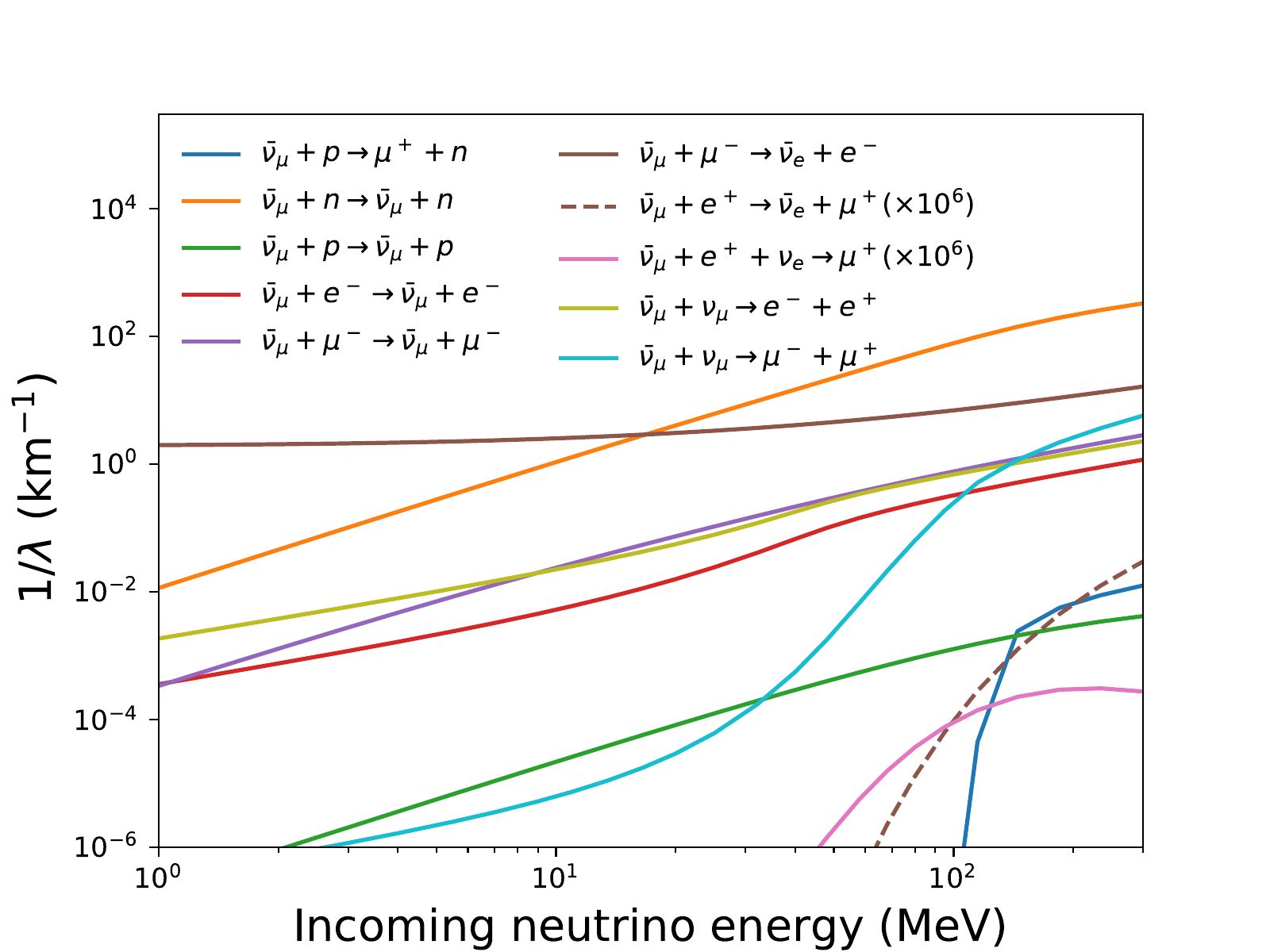}
   \subcaption{}\label{fig:t10s_NuSphere_numub}
  \end{minipage}\\
  \vspace{-0.5cm}
  \begin{minipage}[b]{0.49\linewidth}
    \centering
    \includegraphics[keepaspectratio, scale=0.51]{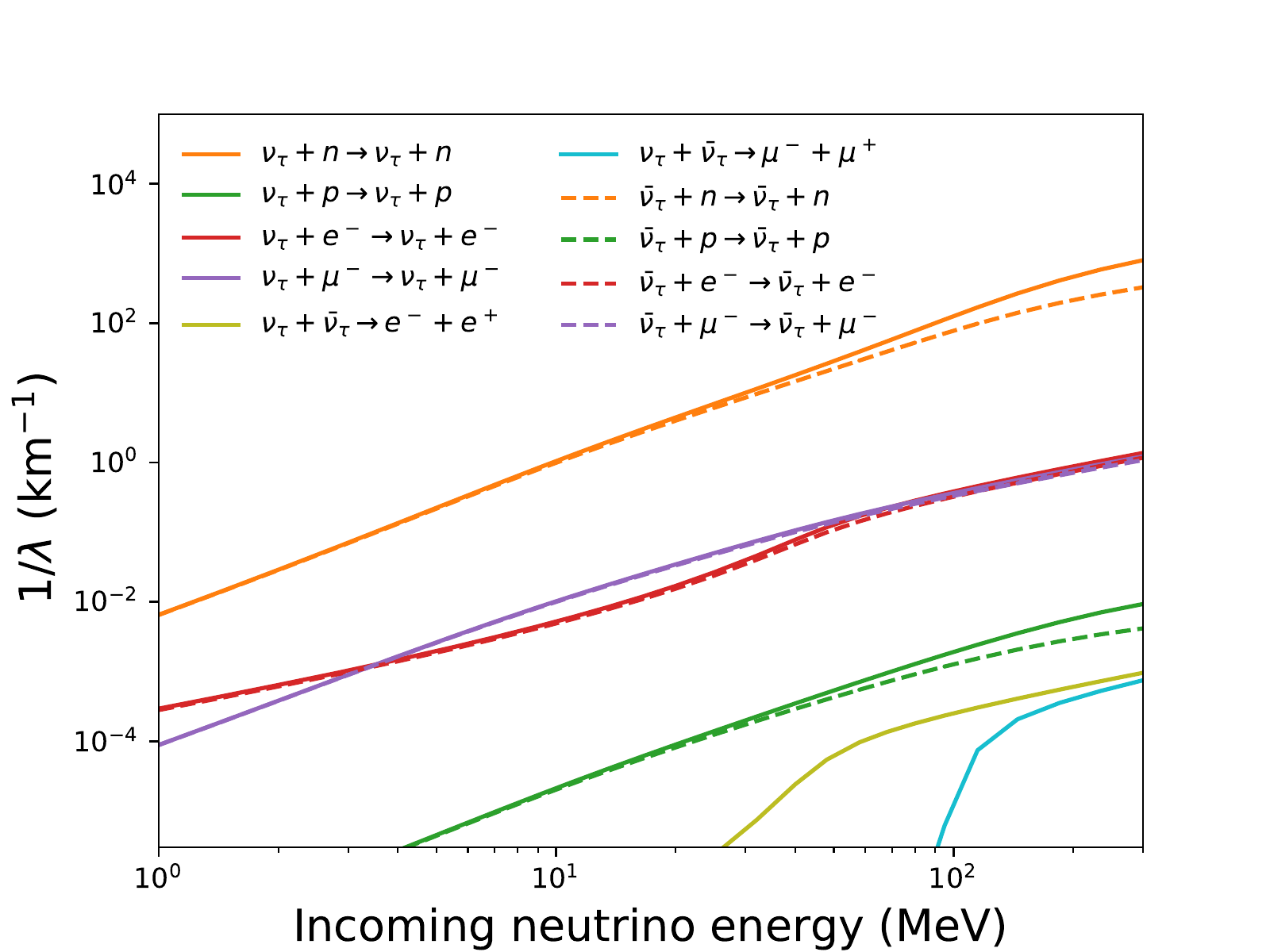}
    \subcaption{}\label{fig:t10s_NuSphere_nutau}
   \end{minipage}
  \caption{Inverse mean free paths for different neutrino flavors in model t10S. Panels (a), (b), (c) and (d) are the results for $\nu_e$, $\bar{\nu}_e$, $\nu_{\mu}$ and $\bar{\nu}_{\mu}$, respectively. Panel (e) shows those for both $\nu_{\tau}$ and $\bar{\nu}_{\tau}$ where the solid and dashed lines are for $\nu_{\tau}$ and $\bar{\nu}_{\tau}$, respectively. Colors denote the different reactions written in the legends in each panel. Note that the values for $ \nu_e + e^+ + \bar{\nu}_{\mu} \rightarrow \mu^+$ in panel (a) are multiplies by $10^{14}$. Note also that the values for $ \bar{\nu}_{\mu} + e^+ + \nu_{e} \rightarrow \mu^+$ in panel (d) are multiplied by $10^{6}$.
  The values for $\bar{\nu}_e + \mu^+ \leftrightarrows \bar{\nu}_{\mu} + e^+$ and $\nu_{\mu} + \mu^+ \leftrightarrows \nu_e + e^+$ are multiplied by the factors given in the legend in each panel.}\label{fig:t10S}
\end{figure*}

Figure \ref{fig:t10S} shows the inverse mean free paths at the neutrino sphere (model t10S) as a function of the energy of the incoming neutrino.
Panel (a) is the results for $\nu_{e}$.
As expected the $\nu_e$ absorption on neutron (blue line), one of CC reactions, is the dominant source of opacity at high energies of the incident neutrino, $E_{\nu_e} \gtrsim 10 \MeV$.
The neutrino scattering on neutron (orange line) is the second contributor at the same energy range.
Since protons are much less abundant \mbox{($Y_p = 4.5 \times 10^{-4}$)} than neutrons ($Y_n = 0.86$) due to the formation of heavy nuclei, the scattering on the neutron is dominant by more than 4 orders compared with the that on proton (green line).
Note that the coherent scatterings on heavy nuclei are not considered here.

At low energies ($E_{\nu_e} \lesssim 10 \MeV$), on the other hand, the flavor exchange reaction \mbox{$\nu_e + \mu^- \rightarrow \nu_{\mu} + e^-$} (brown line) dominates the opacity.
This is one of the muon-related interactions, the main topic in this paper.
It should be mentioned that this reaction is more important than the electron scattering (red line) at almost all energies of relevance \mbox{($E_{\nu_e} \lesssim 100 \MeV$)}.
This is due to the large mass difference between muon and electron.

The scattering on muon (purple line), on the other hand, is always smaller than that on electron, since muons are much less abundant.
Note that the energy dependence is also different between these two reactions, since the muon is much heavier than the electron and the electron is strongly degenerate.
One can see that the rise of the inverse mean free path for the muon scattering with the neutrino energy becomes less steep at $E_{\nu_e} \gtrsim 100 \MeV$ as muons get relativistic.

The pair production of muon and anti-muon (light-blue line) is less efficient than that of electron-positron (yellow line) again owing to the lager mass of muon.
The three body reactions (pink line) are very minor, since they involve positrons, the abundance of which is suppressed by the degeneracy of electrons.
Similarly, the flavor-exchange reaction involving positron and $\mu^+$: $\nu_e + e^+ \rightarrow \nu_{\mu} + \mu^+$ (brown dashed line) is strongly suppressed.

Panel (b) shows the inverse mean free paths for $\bar{\nu}_e$.
This time the absorption via the charged current is quite minor, since protons are scarce and, more importantly, the potential difference between proton and neutron disfavors the reaction, producing an effective threshold in fact.
Note that this is also the threshold for the inverse beta decay of neutron, which occurs essentially below this threshold.

The nucleon scatterings are dominant contributors to the inverse mean free path at high neutrino energies ($E_{\bar{\nu_e}} \gtrsim 10 \MeV$), with the neutron scattering overwhelming the proton scattering for the same reason as for $\nu_e$.
Interestingly, the inverse muon decay is dominant below $ \sim 10 \MeV$.
This happens because there is a large phase space available.
Incidentally, the counterpart for $\nu_e$: \mbox{$\nu_e + e^+ + \bar{\nu}_{\mu} \rightarrow \mu^+$} is pretty minor, since the positron is strongly suppressed.
Other features are common to $\nu_e$.

For $\nu_{\mu}$, the scattering on neutron again gives the dominant contribution in the range of $E_{\nu_{\mu}} = 10 - 100 \MeV$ (panel (c)).
The inverse muon decay comes first at lower energies just as for $\bar{\nu}_e$.
At higher energies, $E_{\nu_{\mu}} + U_n - U_p \gtrsim 100 \MeV$, on the other hand, the neutrino capture on neutron becomes the most important.
The scattering on muon is subdominant but is comparable with or even higher than the scattering on electron because the latter occurs only through the neutral current while the former takes place through both the neutral and charged currents.
The flavor exchange reaction $\nu_{\mu} + e^- \rightarrow \nu_e + \mu^-$ is substantial only at high energies $E_{\nu_{\mu}} \gtrsim 100 \MeV$, since the mass difference between electron and muon disfavors it this time.
The electron-positron and muon-anti-muon pair productions are both very minor.
The flavor exchange reaction $\nu_{\mu} + \mu^+ \rightarrow \nu_e + e^+$ is also negligible due to the strong suppression of $\mu^+$.

As for $\bar{\nu}_{\mu}$ (panel (d)), the scattering on neutron is dominant at \mbox{$E_{\bar{\nu}_{\mu}} \gtrsim 10 \MeV$} while the flavor exchange reaction $\bar{\nu}_{\mu} + \mu^- \rightarrow \bar{\nu}_e + e^-$ becomes most important at lower energies \mbox{$E_{\nu_{\mu}} \lesssim 10 \MeV$}.
The scattering on muon is more important than that on electron and the production of electron-positron pair is comparable to the former.
The production of muon-anti-muon becomes as important at very high energies $E_{\bar{\nu}_{\mu}} \gtrsim 100 \MeV$.
The absorption and scattering on proton are much smaller due to the small abundance of proton.
The inverse beta decay of muon is even smaller, since the positron abundance is more strongly suppressed.

In the presence of muons, the degeneracy between the $\mu$-type and $\tau$-type (anti-)neutrino is resolved.
The muonic reactions for $\tau$-type (anti-)neutrino occur only through the neutral leptonic or nucleonic current and are always subdominant, with the scattering on neutron being dominant (panel (e)).
The difference in opacity between the electron and muon scatterings is originated from the differences in the electron and muon fractions and as well as in the Pauli-blocking factor.
The difference between $\nu_{\tau}$ and $\bar{\nu}_{\tau}$ is pretty minor.
Note, however, that the opacity is always smaller for $\bar{\nu}_{\tau}$ than for $\nu_{\tau}$ as pointed out in \cite{Horowitz2002}.
The difference in the neutrino scatterings on nucleon is ascribed to the weak magnetism in the nucleon current.

\begin{figure*}[htbp]
  \begin{minipage}[b]{0.49\linewidth}
   \centering
   \includegraphics[keepaspectratio, scale=0.51]{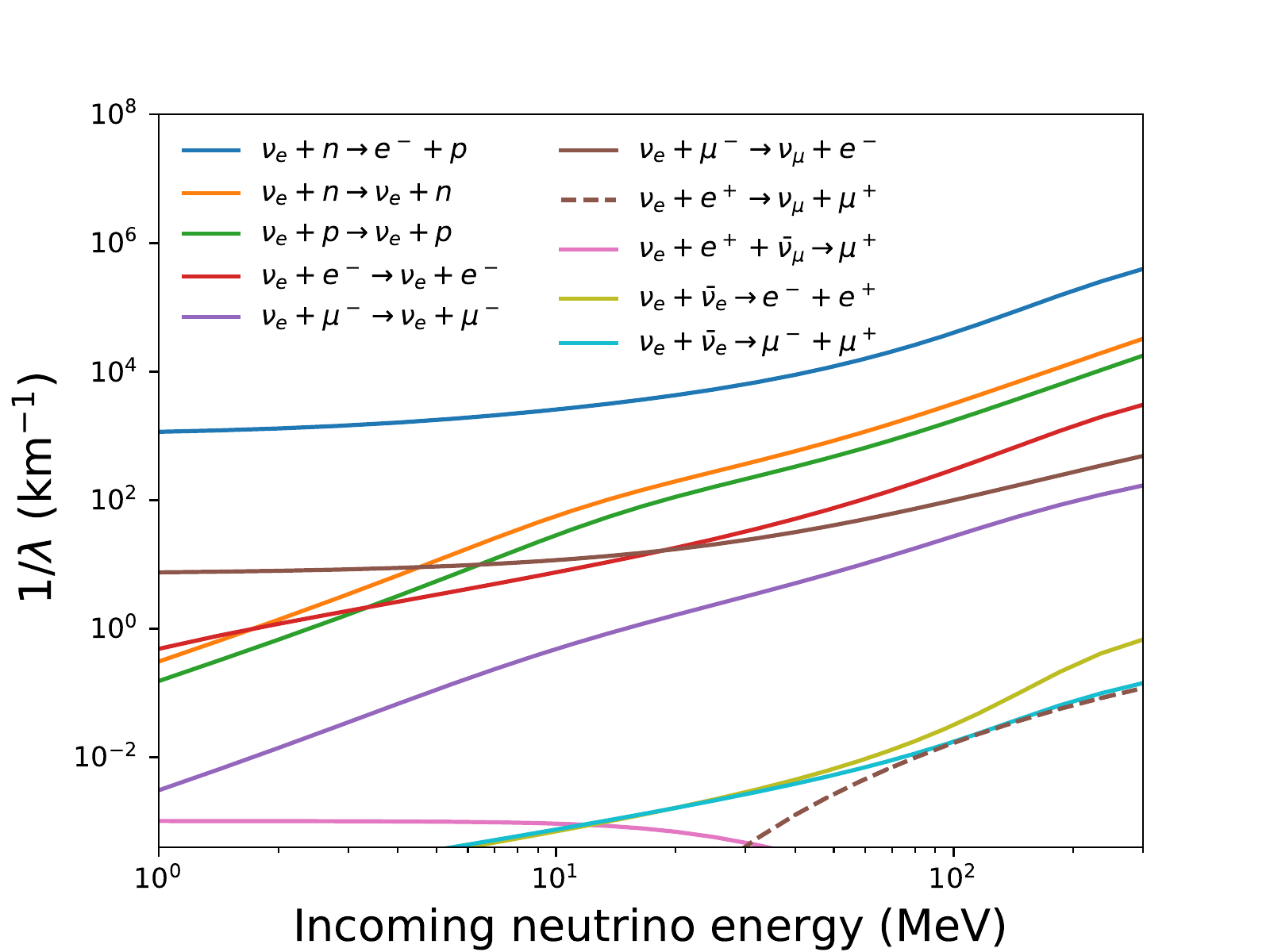}
   \subcaption{}\label{fig:t10d_Deep_nue}
  \end{minipage}
  \begin{minipage}[b]{0.49\linewidth}
   \centering
   \includegraphics[keepaspectratio, scale=0.51]{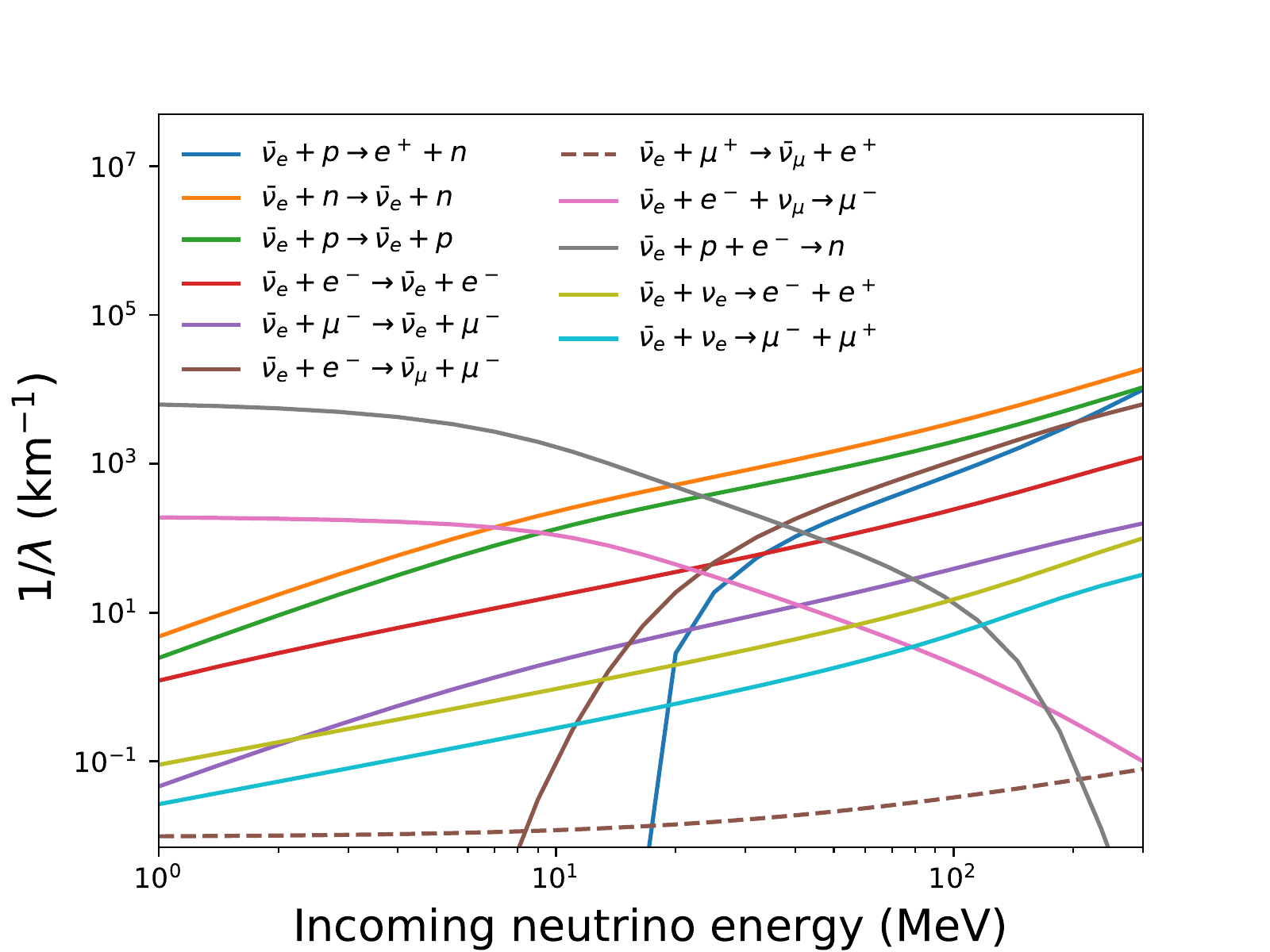}
   \subcaption{}\label{fig:t10d_Deep_nueb}
  \end{minipage}\\
  \begin{minipage}[b]{0.49\linewidth}
   \centering
   \includegraphics[keepaspectratio, scale=0.51]{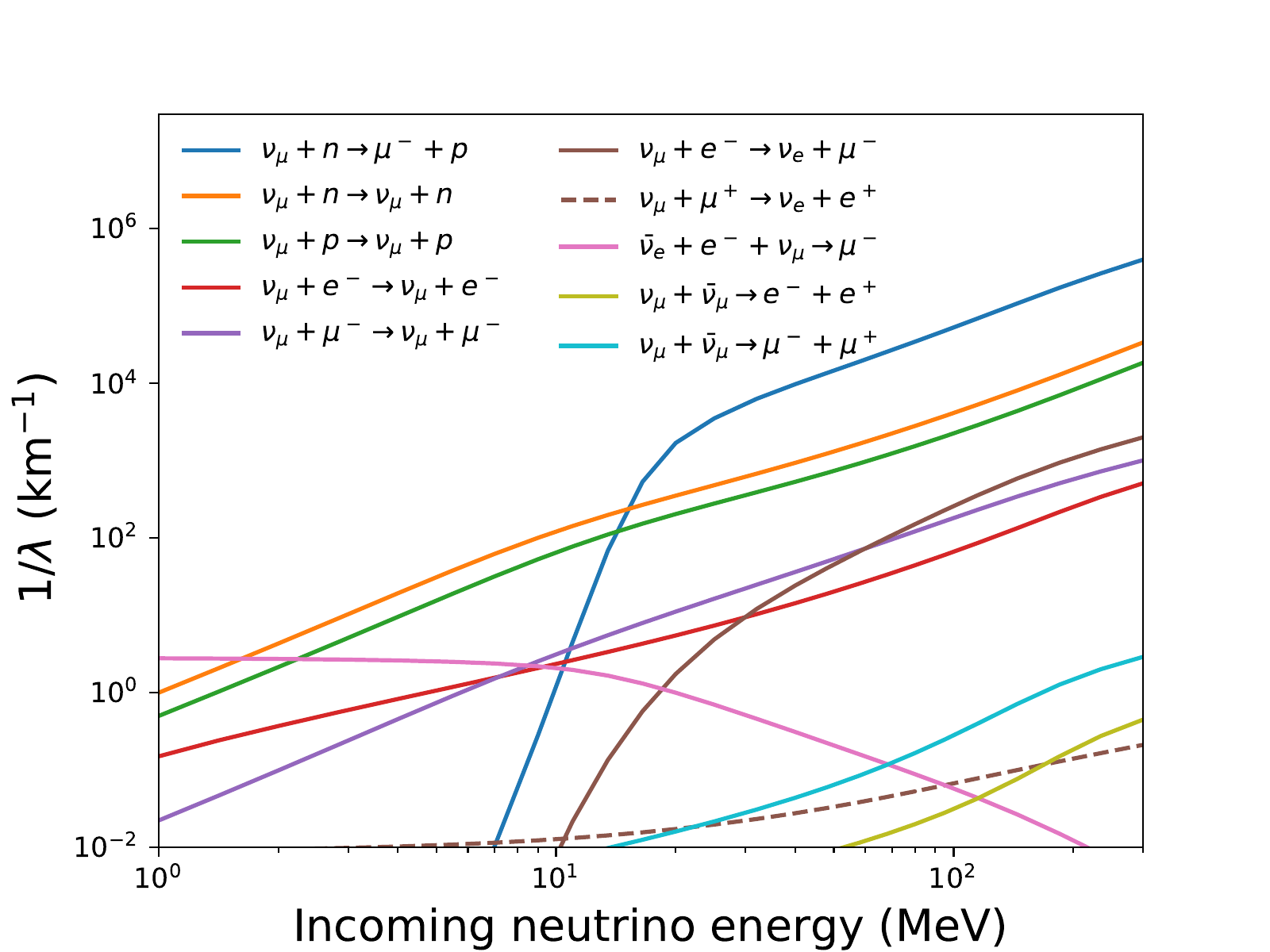}
   \subcaption{}\label{fig:t10d_Deep_numu}
  \end{minipage}
  \begin{minipage}[b]{0.49\linewidth}
   \centering
   \includegraphics[keepaspectratio, scale=0.51]{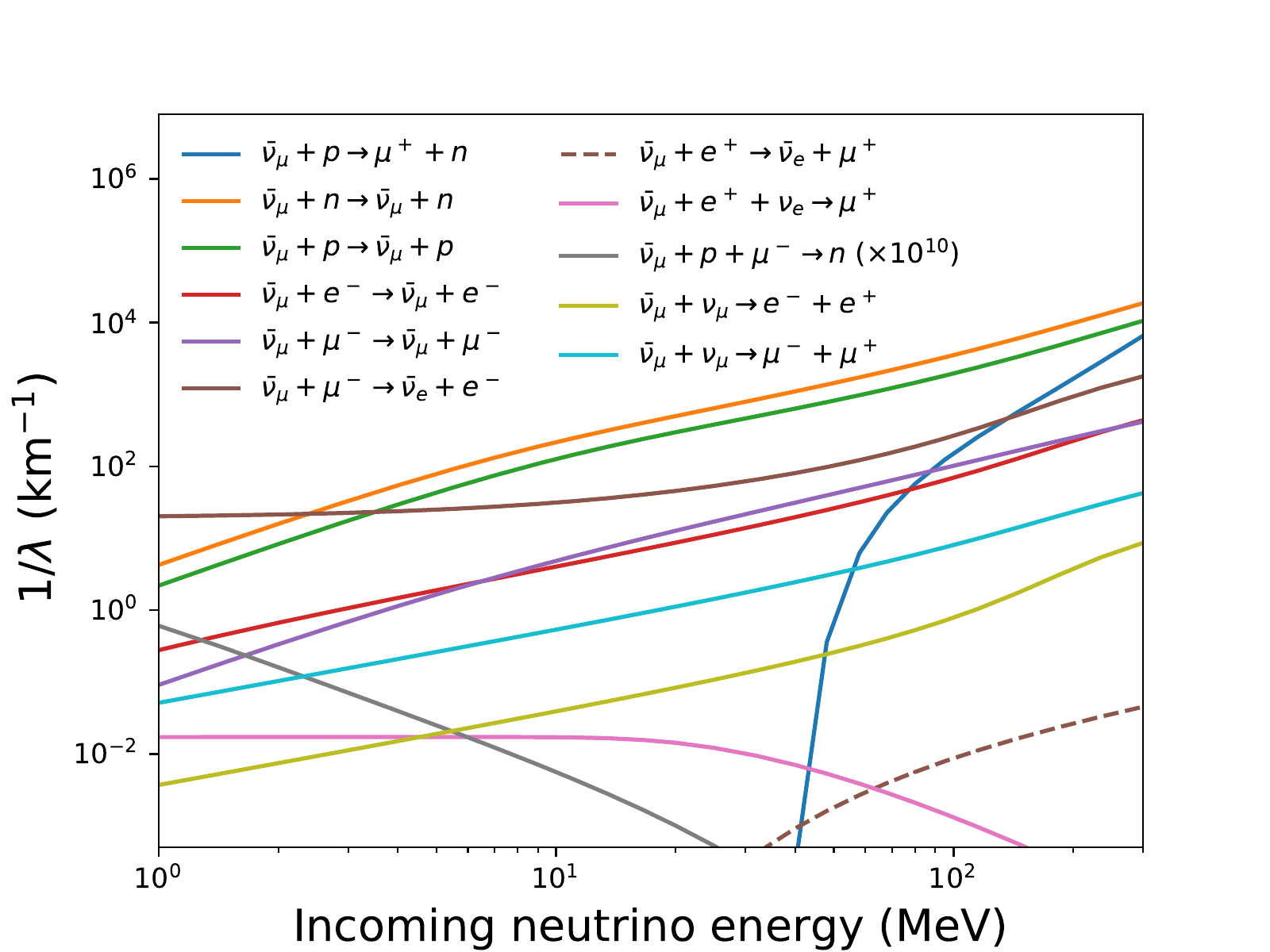}
   \subcaption{}\label{fig:t10d_Deep_numub}
  \end{minipage}\\
  \begin{minipage}[b]{0.49\linewidth}
    \centering
    \includegraphics[keepaspectratio, scale=0.51]{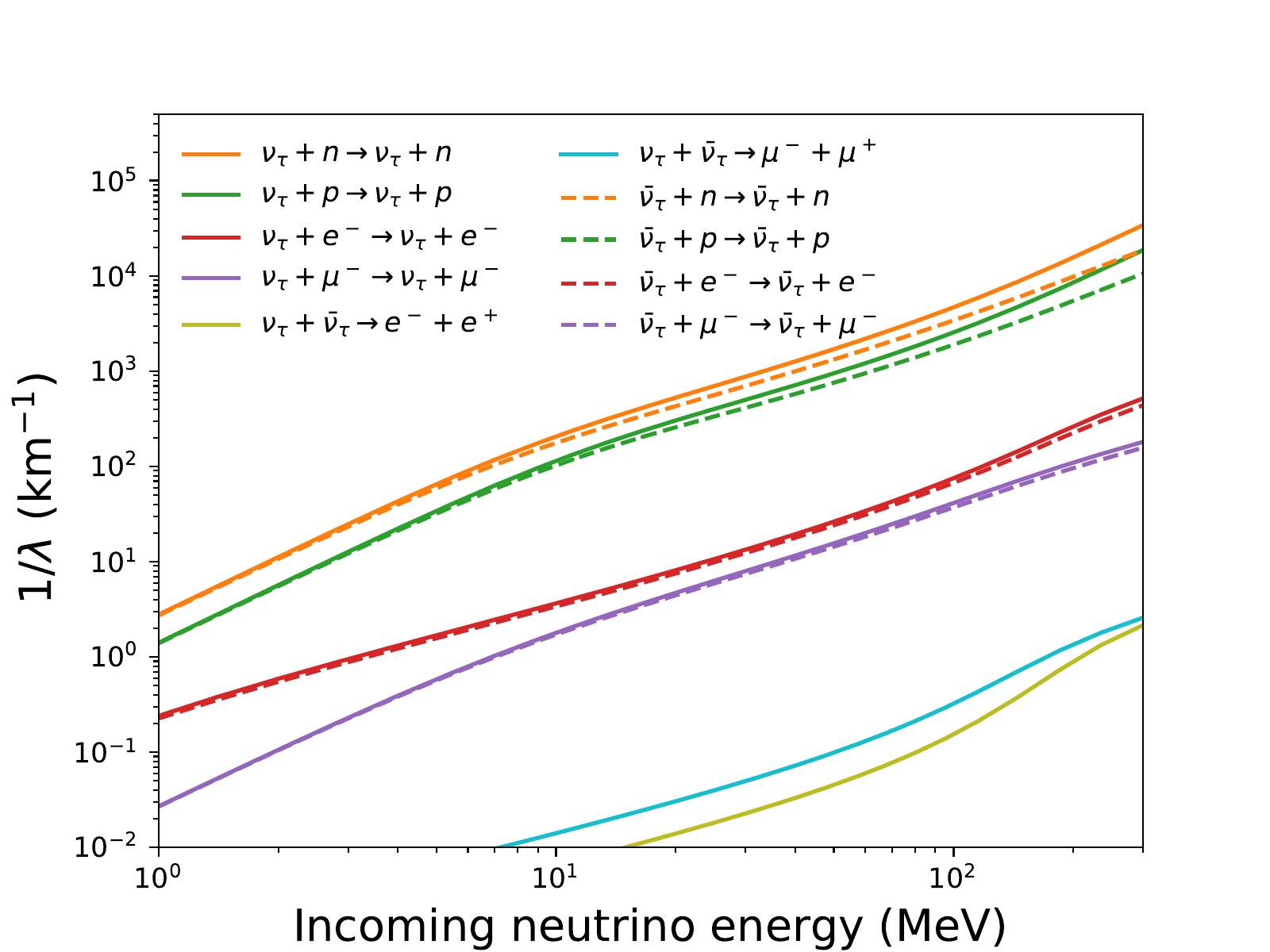}
    \subcaption{}\label{fig:t10d_Deep_nutau}
   \end{minipage}
  \caption{Same as Figure \ref{fig:t10S} but for model t10D. Note that the values for $\bar{\nu}_{\mu} + p + \mu^{-} \rightarrow n$ are multiplied by $10^{10}$.}\label{fig:t10d}
 \end{figure*}

So far we have been concerned with the reaction rates near the neutrino sphere, the region most important for the formation of neutrino signals as observed.
Now we shift our interest to a deeper region with the highest temperature, where muons are expected to be most abundant.
The results are presented in Figure \ref{fig:t10d}.

Panel (a) shows the inverse mean free paths for $\nu_e$.
The absorption on neutron is the most important source of opacity in all range of the incident neutrino energy as expected.
Since the temperature in model t10D is much higher than that in model t10S, the Pauli blocking by electron is more moderate, which is the reason why the CC reaction remains dominant even at low energies.
The nucleon scatterings are the second dominant just as for model t10S (Figure \ref{fig:t10s_NuSphere_nue}). Note that the scattering on proton is much larger compared with model t10S because the free proton is much more abundant in this region ($Y_p = 0.22$) than around the neutrino sphere.

The opacities of the scatterings on neutron, electron and muon are larger by an order than the counterparts in model t10S, since the scattered particles are more numerous.
The flavor exchange reaction $\nu_e + \mu^- \rightarrow \nu_{\mu} + e^-$ is also enhanced compared with model t10S for the same reason.
It is still minor than the CC reaction partly because the chemical potential of muon is overwhelmed by that of electron, which results in a smaller phase space available in the flavor exchange reaction whereas in the CC reaction the difference in the effective potentials between neutron and proton are significant.
As we mentioned, since the Pauli blocking by leptons is weaker in this region, the electron-positron and muon-anti-muon pair processes are allowed even at lower neutrino energies.
The three body reaction and $\nu_e + e^+ \rightarrow \nu_{\mu} + \mu^+$ are very minor as in model t10S.

For $\bar{\nu}_{e}$, the inverse neutron decay is strongly enhanced and is dominant indeed in the energy range of $E_{\bar{\nu}_e} \lesssim 20 \MeV$ (panel (b)).
This happens because the proton number density as well as the potential difference between proton and neutron are larger in model t10D than in model t10S.
The potential difference broadens the available phase space.
The reaction rate of the inverse muon decay is not so large compared with the inverse neutron decay, but is still greater compared with model t10S.
This is because the electron number density is larger and the inequality $\mu_{e^-} > \mu_{\mu^-}$ is satisfied for the chemical potentials at this radius while the opposite inequality $\mu_{e^-} < \mu_{\mu^-}$ holds near the neutrino sphere.
The flavor exchange reaction $\bar{\nu}_e + e^- \rightarrow \bar{\nu}_{\mu} + \mu^-$ is also enhanced from model t10S in the same way as the inverse muon decay is.
The energy threshold for the absorption via the charged current is determined by the potential difference between neutron and proton as in model t10S.
The features of other reactions are common to $\nu_e$.

Panel (c) shows the results for $\nu_{\mu}$.
The neutrino capture on neutron is the greatest source of opacity in a wider energy range $E_{\nu_{\mu}} \gtrsim 10 \MeV$ compared with model t10S due to the larger potential difference between neutron and proton.
The scattering on neutron gives the dominant contribution at $E_{\nu_{\mu}} \lesssim 10 \MeV$.
The flavor exchange reaction $\nu_{\mu} + e^- \rightarrow \nu_e + \mu^-$ and the inverse muon decay are subdominant compared with the scattering and absorption on neutron but are larger than those in model t10S.
The reason is similar to the muon-related reaction for $\bar{\nu}_e$; the electron is more abundant and the chemical potential difference between electron and muon favors these reactions.
The scattering on electron and muon and the pair-processes are minor just as for $\nu_e$ and $\bar{\nu}_e$.

For $\bar{\nu}_{\mu}$, the scattering on neutron gives the dominant contribution except at $E_{\bar{\nu}_{\mu}} \lesssim 2 \MeV$, where the flavor exchange reaction $\bar{\nu}_{\mu} + \mu^- \rightarrow \bar{\nu}_e + e^-$ is the largest contributor, which is similar to model t10S (see panel (d)).
The reaction rate in model t10D, on the other hand, is larger by an order than that in model t10S because the number density of muon is larger by the same factor.
In the similar way, the neutrino capture on proton is enhanced at high energies, since the proton number fraction is greater by an order again compared with model t10S, while the potential difference between proton and neutron prevents the reaction at low energies.
Three body reaction $\bar{\nu}_{\mu} + e^+ + \nu_e \rightarrow \mu^+$ is strongly suppressed as positrons are scarce.
It is mentioned for model t10S that the $\mu$-type (anti-)neutrino scattering on muon is comparable with or higher than that on electron which is also confirmed in model t10D.

The scattering on neutron is again the dominant source of opacity for the $\tau$-type neutrino (panel (e)).
The scattering on proton is smaller owing to the smaller abundance of proton.
As pointed out in model t10S, the opacity of $\bar{\nu}_{\tau}$ is always smaller than $\nu_{\tau}$ for all reactions and at all energies.

\subsection{Some corrections in semi-leptonic reactions \label{subsec:corrections_semilep}}
The semi-leptonic reactions are those reactions that involve nucleons (see Table I).
Since the nucleons are composite particles of three quarks, their weak currents are more involved than those for leptons. In fact, such intricacies as the form factors as well as in-medium modifications of the dispersion relations for nucleons have been studied in the contexts of the CCSN explosion and PNS cooling rather recently.
As a matter of fact, the weak magnetism and the recoil of nucleons were investigated quantitatively both for NC and CC reactions in \cite{Horowitz2002};
the effects of the mean field corrections in the dispersion relations of nucleons have been also explored over the years (see e.g. \cite{Reddy1998}) and \cite{Martinez-Pinedo2012} studied with a non-relativistic treatment possible consequences for the luminosities and energies of neutrinos in the early phase of CCSNe;
in \cite{Roberts2017}, CC reactions were considered in the relativistic framework including the full kinematics and the weak magnetism and in \cite{Fischer2020PRC} their implications both for the supernova explosion and the subsequent PNS cooling up to $\sim 10 \s$ were studied numerically;
the $q^2$ dependence of form factors as well as the pseudoscalar term were also considered rather recently in \cite{Guo2020, Fischer2020}.

Note that all these studies are restricted to $t \lesssim 10 \s$.
As we mentioned in Introduction, the next Galactic supernova is very likely to provide us with an opportunity to probe into the much later phase up to $\sim$ a minute.
We hence think it is important to explore these corrections for the thermodynamic conditions appropriate at this late period.
In this subsection, we numerically evaluate for both NC and CC reactions the mean field effect (MF), the weak magnetism (WM), the $q^2$ dependence of the form factors and the pseudoscalar (PS) term for the thermodynamic conditions considered in the previous subsection and will see when, where and at which neutrino energies they are important.

\begin{figure*}[htbp]
  \begin{minipage}[b]{0.49\linewidth}
    \centering
    \includegraphics[keepaspectratio, scale=0.52]{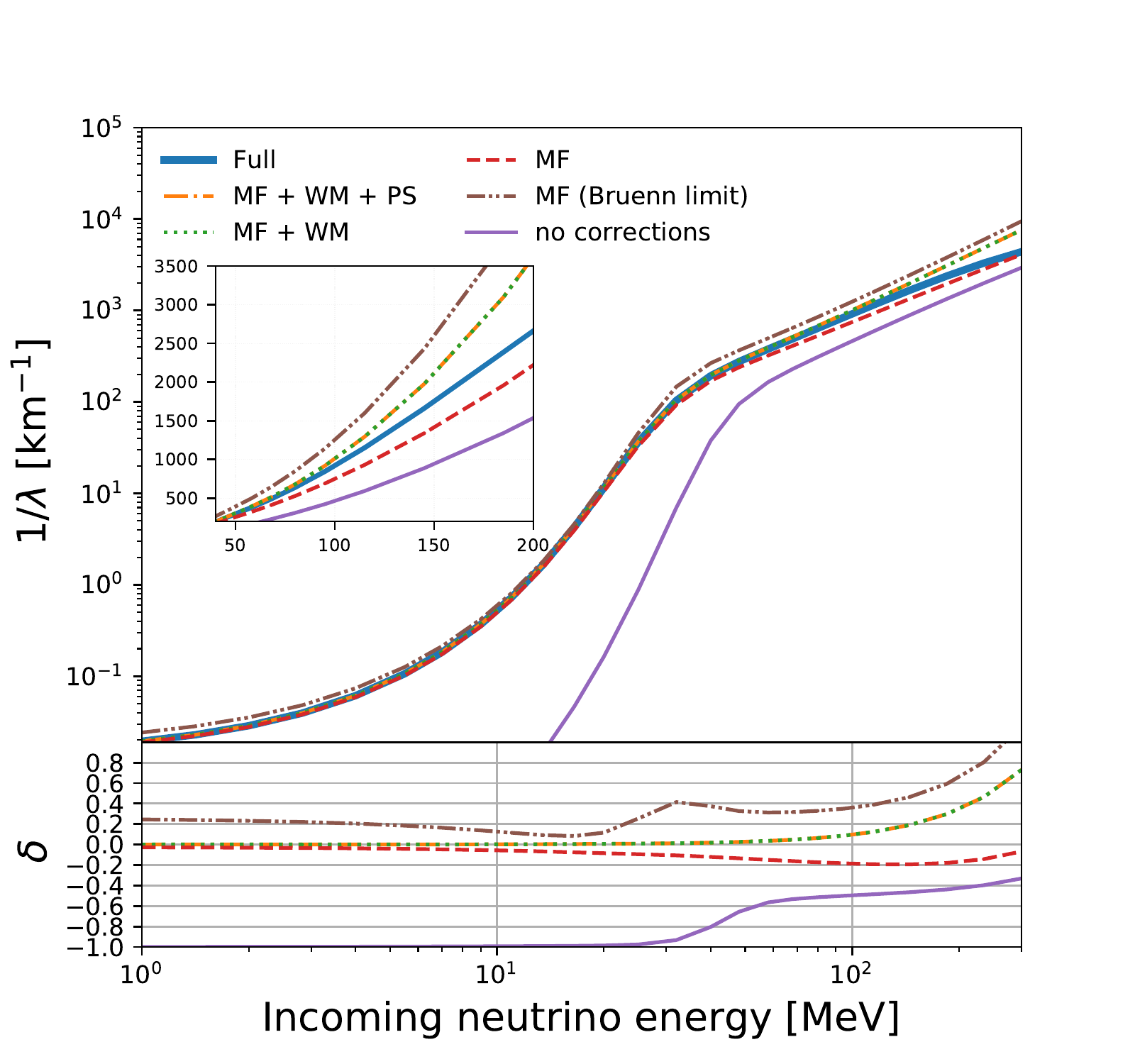}
    \subcaption{}\label{fig:CC_nue}
  \end{minipage}
  \begin{minipage}[b]{0.49\linewidth}
    \centering
    \includegraphics[keepaspectratio, scale=0.52]{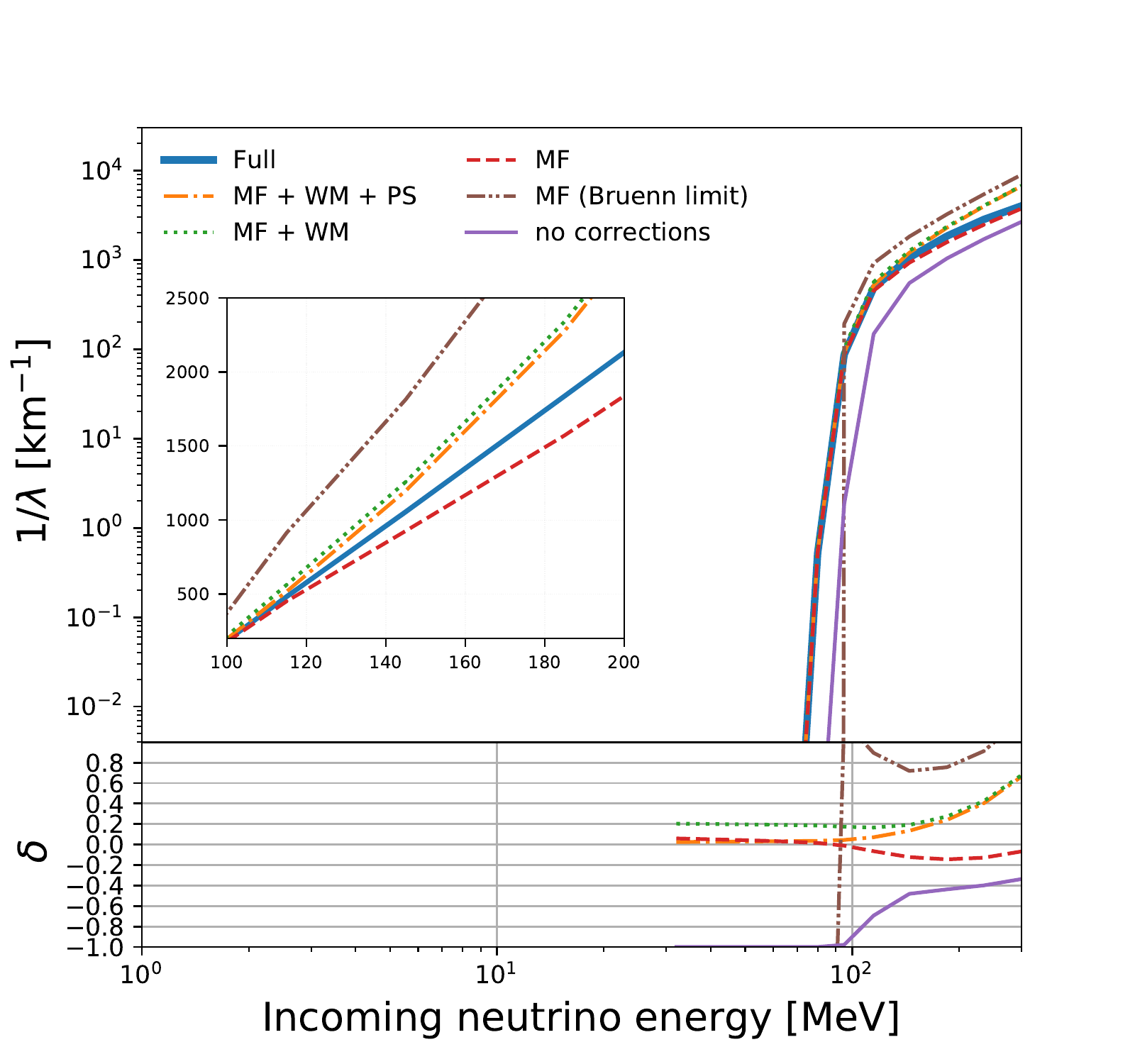}
    \subcaption{}\label{fig:CC_numu}
  \end{minipage}
  \caption{Inverse mean free paths for $\nu_e + n \rightarrow e^- + p$ (panel (a)) and $\nu_{\mu} + n \rightarrow \mu^- + p$ (panel (b)) with and without some corrections to the charged currents of nucleons for model t10S. The label ``Full" (blue line) means that all corrections WM, PS, MF and the $q^2$ dependence of form factors are included;
  the orange line incorporates only WM, PS and MF;
  the green dotted line excludes PS further;
  the red dashed line includes only MF;
  the purple line corresponds to no corrections at all.
  The brown dot-dot-dashed line shows the approximate rate given by Bruenn \cite{Bruenn1985} with the MF being included.
  The relative differences are shown at the bottom of each panel. The insets are the zoom-in on high energy ranges.}\label{fig:CC_effects}
\end{figure*}

Figure \ref{fig:CC_effects} summarizes the results for the neutrino capture on neutron, one of the most important CC reactions of $\nu_e$ and $\nu_{\mu}$ in model t10S.
The label ``Full" (blue line) means that all corrections WM, PS, MF and the $q^2$ dependence of form factors are included;
the orange line incorporates only WM, PS and MF;
the green dotted line excludes PS further;
the red dashed line includes only MF;
the purple line corresponds to no corrections at all.
The brown dot-dot-dashed line shows the approximate rates given by Bruenn \cite{Bruenn1985}, in which they assume that the momentum transfer of the nucleon is approximately zero and the nucleon mass is infinitely large.
We include the MF effect in the Bruenn approximation, which is expressed as
\begin{align}
  \dfrac{1}{\lambda(E_1)}
  &= \dfrac{G_F^2 \cos^2 \theta_C}{\pi} \left( g_V^2 + 3 g_A^2 \right) p_3 E_3 \left[ 1 - f_3(E_3) \right] \dfrac{n_2 - n_4}{1 - \exp \left[ \beta \left(- \mu_2 + \mu_4 \right)\right]},\label{eq:mfp_CC_Bruenn}
\end{align}
where $E_3 = E_1 + U_2 - U_4 + m_2^* - m_4^*$, $p_3 = \sqrt{E_3^2 - m_3^2}$ and $n_2$ and $n_4$ are the number densities of $N_2$ and $N_4$ in \mbox{Eq. (\ref{eq:semi_reaction})}, respectively.
We define the relative deviation of the inverse mean free path for each approximation from the complete one as follows:
\begin{equation}
  \delta_i = \dfrac{1/\lambda_{i} - 1/\lambda_{\text{full}}}{1/\lambda_{\text{full}}}.
\end{equation}

It is obvious that the purple solid line is deviated from all the others substantially.
This is due to the mean field effect as can be most clearly understood from the comparison with the red dashed line.
In fact, the absorptivity is enhanced in general.
This happens for $\nu_e$ (panel (a)) because the energy that the produced electron gains becomes greater by the mean-field-potential difference between proton and neutron and, as a result, the final-state Pauli-blocking of electron is suppressed \cite{Martinez-Pinedo2012}.
For $\nu_{\mu}$ (panel (b)), the reason is essentially the same but the muon rest mass gives the threshold at a higher energy.

We can see how the approximation given by Bruenn \cite{Bruenn1985} behaves by comparing the brown and red dashed lines.
In both $\nu_e$ and $\nu_{\mu}$ captures on neutron, the approximated rates are always larger than the case with full kinematics.
The difference between them depends on the incoming neutrino energy: $\sim 20 \%$ at $ E_{\nu} \lesssim 30 \MeV$ and $\sim 30 \%$ at $E_{\nu} = 100 \MeV$.
Note that in the $\nu_{\mu}$ capture on neutron, the reaction cannot occur when the incoming neutrino is $ E_{\nu} < m_{\mu} - ( U_n - U_p )$ in the approximated rate.

The effect of the weak magnetism can be seen from the comparison between the red dashed line and the green dotted line.
It tends to enhance the absorptivity, the degree of which depends on the momentum transfer $q/M$:
$\sim 10 \%$ at $E_{\nu} \sim 10 \MeV$ and $\sim 30 \%$ at $E_{\nu} \sim 100 \MeV$ for $\nu_e$ whereas for $\nu_{\mu}$ $\sim 20 \%$ at $E_{\nu} \sim 100 \MeV$ as shown at the bottom of each panel.
The difference between the green dotted line and the orange dash-dotted line reflects the effects of pseudoscalar term. Since it is proportional to $m_l^2 \ll M^2$ in the matrix elements, the changes that this term induces are rather small both for $\nu_{e}$ and $\nu_{\mu}$.

The effects of the $q^2$ dependence of form factors are encoded in the difference between the blue solid line and the orange dash-dotted line.
The absorptivity is reduced in general by the inclusion of the dependence and the reduction is larger for higher neutrino energies, scaling with $q^2/M^2$.
This correction reaches $\sim 10 \%$ and $\sim 5 \%$ at $\sim 100 \MeV$ for $\nu_e$ and $\nu_{\mu}$, respectively.

\begin{figure*}[tbp]
  \begin{minipage}[b]{0.49\linewidth}
    \centering
    \includegraphics[keepaspectratio, scale=0.52]{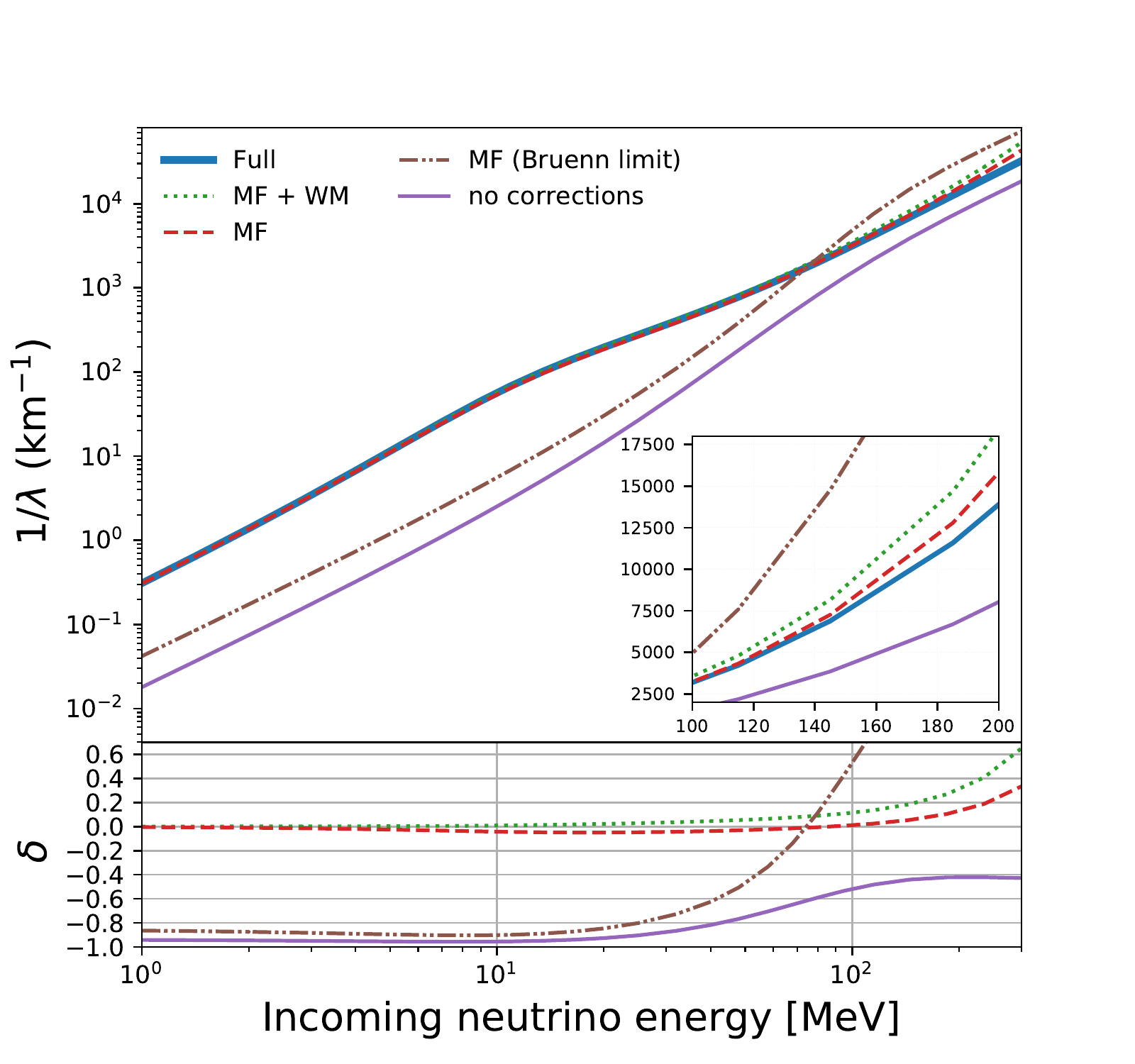}
    \subcaption{}\label{fig:NC_nue}
  \end{minipage}
  \begin{minipage}[b]{0.49\linewidth}
    \centering
    \includegraphics[keepaspectratio, scale=0.52]{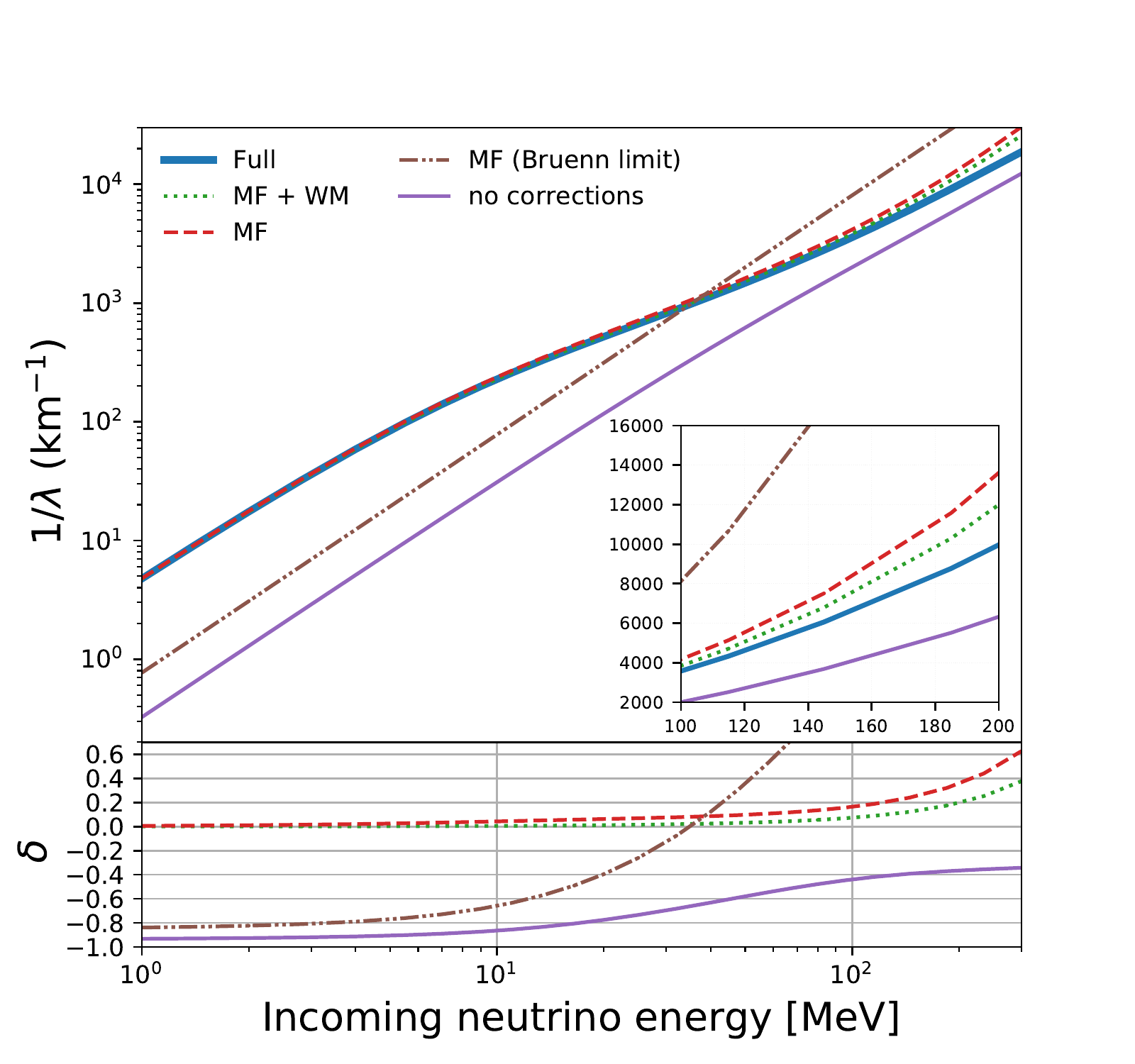}
    \subcaption{}\label{fig:NC_nueb}
  \end{minipage}
  \caption{Inverse mean free paths of the neutrino scattering on neutron for $\nu_e$ (panel (a)) and $\bar{\nu}_e$ (panel (b)) with and without some corrections to the neutral currents of nucleons ofr model t10D.
  The label ``Full" (blue line) means that all corrections WM, MF and the $q^2$ dependence of form factors are included;
  the green dotted line incorporates only WM and MF;
  the red dashed line includes only MF;
  the purple line corresponds to no corrections at all.
  The brown dot-dot-dashed line shows the approximate rate given by Bruenn \cite{Bruenn1985} with the MF being included.
  The relative differences are shown at the bottom of each panel. The insets are the zoom-in on high energy ranges.}\label{fig:NC_effects}
\end{figure*}

Now we shift our interest to the neutrino scattering on neutron via NC.
\mbox{Figure \ref{fig:NC_nue}} shows the effects of the corrections in model t10D.
The effect of MF can be seen from the comparison between the purple solid lines and the red dashed lines, in which the nucleon mass is set to the vacuum value in the former while it is given by the effective mass in the latter.
Note that there is no potential difference in the NC reactions.
In the calculation of the dynamical structure functions of nucleons, the integration ranges that appear in the Fermi integral (for example see Eq. (\ref{eq:I1_before_fermi})) strongly depend on the effective mass.
In fact, for the NC reaction, in which $\kappa = 1$, $\tilde{q}^0 = q^0$ and $\Delta^2 = \left( q^0 \right)^2 - q^2 < 0$ are satisfied, the integration range (see Eqs. (\ref{eq:E2_range}) and (\ref{eq:E^*_2pm})) becomes
\begin{equation}
  E_{\text{min}} = \max \left\{ m_2^*, m_4^* - \tilde{q}^0, E^*_{2,+} \right\}, \ E_{\text{max}} = \infty,
\end{equation}
where $E^*_{2,+}$ is given as
\begin{equation}
  E^*_{2,+} = -\dfrac{q^0}{2} + \dfrac{|q|}{2} \sqrt{1 - \dfrac{4 {m^*_2}^2}{\Delta^2}}. \label{eq:e2_p}
\end{equation}
Since the effective mass of nucleon, $m^*$, in the dense region is substantially smaller than the vacuum mass, the integration ranges are wider in general.
This is the reason why the inverse mean free paths are enhanced by including the effective mass.

The brown line shows the approximation given by Bruenn \cite{Bruenn1985} as follows:
\begin{align}
  &\dfrac{1}{\lambda(E_1)} =  \int
  \dfrac{d^3 \bm{p}_3}{(2 \pi)^3}
  \left[ 1 - f_3(E_3) \right] R_{\text{Bruenn}}, \label{eq:mfp_Bruenn}
\end{align}
with the reaction kernel
\begin{equation}
  R_{\text{Bruenn}} = 2 \pi G_F^2 \eta_{\text{NN}} \delta(E_1 - E_3) \left\{ (G_1^N(0))^2 + 3 (G_A^N(0))^2 + \left[ (G_1^N(0))^2 - (G_A^N(0))^2 \right] \cos \theta  \right\},
\end{equation}
where $\theta$ is an angle between the incoming and outgoing neutrinos and  $\eta_{\text{NN}}$ is defined as
\begin{align}
  \eta_{\text{NN}} = \int \dfrac{2 d^3 \bm{p}_N}{(2 \pi)^3} f_{N}(E_N^*) \left[ 1 - f_N (E_N^*) \right].
\end{align}
From the comparison with red dashed lines, we find that the approximated rate is larger than that including full kinematics for higher energies $ E_{\nu} \gtrsim 70 \MeV$ and
\mbox{$ E_{\nu} \gtrsim 40 \MeV$} for $\nu_e$ and $\bar{\nu}_e$, respectively.
For lower energies, on the other hand, the rate including full kinematics is larger by an order than the approximate one.
We can hence say that the momentum transfer has great influences in the deeper region where the effective mass is reduced.

The WM correction gives opposite modifications for neutrino and anti-neutrino (compare the green and red lines in panels (a) and (b)):
it tends to enhance (suppress) the reaction for neutrino (anti-neutrino).
The degrees of change are a few \% at $E_{\nu} \sim 10 \MeV$ and $\sim 10 \%$ at $E_{\nu} \sim 100 \MeV$ in both cases.
These results are qualitatively consistent with those of the previous study \cite{Horowitz2002}.

The $q^2$ dependence of form factors reduces the inverse mean free paths just as in the CC reactions.
The relative difference it makes is a few percent around $E_{\nu} \sim 10 \MeV$ and reaches a few tens percent at $E_{\nu} = 300 \MeV$.

To summarize, for high-energy neutrinos the corrections in the semi-leptonic reactions are more important than the contributions from the muon-related leptonic reactions to the opacity.
At low energies, on the other hand, some of the muon-related reactions overwhelms the corrections in the semi-leptonic reactions.
Both of them hence need to be taken into account in the PNS cooling.

Note in passing that since the NC reactions do not distinguish the neutrino flavors except their distribution functions, the results mentioned above for the $e$-type neutrino and anti-neutrino are applied also to other flavors of neutrinos and anti-neutrinos.

\subsection{Inverse mean free paths at other times}
We move on to the results for other thermodynamical conditions obtained at $t = 1, 3 \operatorname{s}$ (the earlier phase) and $t = 30, 50 \operatorname{s}$ (the later phase) to infer the time evolutions of the mean free paths in the PNS cooling.

\begin{figure*}[htbp]
  \begin{minipage}[b]{0.49\linewidth}
    \centering
    \includegraphics[keepaspectratio, scale=0.51]{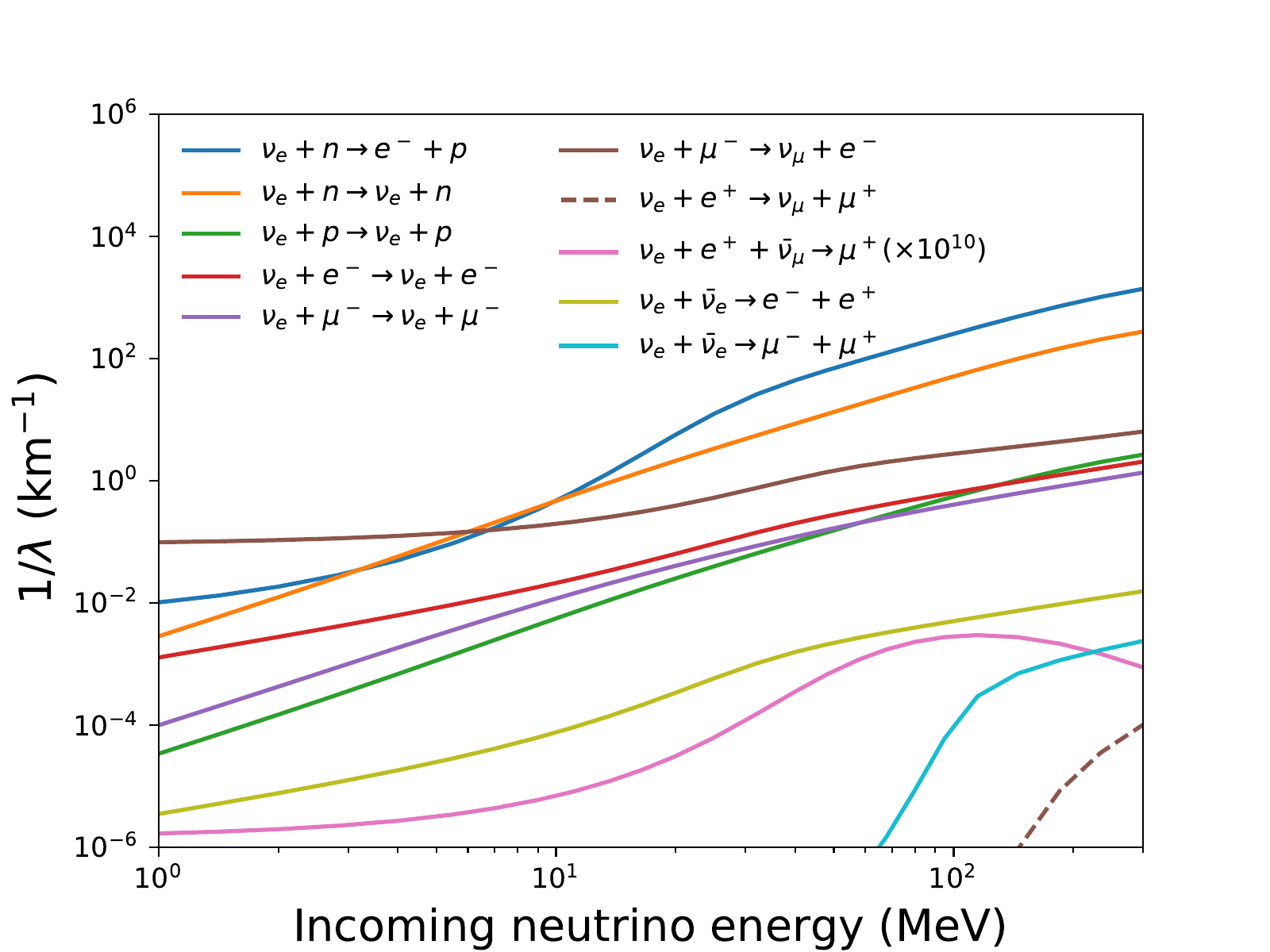}
    \subcaption{}\label{fig:t1s_nue}
  \end{minipage}
  \begin{minipage}[b]{0.49\linewidth}
    \centering
    \includegraphics[keepaspectratio, scale=0.51]{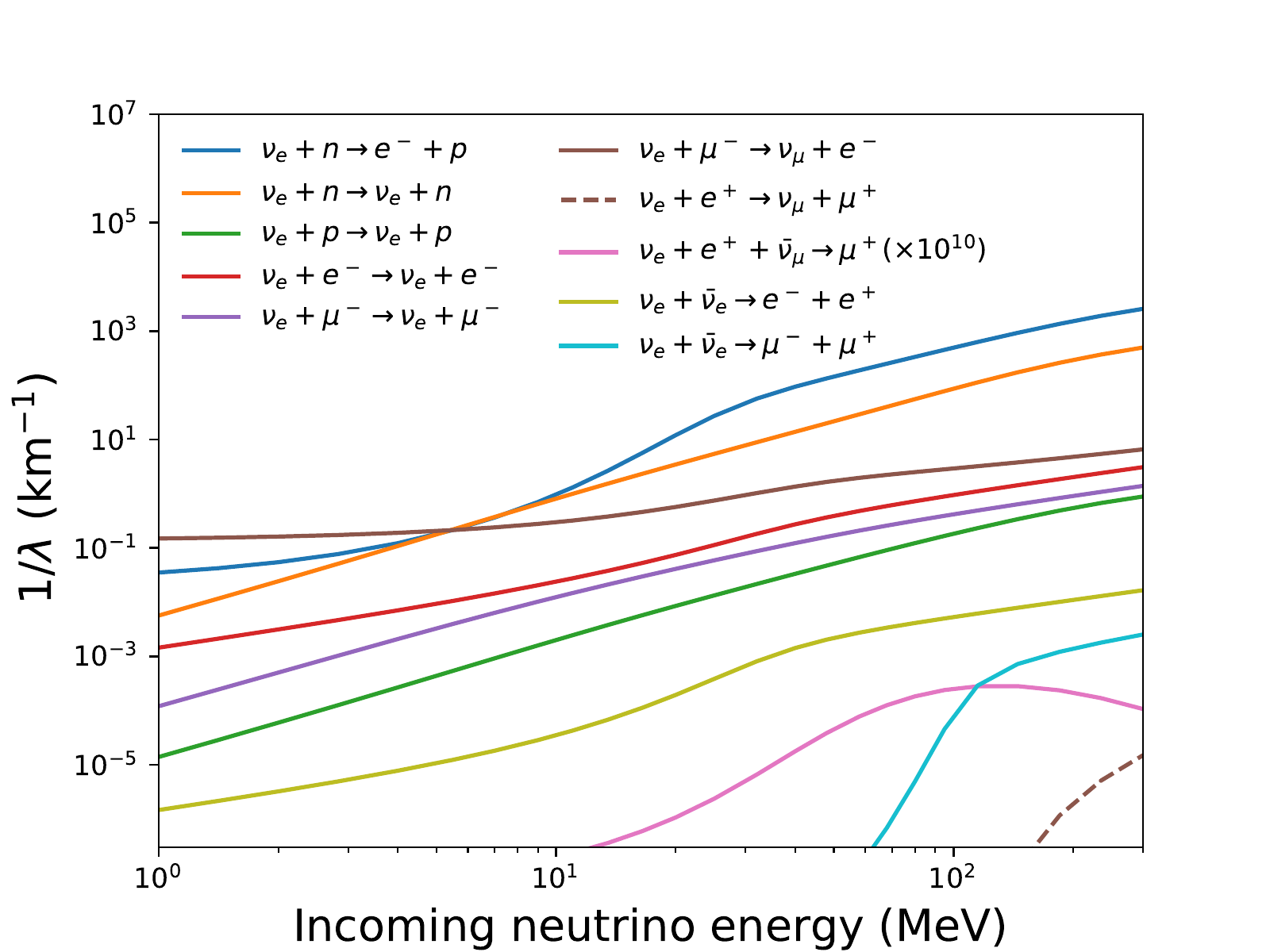}
    \subcaption{}\label{fig:t3s_nue}
  \end{minipage}\\
  \begin{minipage}[b]{0.49\linewidth}
    \centering
    \includegraphics[keepaspectratio, scale=0.51]{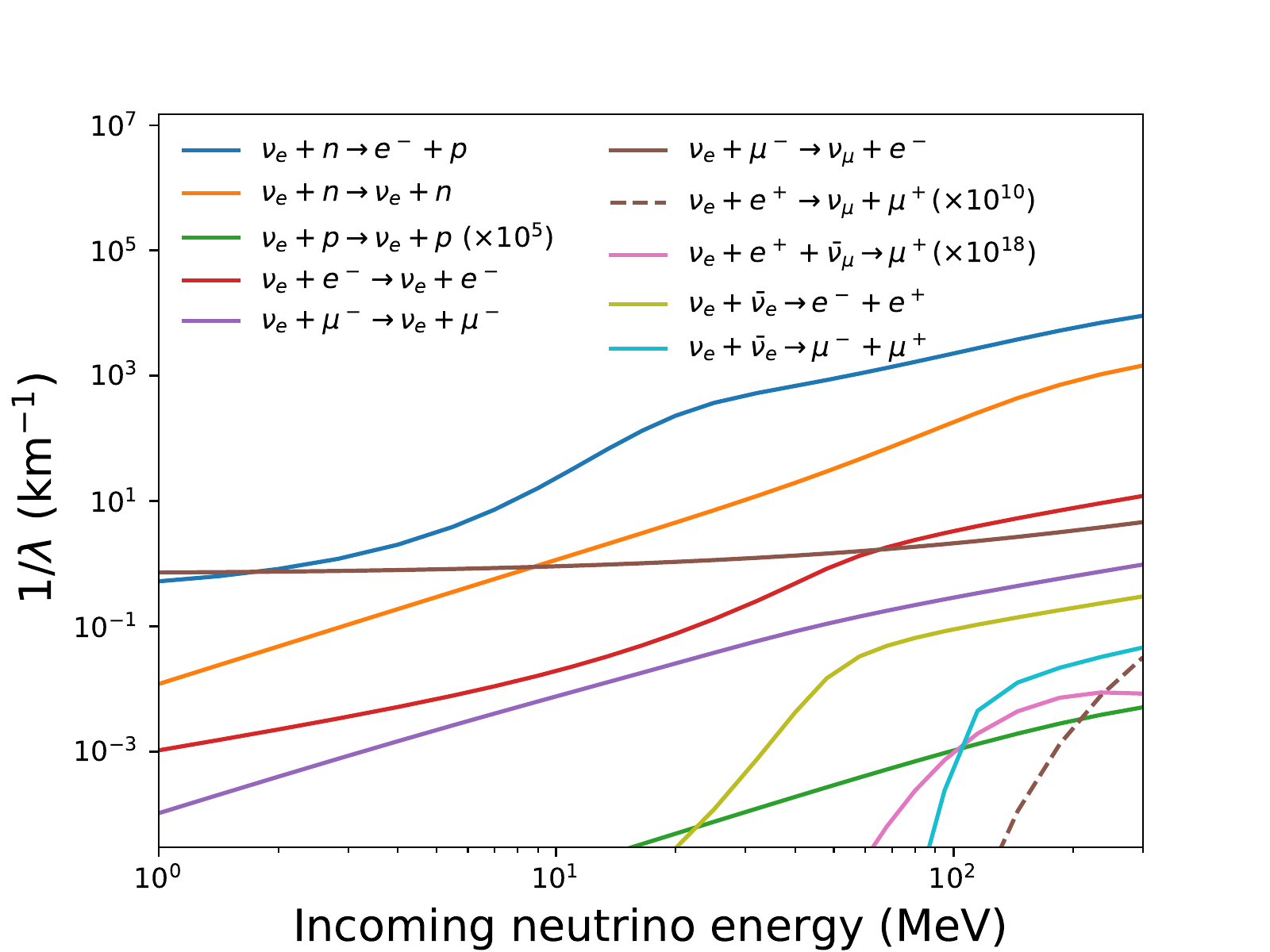}
    \subcaption{}\label{fig:t30s_nue}
  \end{minipage}
  \begin{minipage}[b]{0.49\linewidth}
    \centering
    \includegraphics[keepaspectratio, scale=0.51]{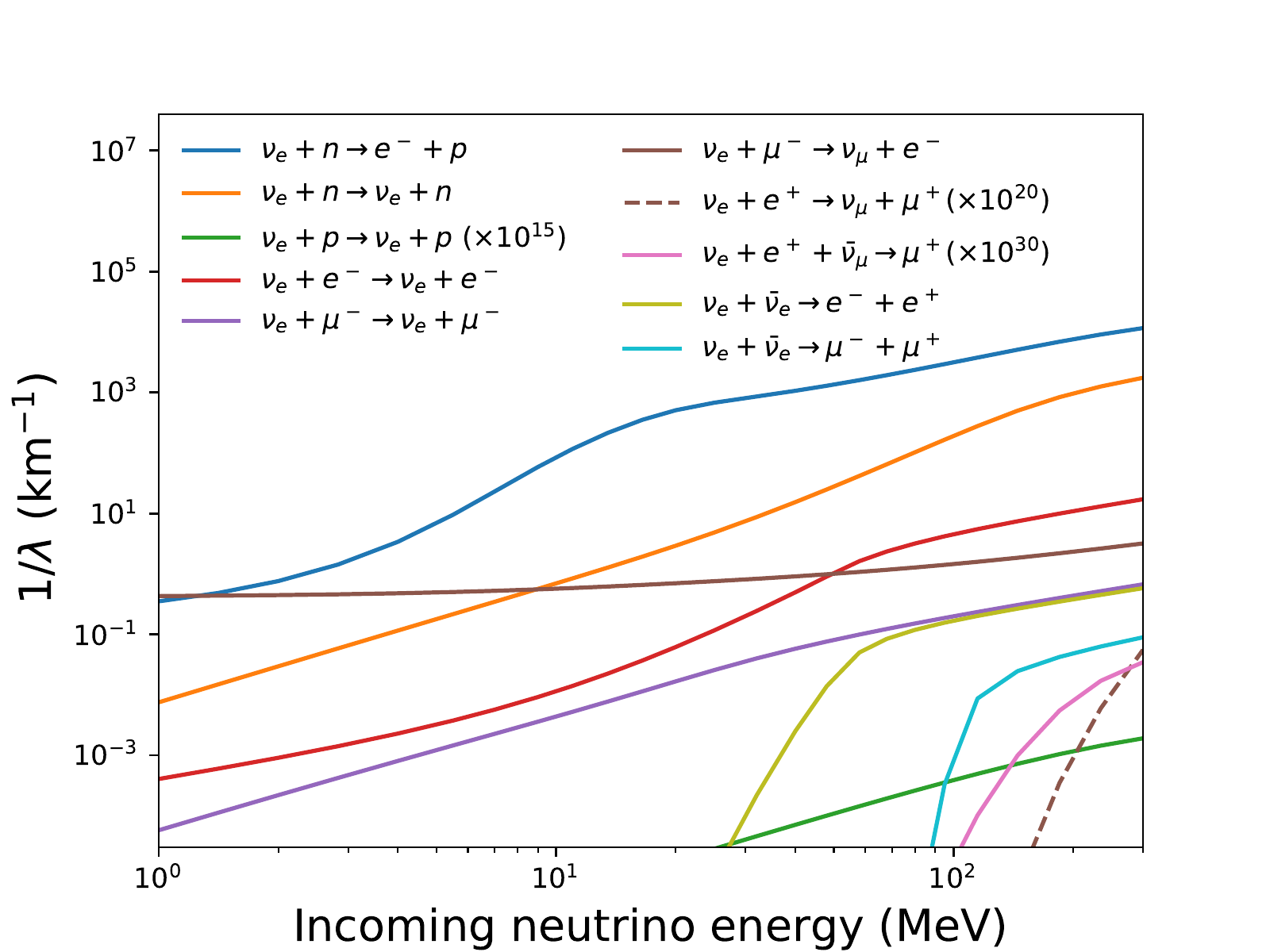}
    \subcaption{}\label{fig:t50s_nue}
  \end{minipage}
  \caption{Inverse mean free paths for $\nu_e$ same as Figure \ref{fig:t10S} (a) but at different times; Panel (a): $t = 1 \operatorname{s}$ (model t1S); Panel (b): $t = 3 \operatorname{s}$ (model t3S); Panel (c): $t = 30 \operatorname{s}$ (model t30S); Panel (d): $t = 50 \operatorname{s}$ (model t50S).
  Colors denote different reactions. Note that the values for $\nu_e + e^+ + \bar{\nu}_{\mu} \rightarrow \mu^+$ and $\nu_e + e^+ \rightarrow \nu_{\mu} + \mu^+$ are multiplied by the factors given in the legend in each panel.}\label{fig:S_nue}
\end{figure*}

The inverse mean free paths for $\nu_e$ at the neutrino sphere are shown in Figure \ref{fig:S_nue}.
Comparing them with Figure \ref{fig:t10s_NuSphere_nue}, one finds that the neutrino capture on neutron is enhanced at later times.
This is due to the widening of the effective potential difference between neutron and proton $U_n - U_p$ as well as to the increase in the number density of neutron.
The inverse mean free paths for the neutrino scattering on neutron and the flavor exchange reaction also rise as the density at the neutrino sphere increases in time.
The scattering on proton, on the other hand, gets strongly suppressed at later times due to the declining proton fraction at the neutrino sphere.
As the temperature lowers, the pair production reactions are suppressed at low energies as we mentioned in \mbox{Figure \ref{fig:t10d}}.

\begin{figure*}[htbp]
  \begin{minipage}[b]{0.49\linewidth}
    \centering
    \includegraphics[keepaspectratio, scale=0.51]{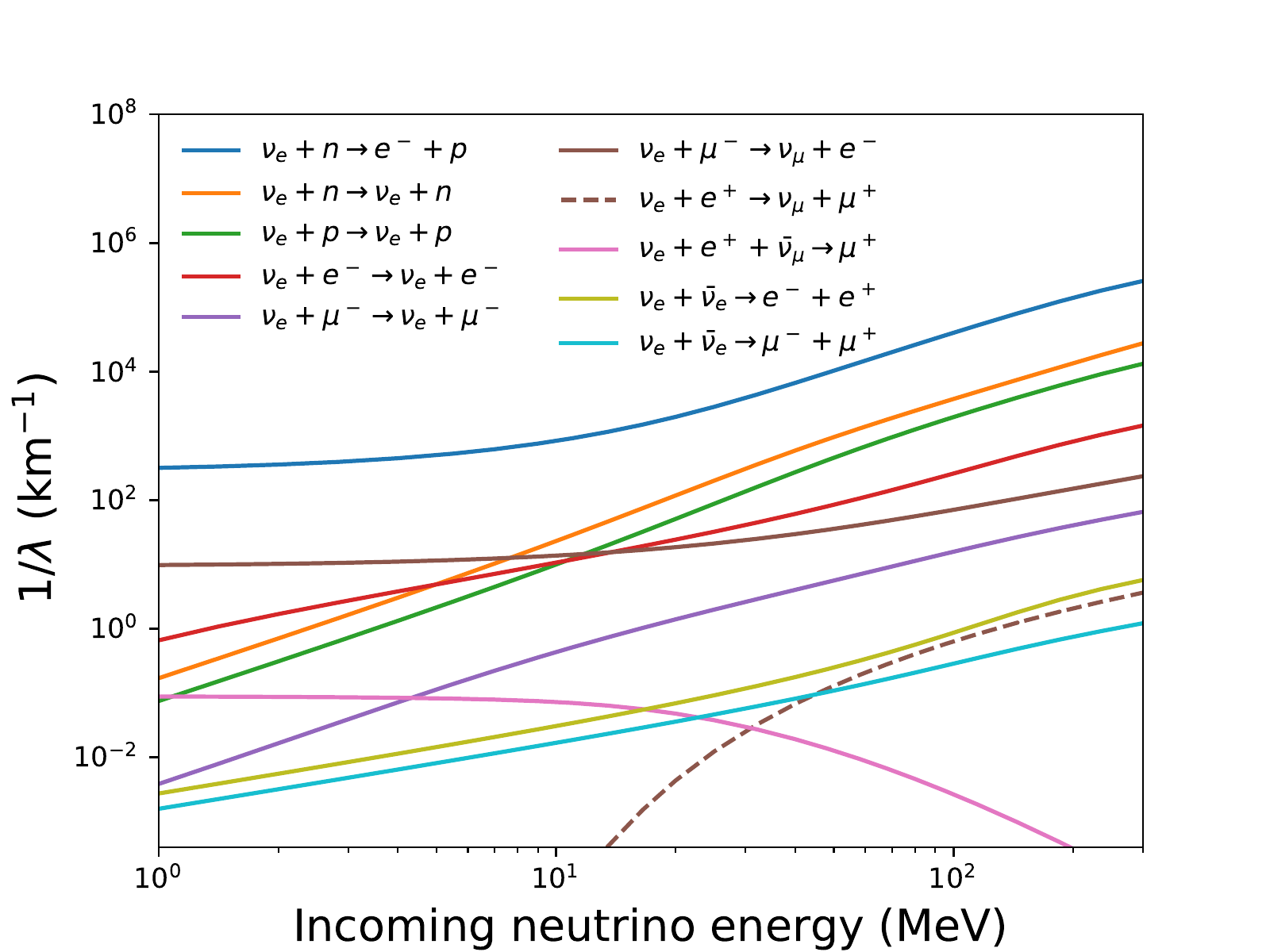}
    \subcaption{}\label{fig:t1d_nue}
  \end{minipage}
  \begin{minipage}[b]{0.49\linewidth}
    \centering
    \includegraphics[keepaspectratio, scale=0.51]{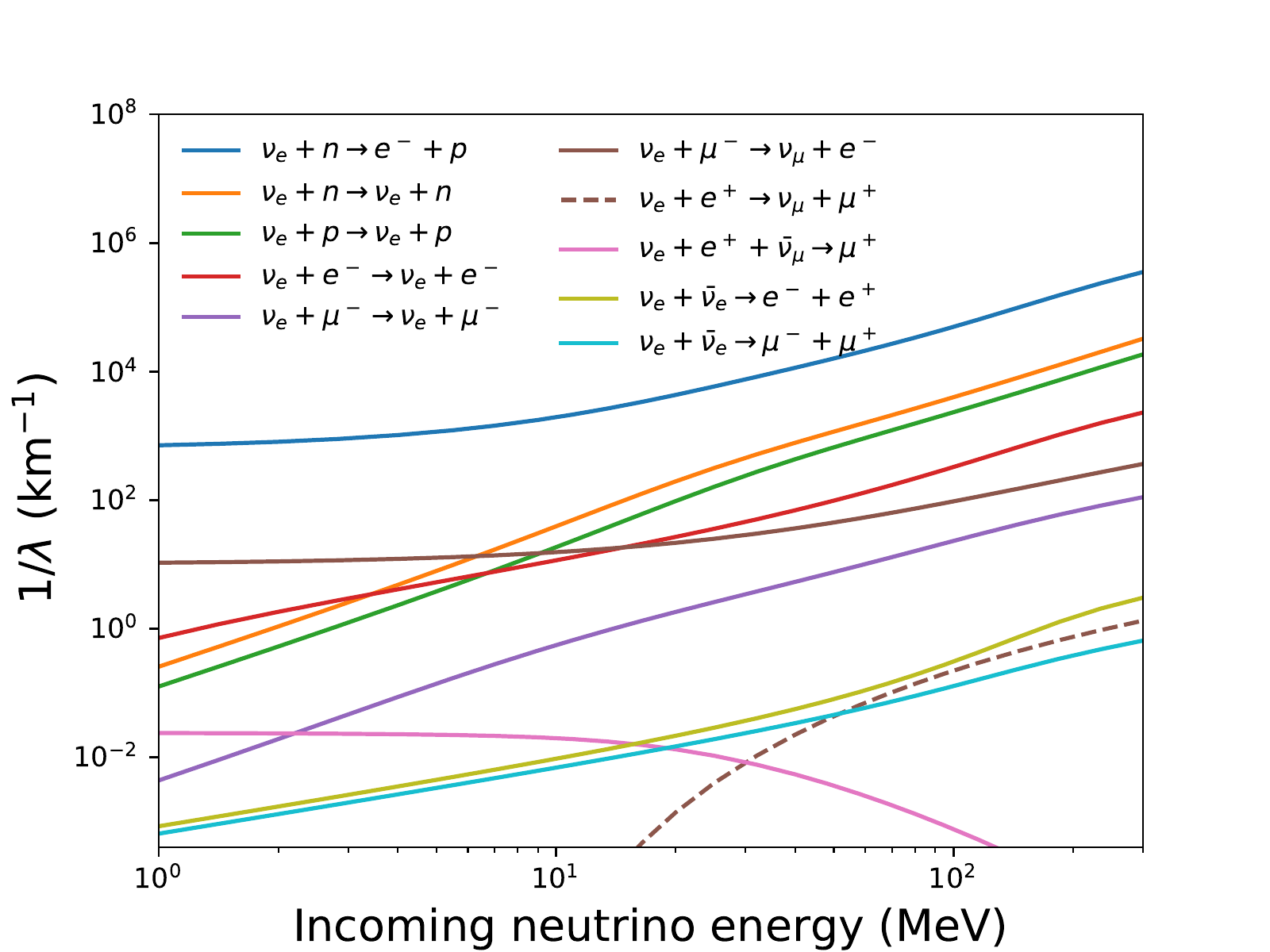}
    \subcaption{}\label{fig:t3d_nue}
  \end{minipage}\\
  \begin{minipage}[b]{0.49\linewidth}
    \centering
    \includegraphics[keepaspectratio, scale=0.51]{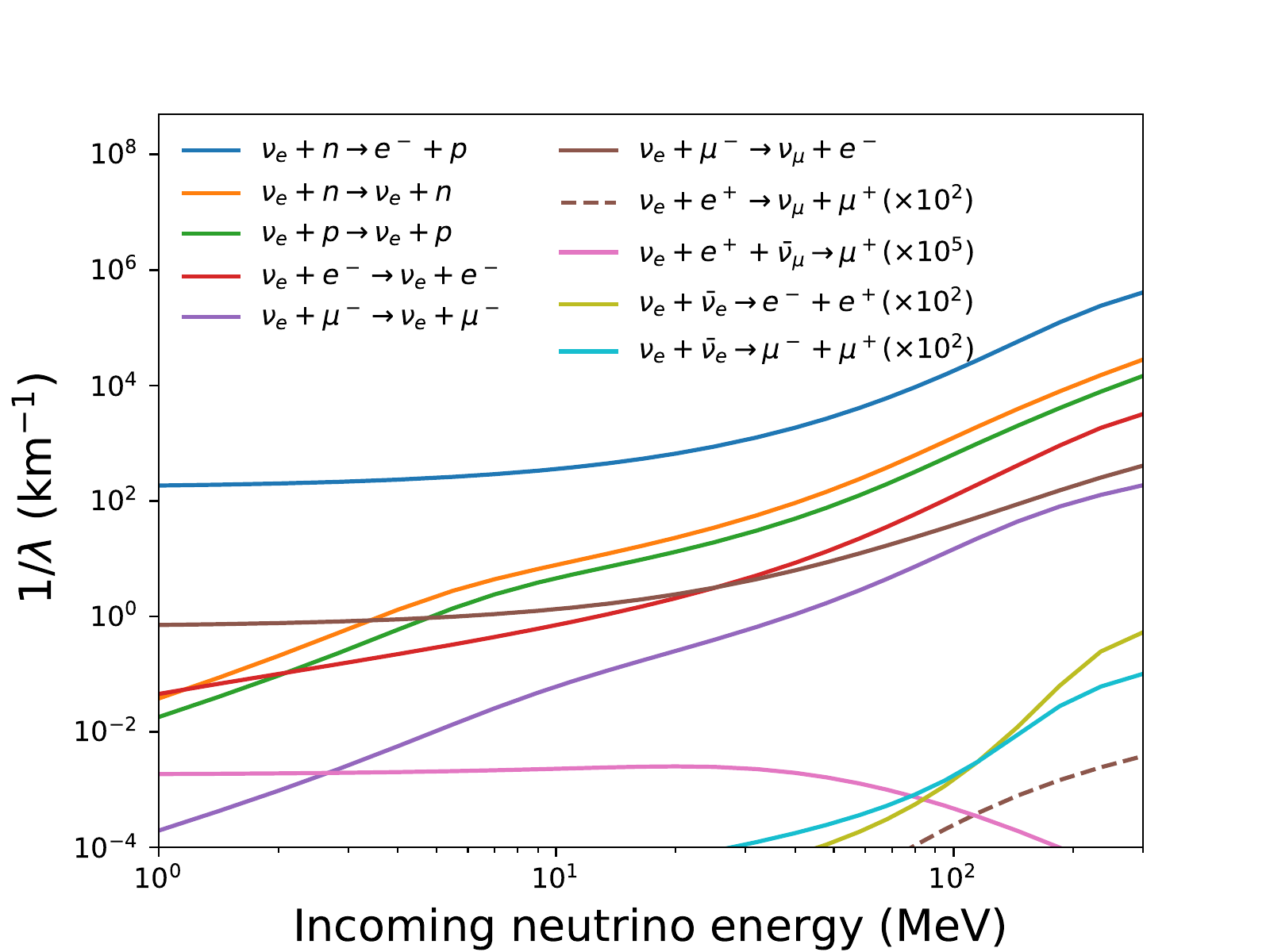}
    \subcaption{}\label{fig:t30d_nue}
  \end{minipage}
  \begin{minipage}[b]{0.49\linewidth}
    \centering
    \includegraphics[keepaspectratio, scale=0.51]{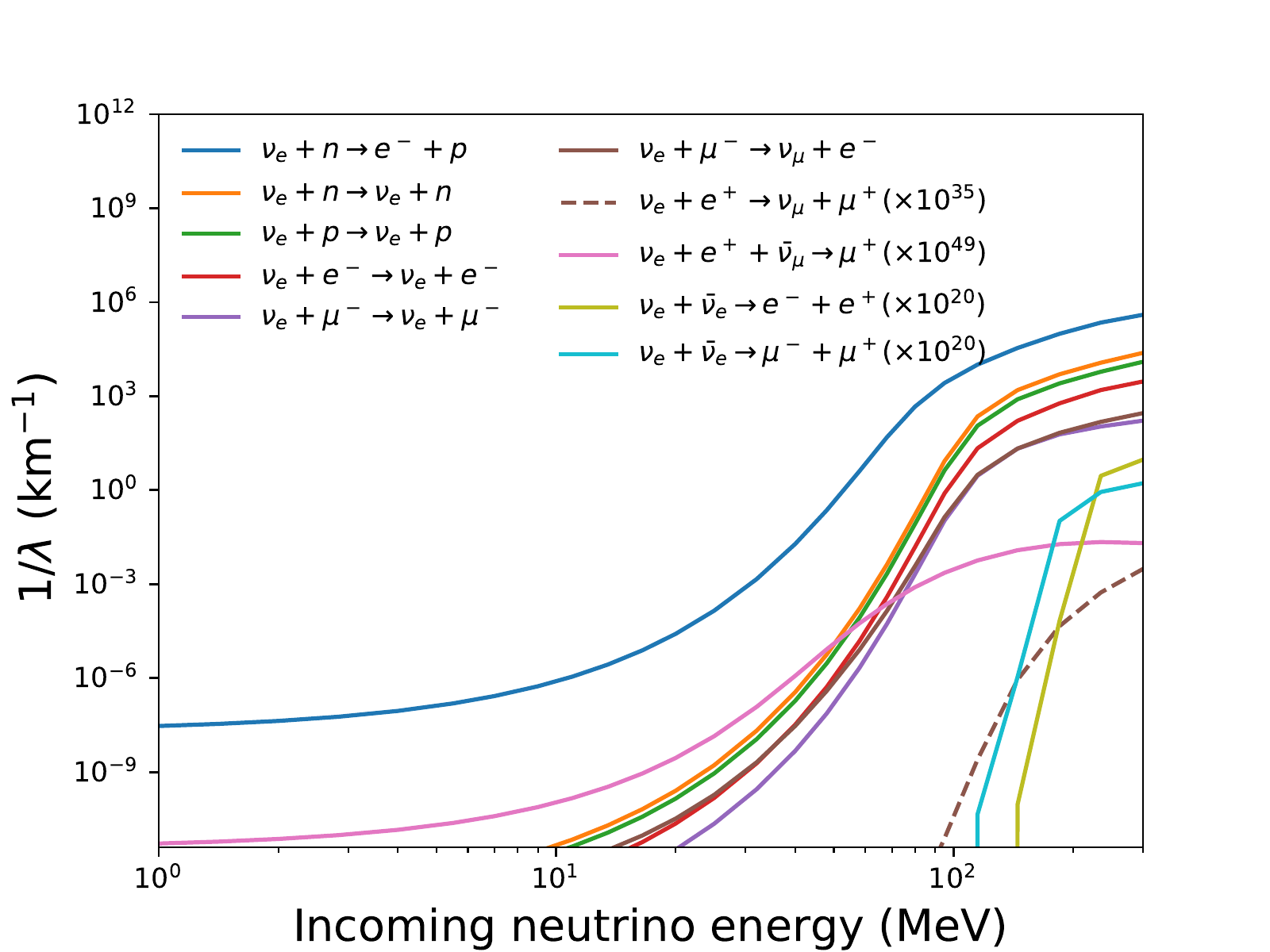}
    \subcaption{}\label{fig:t50d_nue}
  \end{minipage}
  \caption{Same as Figure \ref{fig:S_nue} but for the deeper regions. These figures are same as Figure \ref{fig:t10d} (a) but at different times;
  Panel (a): $t = 1 \operatorname{s}$ (model t1D); Panel (b): $t = 3 \operatorname{s}$ (model t3D); Panel (c): $t = 30 \operatorname{s}$ (model t30D); Panel (d): $t = 50 \operatorname{s}$ (model t50D).
  Colors denote different reactions.
  Note that the values for $\nu_e + e^+ + \bar{\nu}_{\mu} \rightarrow \mu^+$, $\nu_e + e^+ \rightarrow \nu_{\mu} + \mu^+$ and the pair production reactions in models t30D and t50D are multiplied by the factors given in the legend in each panel.}\label{fig:D_nue}
\end{figure*}

Figure \ref{fig:D_nue} is the same as Figure \ref{fig:S_nue} but for the deeper region, where the temperature becomes maximum.
Note that the radius changes in time (see Table \ref{tab:cond}).
The inverse mean free paths of the neutrino capture on neutron is the dominant source of opacity at all times, but it drastically decreases at $t = 50 \operatorname{s}$.
This is because the Pauli blocking by electron at \mbox{$E_{\nu} \lesssim \mu_e - (U_n - U_p)$} is more effective.
This applies also to the suppression of the flavor exchange reaction at $t = 50 \operatorname{s}$.
On the other hand, the decline of the neutrino scatterings on nucleon at $t = 50 \operatorname{s}$ is due to the strong Pauli blocking by $\nu_e$, the Fermi energy of which is \mbox{$\mu_{\nu_e} = \mu_p + \mu_e - \mu_n \simeq 100 \MeV$}.
These results at the very late phase have never been reported so far.

\begin{figure*}[htbp]
  \begin{minipage}[b]{0.49\linewidth}
    \centering
    \includegraphics[keepaspectratio, scale=0.51]{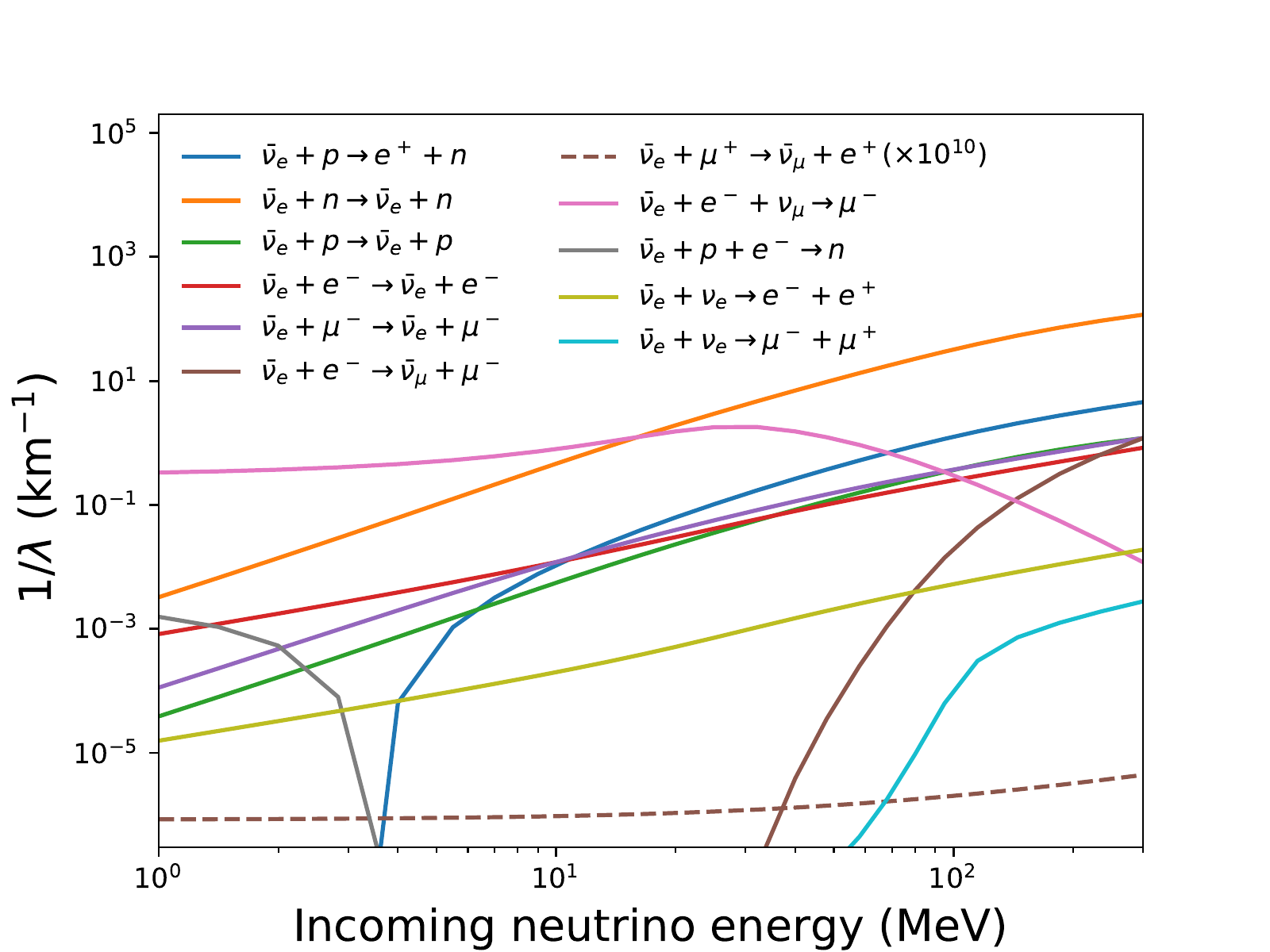}
    \subcaption{}\label{fig:t1s_nueb}
  \end{minipage}
  \begin{minipage}[b]{0.49\linewidth}
    \centering
    \includegraphics[keepaspectratio, scale=0.51]{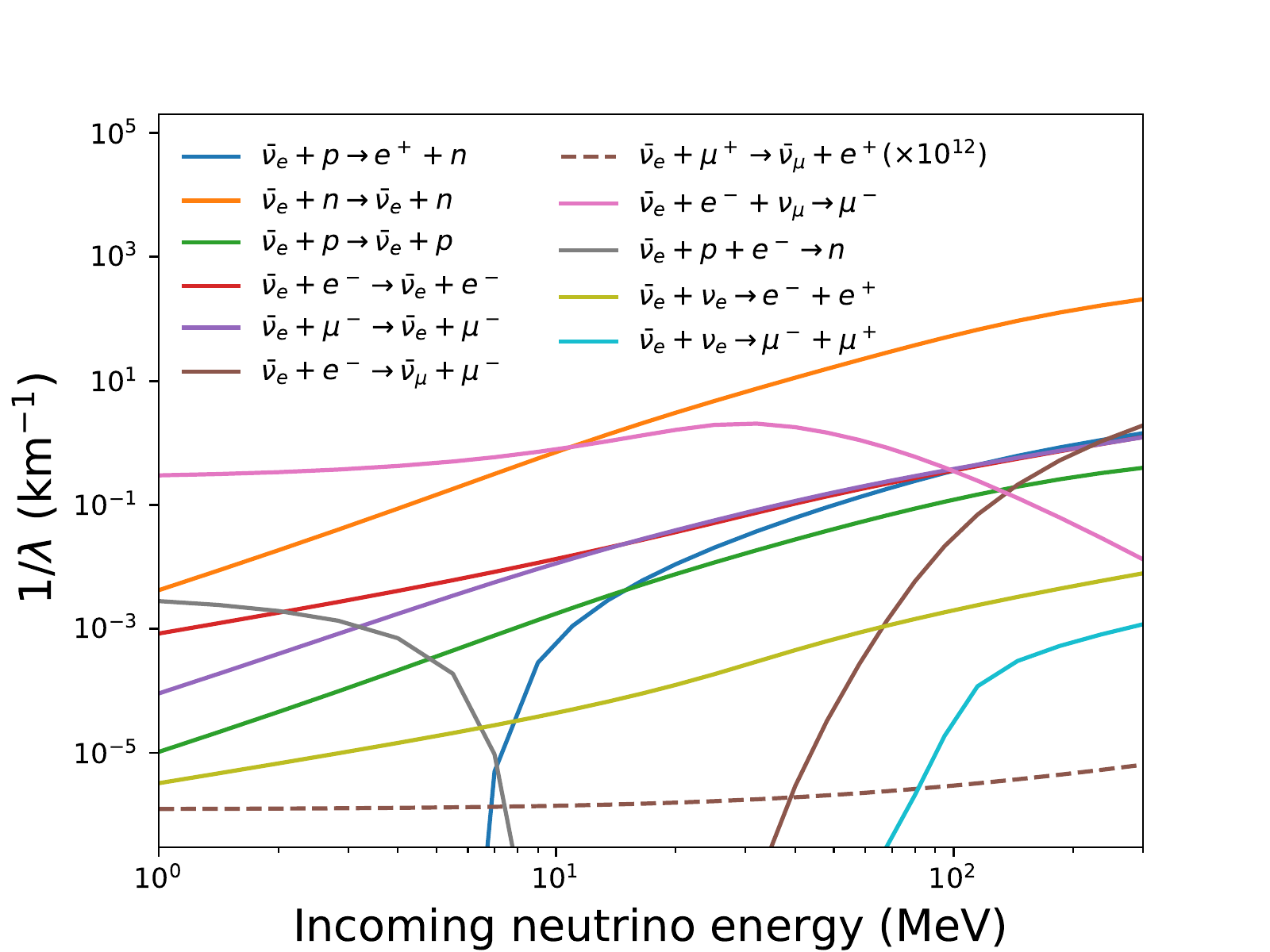}
    \subcaption{}\label{fig:t3s_nueb}
  \end{minipage}\\
  \begin{minipage}[b]{0.49\linewidth}
    \centering
    \includegraphics[keepaspectratio, scale=0.51]{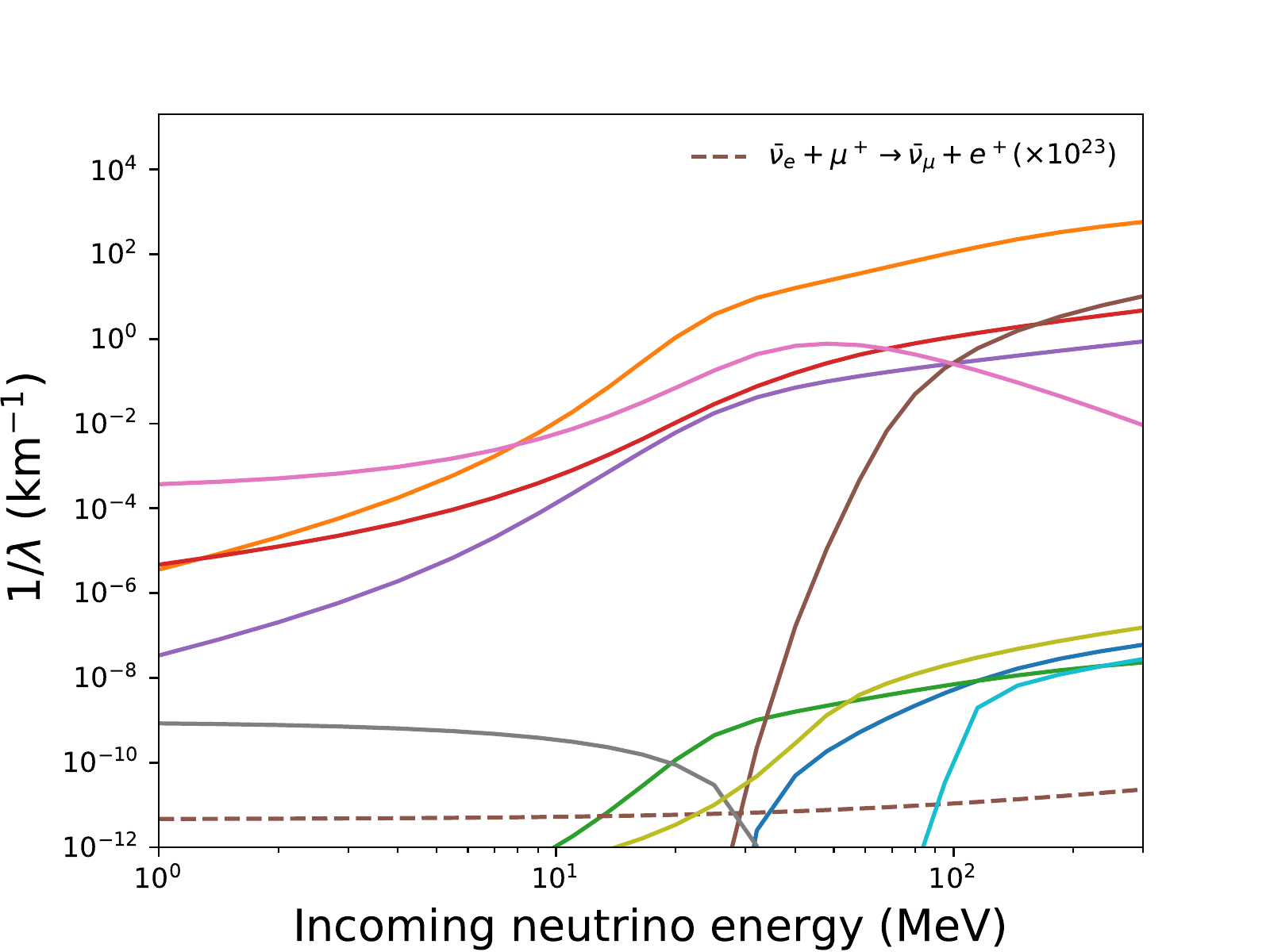}
    \subcaption{}\label{fig:t30s_nueb}
  \end{minipage}
  \begin{minipage}[b]{0.49\linewidth}
    \centering
    \includegraphics[keepaspectratio, scale=0.51]{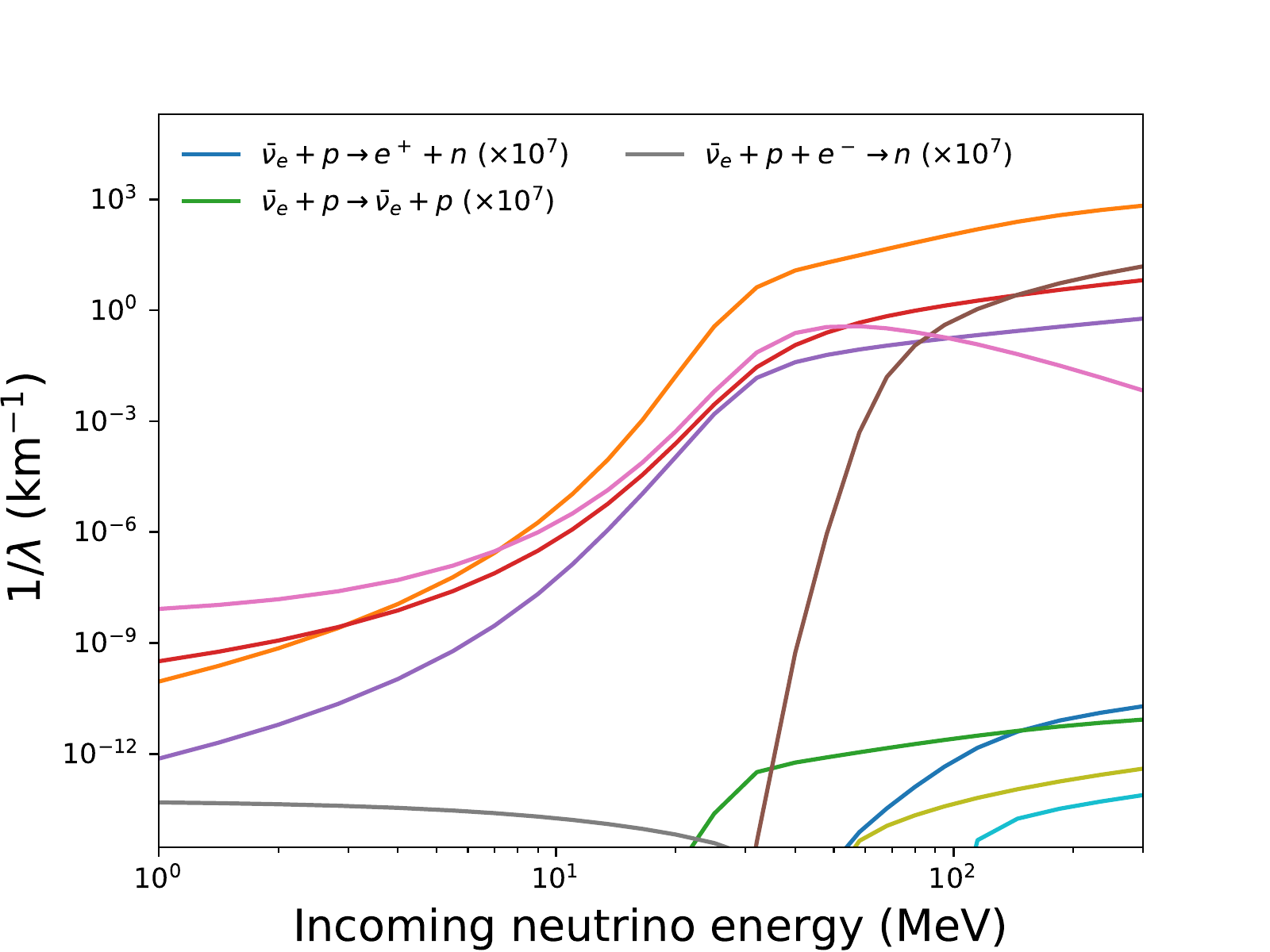}
    \subcaption{}\label{fig:t50s_nueb}
  \end{minipage}
  \caption{Same as Figure \ref{fig:S_nue} but for $\bar{\nu}_e$. These figures are same as Figure \ref{fig:t10S} (b) but at different times. Colors denote the different reactions. The legends are omitted in panels (c) and (d) for visibility but the notations are the same as in panel (a).
  Note that the values for $\bar{\nu}_e + p \rightarrow e^+ + n$, $\bar{\nu}_e + p \rightarrow \bar{\nu}_e + p$ and $\bar{\nu}_e + p + e^- \rightarrow n$ in panel (d) are multiplied by $10^7$.
  The values for $\bar{\nu}_e + \mu^+ \rightarrow \bar{\nu}_{\mu} + e^+$ are multiplied by the factors given in the legend in each panel.
  The flavor exchange reaction $\bar{\nu}_e + \mu^+ \rightarrow \bar{\nu}_{\mu} + e^+$ does not occur at $t = 50 \operatorname{s}$.}\label{fig:S_nueb}
\end{figure*}

\begin{figure*}[htbp]
  \begin{minipage}[b]{0.49\linewidth}
    \centering
    \includegraphics[keepaspectratio, scale=0.51]{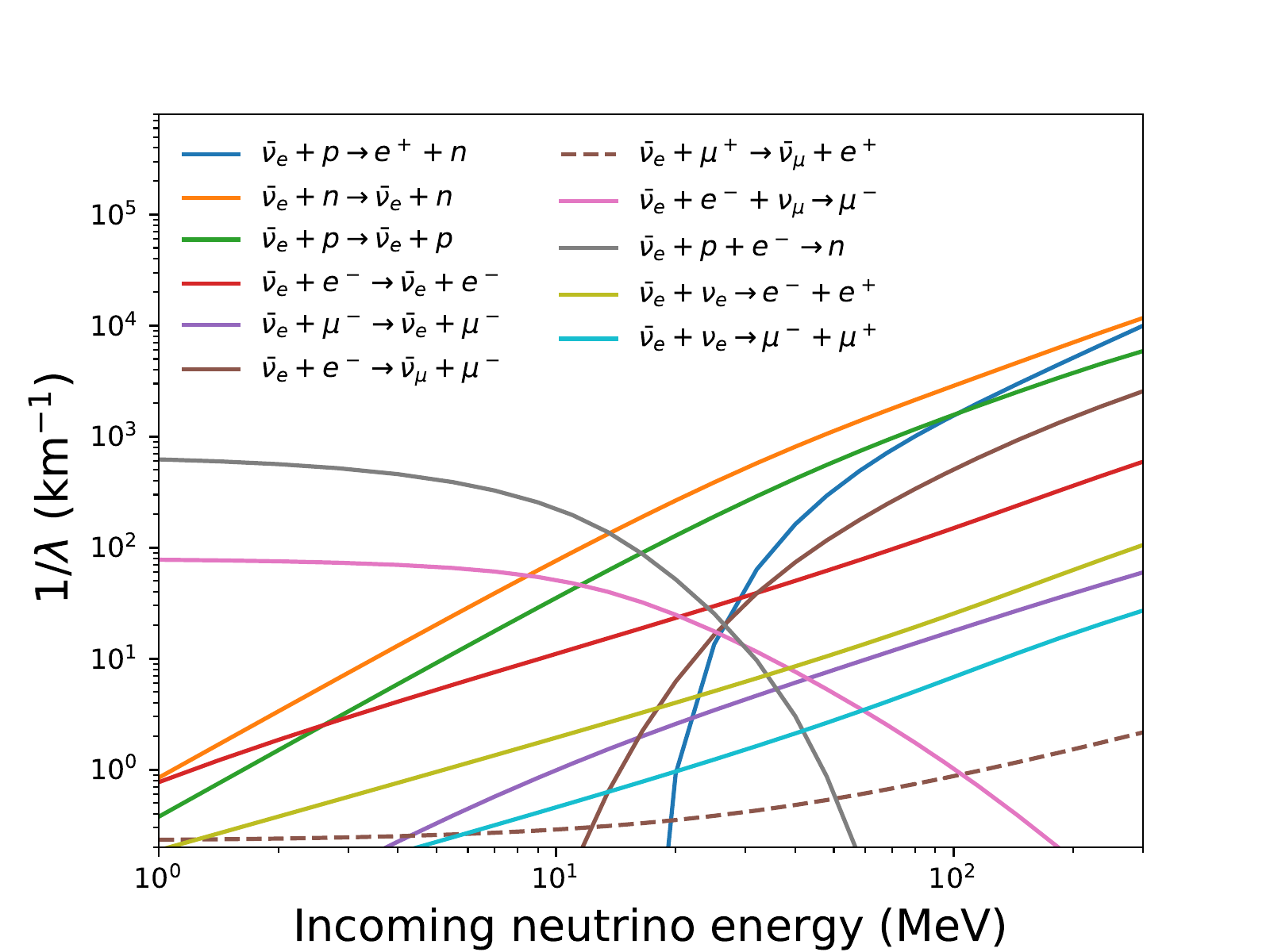}
    \subcaption{}\label{fig:t1d_nueb}
  \end{minipage}
  \begin{minipage}[b]{0.49\linewidth}
    \centering
    \includegraphics[keepaspectratio, scale=0.51]{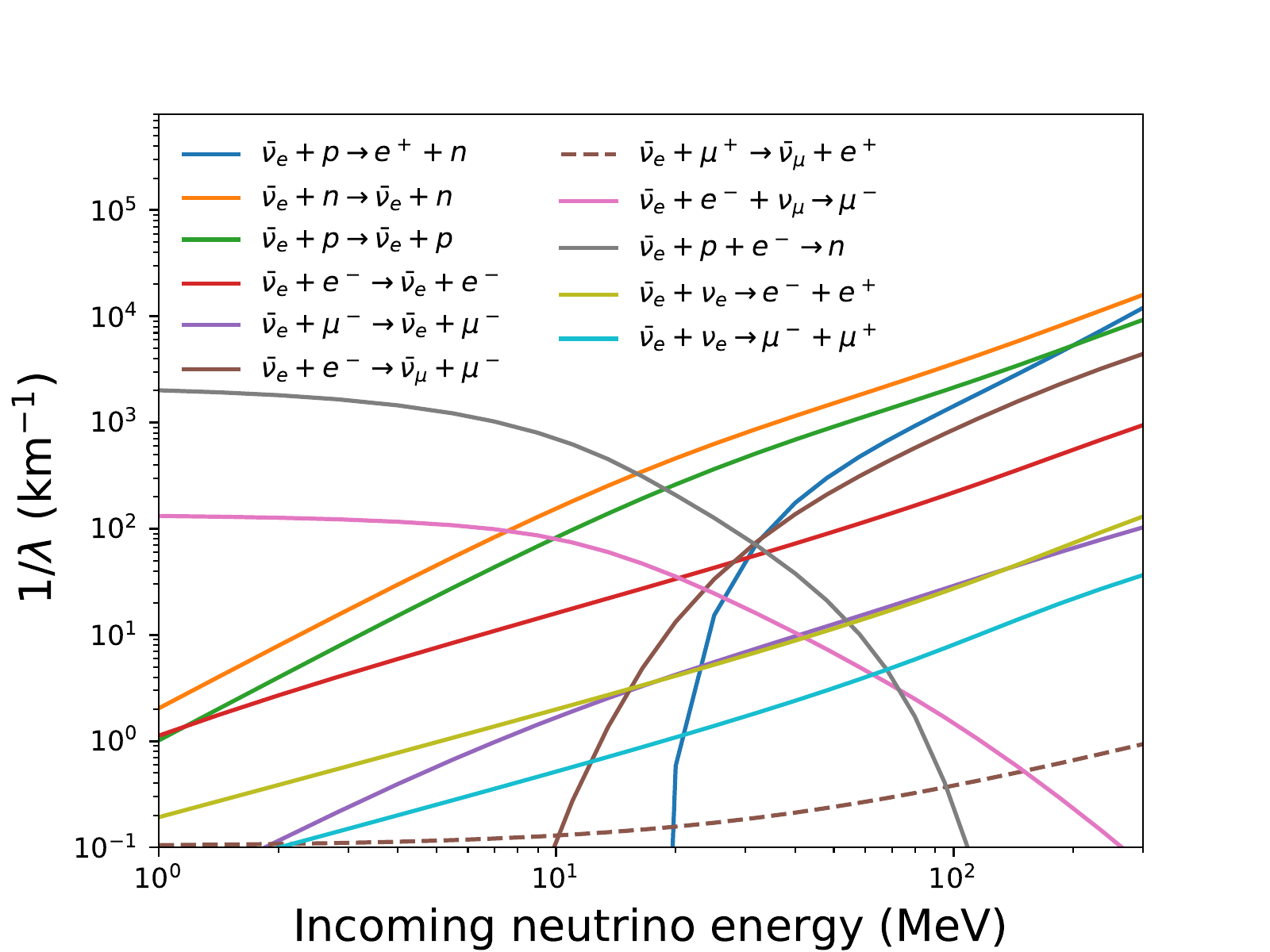}
    \subcaption{}\label{fig:t3d_nueb}
  \end{minipage}\\
  \begin{minipage}[b]{0.49\linewidth}
    \centering
    \includegraphics[keepaspectratio, scale=0.51]{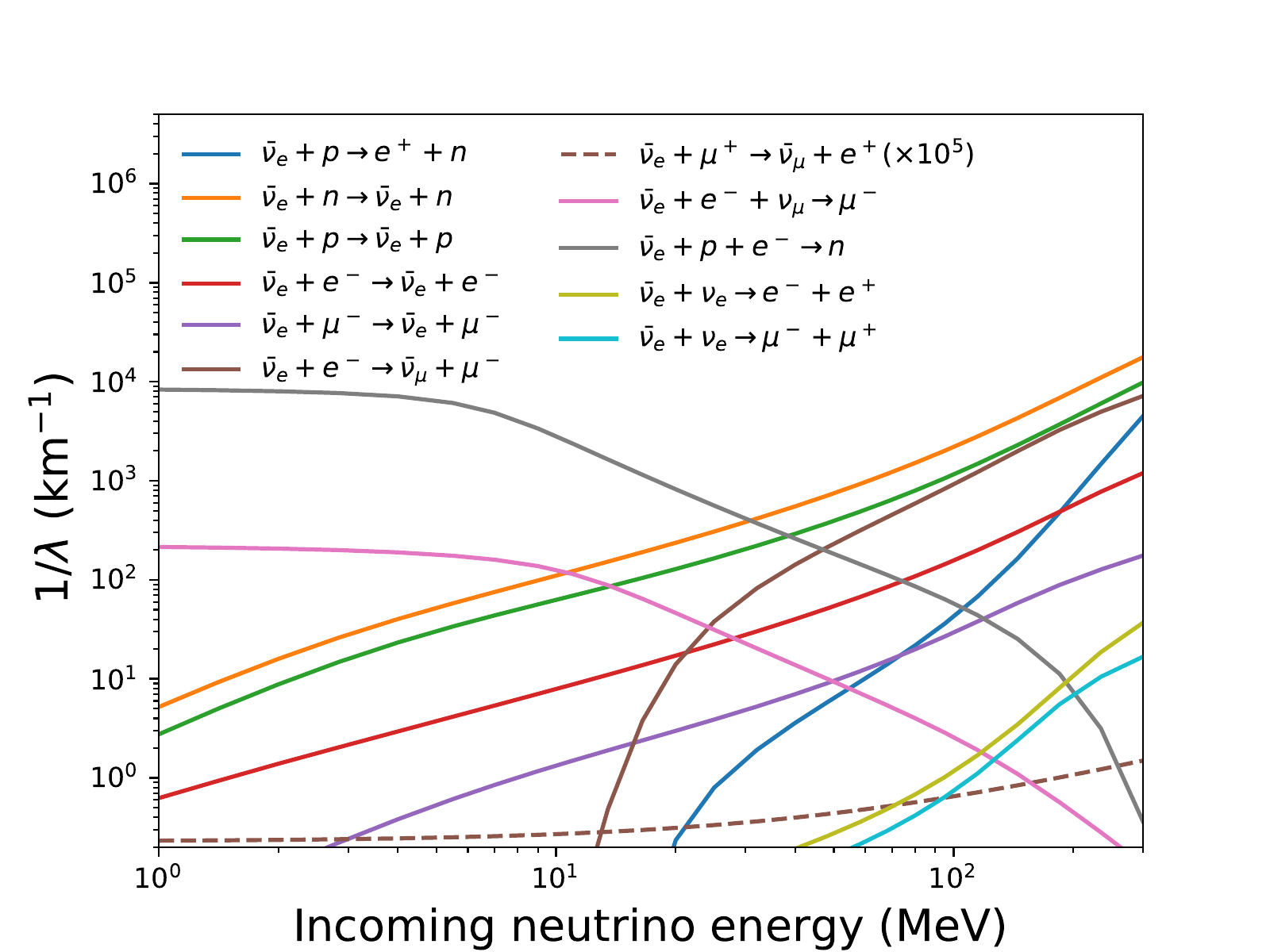}
    \subcaption{}\label{fig:t30d_nueb}
  \end{minipage}
  \begin{minipage}[b]{0.49\linewidth}
    \centering
    \includegraphics[keepaspectratio, scale=0.51]{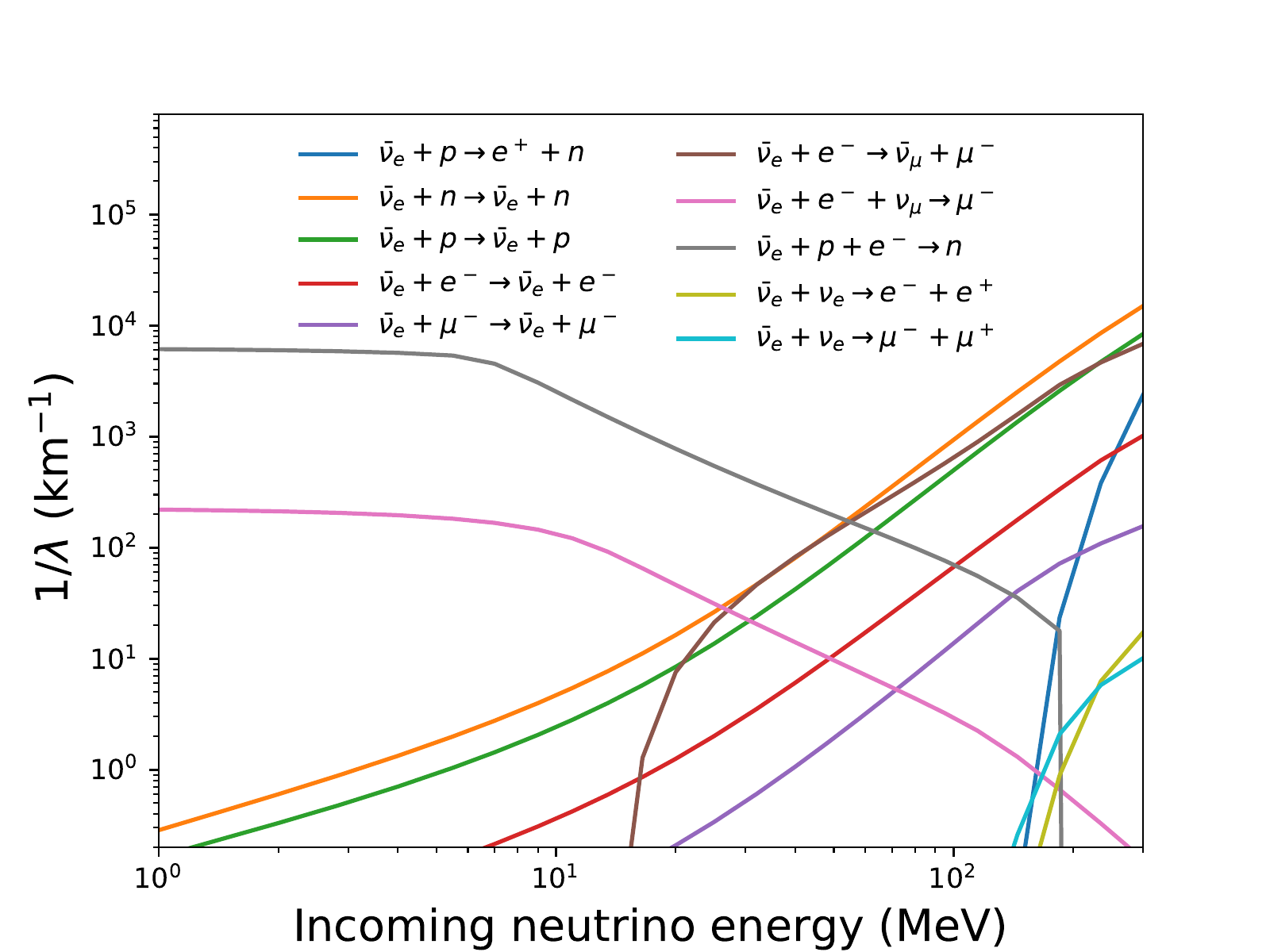}
    \subcaption{}\label{fig:t50d_nueb}
  \end{minipage}
  \caption{Same as Figure \ref{fig:D_nue} but for $\bar{\nu}_e$. These figures are same as Figure \ref{fig:t10d} (b) but at different times. Colors denote different reactions.
  Note that the values for $\bar{\nu}_e + \mu^+ \rightarrow \bar{\nu}_{\mu} + e^+$ in panel (c) are multiplied by $10^5$.
  The flavor exchange reaction $\bar{\nu}_e + \mu^+ \rightarrow \bar{\nu}_{\mu} + e^+$ does not occur at $t = 50 \operatorname{s}$.}\label{fig:D_nueb}
\end{figure*}

Figures \ref{fig:S_nueb} and \ref{fig:D_nueb} are the inverse mean free paths for
$\bar{\nu}_{e}$ at the neutrino sphere and at the deeper region, respectively, and the four panels in each figure correspond to different times as in Figures \ref{fig:S_nue} and \ref{fig:D_nue}.
These results have similar features as those for models t10S and t10D at $t = 10 \operatorname{s}$.
At low energies, the inverse muon decay is dominant at the neutrino sphere, and is second dominant at the deeper region.
In model t50S (see panel \ref{fig:t50s_nueb}), however, its rate is strongly suppressed due to the Pauli blocking by muon.
In the deeper region, the inverse neutron decay is dominant at low energies in all phases just as in model t10D.
The neutrino capture on proton is strongly suppressed due to the effective potential difference between neutron and proton also as in models t10S and t10D.
The neutrino scattering on neutron is the dominant source of opacity at high neutrino energies, which is also similar to the result at $t = 10 \operatorname{s}$.
The inverse mean free paths of the scatterings at $t = 50 \operatorname{s}$ are smaller by an order compared with those at $t = 30 \operatorname{s}$ because the temperature decreases by an order.
Note that $J_i$'s that appear in the dynamical structure functions (Eqs. (\ref{eq:N_J0})-(\ref{eq:N_J2})) are proportional to $T^{i + 1}$.
Again these findings at this very late time are new.

\begin{figure*}[htbp]
  \begin{minipage}[b]{0.49\linewidth}
    \centering
    \includegraphics[keepaspectratio, scale=0.51]{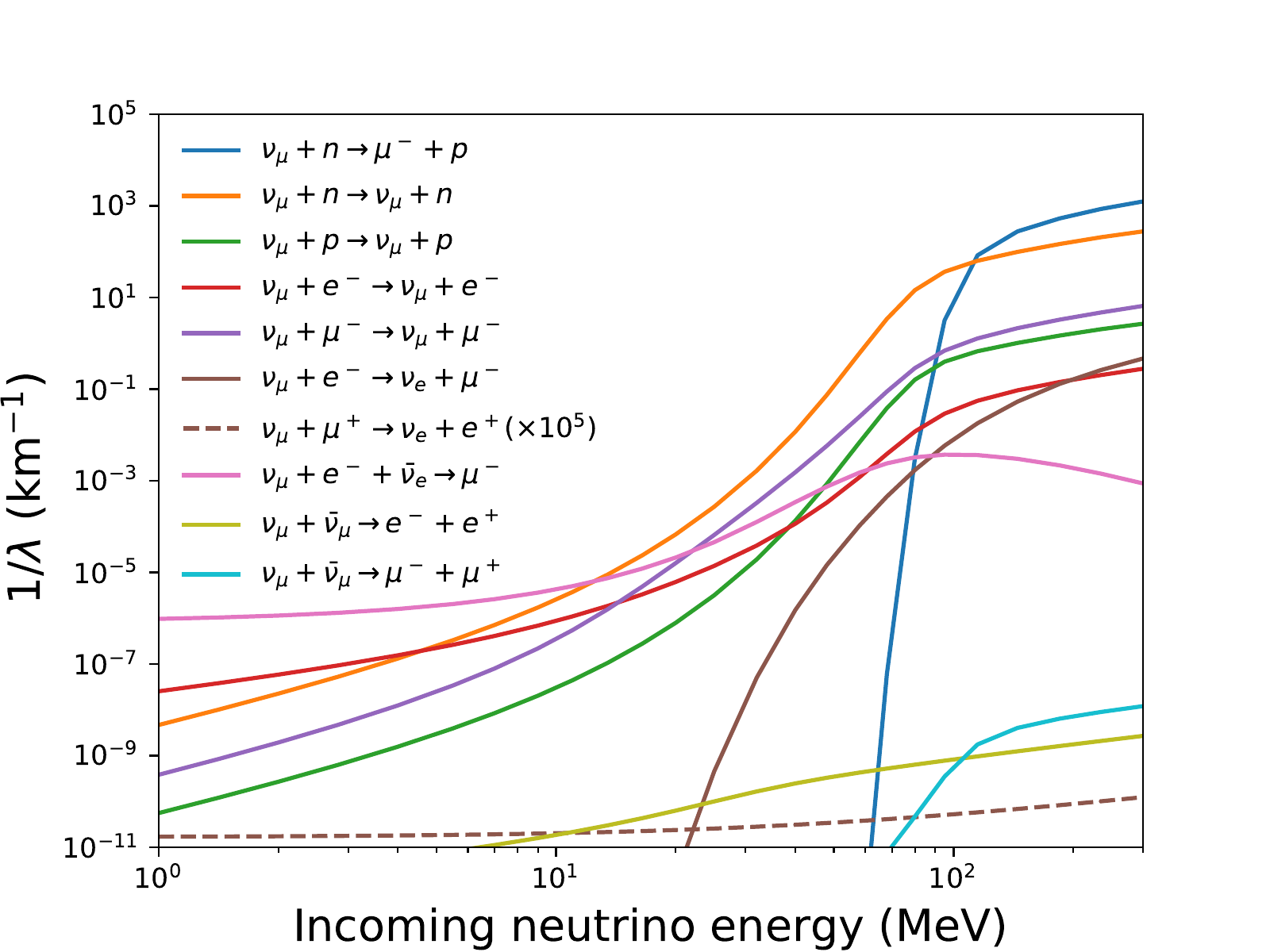}
    \subcaption{}\label{fig:t1s_numu}
  \end{minipage}
  \begin{minipage}[b]{0.49\linewidth}
    \centering
    \includegraphics[keepaspectratio, scale=0.51]{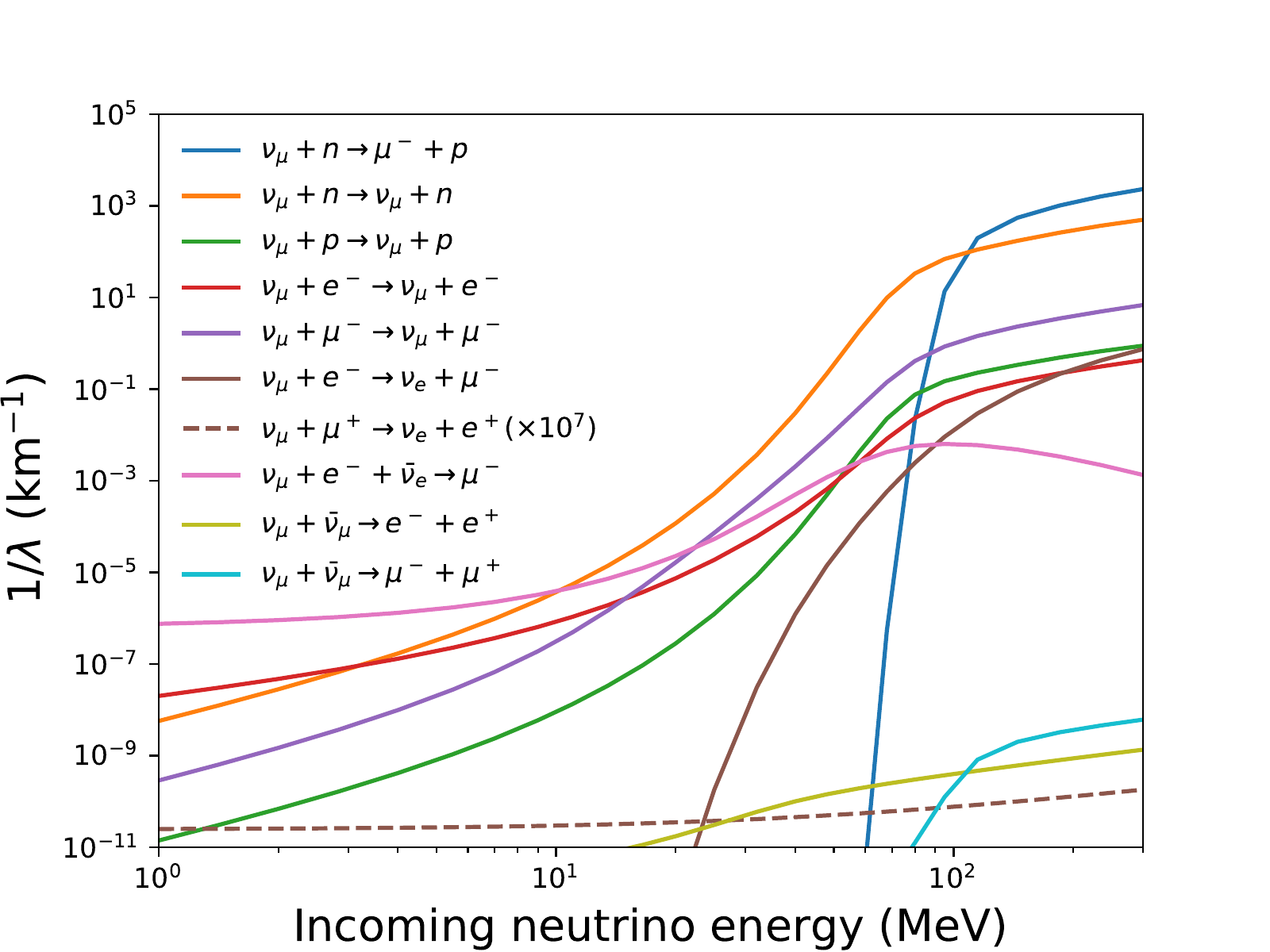}
    \subcaption{}\label{fig:t3s_numu}
  \end{minipage}\\
  \begin{minipage}[b]{0.49\linewidth}
    \centering
    \includegraphics[keepaspectratio, scale=0.51]{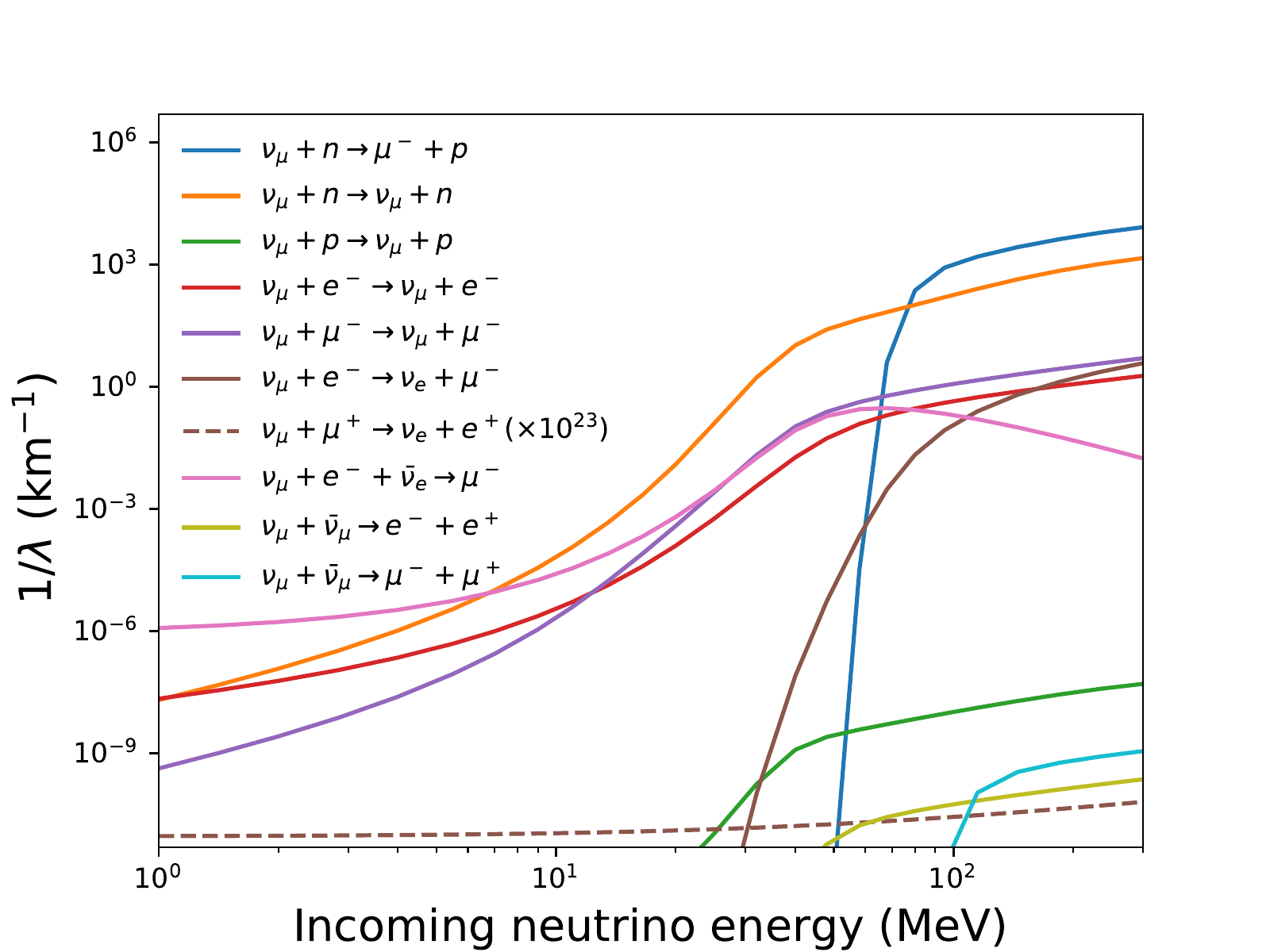}
    \subcaption{}\label{fig:t30s_numu}
  \end{minipage}
  \begin{minipage}[b]{0.49\linewidth}
    \centering
    \includegraphics[keepaspectratio, scale=0.51]{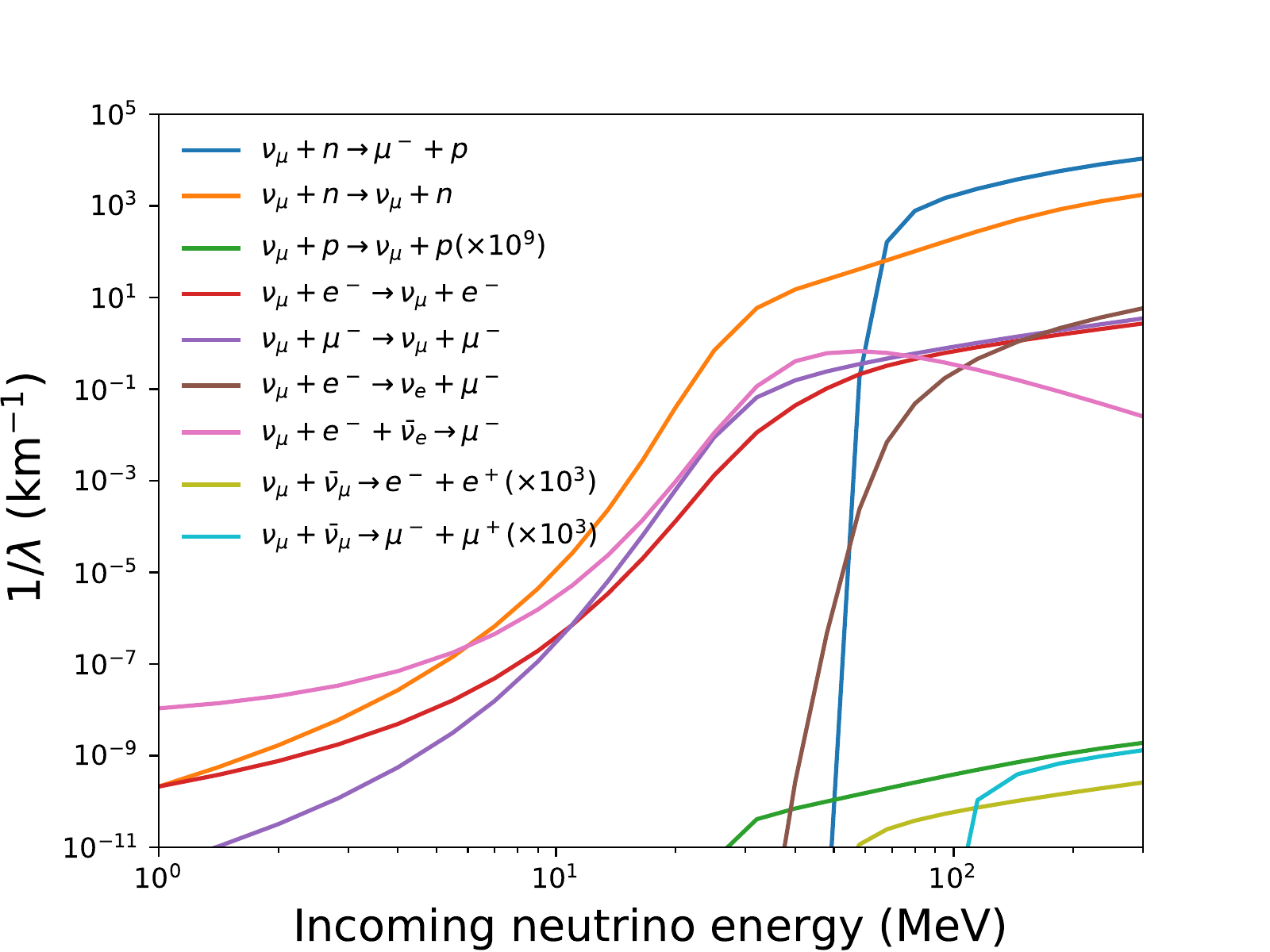}
    \subcaption{}\label{fig:t50s_numu}
  \end{minipage}
  \caption{Same as Figure \ref{fig:S_nue} but for $\nu_{\mu}$. These figures are same as Figure \ref{fig:t10S} (c) but at different times. Colors denote different reactions. Not that the values of the pair production reactions in panel (d) are multiplied by $10^3$.
  The values for $\nu_{\mu} + \mu^+ \rightarrow \nu_e + e^+$ are multiplied by the factors given in the legend in each panel.
  The flavor exchange reaction $\nu_{\mu} + \mu^+ \rightarrow \nu_e + e^+$ does not occur at $t = 50 \operatorname{s}$.}\label{fig:S_numu}
\end{figure*}

\begin{figure*}[htbp]
  \begin{minipage}[b]{0.49\linewidth}
    \centering
    \includegraphics[keepaspectratio, scale=0.51]{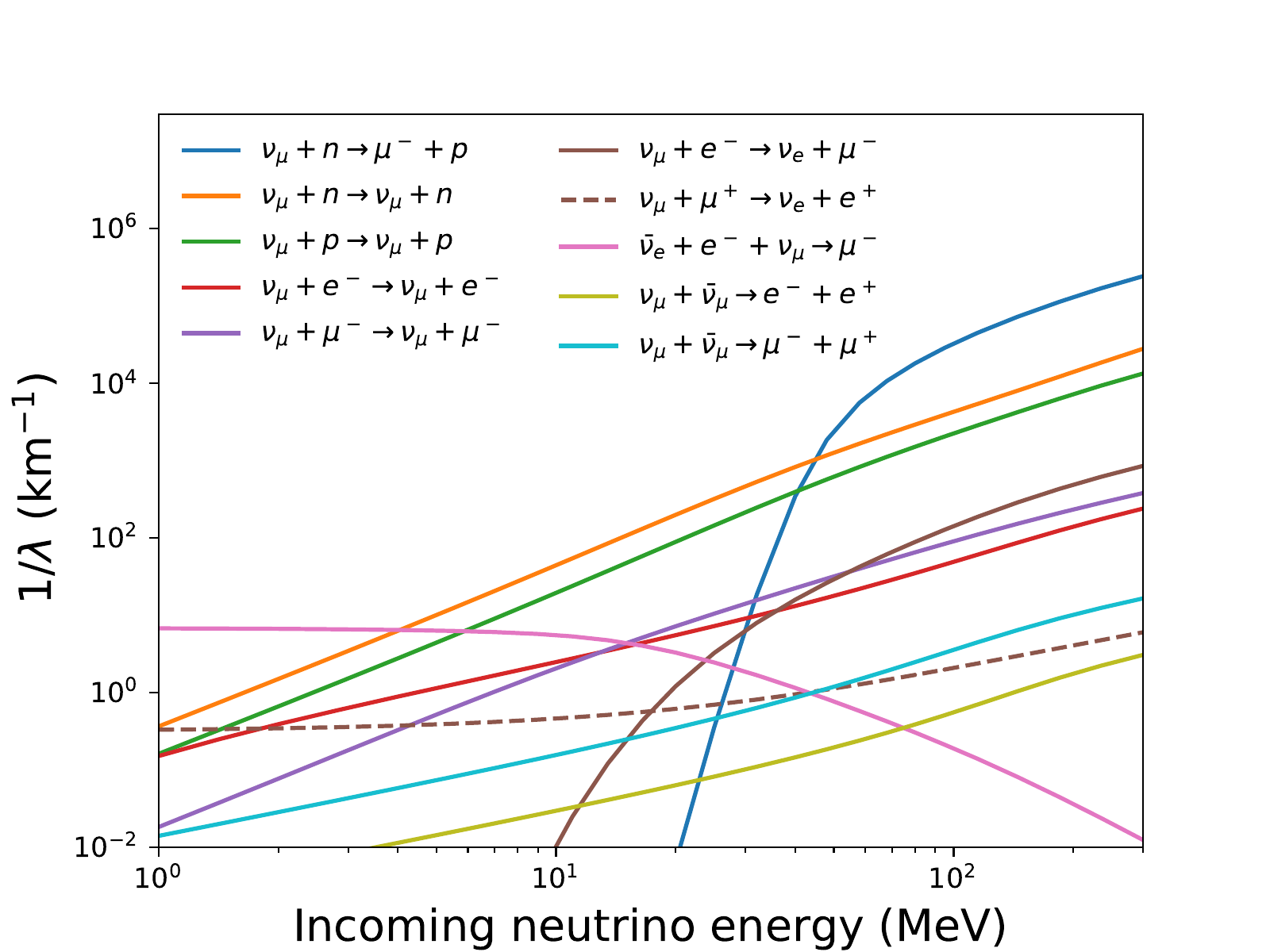}
    \subcaption{}\label{fig:t1d_numu}
  \end{minipage}
  \begin{minipage}[b]{0.49\linewidth}
    \centering
    \includegraphics[keepaspectratio, scale=0.51]{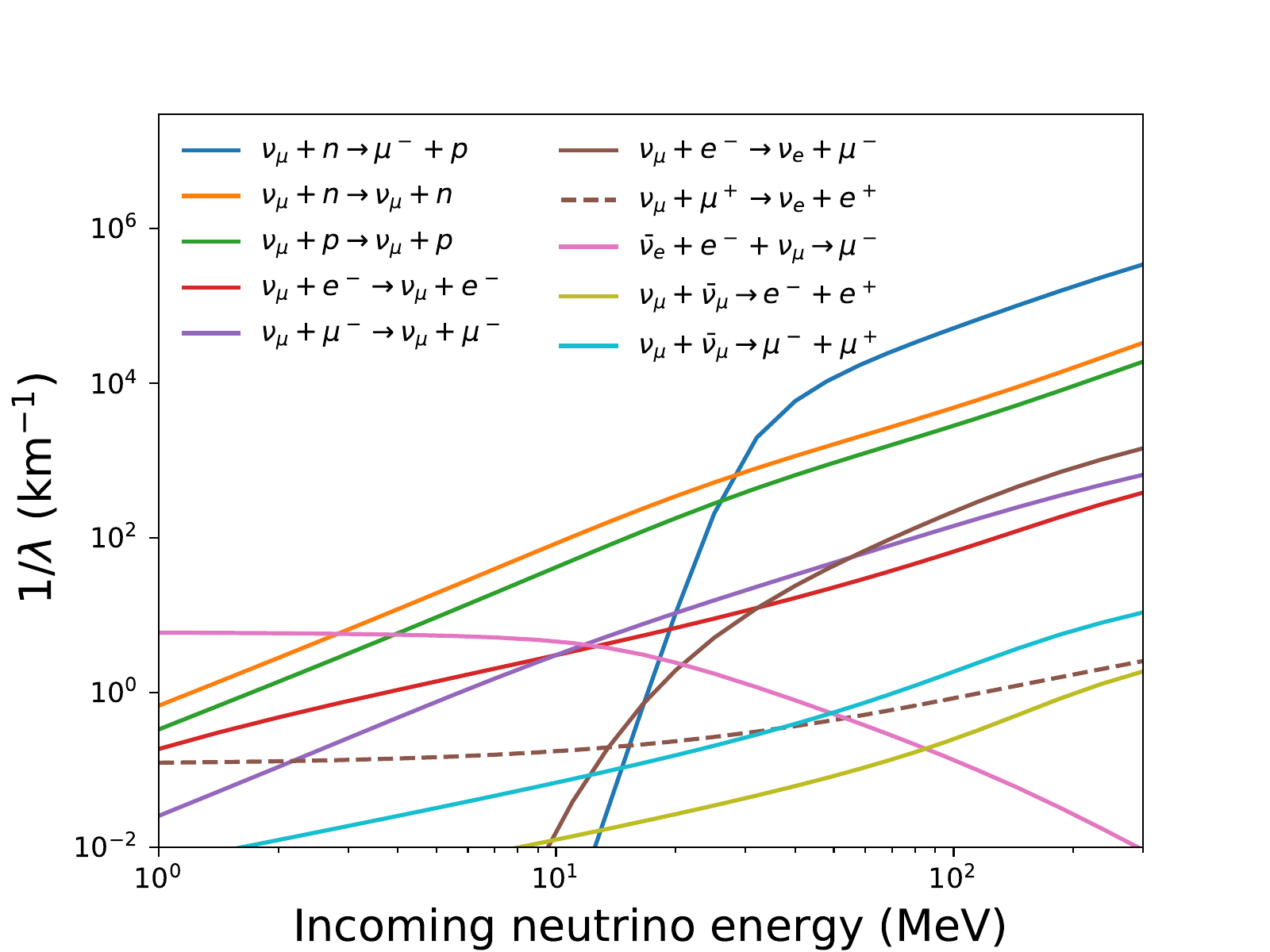}
    \subcaption{}\label{fig:t3d_numu}
  \end{minipage}\\
  \begin{minipage}[b]{0.49\linewidth}
    \centering
    \includegraphics[keepaspectratio, scale=0.51]{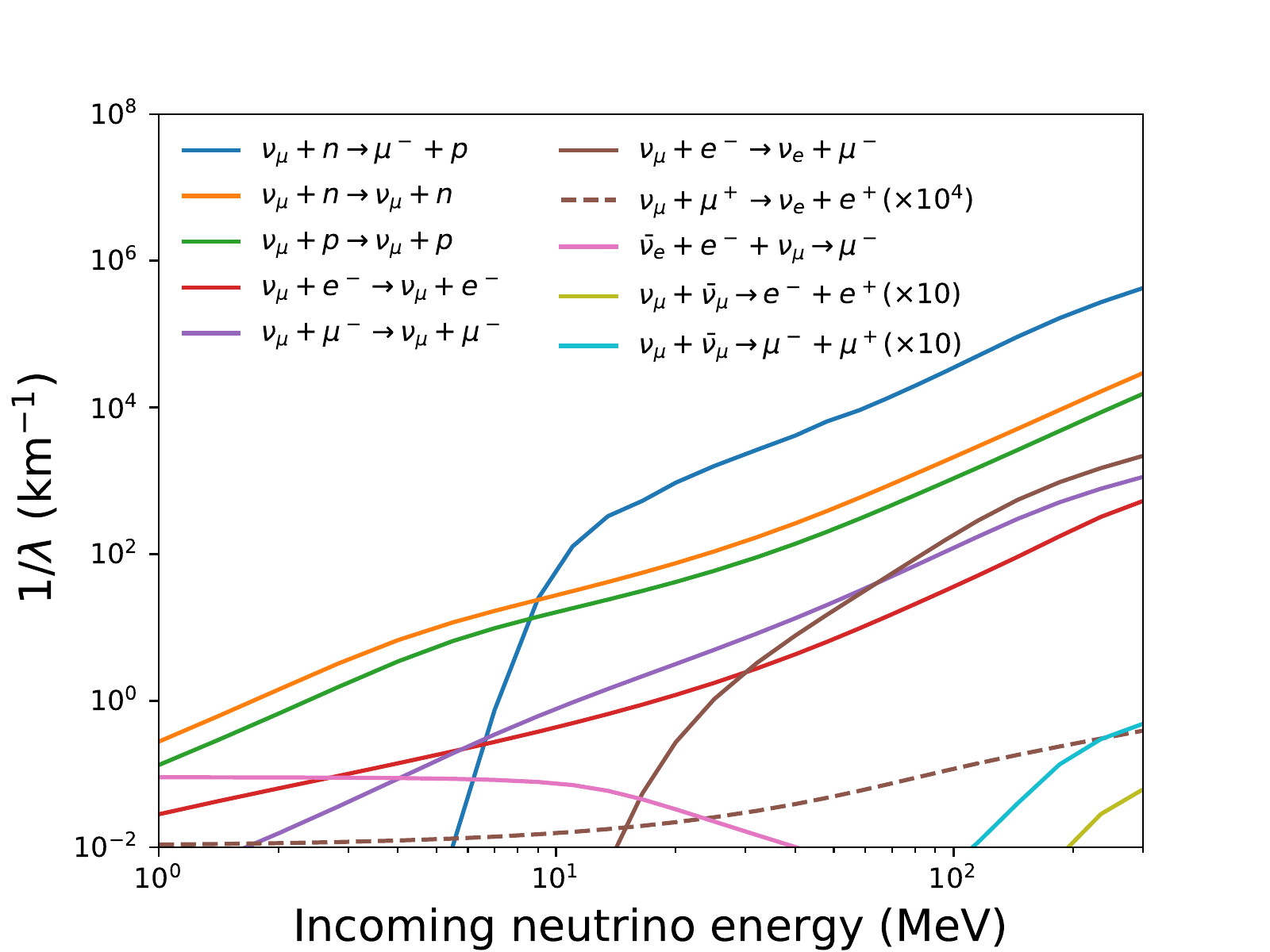}
    \subcaption{}\label{fig:t30d_numu}
  \end{minipage}
  \begin{minipage}[b]{0.49\linewidth}
    \centering
    \includegraphics[keepaspectratio, scale=0.51]{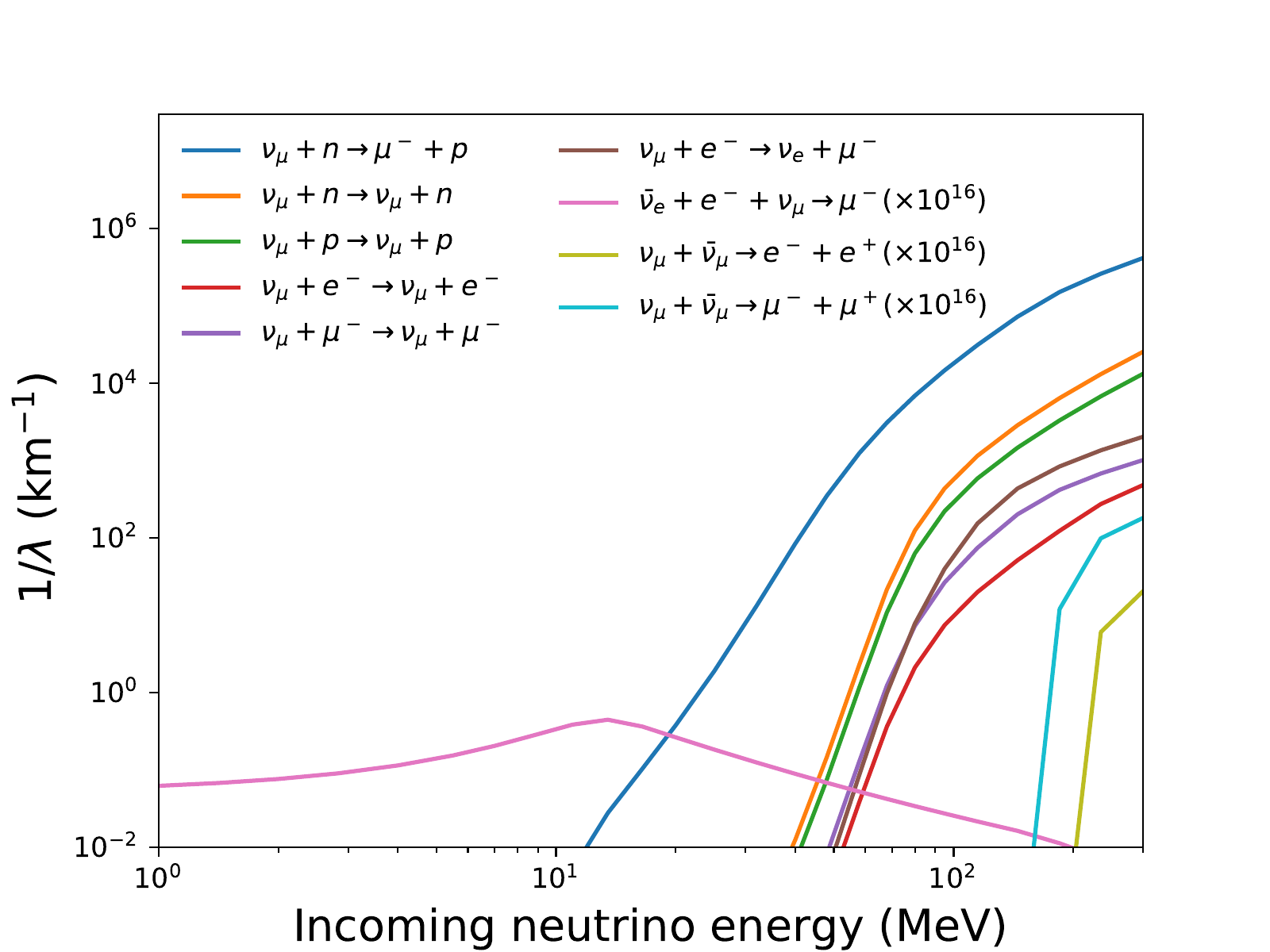}
    \subcaption{}\label{fig:t50d_numu}
  \end{minipage}
  \caption{Same as Figure \ref{fig:D_nue} but for $\nu_{\mu}$. These figures are same as Figure \ref{fig:t10d} (c) but at different times. Colors denote different reactions.
  Note that the values for $\nu_{\mu} + e^- + \bar{\nu}_e \rightarrow \mu^-$ and the pair production reactions in panels (c) and (d) are multiplied by the factors given in the legend in each panel.
  The values for $\nu_{\mu} + \mu^+ \rightarrow \nu_e + e^+$ are multiplied by the factors given in the legend in each panel.
  The flavor exchange reaction $\nu_{\mu} + \mu^+ \rightarrow \nu_e + e^+$ does not occur at $t = 50 \operatorname{s}$.}\label{fig:D_numu}
\end{figure*}

The results for $\nu_{\mu}$ are exhibited in Figures \ref{fig:S_numu} and \ref{fig:D_numu}.
At the neutrino sphere (see Figure \ref{fig:S_numu}), the time evolutions of the inverse mean free paths are not so remarkable: the inverse muon decay, the neutrino scattering on neutron and the neutrino absorption on neutron are the dominant sources of opacity at low, middle and high neutrino energies, respectively.
However, in the deeper region at the later phase, the inverse muon decay $\nu_{\mu} + e^- + \bar{\nu}_{e} \rightarrow \mu^-$ remains subdominant even at low neutrino energies because the number density of $\bar{\nu}_e$ declines due to large negative values of $\mu_{\bar{\nu}_e}$ in beta-equilibrium.

\begin{figure*}[htbp]
  \begin{minipage}[b]{0.49\linewidth}
    \centering
    \includegraphics[keepaspectratio, scale=0.51]{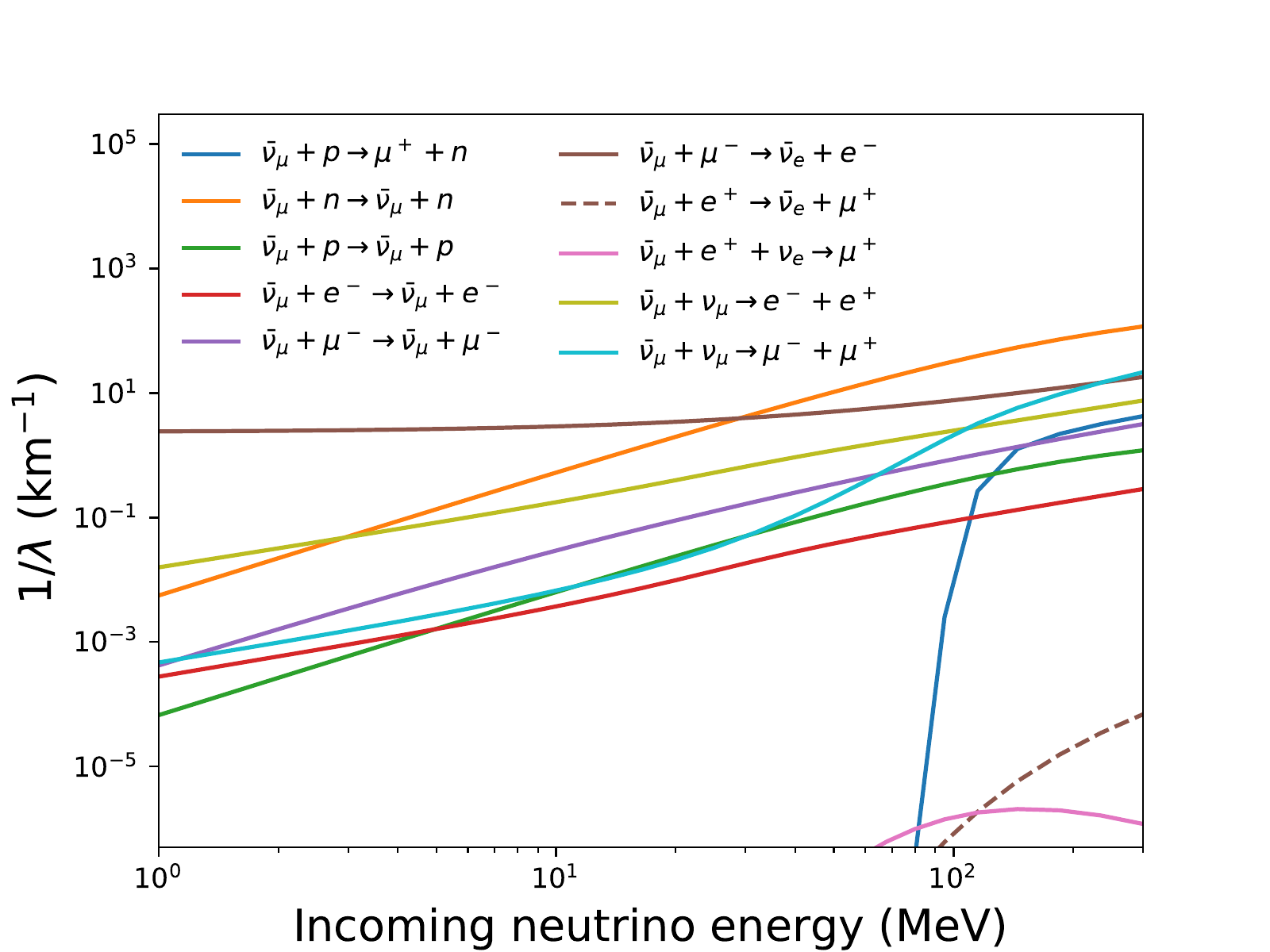}
    \subcaption{}\label{fig:t1s_numub}
  \end{minipage}
  \begin{minipage}[b]{0.49\linewidth}
    \centering
    \includegraphics[keepaspectratio, scale=0.51]{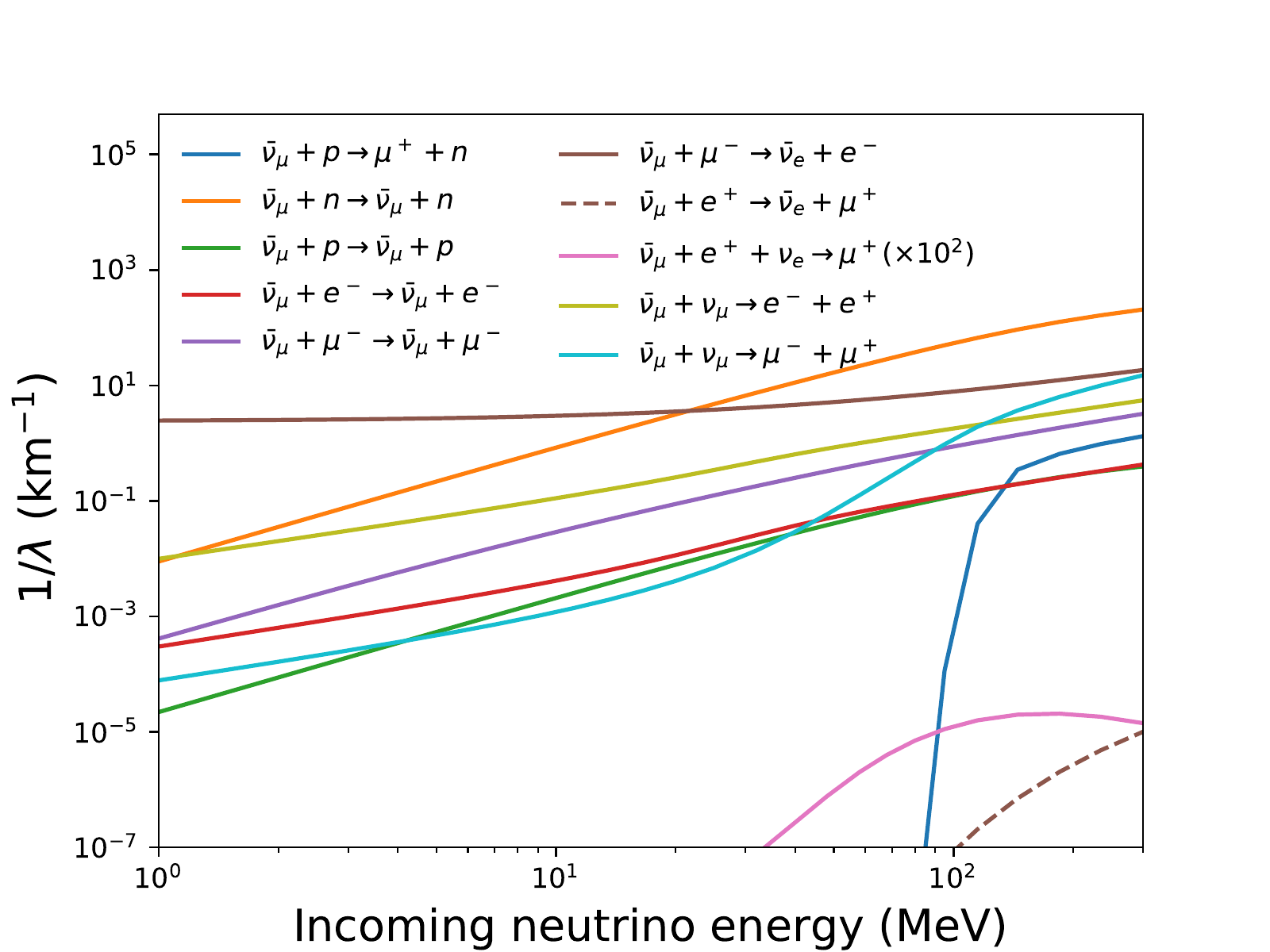}
    \subcaption{}\label{fig:t3s_numub}
  \end{minipage}\\
  \begin{minipage}[b]{0.49\linewidth}
    \centering
    \includegraphics[keepaspectratio, scale=0.51]{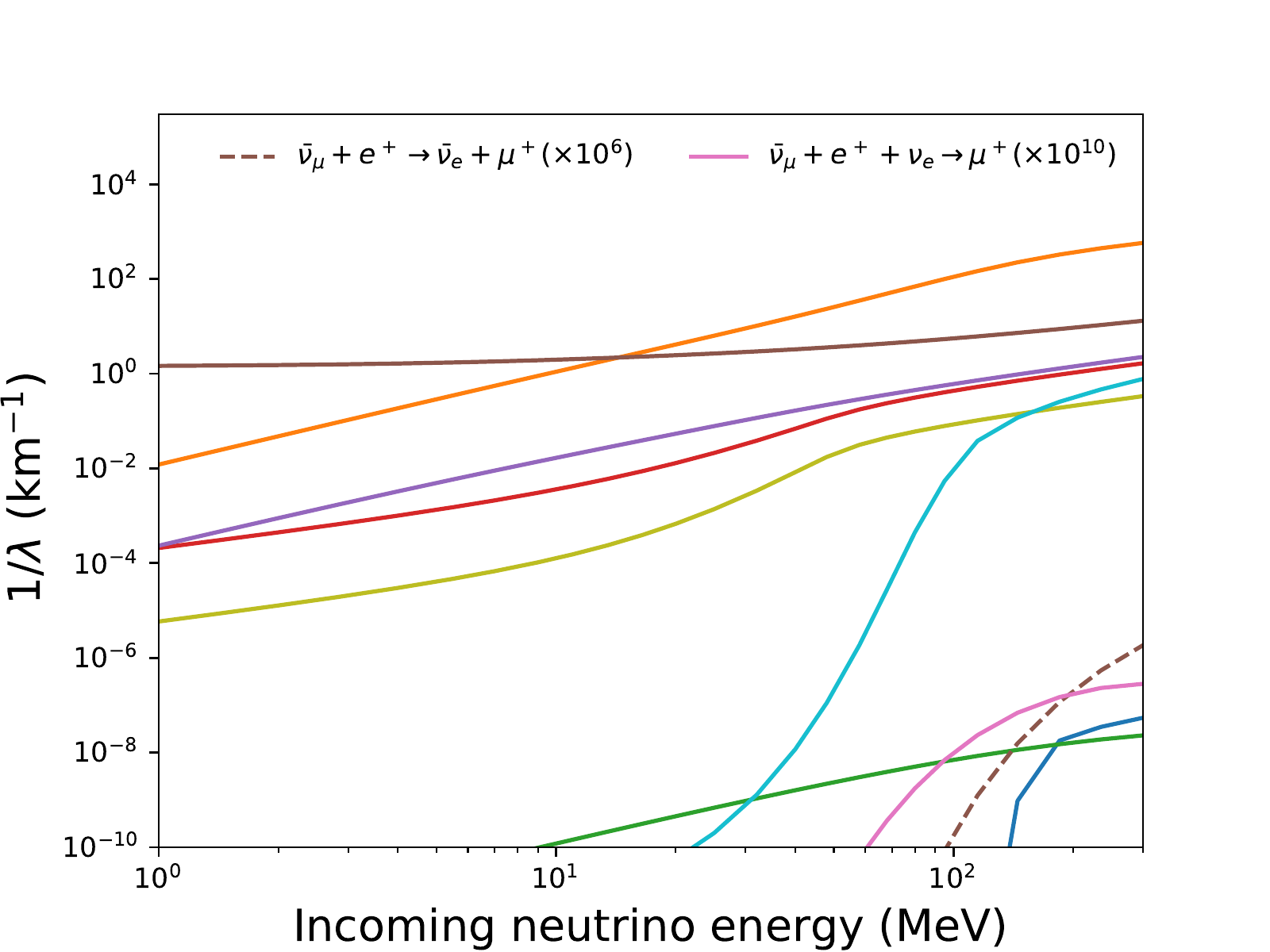}
    \subcaption{}\label{fig:t30s_numub}
  \end{minipage}
  \begin{minipage}[b]{0.49\linewidth}
    \centering
    \includegraphics[keepaspectratio, scale=0.51]{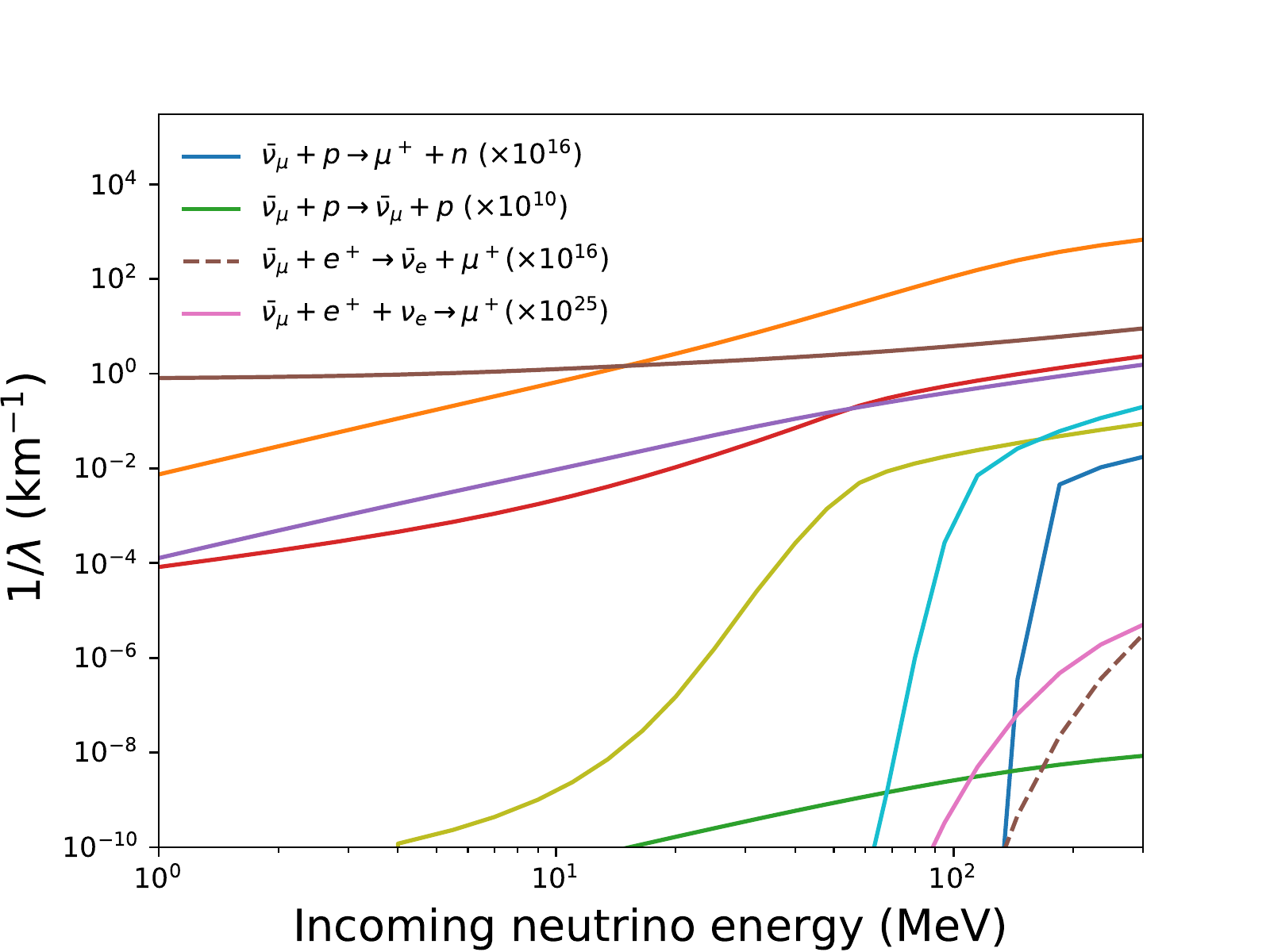}
    \subcaption{}\label{fig:t50s_numub}
  \end{minipage}
  \caption{Same as Figure \ref{fig:S_nue} but for $\bar{\nu}_{\mu}$. These figures are same as Figure \ref{fig:t10S} (d) but at different times. Colors denote different reactions.
  The legends are omitted in panels (c) and (d) but the notations are the same as in panel (a).
  Note that the values for $\bar{\nu}_{\mu} + p \rightarrow \mu^{+} + n$, $\bar{\nu}_{\mu} + e^+ \rightarrow \bar{\nu}_e + \mu^+$ and the scattering on proton in panel (d) and $\bar{\nu}_{\mu} + e^+ + \nu_e \rightarrow \mu^+ $ in panels (b), (c) and (d) are multiplied by the factors given in the legend in each panel.}\label{fig:S_numub}
\end{figure*}

\begin{figure*}[htbp]
  \begin{minipage}[b]{0.49\linewidth}
    \centering
    \includegraphics[keepaspectratio, scale=0.51]{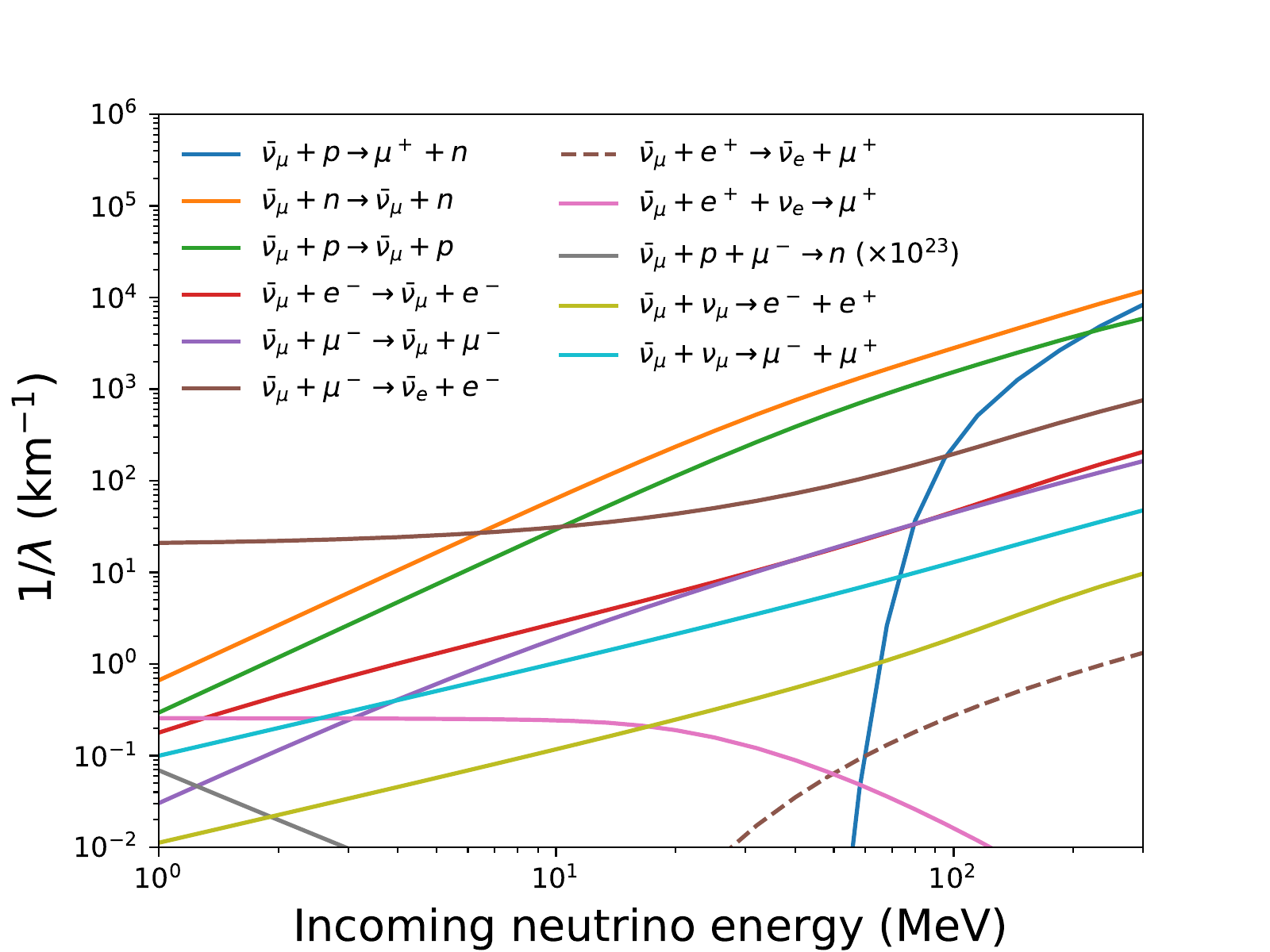}
    \subcaption{}\label{fig:t1d_numub}
  \end{minipage}
  \begin{minipage}[b]{0.49\linewidth}
    \centering
    \includegraphics[keepaspectratio, scale=0.51]{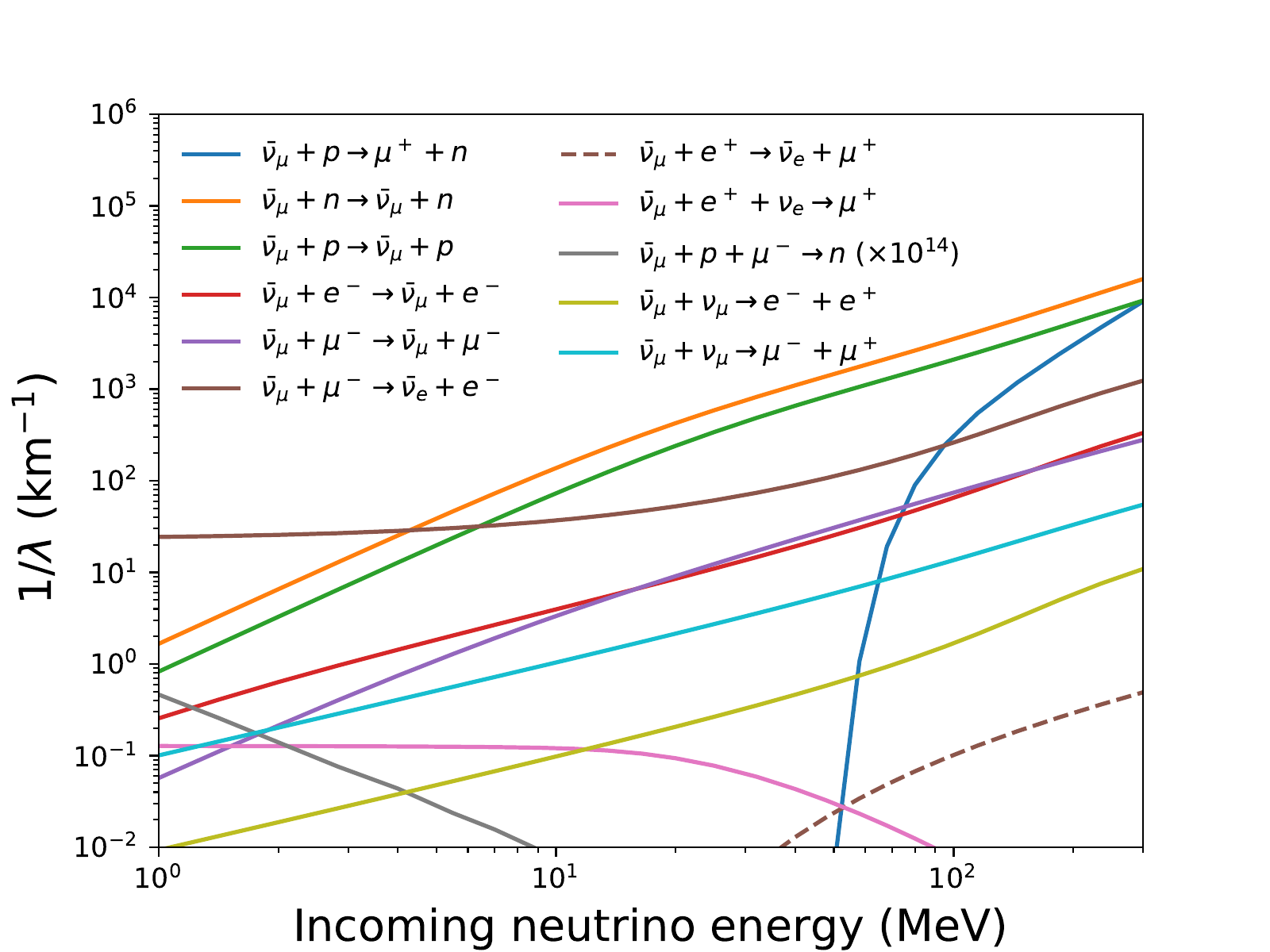}
    \subcaption{}\label{fig:t3d_numub}
  \end{minipage}\\
  \begin{minipage}[b]{0.49\linewidth}
    \centering
    \includegraphics[keepaspectratio, scale=0.51]{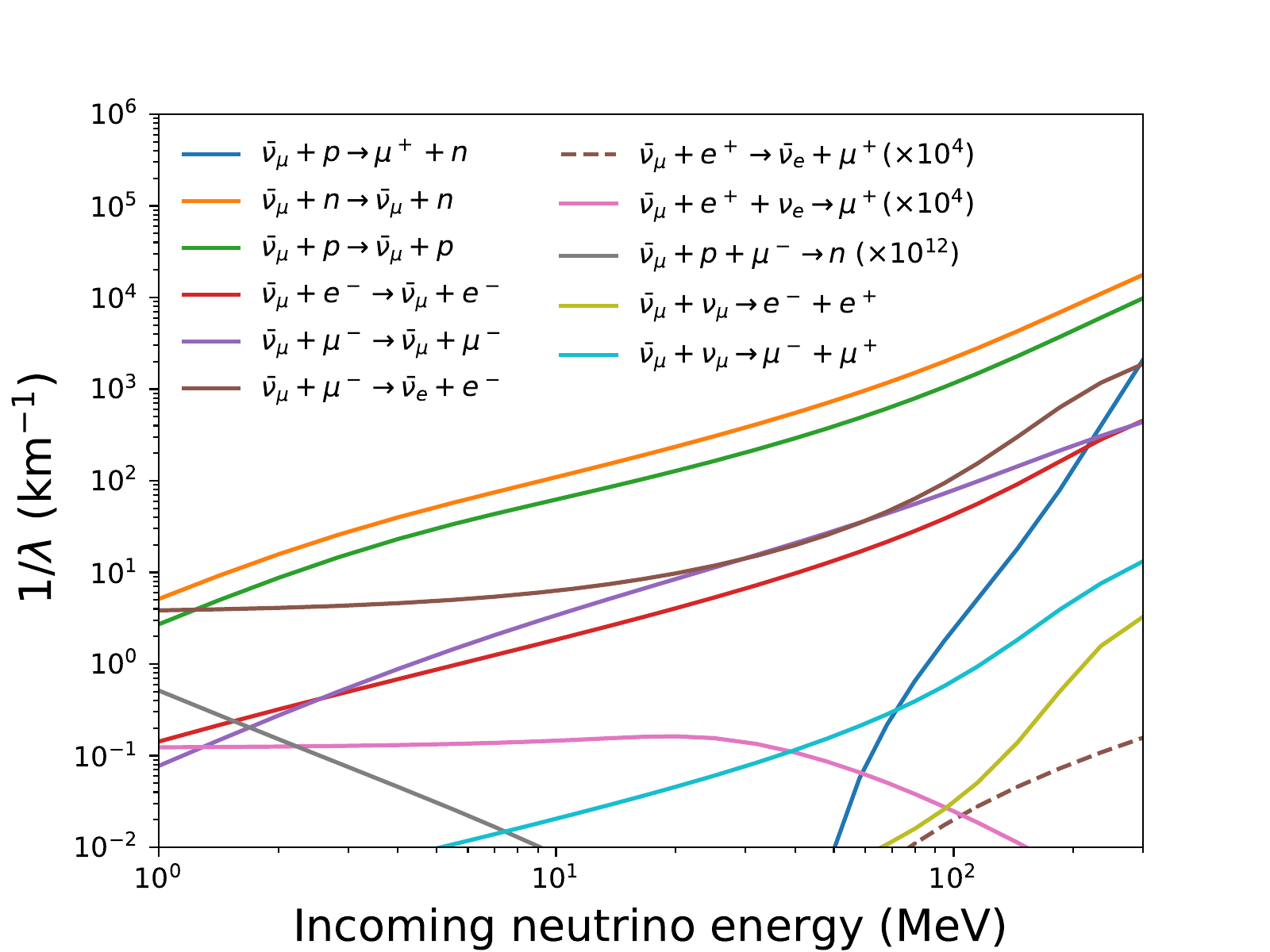}
    \subcaption{}\label{fig:t30d_numub}
  \end{minipage}
  \begin{minipage}[b]{0.49\linewidth}
    \centering
    \includegraphics[keepaspectratio, scale=0.51]{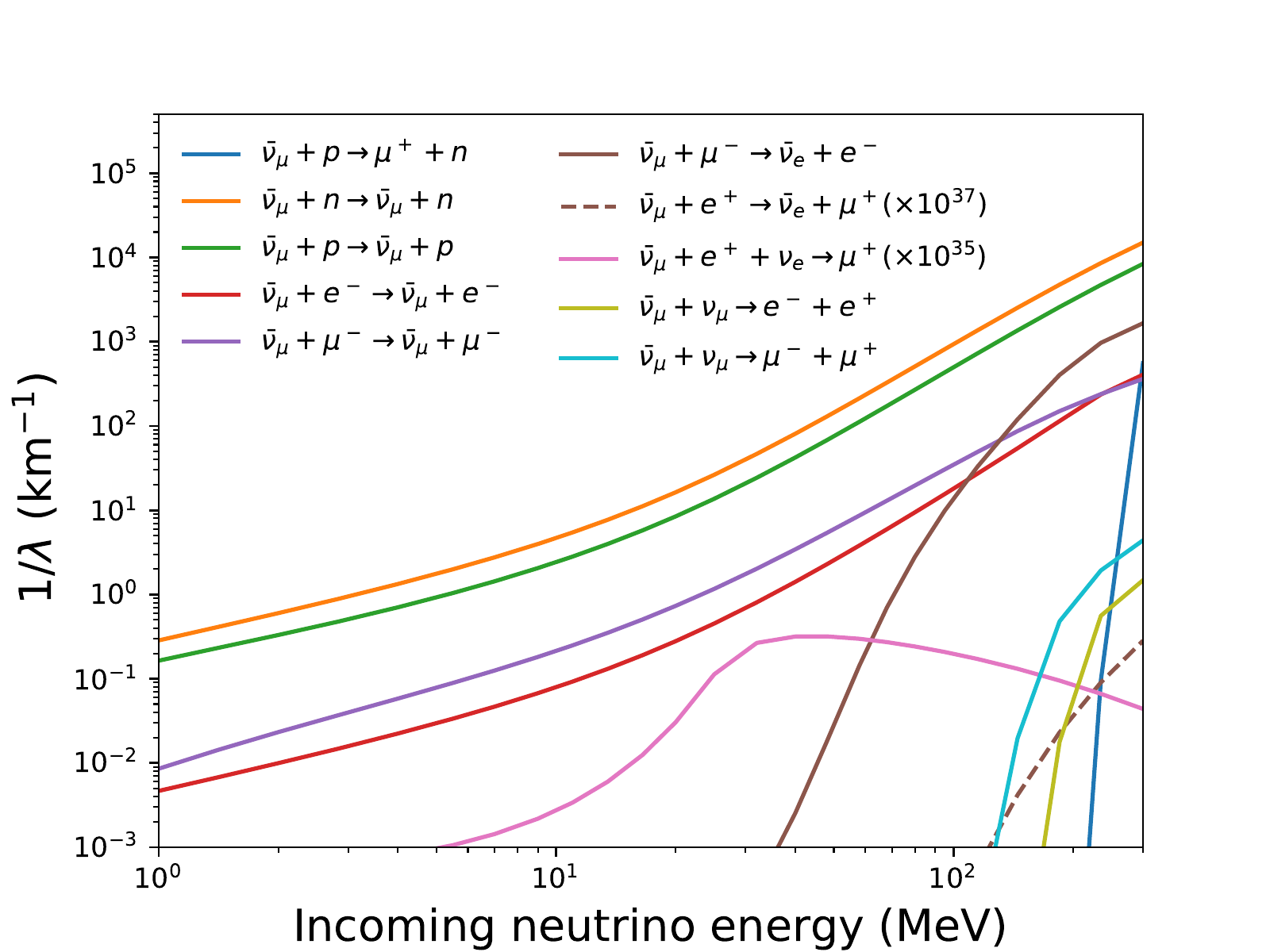}
    \subcaption{}\label{fig:t50d_numub}
  \end{minipage}
  \caption{Same as Figure \ref{fig:D_nue} but for $\bar{\nu}_{\mu}$. These figures are same as Figure \ref{fig:t10d} (d) but at different times. Colors denote different reactions. Note that the values for $\bar{\nu}_{\mu} + e^+ + \nu_e \rightarrow \mu^+ $, $\bar{\nu}_{\mu} + e^+ \rightarrow \bar{\nu}_e + \mu^+$ and $\bar{\nu}_{\mu} + p + \mu^- \rightarrow n$ are multiplied by the factors given in the legends in each panel.}\label{fig:D_numub}
\end{figure*}

Finally, figures \ref{fig:S_numub} and \ref{fig:D_numub} present the inverse mean free paths for $\bar{\nu}_{\mu}$.
At the neutrino sphere, the neutrino scattering on neutron and the flavor exchange reaction $\bar{\nu}_{\mu} + \mu^- \rightarrow \bar{\nu}_e + e^-$ are the dominant sources of opacity at high and low energies, respectively.
The electron-positron pair production comes next except at very high energies, where muon-anti-muon production takes its place in the earlier phase because the incoming neutrino energy is so high that muons can be produced, but gets suppressed in the later phase due to the strong Pauli blocking by electron and muon at low temperatures.
In the deeper region, the flavor exchange reaction is dominant at low energies in the early phase just as in model t10D, while it is suppressed in the later phase (especially in model t50D) because the temperature is so low that the Pauli blocking by muon cannot be overcome.
In these conditions, the scattering on proton is the second dominant reaction.


Before closing this subsection, we give a brief comment on the comparison with the previous work \cite{Guo2020}, which focused on the muonization of matter and the muon-related reactions of $\bar{\nu}_{e}$ and $\nu_{\mu}$ in the very early phase of PNS cooling, or more appropriately the post-bounce phase of CCSNe.
The thermal conditions in the earlier phase, especially for model t1D, are similar to condition A in \cite{Guo2020}.
Our results, Figures \ref{fig:D_nueb}(a) and \ref{fig:D_numu}(a), are perfectly consistent with their results: condition A in figures 5 and 6 in their paper except for minor deviations induced inevitably by the use of different EoS and the values of the mean field parameters.
Condition B in \cite{Guo2020} is similar to our model t10S except that the temperature and the electron fraction in the latter are lower than those in the former due to the advanced cooling and neutronization in our model.
The trends in the inverse mean free paths are qualitatively the same: for $\nu_{\mu}$, for example, the inverse muon decay and the neutrino scattering on neutron are dominant at low and high energies, respectively.


\subsection{Reaction kernels of the flavor-exchange reaction and the inverse muon decay}
So far we have looked at the inverse mean free path alone, the quantity integrated over the energy and angle.
In this section we will look into more details, i.e., the energy- and angular dependences of the reaction kernels for some muon-related reactions that become significant as an opacity source at some energies: the flavor-exchange reaction \mbox{$\nu_e + \mu^- \leftrightarrows \nu_{\mu} + e^-$} and the inverse muon decay $\bar{\nu}_e + \nu_{\mu} + e^- \rightarrow \mu^- $.
We show the results at $t = 10 \s$.
These are relevant information for detailed neutrino transport calculations but have not been presented so far.

\begin{figure*}[htbp]
  \begin{minipage}[b]{0.32\linewidth}
    \centering
    \includegraphics[keepaspectratio, scale=0.4]{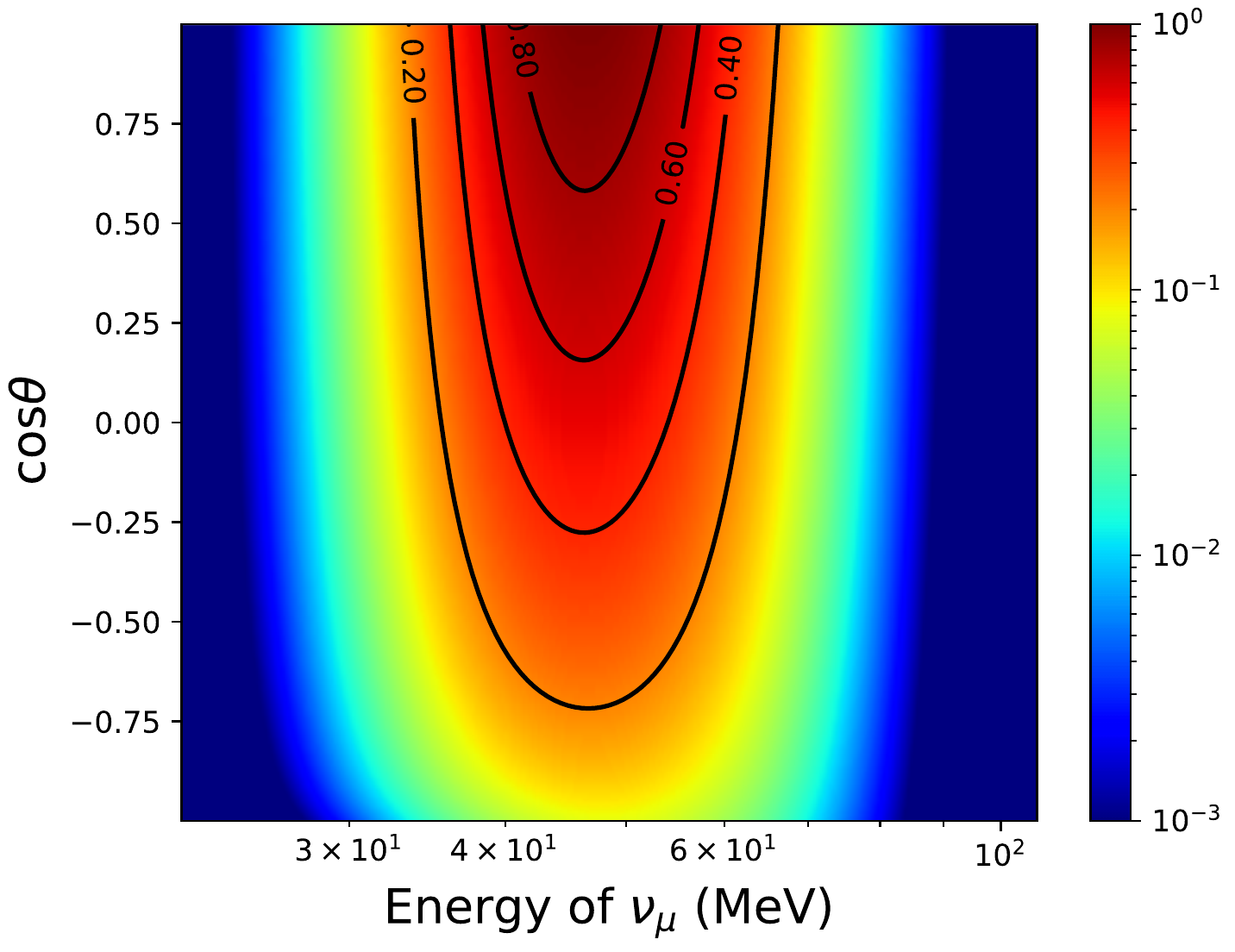}
    \subcaption{$E_{\nu_e} = 1 \MeV$}\label{fig:kernel_flex_1_e1s}
  \end{minipage}
  \begin{minipage}[b]{0.32\linewidth}
    \centering
    \includegraphics[keepaspectratio, scale=0.4]{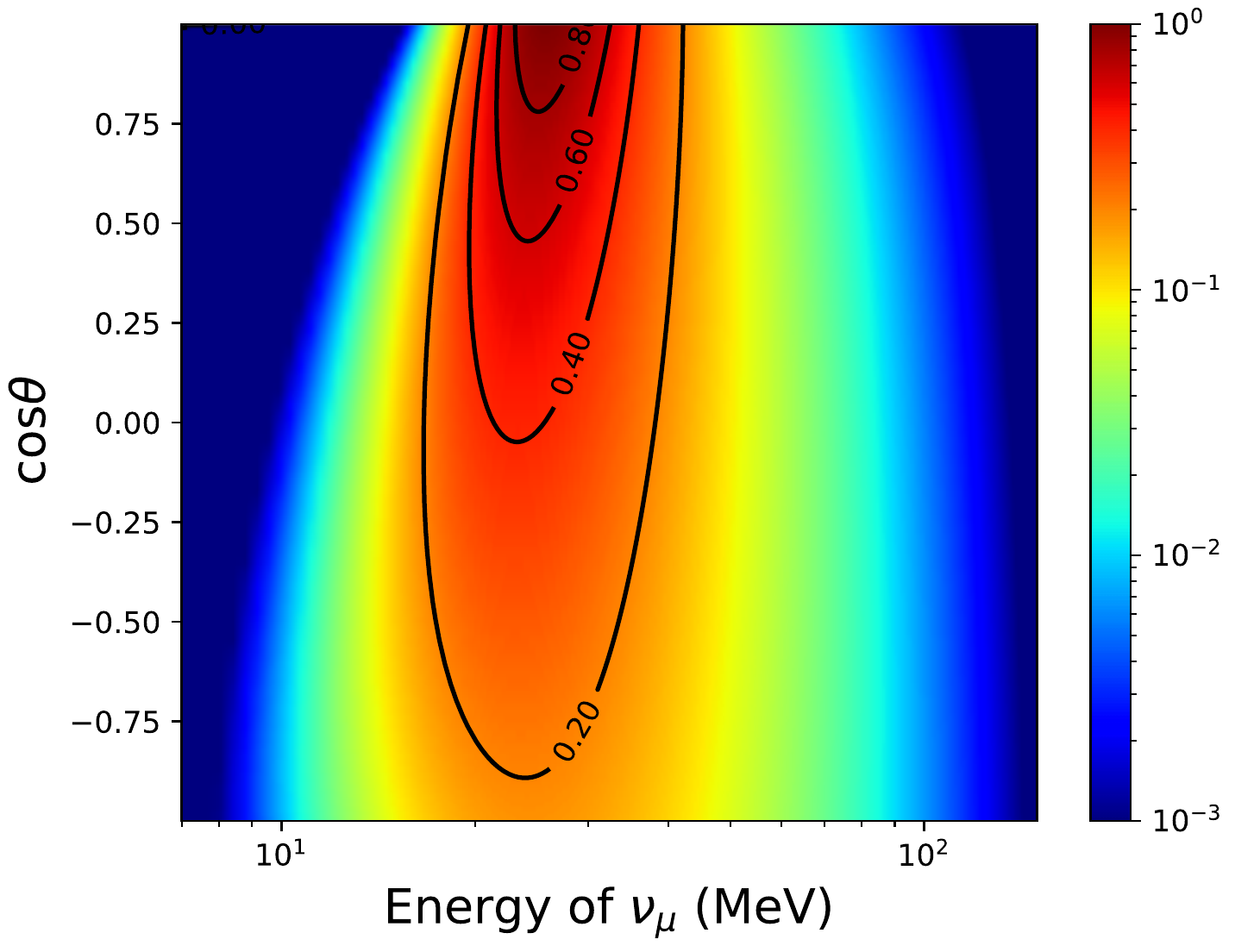}
    \subcaption{$E_{\nu_e} = 10 \MeV$}\label{fig:kernel_flex_1_e10s}
  \end{minipage}
  \begin{minipage}[b]{0.32\linewidth}
    \centering
    \includegraphics[keepaspectratio, scale=0.4]{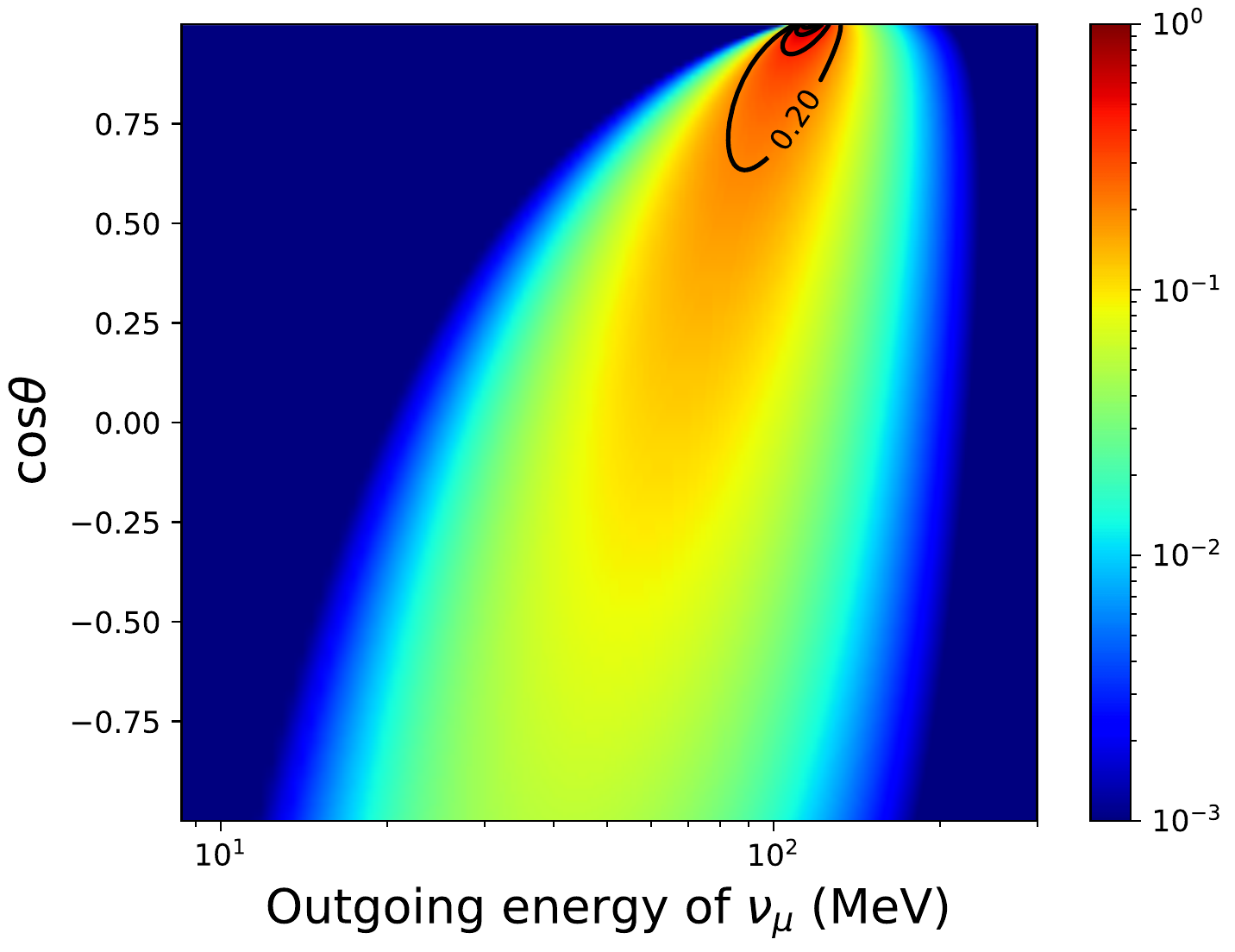}
    \subcaption{$E_{\nu_e} = 100 \MeV$}\label{fig:kernel_flex_1_e100s}
  \end{minipage}\\
  \begin{minipage}[b]{0.32\linewidth}
    \centering
    \includegraphics[keepaspectratio, scale=0.4]{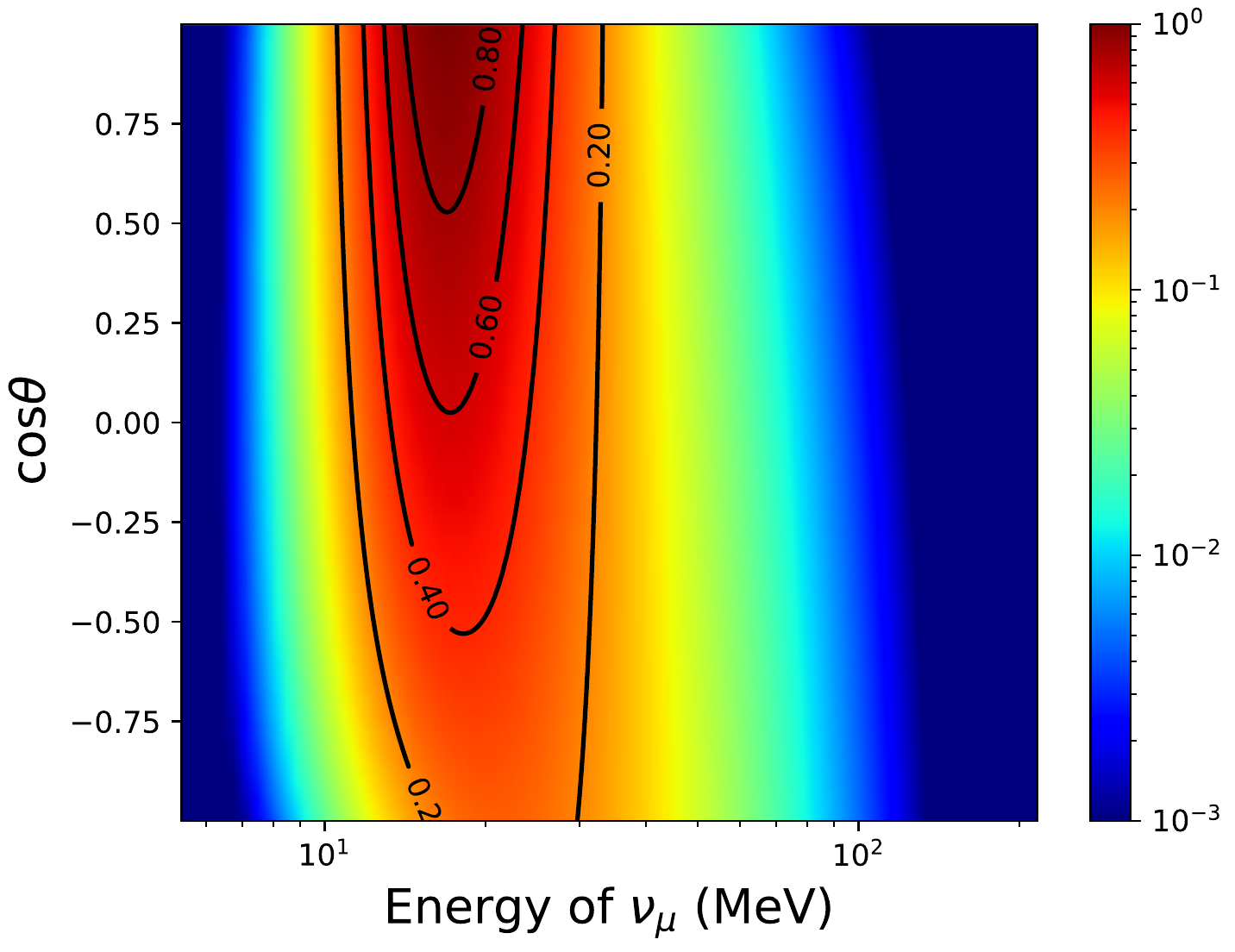}
    \subcaption{$E_{\nu_e} = 1 \MeV$}\label{fig:kernel_flex_1_e1d}
  \end{minipage}
  \begin{minipage}[b]{0.32\linewidth}
    \centering
    \includegraphics[keepaspectratio, scale=0.4]{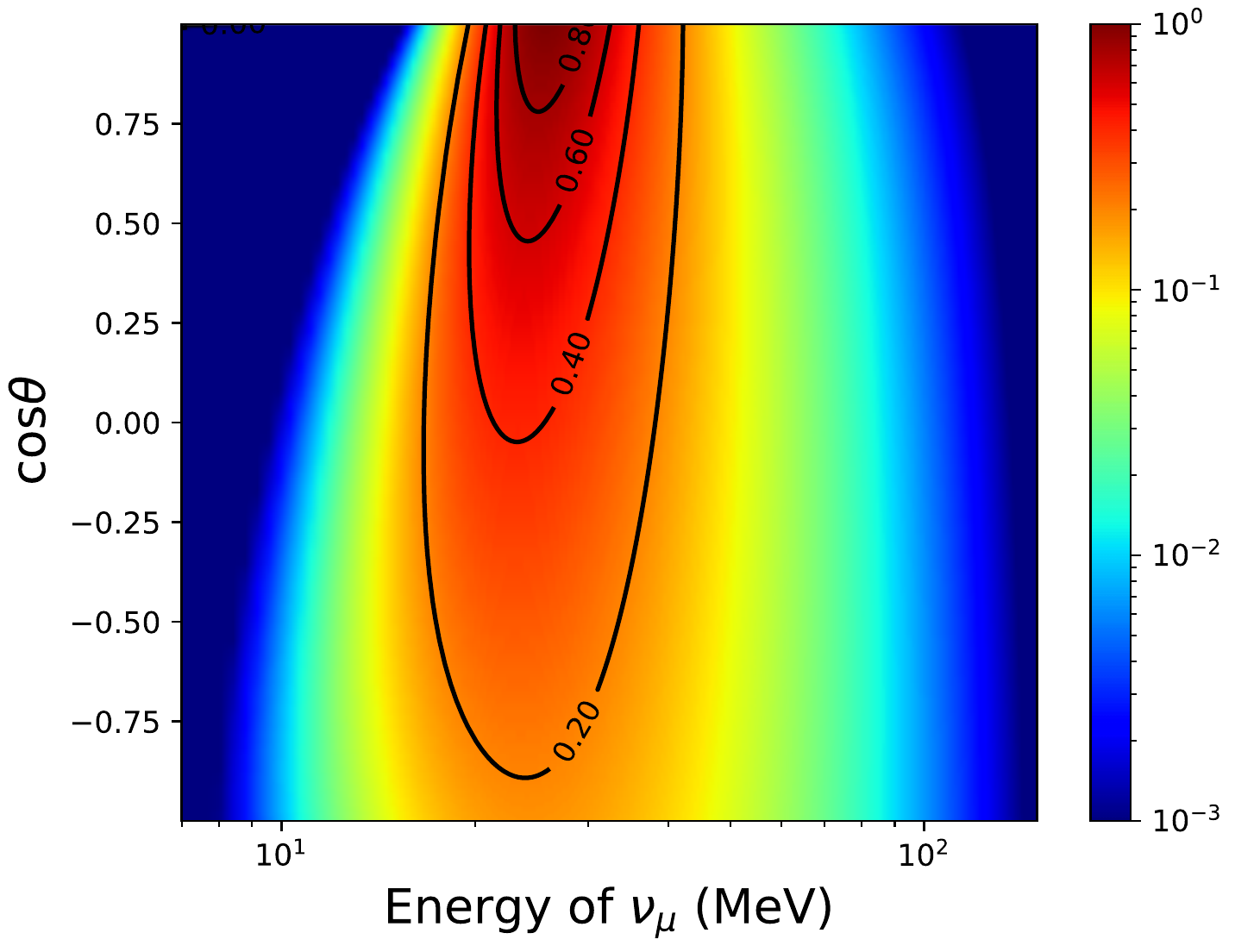}
    \subcaption{$E_{\nu_e} = 10 \MeV$}\label{fig:kernel_flex_1_e10d}
  \end{minipage}
  \begin{minipage}[b]{0.32\linewidth}
    \centering
    \includegraphics[keepaspectratio, scale=0.4]{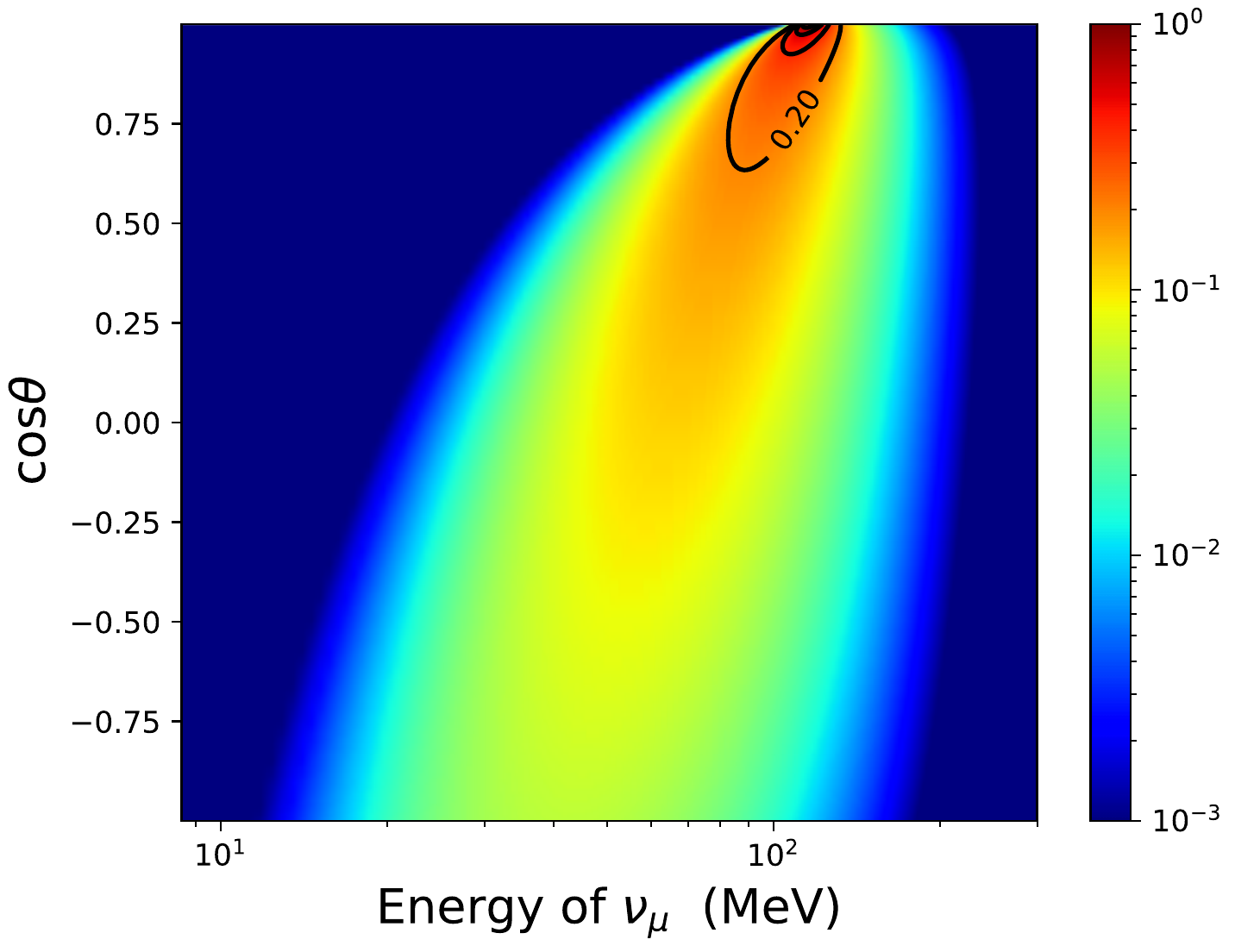}
    \subcaption{$E_{\nu_e} = 100 \MeV$}\label{fig:kernel_flex_1_e100d}
  \end{minipage}
  \caption{The reaction kernel $R_{\nu_e}^{\text{in}}$ of the flavor-exchange reaction $\nu_e + \mu^- \rightarrow \nu_{\mu} + e^-$ for models t10S (panels (a), (b) and (c)) and t10D (panels (d), (e) and (f)) as a function of the energy and angle of outgoing $\nu_{\mu}$.
  The angle is measured from the flight direction of $\nu_e$.
  The energy of $\nu_e$ is fixed to the value show in each panel.
  The value of the kernel is normalized by its maximum.}\label{fig:kernel_flex_1}
\end{figure*}

Figure \ref{fig:kernel_flex_1} exhibits as a color contour the reaction kernel $R_{\nu_e}^{\text{in}}$ for the flavor exchange reaction, $\nu_e + \mu^- \rightarrow \nu_{\mu} + e^-$, (see Eq. (\ref{eq:mfp_GroupA})) as a function of the energy and the angle of the outgoing $\nu_{\mu}$ with the energy of the incoming $\nu_e$ being fixed.
The angle is measured from the flight direction of the incident $\nu_e$.
The upper panels (a), (b) and (c) are the results for model t10S while panels (d), (e) and (f) are for model t10D.
The energy of the incoming $\nu_e$ is set to $1$, $10$ and $100 \MeV$ for the left, middle and right columns, respectively.

It is observed that the energy of $\nu_{\mu}$ is larger than the energy of $\nu_e$ thanks to the large difference of the rest masses between muon and electron.
It is also clear that $\nu_{\mu}$ is preferentially emitted in the forward direction irrespective of the incident energy of $\nu_e$.
As the energy $\nu_e$ gets larger, the forward peak becomes more remarkable whereas the energy gain gets smaller.
From the comparison between models t10S and t10D, we find that these features are shared not only qualitatively but also quantitatively although the thermodynamic conditions are fairly different and the energy of $\nu_{\mu}$, at which the kernel attains the maximum, is somewhat different.
Since this is true also for the other reactions, we will focus on model t10S in the following.

\begin{figure*}[htbp]
  \begin{minipage}[b]{0.32\linewidth}
    \centering
    \includegraphics[keepaspectratio, scale=0.4]{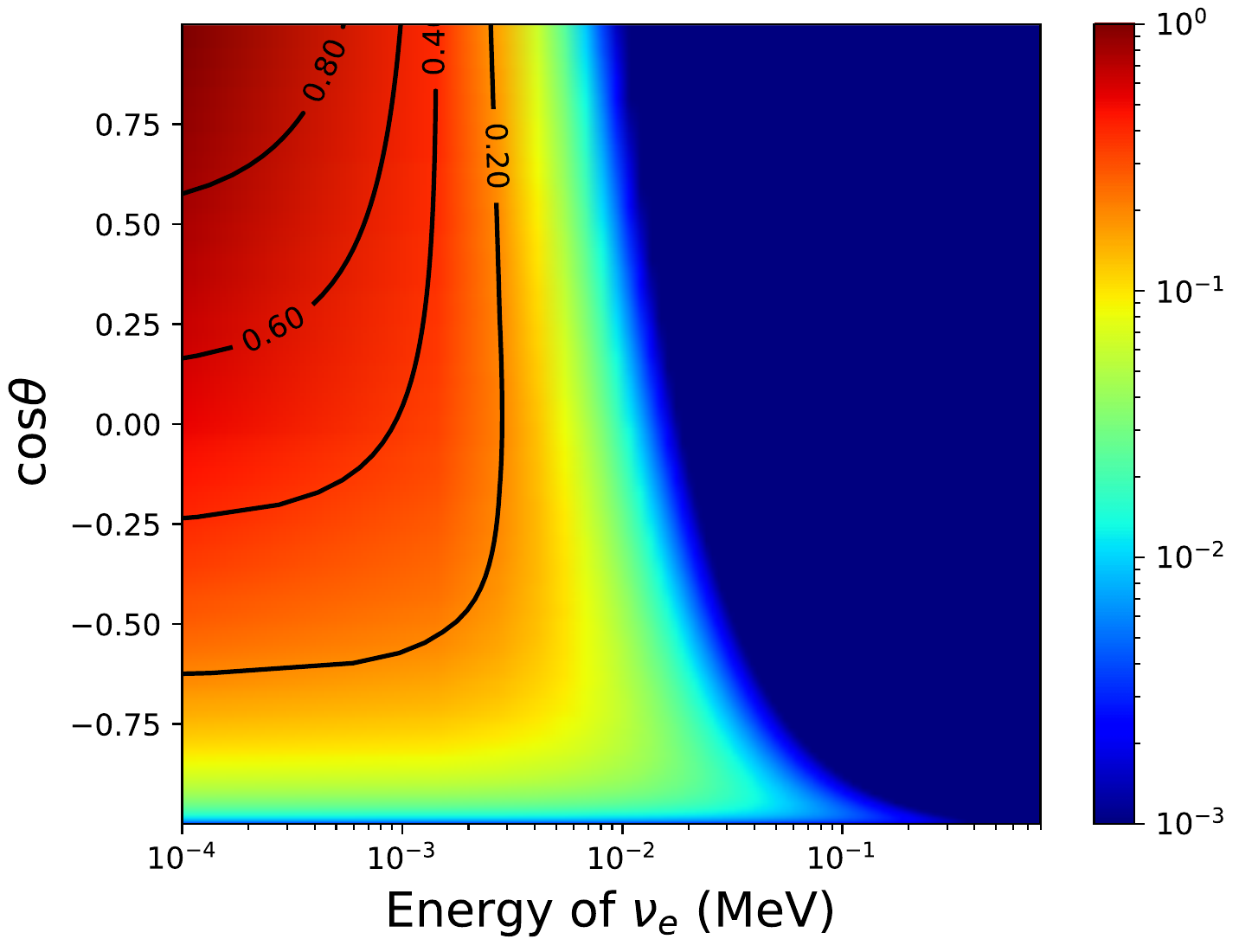}
    \subcaption{$E_{\nu_{\mu}} = 1 \MeV$}\label{fig:kernel_flex_3_e1s}
  \end{minipage}
  \begin{minipage}[b]{0.32\linewidth}
    \centering
    \includegraphics[keepaspectratio, scale=0.4]{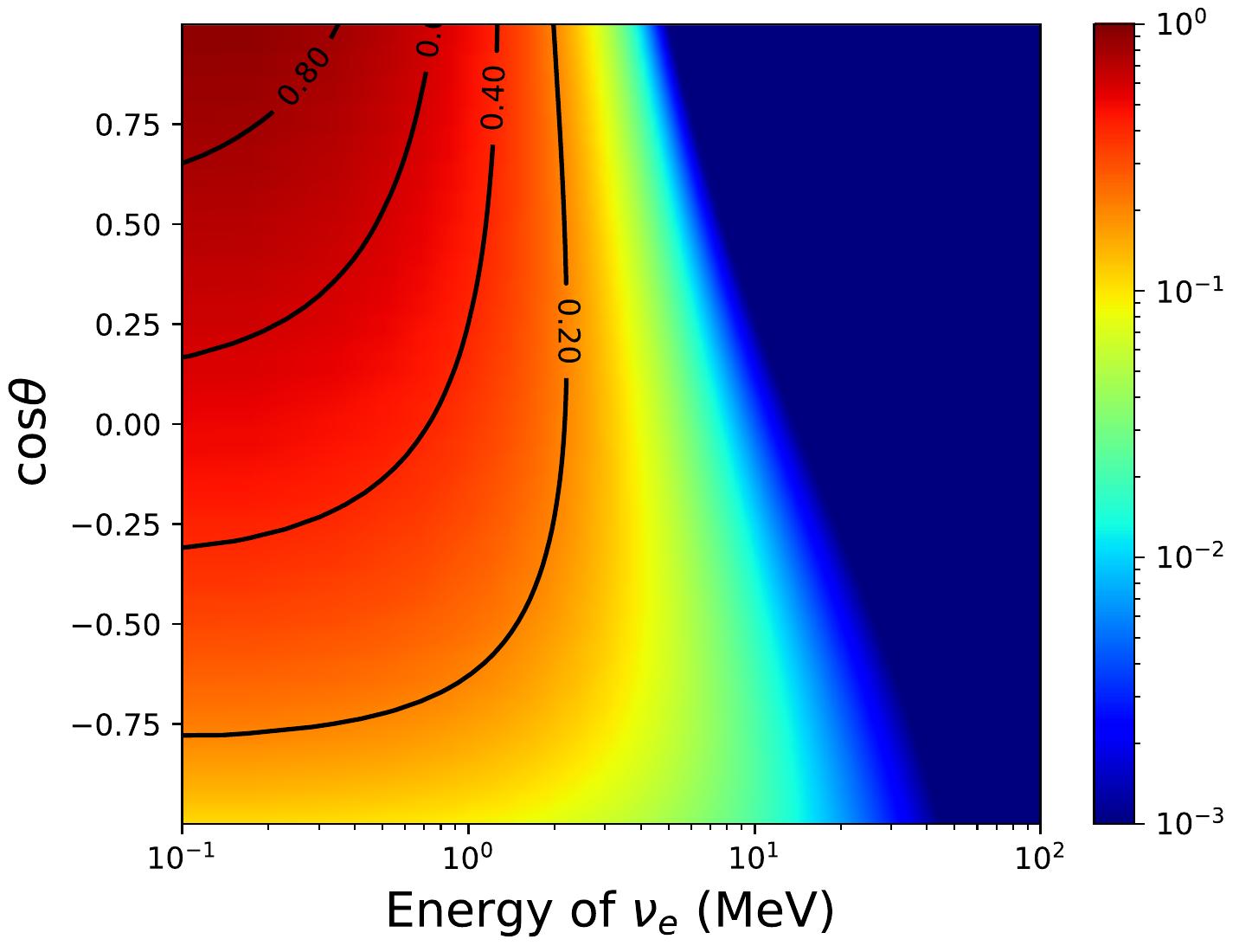}
    \subcaption{$E_{\nu_{\mu}} = 10 \MeV$}\label{fig:kernel_flex_3_e10s}
  \end{minipage}
  \begin{minipage}[b]{0.32\linewidth}
    \centering
    \includegraphics[keepaspectratio, scale=0.4]{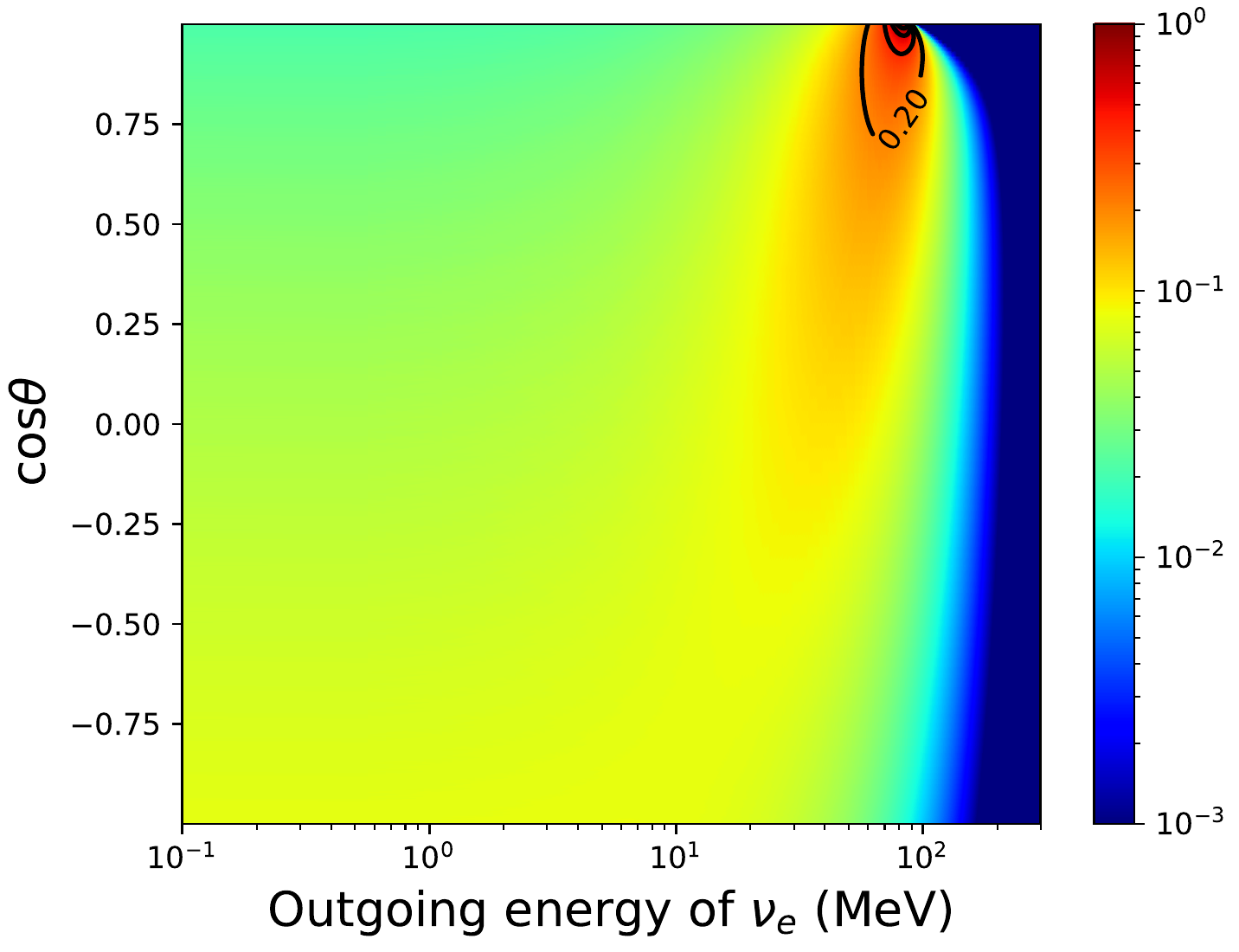}
    \subcaption{$E_{\nu_{\mu}} = 100 \MeV$}\label{fig:kernel_flex_3_e100s}
  \end{minipage}
  \caption{Same as Figure \ref{fig:kernel_flex_1} but the kernel $R_{\nu_{\mu}}^{\text{in}}$ for $\nu_{\mu} + e^- \rightarrow \nu_e + \mu^-$ as a function of the energy and angle of outgoing $\nu_{e}$ for model t10S.}\label{fig:kernel_flex_3}
\end{figure*}

Figure \ref{fig:kernel_flex_3} presents the reaction kernel $R_{\nu_{\mu}}^{\text{in}} \left( = R_{\nu_e}^{\text{out}} \right)$ for the inverse process of the flavor-exchange reaction discussed above: $\nu_{\mu} + e^- \rightarrow \nu_e + \mu^-$.
This time the energy of the incident $\nu_{\mu}$ is fixed to $1$, $10$ and $100 \MeV$ in panels (a), (b) and (c), respectively, and the reaction kernel is regarded as a function of the energy and angle of the outgoing $\nu_e$.
For rather low incident energies (see panels (a) and (b)), the outgoing $\nu_e$ has very low energies, since most of the energy is exhausted to generate the muon mass.
We can see again the outgoing neutrinos are emitted preferentially in the forward direction.
For the high incident energy, the energy of outgoing neutrino rises accordingly and the forward peak gets pronounced.
These results simply reflect the detailed balance expected as Eq. (\ref{eq:DetailedBalance_lsc})

\begin{figure*}[htbp]
  \begin{minipage}[b]{0.32\linewidth}
    \centering
    \includegraphics[keepaspectratio, scale=0.4]{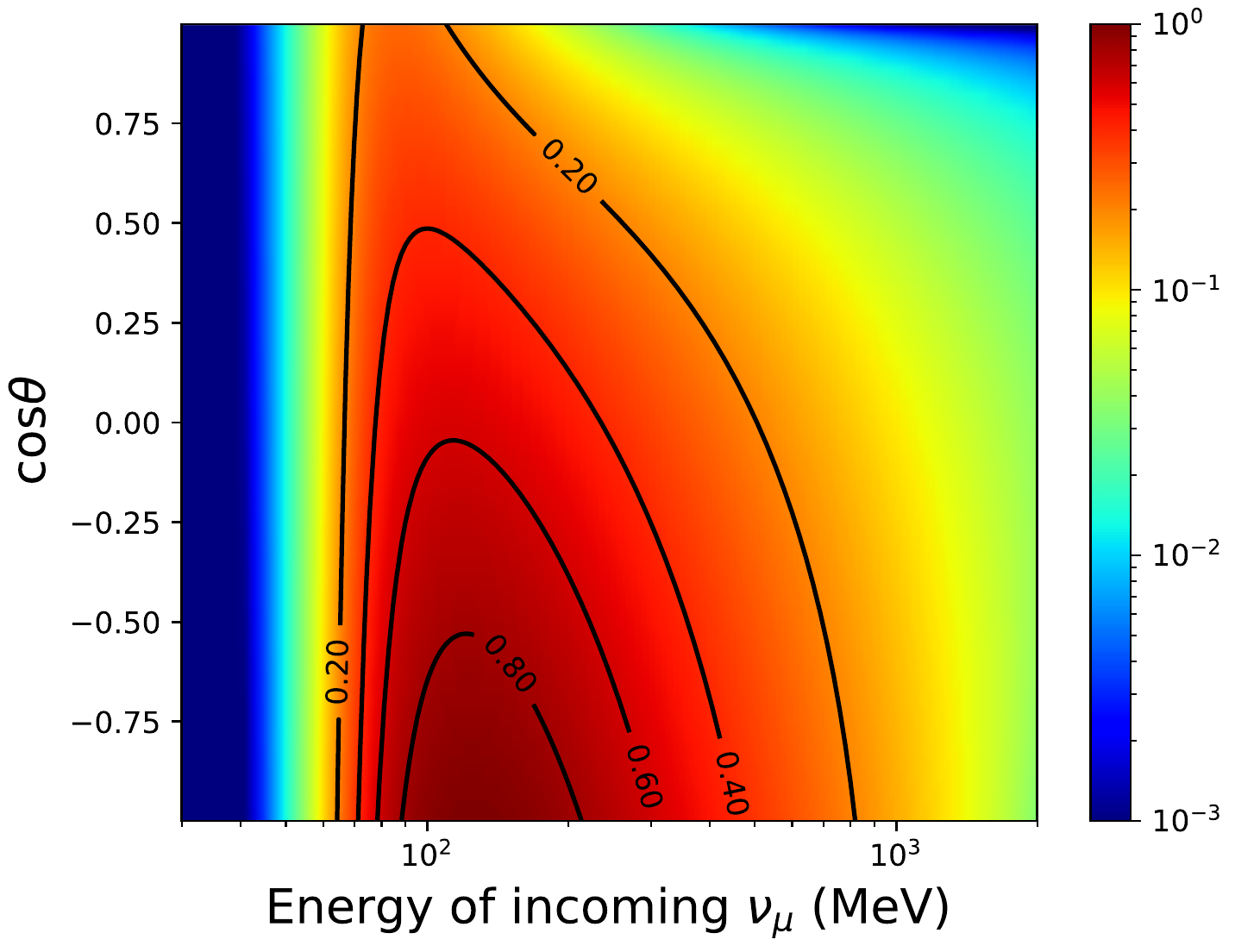}
    \subcaption{$E_{\bar{\nu}_e} = 1 \MeV$}\label{fig:kernel_mudecay_2_e1s}
  \end{minipage}
  \begin{minipage}[b]{0.32\linewidth}
    \centering
    \includegraphics[keepaspectratio, scale=0.4]{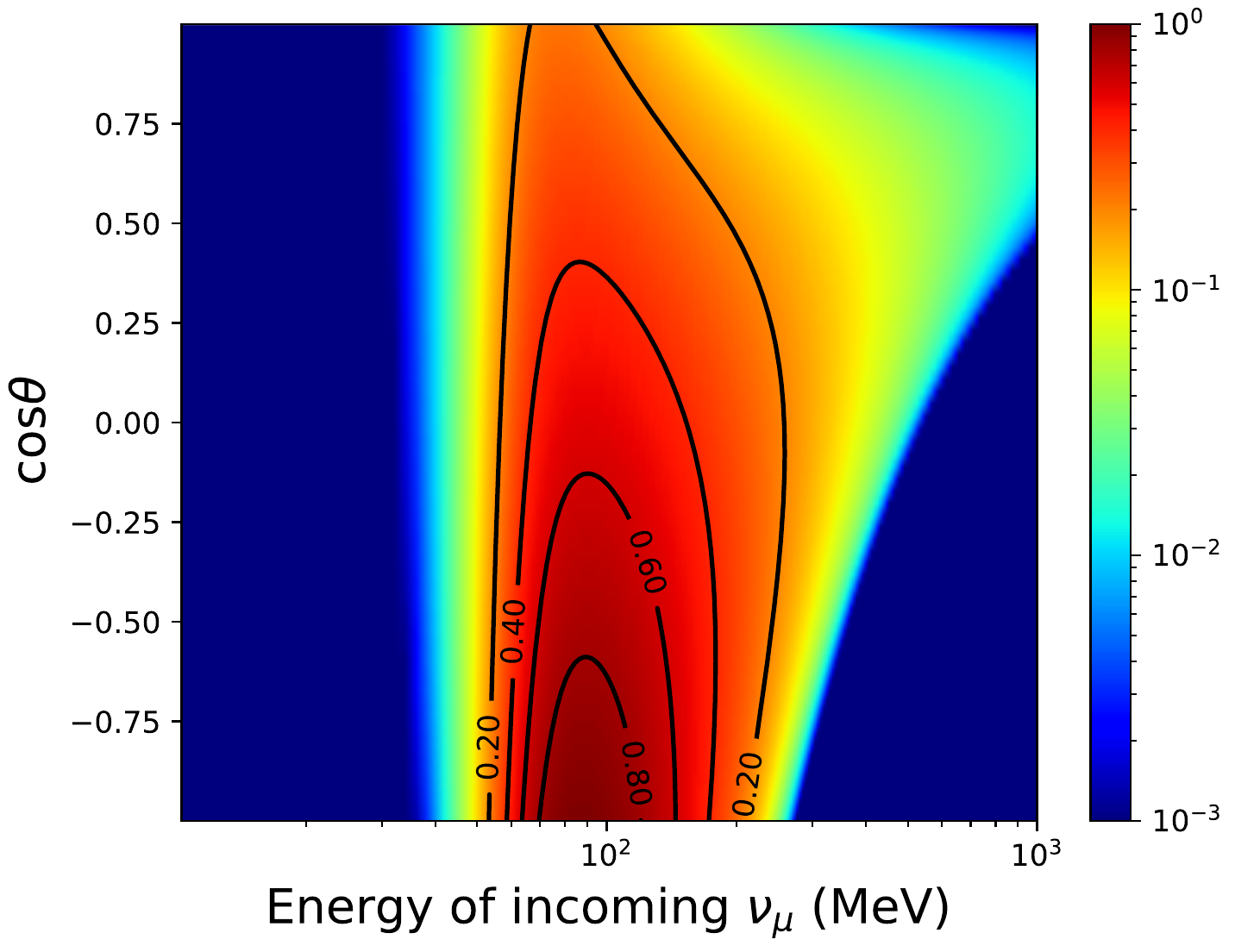}
    \subcaption{$E_{\bar{\nu}_e} = 10 \MeV$}\label{fig:kernel_mudecay_2_e10s}
  \end{minipage}
  \begin{minipage}[b]{0.32\linewidth}
    \centering
    \includegraphics[keepaspectratio, scale=0.4]{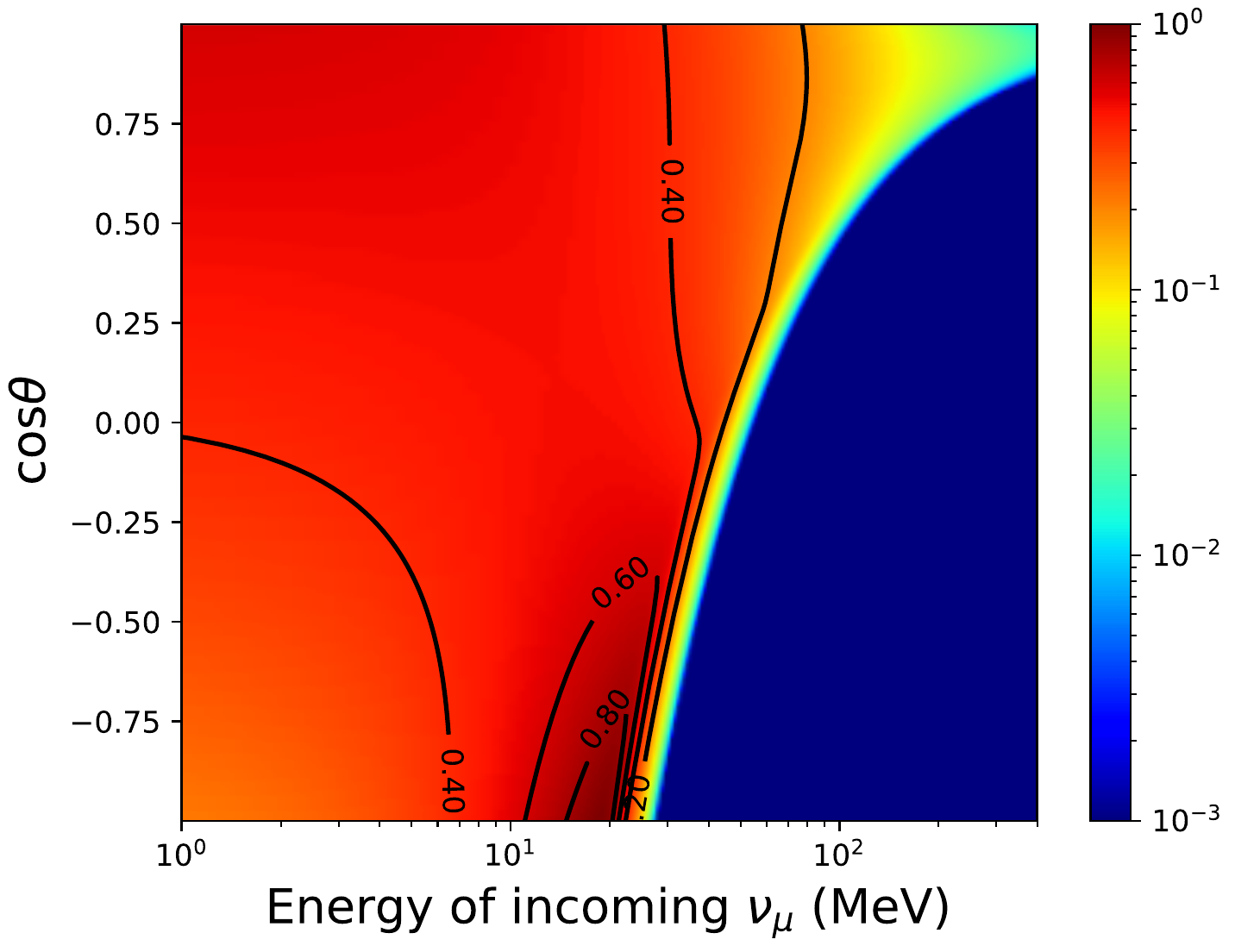}
    \subcaption{$E_{\bar{\nu}_e} = 100 \MeV$}\label{fig:kernel_mudecay_2_e100s}
  \end{minipage}
  \caption{The reaction kernel $R_{\bar{\nu}_e}^{\text{in}}$ of the inverse muon decay $\bar{\nu}_e + \nu_{\mu} + e^- \rightarrow \mu^-$ for model t10S as a function of energy and angle of incoming $\nu_{\mu}$.
  The energy of $\bar{\nu}_{e}$ is fixed to \mbox{$ E_{\bar{\nu}_e} = 1, 10, 100 \MeV$} for panels (a), (b) and (c), respectively.
  The values of the kernel is normalized by its maximum.}\label{fig:kernel_mudecay_2}
\end{figure*}

\begin{figure*}[htbp]
  \begin{minipage}[b]{0.32\linewidth}
    \centering
    \includegraphics[keepaspectratio, scale=0.4]{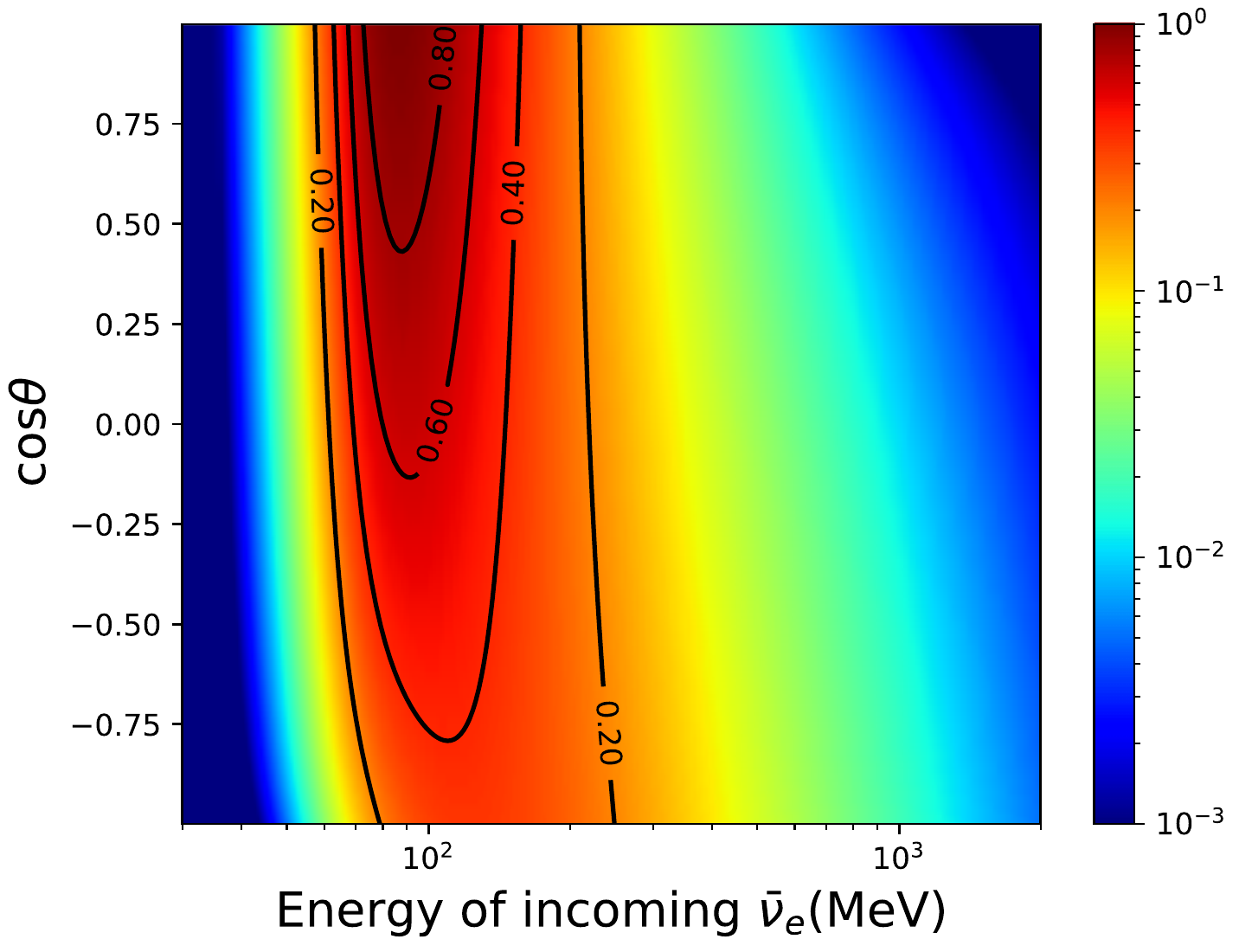}
    \subcaption{$E_{\nu_{\mu}} = 1 \MeV$}\label{fig:kernel_mudecay_2_e1s}
  \end{minipage}
  \begin{minipage}[b]{0.32\linewidth}
    \centering
    \includegraphics[keepaspectratio, scale=0.4]{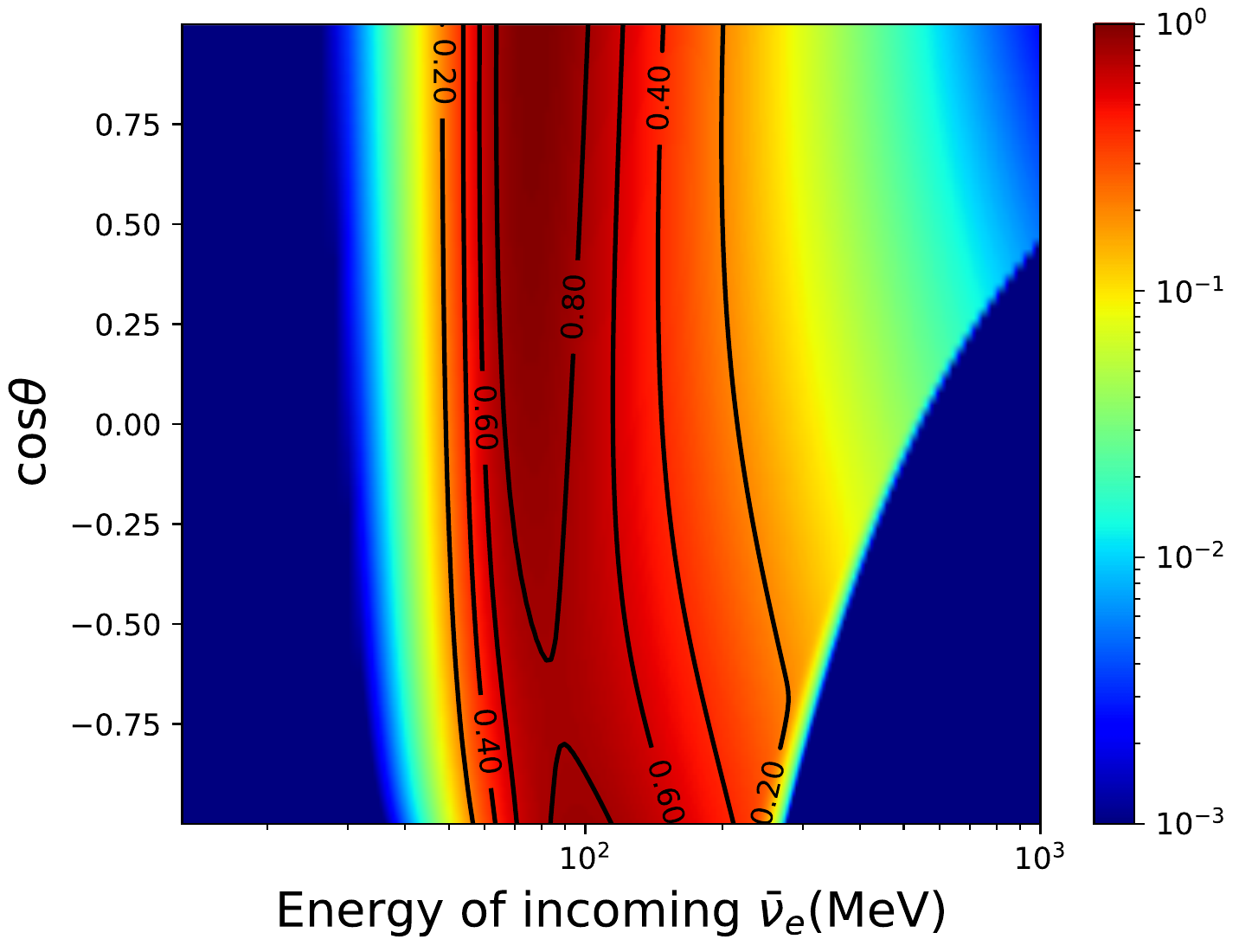}
    \subcaption{$E_{\nu_{\mu}} = 10 \MeV$}\label{fig:kernel_mudecay_2_e10s}
  \end{minipage}
  \begin{minipage}[b]{0.32\linewidth}
    \centering
    \includegraphics[keepaspectratio, scale=0.4]{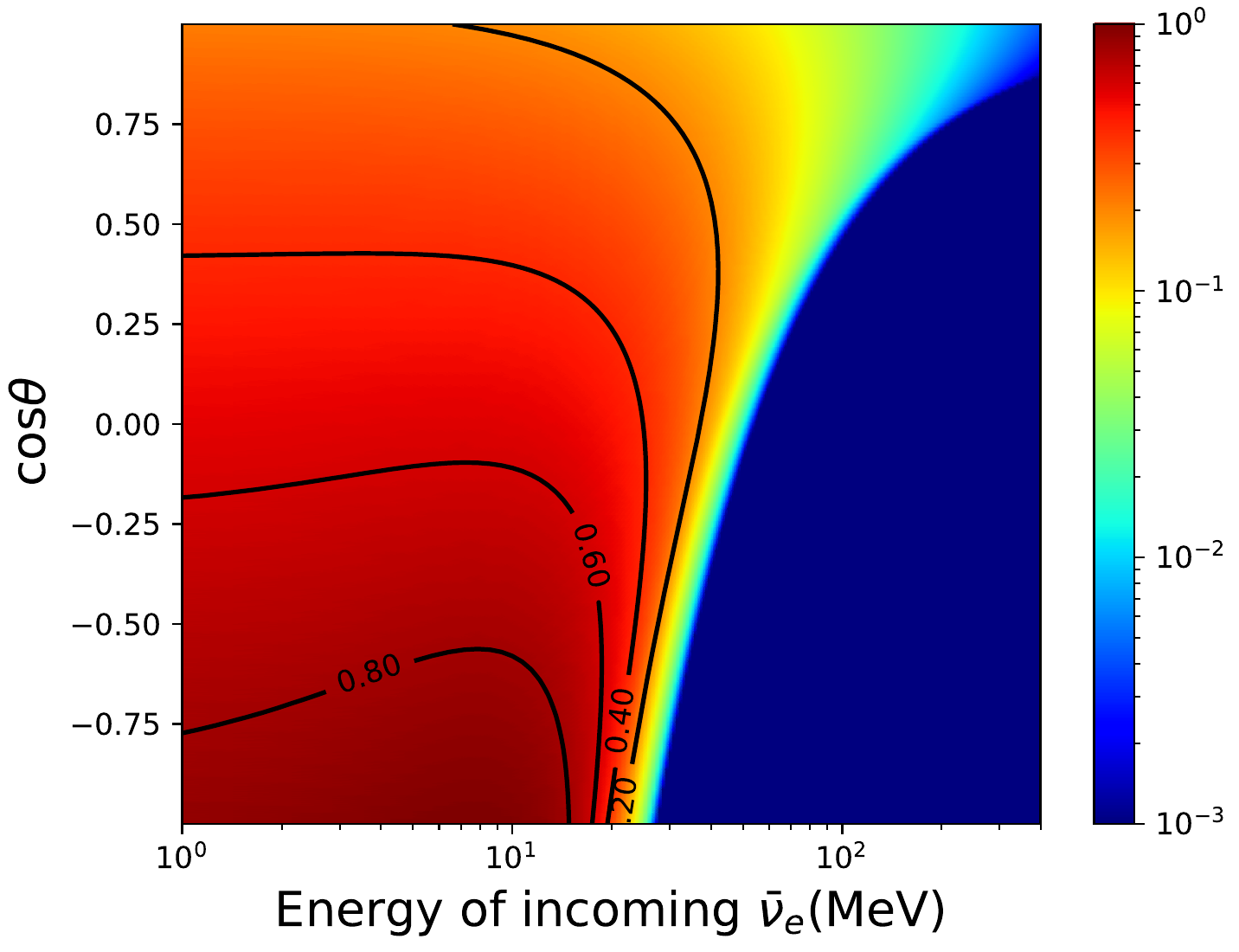}
    \subcaption{$E_{\nu_{\mu}} = 100 \MeV$}\label{fig:kernel_mudecay_2_e100s}
  \end{minipage}
  \caption{Same as Figure \ref{fig:kernel_mudecay_2} but as a function of the energy and angle of incoming $\bar{\nu}_{e}$.
  The energy of $\nu_{\mu}$ is fixed to $ E_{\nu_{\mu}} = 1, 10, 100 \MeV$ for panels (a), (b) and (c), respectively.}\label{fig:kernel_mudecay_3}
\end{figure*}

Now we move on to the reaction kernel for the inverse muon decay, $\bar{\nu}_e + \nu_{\mu} + e^- \rightarrow \mu^-$.
Figure \ref{fig:kernel_mudecay_2} shows $R_{\bar{\nu}_e}^{\text{in}}$ (see Eq. (\ref{eq:kernel_GroupB})) as a function of the energy and angle of $\nu_{\mu}$ with the energy of incoming $\bar{\nu}_e$ being fixed to either $1$, $10$ or $100 \MeV$ (corresponding to panels (a), (b) and (c), respectively).
The angle is measured from the flight direction of the other neutrino, $\bar{\nu}_e$.
It is found that the kernel is largest when the sum of energies, $E_{\bar{\nu}_e} + E_{\nu_{\mu}}$ is close to the rest mass of muon and the neutrinos collide head on, i.e., $\cos \theta = -1$.
When one looks at this reaction from the standpoint of $\nu_{\mu}$, on the other hand, the picture could be different.
This can be seen in Figure \ref{fig:kernel_mudecay_3}, in which we display the same kernel as a function of the energy and angle of $\bar{\nu}_e$ for some fixed energies of $\nu_{\mu}$: $1$, $10$ or $100 \MeV$ for panels (a), (b) and (c).
For the low energy of $1 \MeV$ (see panel (a)), the kernel reaches the maximum at $E_{\bar{\nu}_e} + E_{\nu_{\mu}} \approx m_{\mu}$ again but when the two neutrinos move in the same direction, i.e., $\cos \theta = 1$.
As the incident energy becomes higher, however, the head-on collision is preferred again as is obvious in panels (b) and (c).

\subsection{Possible implications for the cooling timescale\label{subsection:dc}}
Although we have so far looked into the inverse mean free paths and reaction kernels for individual reactions,
what determines the thermal history of PNS is the sum of all these and inverse reactions as well as the advection of all species of neutrinos in the PNS.
Except near the PNS surface, matter is so dense in the PNS interior that neutrinos are almost in thermal and chemical equilibria with matter.
In such circumstances, each reaction is almost balanced with its inverse reaction.
Then the neutrino distributions become nearly isotropic in momentum space and can be expanded as $f_{\nu}(\varepsilon , \cos \theta) = f_{\nu}^{(0)}(\varepsilon) + \cos \theta f_{\nu}^{(1)}(\varepsilon)$.
This is nothing but the diffusion approximation, in which the energy flux of $\nu_i$ can be written as \cite{Pons1999, Roberts2017a}
\begin{align}
  F_{\nu_i} &= \dfrac{1}{6 \pi^2} \int f_{\nu_i}^{(1)}(\varepsilon) \varepsilon^3 d\varepsilon \notag\\
  &= - \dfrac{\Gamma T^3}{\alpha 6 \pi^2} \left[ D_{4,\nu_i} \dfrac{\partial \alpha T}{\partial r} + D_{3, \nu_i} \left( \alpha T \right) \dfrac{\partial \eta_{\nu_i}}{\partial r}\right] \label{eq:diffusion_flux},
\end{align}
where $\eta_{\nu_i} = \mu_{\nu_i} / T$, $\Gamma = \sqrt{1 - 2 G M_g/r}$ and $\alpha$ is the lapse function, or the $(00)$-component of the metric.
Here we assume that the PNS is spherically symmetric.
The diffusion coefficients in this expression are given as
\begin{align}
  D_{n, \nu_i} = \int_0^{\infty} d \varepsilon \dfrac{\varepsilon^n}{T^{n+1}}  \dfrac{1}{\kappa^*_{\nu_i}} f^{(0)}_{\nu_i}(\varepsilon) \left[ 1 - f^{(0)}_{\nu_i}(\varepsilon)\right], \label{eq:Diffusion_coefficient}
\end{align}
where $\kappa_{\nu_i}^*$ is the total absorption opacity defined as
\begin{equation}
  \kappa_{\nu_i}^* = \dfrac{1}{\lambda_{\nu_i}^{\text{tot}}} + j_{\nu_i}^{\text{tot}}.
\end{equation}
Since the second term in Eq. (\ref{eq:diffusion_flux}) is normally negligible in the PNS cooling \cite{Roberts2017a}, the energy flux is reduced to
\begin{equation}
  F_{\nu_i} \simeq - \dfrac{\Gamma T^3}{\alpha 6 \pi^2} D_{4,\nu_i} \dfrac{\partial \alpha T}{\partial r}  \label{eq:diffusion_flux_simple}.
\end{equation}
It is then obvious that the diffusion coefficient $D_{4, \nu_i}$ is the key factor to determine the cooling timescale through \mbox{Eq. (\ref{eq:KH_timescale})}.
We hence investigate this coefficient, in particular, the contribution of each reaction:
\begin{equation}
  D_{4, \nu_i}^r = \int_0^{\infty} d \varepsilon \dfrac{\varepsilon^4}{T^{5}}  \dfrac{1}{\kappa^r_{\nu_i}} f^{(0)}_{\nu_i}(\varepsilon) \left[ 1 - f^{(0)}_{\nu_i}(\varepsilon)\right], \label{eq:Diffusion_coefficient_r}
\end{equation}
where $r$ specifies the reaction we consider and $\kappa^r_{\nu_i} = 1/\lambda_{\nu_i}^r + j_{\nu_i}^r$.
Although $1/D_{4, \nu_i}$ is not the sum of $1/D_{4, \nu_i}^r$, we use it instead of $D_{4, \nu_i}$, since it corresponds to the inverse mean free path.

\begin{figure*}[htbp]
  \begin{minipage}[b]{0.49\linewidth}
    \centering
    \includegraphics[keepaspectratio, scale=0.51]{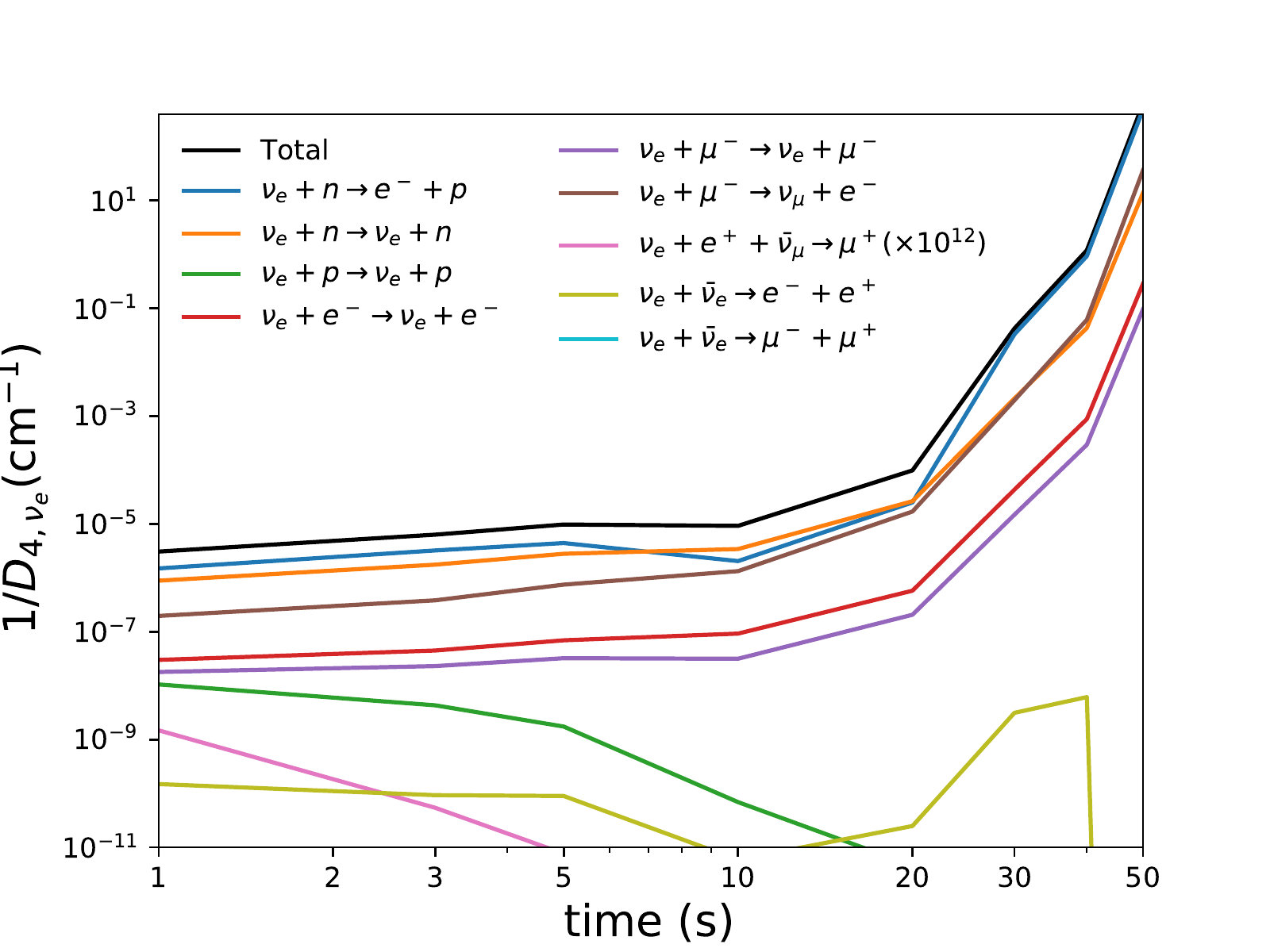}
    \subcaption{}\label{fig:DCS_nue}
  \end{minipage}
  \begin{minipage}[b]{0.49\linewidth}
    \centering
    \includegraphics[keepaspectratio, scale=0.51]{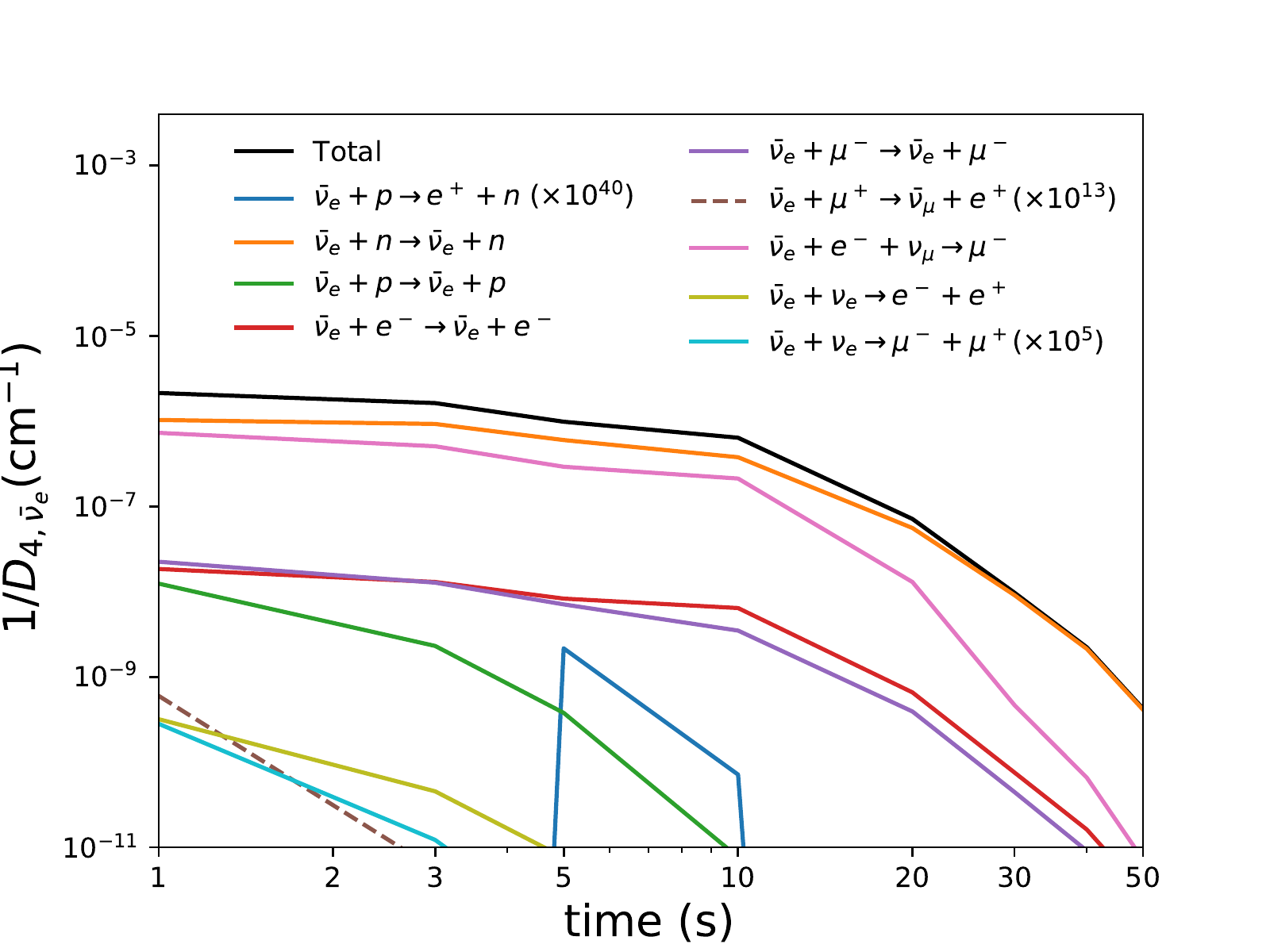}
    \subcaption{}\label{fig:DCS_nueb}
  \end{minipage}\\
  \begin{minipage}[b]{0.49\linewidth}
    \centering
    \includegraphics[keepaspectratio, scale=0.51]{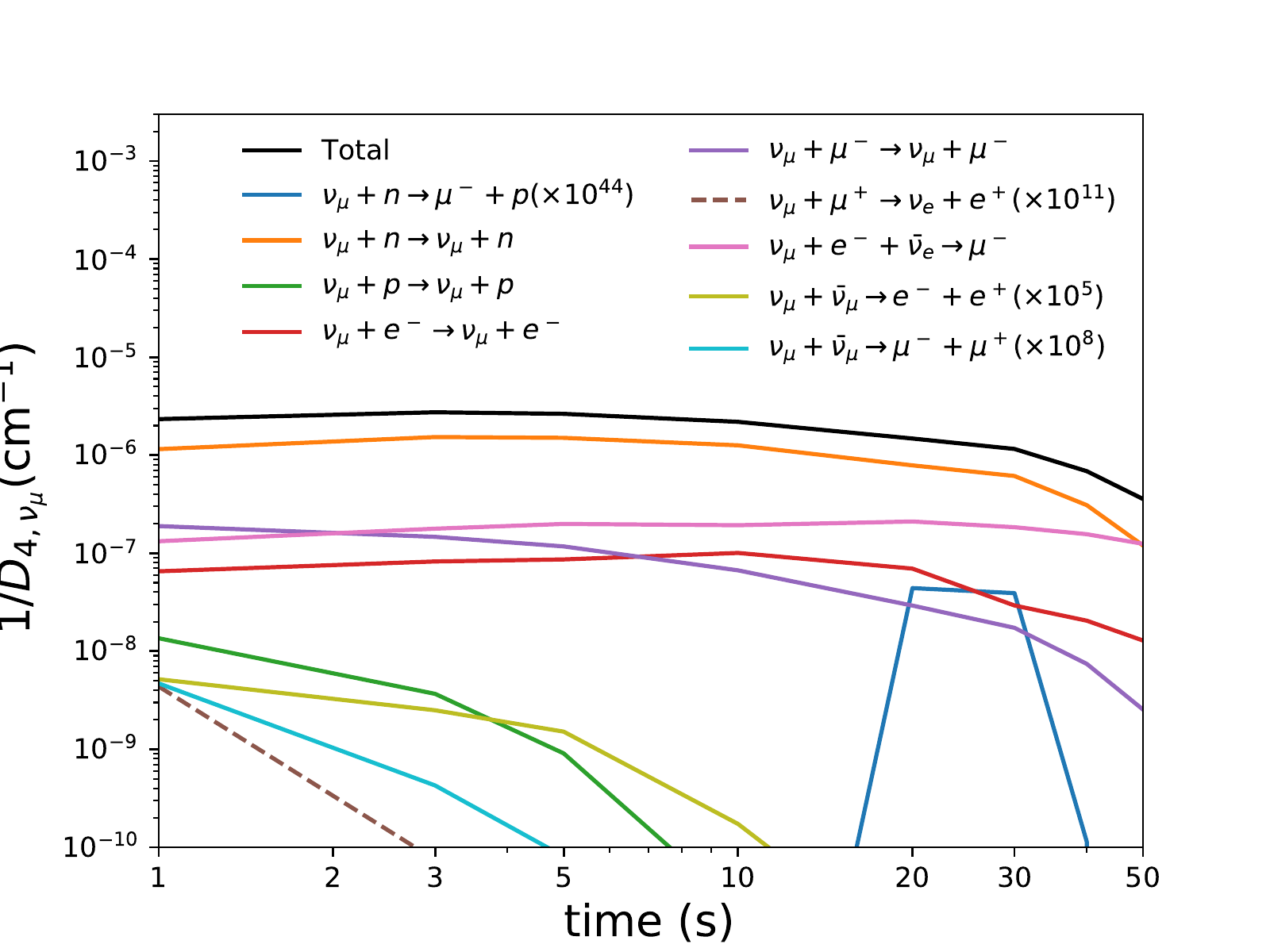}
    \subcaption{}\label{fig:DCS_numu}
  \end{minipage}
  \begin{minipage}[b]{0.49\linewidth}
    \centering
    \includegraphics[keepaspectratio, scale=0.51]{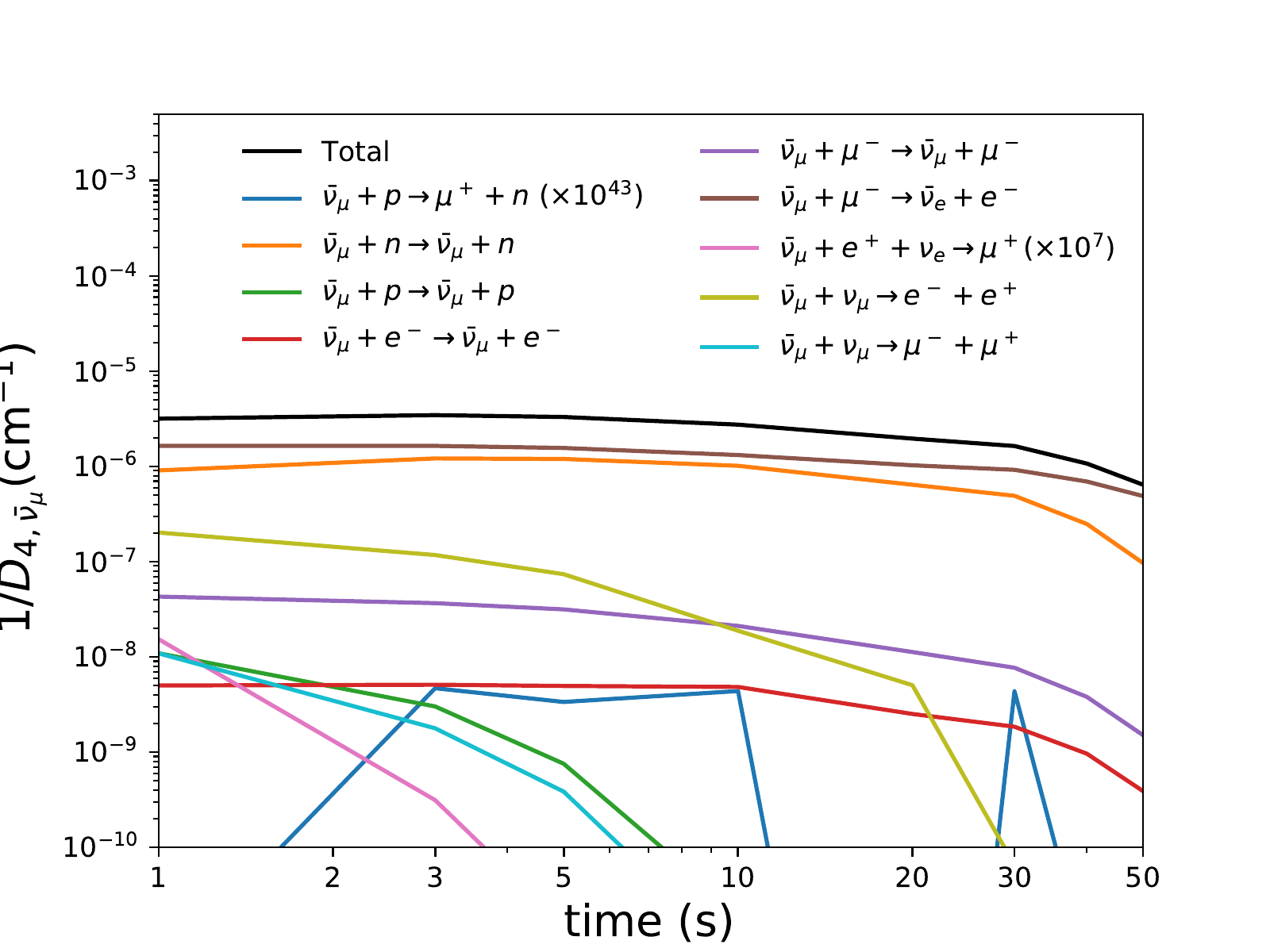}
    \subcaption{}\label{fig:DCS_numub}
  \end{minipage}\\
  \begin{minipage}[b]{0.49\linewidth}
    \centering
    \includegraphics[keepaspectratio, scale=0.51]{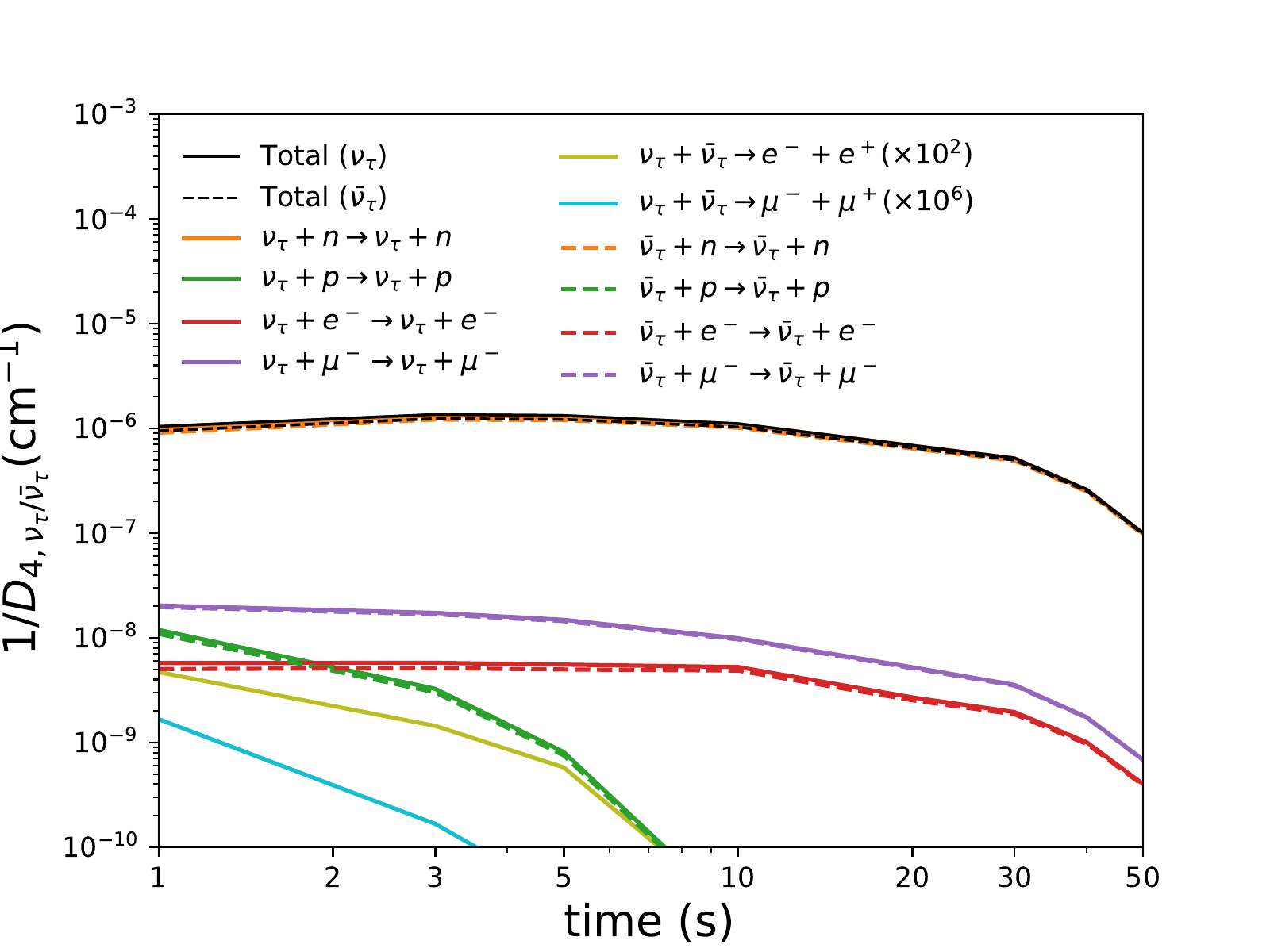}
    \subcaption{}\label{fig:DCS_nutau}
  \end{minipage}
  \caption{Inverse of the diffusion coefficients $D_4$'s at the neutrino sphere for different reactions as a function of time. Panels (a), (b), (c) and (d) are the results for $\nu_e$, $\bar{\nu}_e$, $\nu_{\mu}$ and $\bar{\nu}_{\mu}$, respectively. Panel (e) shows those for both $\nu_{\tau}$ (solid line) and $\bar{\nu}_{\tau}$ (dashed line), respectively. Colors denote different reactions (see the legends in each panel). }\label{fig:Diffusion_coefficient_S}
\end{figure*}

\begin{figure*}[htbp]
  \begin{minipage}[b]{0.49\linewidth}
    \centering
    \includegraphics[keepaspectratio, scale=0.51]{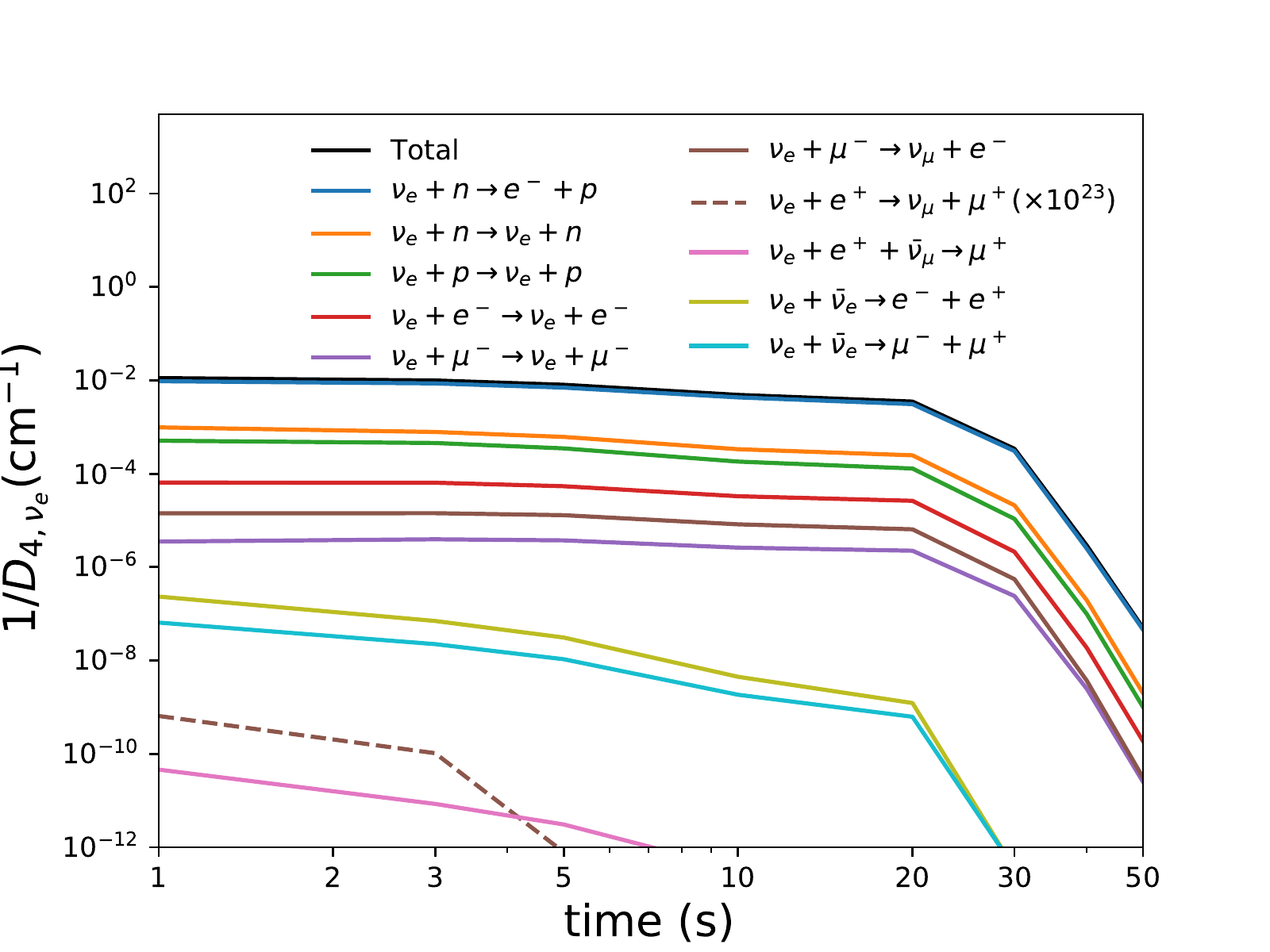}
    \subcaption{}\label{fig:DCD_nue}
  \end{minipage}
  \begin{minipage}[b]{0.49\linewidth}
    \centering
    \includegraphics[keepaspectratio, scale=0.51]{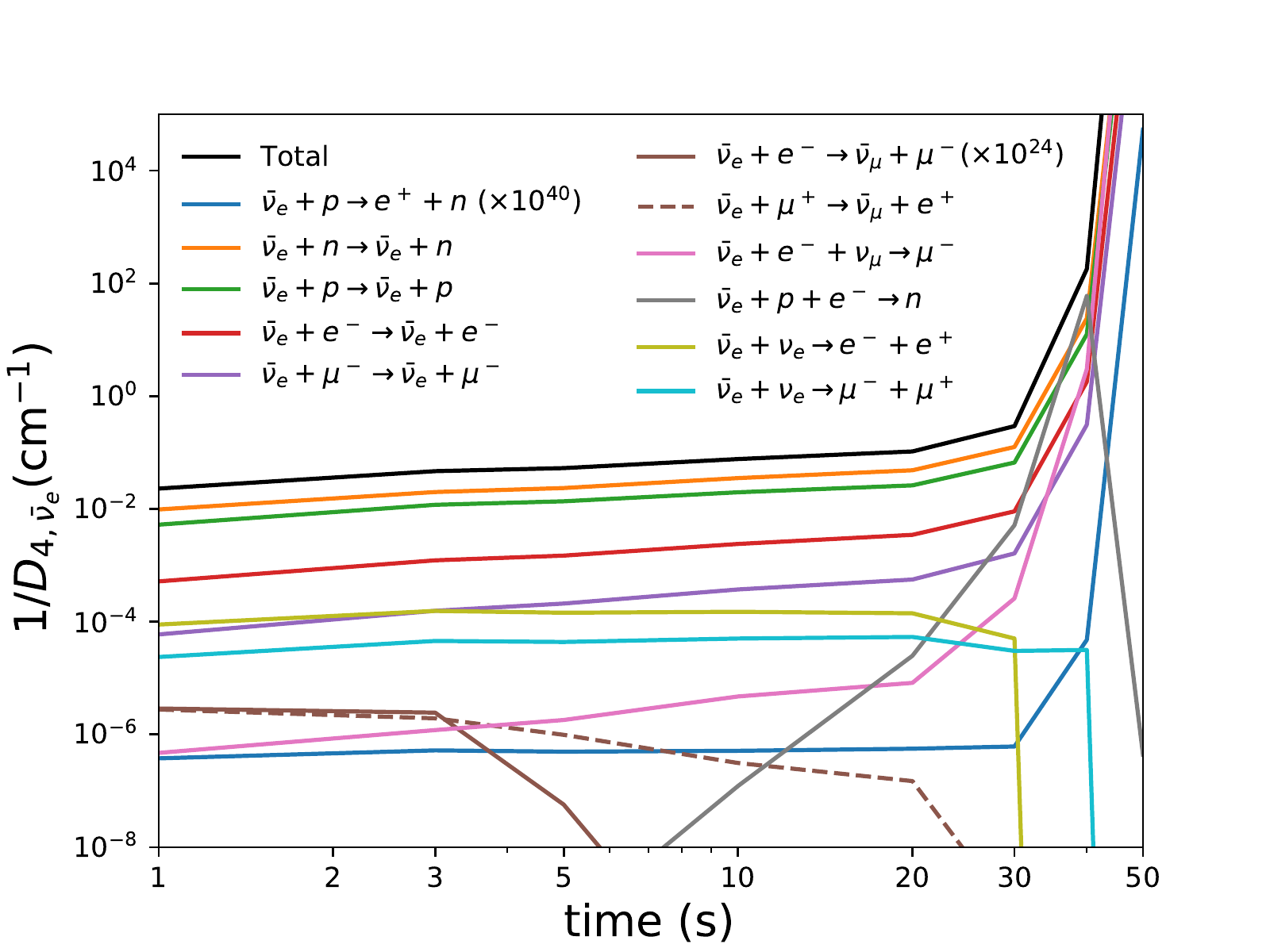}
    \subcaption{}\label{fig:DCD_nueb}
  \end{minipage}\\
  \begin{minipage}[b]{0.49\linewidth}
    \centering
    \includegraphics[keepaspectratio, scale=0.51]{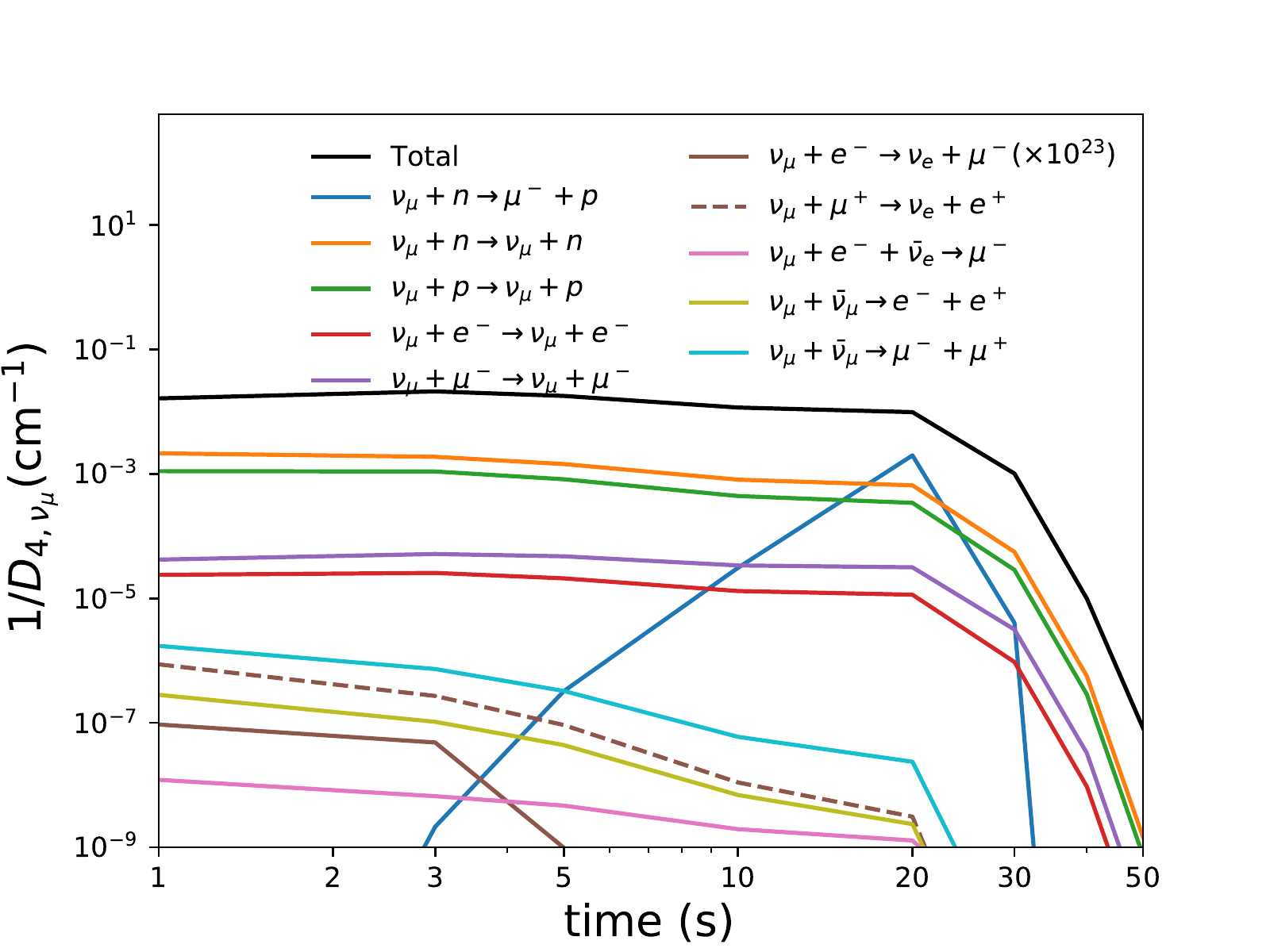}
    \subcaption{}\label{fig:DCD_numu}
  \end{minipage}
  \begin{minipage}[b]{0.49\linewidth}
    \centering
    \includegraphics[keepaspectratio, scale=0.51]{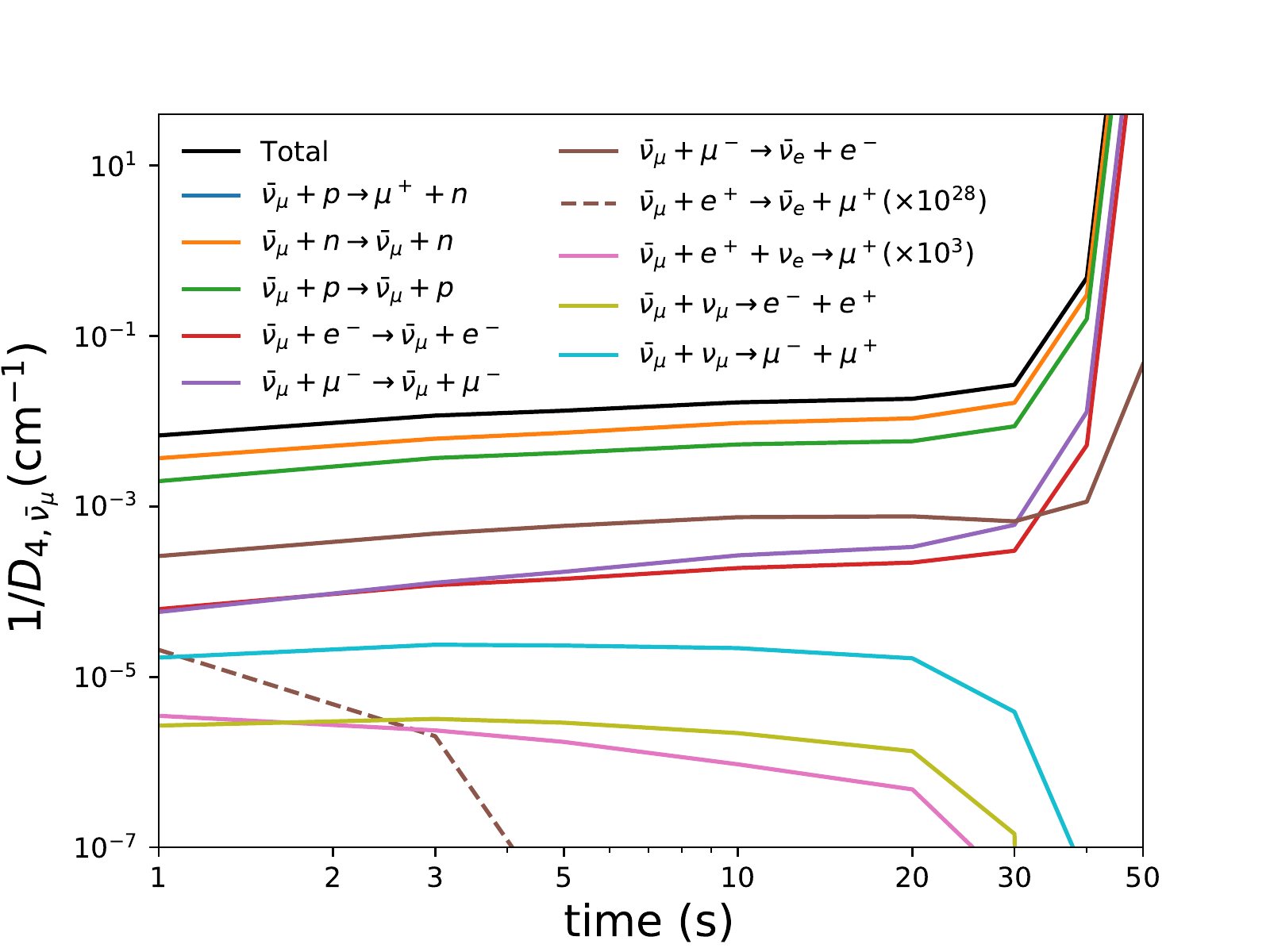}
    \subcaption{}\label{fig:DCD_numub}
  \end{minipage}\\
  \begin{minipage}[b]{0.49\linewidth}
    \centering
    \includegraphics[keepaspectratio, scale=0.51]{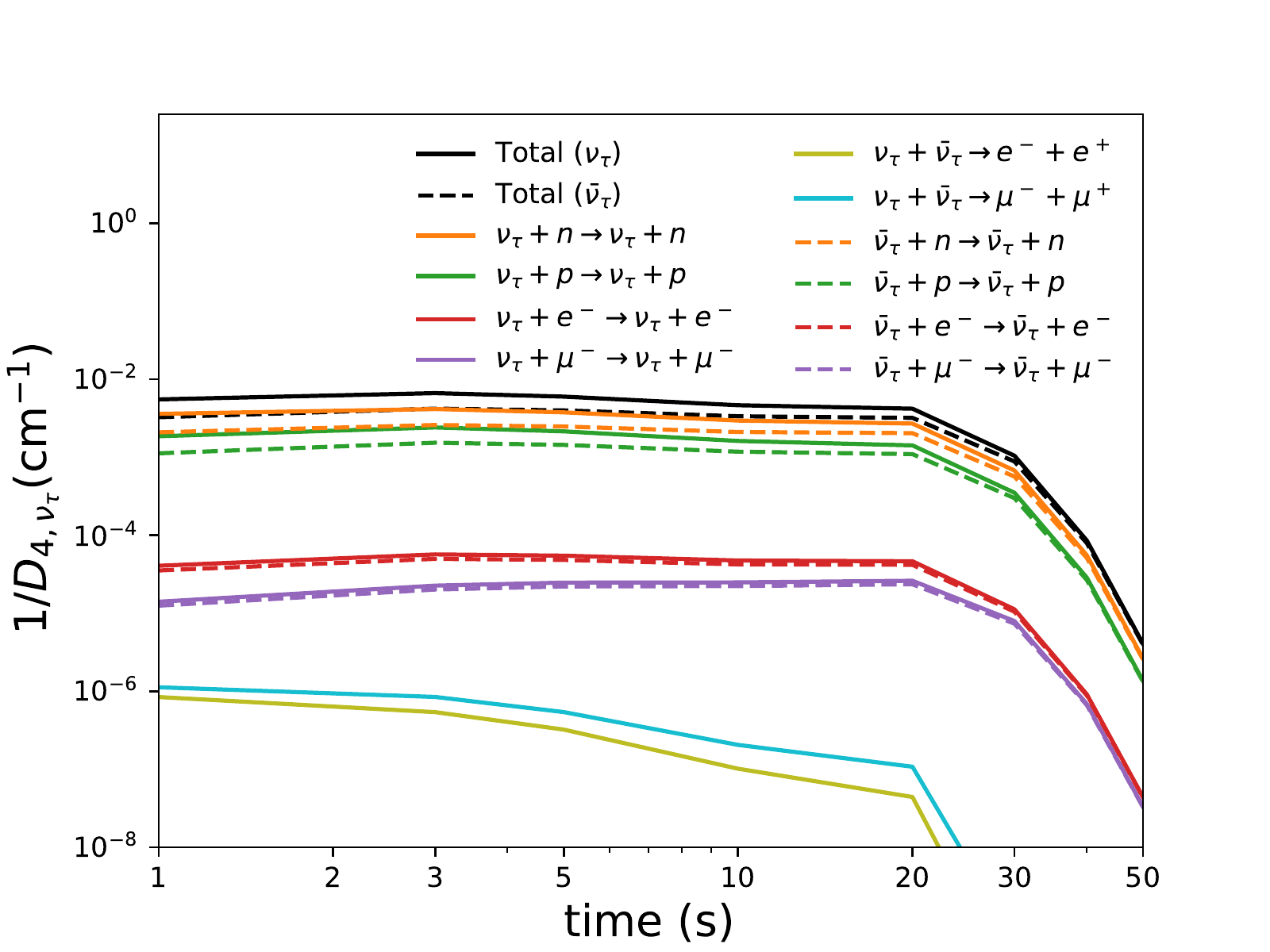}
    \subcaption{}\label{fig:DCD_nutau}
  \end{minipage}
  \caption{Same as Figure \ref{fig:Diffusion_coefficient_S} but for the deeper region.}\label{fig:Diffusion_coefficient_D}
\end{figure*}

Figures \ref{fig:Diffusion_coefficient_S} and \ref{fig:Diffusion_coefficient_D} show the values of $1/D_{4, \nu_i}^r$ for different reactions again at the neutrino sphere (Figure \ref{fig:Diffusion_coefficient_S}) and at the radius where the temperature is highest and hence muons are expected to be most abundant (Figure \ref{fig:Diffusion_coefficient_D}), as a function of time.
We can see from Figure \ref{fig:Diffusion_coefficient_S} that at the neutrino sphere, the neutrino scattering on neutron is dominant for $\bar{\nu}_e$, $\nu_{\mu}$, $\nu_{\tau}$ and $\bar{\nu}_{\tau}$.
For $\nu_e$, the capture on neutron is almost comparable.
For $\bar{\nu}_{\mu}$, on the other hand, the flavor exchange reaction exceeds the neutron scattering at all times, which suggests that the $\bar{\nu}_{\mu}$ flux may be most affected by the muon existence.
The flavor exchange reaction and the muon decay are smaller but not negligible also for $\nu_e$, $\bar{\nu}_e$ and $\nu_{\mu}$.
In the deeper region, on the other hand, the flavor exchange reaction as well as the muon decay is suppressed.
We find for $\nu_{\mu}$ that the CC reaction dominates all other reactions around $t = 20 \operatorname{s}$ due to the reduction of effective mass, which makes a wider region in the phase space available just as for the enhancement of the neutrino scattering rate discussed in subsection \ref{subsec:corrections_semilep} (see Eq. (\ref{eq:e2_p})).
In the later phase, this CC reaction is suppressed due to the strong Pauli blocking of muon in the cooled PNS.

If one looks at these figures more closely, one recognizes that the inverse diffusion coefficients increase rather rapidly at late times $t \gtrsim 30 \operatorname{s}$ in Figures \ref{fig:DCS_nue}, \ref{fig:DCD_nueb} and \ref{fig:DCD_numub}.
This behavior is mainly originated from the factor $f^{(0)}_{\nu} (1 - f^{(0)}_{\nu})$ in Eq. (\ref{eq:Diffusion_coefficient}), which decrease quickly at these times.
In fact, in Figure \ref{fig:DCS_nue}, $\mu_{\nu_e} (= \mu_p + \mu_e - \mu_n)$ get smaller at the neutrino sphere as the proton fraction (and hence the proton chemical potential) becomes lower at late times since free protons get depleted there, being incorporated into nuclei as the temperature decreases:
for instance, $\mu_{\nu_e} = -5.1 \MeV$ at $t = 10 \s$ whereas $\mu_{\nu_e} = -28.4 \MeV$ at $t = 50 \s$.
This decline of  $f_{\nu_e}$ leads to that of diffusion coefficients, and equivalently to the rise of their inverse.
On the other hand, in Figures \ref{fig:DCD_nueb} and \ref{fig:DCD_numub}, $\mu_{{\nu}_{e,\mu}} = \mu_p + \mu_{e,\mu} - \mu_n$ both take large positive values because electrons and muons are highly degenerate in the deeper region:
for instance, at $t = 50 \s$, $\mu_{\nu_{e}} = 101.5 \MeV$ and $\mu_{\nu_{\mu}} = 60.3 \MeV$, which are much lager than the local temperature $T = 2.8 \MeV$ there.
Since $\mu_{\bar{\nu}_{e,\mu}} = - \mu_{{\nu}_{e,\mu}}$ in beta-equilibrium, these anti-neutrinos get more strongly suppressed at  these late times and the factor $f^{(0)}_{\bar{\nu}_{e, \mu}} (1 - f^{(0)}_{\bar{\nu}_{e, \mu}})$ becomes smaller, resulting in the rapid  rise of the inverse diffusion coefficients.

The inverse diffusion coefficients for $\nu_e$, $\nu_{\mu}$, $\nu_{\tau}$ and $\bar{\nu}_{\tau}$ decline rather quickly in Figures
\ref{fig:DCD_nue}, \ref{fig:DCD_numu} and \ref{fig:DCD_nutau} at the late times, on the other hand.
This stems from two factors:
(1) the temperature, which appears in the denominator in Eq.(\ref{eq:Diffusion_coefficient_r}), gets lower;
(2) the opacities themselves become smaller.
The first factor is common to all the species of neutrinos whereas the second one is important for $\nu_e$ and $\nu_{\mu}$, since they get strongly degenerate and the Pauli blocking in the final state tends to suppress their reactions.
Incidentally, matter becomes transparent for neutrinos eventually at some point in the late phase.
Then, the diffusion approximation is no longer valid and the diffusion coefficients are not a good measure for the cooling timescale.

\begin{figure}[htbp]
   \centering
   \includegraphics[keepaspectratio, scale=0.58]{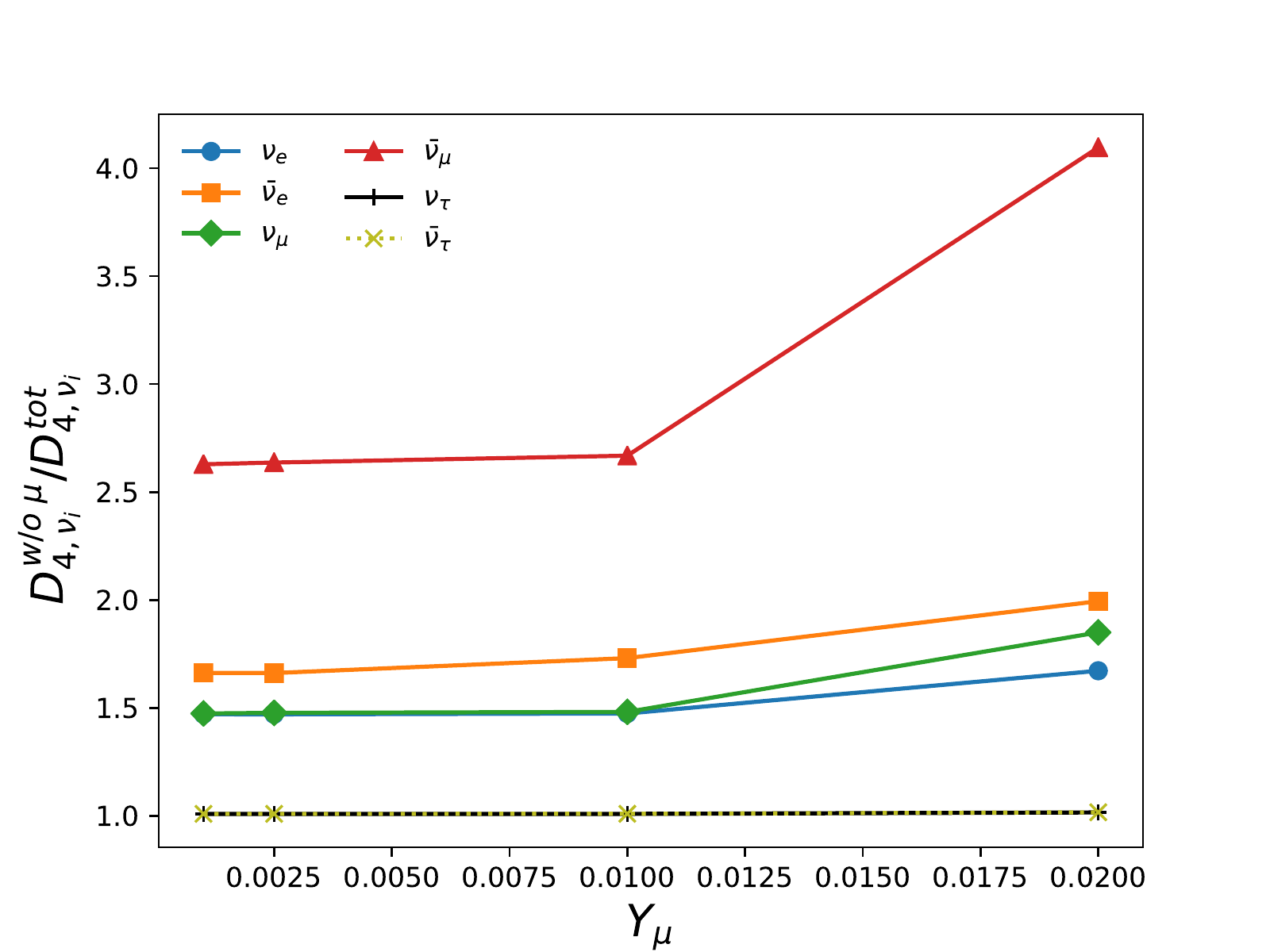}
  \caption{The ratio of the diffusion coefficient without the contribution from the muon-related reactions to the one with their contributions at $t = 10 \s$. Different colors indicate the neutrino flavors.}\label{fig:Diffution_coefficient}
 \end{figure}

In Figure \ref{fig:Diffution_coefficient}, we show $D_{4, \nu_i}^{\text{w/o} \ \mu}/D_{4, \nu_i}^{\text{tot}}$, the ratios of the diffusion coefficients not including the contributions from the muon-related reactions to that
including the contributions from all reactions, to see to what extent the muon-related reactions affect the energy flux at the neutrino sphere.
In so doing we change the muon fraction rather arbitrarily.
We can see from the figure that the ratios are enhanced for neutrinos and anti-neutrinos other than the $\tau$-type of them as the muon fraction gets larger, i.e., the diffusion of these neutrinos becomes slower.
To these changes the following reactions $\nu_e + \mu^- \rightarrow \nu_{\mu} + e^-$ and $\bar{\nu}_{\mu} + \mu^- \rightarrow \bar{\nu}_e + e^-$ give the greatest contributions for $\nu_e$ and $\bar{\nu}_{\mu}$, respectively,
whereas the inverse muon decay $\bar{\nu}_e + \nu_{\mu} + e^- \rightarrow \mu^-$ is the most important for $\bar{\nu}_e$ and $\nu_{\mu}$.
These results may imply that muons behave as a reservoir of these neutrinos, disturbing their diffusions in the PNS.
This is expected in turn to lead to the prolongation of the cooling timescale of PNS.
It is an issue of quantitative investigation, however.
As a matter of fact, the simulation of PNS cooling reported in \cite{Fischer2021Axion} did not find a discernible change in the neutrino luminosities and energies up to $10 \operatorname{s}$ by incorporating the muon-related reactions.
They ascribe this result to the rather small abundance of muons around the neutrino sphere at this time (see Appendix in this paper).
This may change as the time passes further, though, since the muon will be more populated later.
We need to await a quantitative calculation of PNS cooling up to the very late phase with appropriate physics incorporated.

\section{Summary \label{sec:Summary}}
In this paper, we have numerically evaluated the rates of muon-related weak interactions of all species of neutrinos that are relevant in the cooling phase of PNS.
We are particularly interested in the late phase of PNS cooling, $t \gtrsim 10 \operatorname{s}$, which may be accessible for the next Galactic supernova \cite{Suwa2019, Li2021} but lacks the basic information on the these reactions so far in the literature.
For the semi-leptonic interactions, we have taken fully into account the relativistic kinematics of nucleon as well as the weak magnetism, the pseudoscalar term, and the $q^2$ dependence of the form factors for nucleon; we have also considered the corrections to the dispersion relations of nucleons from nuclear interactions on the mean field level.

We have compared the inverse mean free paths of these reactions at different times.
At the neutrino sphere, the inverse muon decay $\bar{\nu}_e + \nu_{\mu} + e^-  \rightarrow \mu^-$ is the dominant source of opacity for $\bar{\nu}_e$ and $\nu_{\mu}$ at low incoming neutrino energies whereas the flavor exchanging reactions: $\nu_e + \mu^- \rightarrow \nu_{\mu} + e^-$ and $\bar{\nu}_{\mu} + \mu^- \rightarrow \bar{\nu}_e + e^-$ give the greatest contributions for $\nu_e$ and $\bar{\nu}_{\mu}$ also at low neutrino energies.
At high energies, on the other hand, the neutrino capture on neutron is dominant for $\nu_e$ and $\nu_{\mu}$ whereas the scattering on neutron dominates the opacity for $\bar{\nu}_e$ and $\bar{\nu}_{\mu}$.
In the deeper region, the muon-related reactions are suppressed compared with the semi-leptonic reactions although general features are similar to what we have found at the neutrino sphere.

In the exploration of the semi-leptonic reactions, we have observed that the weak-magnetism enhances (suppresses) the opacities for neutrino (anti-neutrino) both in the CC and NC reactions;
the $q^2$ dependence of form factors tends to reduce the opacities both via CC and NC.
We have confirmed that the pseudoscalar term gives only minor corrections even in the muon-related CC semi-leptonic reactions.

The difference of the effective potentials between neutron and proton, which we incorporate on the mean field level, shifts the threshold of the CC semi-leptonic reactions and, as a results, enhances the neutrino capture on neutron as well as the inverse neutron decay, which is one of the dominant sources of opacity in the deeper region.
The neutrino scattering on neutron is enhanced, on the other hand, because the effective mass of nucleons becomes smaller in the dense region.

The energy and angular dependences of the flavor-exchange reaction $\nu_e + \mu^- \leftrightarrows \nu_{\mu} + e^- $ and the inverse muon decay  $\bar{\nu}_e + \nu_{\mu} + e^- \rightarrow \mu^- $, both of which could make a dominant contribution to the opacity in some cases, have been inspected as more detailed information.
In fact, they are important for neutrino transport calculations beyond the flux-limited diffusion approximation.
In the former case, $\nu_{\mu}$ is emitted preferentially in the forward direction with a much larger energy than the incident $\nu_e$ thanks to the mass difference between muon and electron.
As expected from the detailed balance, $\nu_e$ is emitted also in the forward direction preferentially with smaller energies in the inverse process.
In the latter case of the inverse muon decay, on the other hand, the reaction kernel as a function of the energy and angle of $\nu_{\mu}$ becomes largest when two neutrinos collide head on in the laboratory frame and the sum of the energies of two incoming neutrinos is close to the rest mass of muon.
If it is regarded as a function of the energy and angle of $\nu_e$ instead, we have observed that the angular dependence is opposite between the low and high incident energies.
The importance of these findings will be assessed later quantitatively with detailed transport calculations.

We have finally investigated the diffusion coefficients for neutrinos, which are relevant for neutrino transport in the optically thick regime.
We have paid particular attention to the changes that the existence of muon will make.
We have found that muons may play a role of reservoir for the e- and $\mu$-type neutrinos by disturbing their diffusion in the PNS interior.
Although we expect that this will make the cooling timescale longer particularly at later times, we need a detailed calculation of PNS cooling up to the very late phase to confirm this.
In fact, the previous simulation up to $10 \operatorname{s}$ \cite{Fischer2021Axion} did not find an appreciable change by incorporating the muon-related reactions.

Our eventual goal is to employ the reaction rates obtained here in calculations of the PNS cooling with the Boltzmann neutrino-transport code we have developed over the years \cite{Nagakura2018ApJ, Iwakami2020, Akaho2021}, possibly in multi-spatial dimensions, and explore quantitatively their influences on the thermal history and neutrino emissions of PNS.
It will be also interesting to apply them to CCSN simulations.
The reaction rates obtained in this paper are admittedly imperfect and there is an ample room for improvement.
For example, it is well known that the corrections to the dispersion relations of nucleons should be accompanied by the corrections to the vertex.
On the mean field level this corresponds to RPA to the structure functions.
Note that the relativistic mean field theory, which we adopt for EoS is particularly convenient for the RPA calculations \cite{Yamada&Toki1999}.
Pions are another interesting particles that may be populated in PNS and have an impact on the the opacity of $\nu_{\mu}$ \cite{BryceFore2020}.
Implementing these effects in the calculations of PNS cooling is certainly the future task.


\section*{Acknowledgements}
We would like to thank, H. Suzuki and K. Nakazato for helpful discussion on the numerical calculation of PNS cooling.
K.Sugiura is supported by JSPS Grant-in-Aid for JSPS Fellows (No. 19J21669) from the Ministry of Education, Culture, Sports, Science and Technology (MEXT), Japan.
This work is supported by Grant-in-Aid for Scientific Research (19K03837, 20H01905) and Grant-in-Aid for Scientific Research on Innovative areas "Gravitational wave physics and astronomy:Genesis" (17H06357, 17H06365) from MEXT, Japan.
S. Furusawa was supported by JSPS KAKENHI (Grant Numbers JP 19K14723, 20H01905).
K. Sumiyoshi acknowledges Computing Research Center at KEK, JLDG, RCNP, Osaka University, YTIP, Kyoto University, and University of Tokyo for the usage of high performance computing resources.
This work was partly supported by the Particle, Nuclear and Astro Physics Simulation Program (Nos. 2020-004, 2021-004) of Institute of Particle and Nuclear Studies, High Energy Accelerator Research Organization (KEK).
A part of the numerical calculations was conducted on the PC cluster at the Center for Computational Astrophysics, National Astronomical Observatory of Japan.
S. Y. is supported by the Institute for Advanced Theoretical and Experimental Physics, Waseda University and the Waseda University Grant for Special Research Projects (project number: 2020C-273, 2021C-197, 2022C-140).

\bibliographystyle{ptephy}
\bibliography{library,muon}

\appendix

\section{Details of the derivation of leptonic reaction rates}
In this appendix, we present the detailed calculations of reaction kernels for the leptonic reactions listed in Table \ref{tab:mu_reaction}.

\subsection{Scatterings: $\nu + l \leftrightarrows \nu + l$\label{appendix:lepsca}}
The spin-averaged matrix elements squared for this type of reactions are given in Eq. (\ref{eq:M_lepsca}) with the coefficients $\beta_i$ being listed in Table \ref{tab:beta_lepsca}.
The corresponding reaction kernel is written as
\begin{align}
  R^{\text{in}}_{\nu} \left( E_1, E_2, \cos \theta \right) &= \iint
  \dfrac{d^3 \bm{p}_1}{(2 \pi)^3}
  \dfrac{d^3 \bm{p}_2}{(2 \pi)^3} \dfrac{ \langle |\mathcal{M}|^2 \rangle_{\text{lsc}}}{16 E_1 E_2 p_1^0 p_2^0} \, 2 f_l\left(p_1^0\right) \left[ 1 -  f_{l}\left(p_2^0\right) \right]
  \left( 2 \pi \right)^4 \delta^{(4)} \left(q_1^{\alpha}+p_1^{\alpha} - q_2^{\alpha}-p_2^{\alpha}\right) \\
  &= \dfrac{G_F^2}{(2 \pi)^2 E_1 E_2} \left[ \beta_1 I_1\left( E_1, E_2, \cos \theta \right) + \beta_2 I_2\left( E_1, E_2, \cos \theta \right)  + \beta_3 I_3\left( E_1, E_2, \cos \theta \right) \right], \label{eq:rk_lepsca}
\end{align}
where the three functions $I_1$ through $I_3$ are given as
\begin{align}
  I_1\left( E_1, E_2, \cos \theta \right) &= \iint d^3 \bm{p}_1 d^3 \bm{p}_2 \dfrac{1}{p_1^0 p_2^0} \delta^{(4)}\left( q_1^{\alpha}+p_1^{\alpha} - q_2^{\alpha}-p_2^{\alpha} \right) f_l\left(p_1^0\right)  \left[ 1 -  f_{l}\left(p_2^0\right) \right] \left( q_1 \cdot p_1 \right)\left( q_2 \cdot p_2 \right), \\
  I_2\left( E_1, E_2, \cos \theta \right) &= \iint d^3 \bm{p}_1 d^3 \bm{p}_2 \dfrac{1}{p_1^0 p_2^0} \delta^{(4)}\left( q_1^{\alpha}+p_1^{\alpha} - q_2^{\alpha}-p_2^{\alpha} \right) f_l\left(p_1^0\right)  \left[ 1 -  f_{l}\left(p_2^0\right) \right] \left( q_1 \cdot p_2 \right)\left( q_2 \cdot p_1 \right), \\
  I_3\left( E_1, E_2, \cos \theta \right) &= \iint d^3 \bm{p}_1 d^3 \bm{p}_2 \dfrac{1}{p_1^0 p_2^0} \delta^{(4)}\left( q_1^{\alpha}+p_1^{\alpha} - q_2^{\alpha}-p_2^{\alpha} \right) f_l\left(p_1^0\right)  \left[ 1 -  f_{l}\left(p_2^0\right) \right] m_l^2 \left( q_1 \cdot q_2 \right).
\end{align}
The integrations in $I_1, I_2$ and $I_3$ can be analytically done following the previous works \cite{Yueh1977, Mezzacappa1993}:
\begin{align}
  I_1\left( E_1, E_2, \cos \theta \right) &= \dfrac{2 \pi E_1^2 E_2^2}{\Delta^5} \left( 1 - \cos \theta \right)^2
  \int_{\epsilon_{\text{min}}}^{\infty} d \epsilon_l f_l\left(\epsilon_l\right)  \left[ 1 -  f_{l} \left(\epsilon_l + E_1 - E_2 \right) \right] \left( A_1 \epsilon_l^2 + B_1 \epsilon_l + C_1 \right),\\
  A_1 &= E_1^2 + E_2^2 + E_1 E_2 \left( 3 + \cos \theta \right), \\
  B_1 &= E_1 \left[ 2E_1^2 + E_1 E_2 \left( 3 - \cos \theta \right) - E_2^2 \left( 1 + 3 \cos \theta \right)\right], \\
  C_1 &= E_1^2 \left[ E_1^2 - 2 E_1 E_2 \cos \theta + \dfrac{1}{2} E_2^2 \left( 3 \cos^2 \theta - 1 \right) - \dfrac{m_l^2}{2} \dfrac{1 + \cos \theta}{1 - \cos \theta} \dfrac{\Delta^2}{E_1^2} \right] , \\
  \Delta &= \sqrt{E_1^2 + E_2^2 - 2 E_1 E_2 \cos \theta}, \\
  \epsilon_{\text{min}} &= \max \left\{ m_l, \epsilon_{-}, -(E_1 - E_2) \right\}, \\
  \epsilon_{-} &= - \dfrac{E_1 - E_2}{2} + \dfrac{\Delta}{2} \sqrt{1 + \dfrac{2 m_l^2}{E_1 E_2 (1 - \cos \theta)}};
\end{align}
the remaining integral over $\epsilon_l$ can be reduced to the Fermi-Dirac integrals:
\begin{align}
  J_0 &= \int_{\epsilon_{\text{min}}}^{\infty} d \epsilon_l f_l\left(\epsilon_l\right)  \left[ 1 -  f_{l}\left(\epsilon_l + E_1 - E_2 \right) \right] \notag \\
      &= \dfrac{1}{e^{- \beta (E_1 - E_2)} - 1 } \int_{\epsilon_{\text{min}}}^{\infty} d \epsilon_l \left[ f_{l}\left(\epsilon_l + E_1 - E_2 \right) - f_l\left(\epsilon_l\right)  \right] \notag \\
      &= \dfrac{1}{\beta} \dfrac{1}{e^{- \beta (E_1 - E_2)} - 1 } \int_{0}^{\infty} dz
      \left\{ \dfrac{1}{1 + \exp \left[ z - \beta \left( \mu_l - \left(E_1 - E_2\right) - \epsilon_{\text{min}} \right) \right]} - \dfrac{1}{1 + \exp \left[z - \beta \left( \mu_l - \epsilon_{\text{min}} \right) \right] } \right\} \notag \\
      &= \dfrac{T}{e^{(E_2 - E_1)/T} - 1} G_0\left( \beta \epsilon_{\text{min}} \right) , \label{eq:G0} \\
  J_1 &= \int_{\epsilon_{\text{min}}}^{\infty} d \epsilon_l f_l\left(\epsilon_l\right)  \left[ 1 -  f_{l}\left(\epsilon_l + E_1 - E_2 \right) \right] \epsilon_l  \notag\\
      &= \dfrac{T^2}{e^{(E_2 - E_1)/T} - 1} \left[ G_1\left(\beta \epsilon_{\text{min}}\right) + \beta \epsilon_{\text{min}} G_0\left( \beta \epsilon_{\text{min}} \right)\right], \label{eq:G1}\\
  J_2 &= \int_{\epsilon_{\text{min}}}^{\infty} d \epsilon_l f_l\left(\epsilon_l\right)  \left[ 1 -  f_{l}\left(\epsilon_l + E_1 - E_2 \right) \right] \epsilon_l^2  \notag\\
      &= \dfrac{T^3}{e^{(E_2 - E_1)/T} - 1} \left[ G_2\left(\beta \epsilon_{\text{min}}\right) + 2 \beta \epsilon_{\text{min}} G_1\left(\beta \epsilon_{\text{min}}\right) + \left( \beta \epsilon_{\text{min}} \right)^2 G_0\left( \beta \epsilon_{\text{min}} \right)\right], \label{eq:G2}
\end{align}
where $\beta = 1/T$ and $\mu_l$ is the chemical potential of lepton $l$ and $G_i$ is defined as
\begin{align}
  &G_i(\beta \epsilon) = F_i\left[ \beta \left( \mu_l - \left(E_1 - E_2\right) - \epsilon \right)\right] - F_i\left[ \beta \left( \mu_l - \epsilon \right)\right]
\end{align}
with $F_i[\eta]$ being the Fermi-Dirac integrals defined as
\begin{align}
  F_i[\eta] = \int_0^{\infty} \dfrac{x^i}{e^{x - \eta} + 1} dx \ \ (\text{for} \ i \geq 0).
\end{align}
Note that $J_0$ through $J_2$ are not divergent at $E_1 = E_2$ and are given as
\begin{align}
  J_0 &= T F_{-1}\left[ \beta\left(\mu_l - \epsilon_{\text{min}}\right) \right] , \\
  J_1 &= T^2 \left\{ F_0\left[\beta \left(\mu_l - \epsilon_{\text{min}}\right) \right] + \beta \epsilon_{\text{min}} F_{-1}\left[ \beta \left(\mu_l - \epsilon_{\text{min}}\right) \right] \right\},\\
  J_2 &= T^3 \left\{ 2 F_1\left[\beta \left(\mu_l - \epsilon_{\text{min}}\right) \right]
  + 2 \beta \epsilon_{\text{min}} F_0\left[\beta \left(\mu_l - \epsilon_{\text{min}}\right) \right] + \left( \beta \epsilon_{\text{min}} \right)^2 F_{-1}\left[ \beta \left(\mu_l - \epsilon_{\text{min}}\right) \right] \right\},
\end{align}
with
\begin{equation}
  F_{-1}[\eta] = \dfrac{1}{e^{- \eta} + 1}.
\end{equation}
The Fermi-Dirac integrals included in $J_i$ (see equations (\ref{eq:G0})--(\ref{eq:G2})) are numerically evaluated \cite{JosepM.Aparicio}.
To summarize, we get
\begin{align}
  I_1\left( E_1, E_2, \cos \theta \right) = \dfrac{2 \pi E_1^2 E_2^2}{\Delta^5} \left( 1 - \cos \theta \right)^2 \left( A_1 J_2 + B_1 J_1 + C_1 J_0 \right).
\end{align}
The other two integrals $I_2$ and $I_3$ are calculated in a similar way as
\begin{align}
  I_2\left( E_1, E_2, \cos \theta \right) &= \int d^3 \bm{p}_1 d^3 \bm{p}_2 \dfrac{1}{p_1^0 p_2^0} \, \delta^{(4)} \bm{(} \left(- q_2^{\alpha}\right)+p_1^{\alpha} - \left(-q_1^{\alpha}\right)-p_2^{\alpha} \bm{)} \, f_l\left(p_1^0\right)  \left[ 1 -  f_{l}\left(p_2^0\right) \right] \notag \\
  & \ \ \ \ \ \ \ \ \ \ \ \ \ \ \ \ \ \ \times \left( \left(-q_1\right) \cdot p_2 \right)\left( \left(-q_2\right) \cdot p_1 \right) \notag\\
  &= I_1\left( -E_2, -E_1, \cos \theta \right), \\
  I_3\left( E_1, E_2, \cos \theta \right) &= \dfrac{2 \pi m_l^2 E_1 E_2}{\Delta} \left(1 - \cos \theta\right) \int_{\epsilon_{\text{min}}}^{\infty} d \epsilon_l f_l\left(\epsilon_l\right)  \left[ 1 -  f_{l}\left(\epsilon_l + E_1 - E_2 \right) \right] \notag \\
  &= \dfrac{2 \pi m_l^2 E_1 E_2}{\Delta} \left(1 - \cos \theta\right) J_0.
\end{align}
The other kernel $R^{\text{out}_{\nu}}$ is obtained from the detailed balance, Eq. (\ref{eq:DetailedBalance_lsc}).

\subsection{ Lepton flavor exchange/conversion reactions \label{appendix:flex}}
The following 4 reactions
\begin{align}
  &\nu_e + \mu^- \leftrightarrows \nu_{\mu} + e^-,\\
  &\bar{\nu}_e + \mu^+ \leftrightarrows \bar{\nu}_{\mu} + e^+, \\
  &\bar{\nu}_{\mu} + \mu^- \leftrightarrows \bar{\nu}_e + e^-, \\
  &\nu_{\mu} + \mu^+\leftrightarrows \nu_e + e^+,
\end{align}
are collectively expressed in this subsection as
\begin{equation}
  \nu_1 + \mu \leftrightarrows \nu_2 + e,
\end{equation}
and the 4-momenta of $\nu_1$, $\nu_2$, $\mu$ and $e$ are denoted by $q_1^{\alpha}$, $q_2^{\alpha}$, $p_{\mu}^{\alpha}$ and $p_e^{\alpha}$, respectively.
The spin-averaged matrix elements squared for these reactions are given in Eq. (\ref{eq:M_flex}) and the coefficients $\alpha_i$ are listed in Table \ref{tab:alpha_flex}.
We present here the detailed expression of $R^{\text{out}}_{\nu_1}$ in Eq. (\ref{eq:emi_GroupA}), which is similar to that of the lepton scattering counterpart:
\begin{align}
R^{\text{out}}_{\nu_1} \left( E_1, E_2, \cos \theta \right) &= \iint
\dfrac{d^3 \bm{p_1}}{(2 \pi)^3}
\dfrac{d^3 \bm{p_2}}{(2 \pi)^3} \dfrac{ \langle |\mathcal{M}|^2 \rangle_{\text{flex}}}{16 E_1 E_2 p_1^0 p_2^0} \, 2 f_{e}\left(p_{e}^0\right) \left[ 1 -  f_{\mu}\left(p_{\mu}^0\right) \right]
\left( 2 \pi \right)^4 \delta^{(4)} \left(q_1^{\alpha}+p_{\mu}^{\alpha} - q_2^{\alpha}-p_e^{\alpha}\right) \\
&= \dfrac{ 8 G_F^2}{(2 \pi)^2 E_1 E_2} \left[ \alpha_1 I_1\left( E_1, E_2, \cos \theta \right) + \alpha_2 I_2\left( E_1, E_2, \cos \theta \right) \right], \label{eq:kernel_flex_out}
\end{align}
where $I_1$ is expressed as
\begin{align}
  I_1\left( E_1, E_2, \cos \theta \right) &= \iint d^3 \bm{p}_{\mu} d^3 \bm{p}_e \dfrac{1}{p_{\mu}^0 p_{e}^0} \, \delta^{(4)}\left( q_1^{\alpha}+p_{\mu}^{\alpha} - q_2^{\alpha}-p_{e}^{\alpha} \right) f_e\left(p_{e}^0\right)  \left[ 1 -  f_{\mu}\left(p_{\mu}^0\right) \right] \left( q_1 \cdot p_{e} \right)\left( q_2 \cdot p_{\mu} \right) \\
  &= \dfrac{2 \pi E_1 E_2}{\Delta^5} \left( A_1 J_2 + B_1 J_1 + C_1 J_0 \right), \\
  A_1 &= E_1 E_2 \left(1 - \cos \theta \right)^2 \left[ E_1^2 + E_2^2 + E_1 E_2 \left( 3 + \cos \theta \right) \right], \\
  B_1 &= E_1 E_2^2 \left(1 - \cos \theta \right)^2 \left[ E_1^2 \left( 1 + 3 \cos \theta \right) - E_1 E_2 \left( 3 - \cos \theta \right) - 2E_2^2 \right] \notag \\
  & \ \ \ + Q \left(E_2 - E_1\right) \left( 1 - \cos \theta \right) \left[ E_1^2 + E_1 E_2 \left( 3 + \cos \theta \right) + E_2^2 \right], \\
  C_1 &= E_1 E_2^3 \left(1 - \cos \theta \right)^2
  \left[ \dfrac{1}{2} E_1^2 \left( 3 \cos^2 \theta -1 \right) - 2 E_1 E_2 \cos \theta + E_2^2  \right]
  - \dfrac{m_{\mu}^2}{2}  E_1 E_2 \left( 1 - \cos^2 \theta \right) \Delta^2 \notag \\
  & \ \ \ - Q E_2 \left(1 - \cos \theta \right) \left[ E_1^3 \cos \theta - E_1^2 E_2 \left( 2 - \cos^2 \theta \right) - E_1 E_2^2 \cos \theta + E_2^3\right] \notag \\
  & \ \ \ + Q^2 \left[ E_1^2 \cos \theta - \dfrac{1}{2}E_1 E_2 \left( 3 + \cos^2 \theta \right) + E_2^2 \cos \theta \right], \\
  \Delta &= \sqrt{E_1 ^2 + E_2^2 - 2 E_1 E_2 \cos \theta}, \\
  Q &= \dfrac{1}{2}\left(m_{\mu}^2 - m_e^2 \right).
\end{align}

In the above expression, $J_i$'s are again written with the Fermi-Dirac integrals:
\begin{align}
  J_0 &= \dfrac{T}{1 - e^{\left[\mu_{\mu} - \left( \mu_e + E_2 - E_1 \right) \right] /T}} G_0\left( \beta \epsilon_{\text{min}} \right), \\
  J_1 &= \dfrac{T^2}{1 - e^{\left[\mu_{\mu} - \left( \mu_e + E_2 - E_1 \right) \right] /T}} \left[ G_1\left(\beta \epsilon_{\text{min}}\right) + \beta \epsilon_{\text{min}} G_0\left( \beta \epsilon_{\text{min}} \right)\right], \\
  J_2 &=\dfrac{T^3}{1 - e^{\left[\mu_{\mu} - \left( \mu_e + E_2 - E_1 \right) \right] /T}} \left[ G_2\left(\beta \epsilon_{\text{min}}\right) + 2 \beta \epsilon_{\text{min}} G_1\left(\beta \epsilon_{\text{min}}\right) + \left( \beta \epsilon_{\text{min}} \right)^2 G_0\left( \beta \epsilon_{\text{min}} \right)\right], \\
  \Delta &= \sqrt{E_1 ^2 + E_2^2 - 2 E_1 E_2 \cos \theta}, \\
  \epsilon_{\text{min}} &= \max \left\{ m_{\mu}, \epsilon_{+}, m_e + E_2 - E_1 \right\}, \\
  \epsilon_{+} &= \dfrac{E_2 - E_1}{2}\left( 1 - \kappa \right) + \dfrac{\Delta}{2} \sqrt{ \left( 1 - \kappa \right)^2 + \dfrac{2 m_{\mu}^2}{E_1 E_2 (1 - \cos \theta)}}, \\
  \kappa &= \dfrac{Q}{E_1 E_2 \left( 1 - \cos \theta \right)},
\end{align}
where $\mu_{\mu}$ and $\mu_e$ are the chemical potentials of muon and electron, respectively, and $G_i$'s are given as
\begin{equation}
  G_i\left(\beta \epsilon\right) = F_i\left[ \beta \left( \mu_e - \left( E_1 - E_2\right) - \epsilon \right)\right] - F_i\left[ \beta \left( \mu_{\mu} - \epsilon \right)\right].
\end{equation}
There is no divergence at $\mu_{\mu} = \mu_e + E_2 - E_1$, and $J_0$ through $J_2$ are obtained as
\begin{align}
  J_0 &= T F_{-1}\left[ \beta\left(\mu_{\mu} - \epsilon_{\text{min}}\right) \right] , \\
  J_1 &= T^2 \left\{ F_0\left[\beta \left(\mu_{\mu} - \epsilon_{\text{min}}\right) \right] + \beta \epsilon_{\text{min}} F_{-1}\left[ \beta \left(\mu_{\mu} - \epsilon_{\text{min}}\right) \right] \right\},\\
  J_2 &= T^3 \left\{ 2 F_1\left[\beta \left(\mu_{\mu} - \epsilon_{\text{min}}\right) \right]
  + 2 \beta \epsilon_{\text{min}} F_0\left[\beta \left(\mu_{\mu} - \epsilon_{\text{min}}\right) \right] + \left( \beta \epsilon_{\text{min}} \right)^2 F_{-1}\left[ \beta \left(\mu_{\mu} - \epsilon_{\text{min}}\right) \right] \right\}.
\end{align}
The remaining $I_2$ can be obtained from the following relation:
\begin{equation}
  I_2\left( E_1, E_2, \cos \theta \right) = I_1\left( -E_2, -E_1, \cos \theta \right).
\end{equation}
The other kernel $R^{\text{in}}_{\nu_1}$ is obtained from the detailed balance,  Eq. (\ref{eq:DetailedBalance_lsc}).

\subsection{Muon decays \label{appendix:mu_decay}}
In this subsection, we give only the reaction rate of $\mu^- \leftrightarrows e^- + \bar{\nu}_e + \nu_{\mu}$.
The counterpart for $\mu^+ \leftrightarrows e^+ + \nu_e + \bar{\nu}_{\mu}$ can be obtained just by changing the signature of the chemical potentials of charged leptons and exchanging the neutrino and anti-neutrino of the same flavor.
The calculation of $R^{\text{out}}_{\bar{\nu}_e}$ in Eq. (\ref{eq:emi_GroupB}) proceeds in a similar way to those for the above two reactions:
\begin{align}
&R^{\text{out}}_{\bar{\nu}_e} \left( E_{\bar{\nu}_e}, E_{\nu_{\mu}}, \cos \theta \right) \notag \\
&= \iint \dfrac{d^3 \bm{p}_{\mu}}{(2 \pi)^3} \dfrac{d^3 \bm{p}_{e}}{(2 \pi)^3}
\dfrac{ \langle |\mathcal{M}|^2 \rangle_{\mu\text{decay}}} {16 E_{\bar{\nu}_e} E_{\nu_{\mu}} p_{\mu}^0 p_e^0}
\, 2 f_{\mu}\left(p_{\mu}^0\right) \left[ 1 -  f_{e}\left(p_{e}^0\right) \right]
\left( 2 \pi \right)^4 \delta^{(4)} \left(p_{\mu}^{\alpha} - p_e^{\alpha} - q_{\bar{\nu}_e}^{\alpha} - q_{\nu_{\mu}}^{\alpha}\right) \\
&= \dfrac{ 8 G_F^2}{(2 \pi)^2 E_{\bar{\nu}_e} E_{\nu_{\mu}}} I_1\left( E_{\bar{\nu}_e}, E_{\nu_{\mu}}, \cos \theta \right) , \label{eq:rk_mudecay}
\end{align}
in which the spin-averaged matrix element squared is given in Eq. (\ref{eq:M_mudecay}) and $I_1$ is written as
\begin{align}
  &I_1\left( E_{\bar{\nu}_e}, E_{\nu_{\mu}}, \cos \theta \right) \notag \\
  &= \iint d^3 \bm{p}_{\mu} d^3 \bm{p}_e \dfrac{1}{p_{\mu}^0 p_{e}^0}
  \delta^{(4)}\left( p_{\mu}^{\alpha} - p_e^{\alpha} - q_{\bar{\nu}_e}^{\alpha} - q_{\nu_{\mu}}^{\alpha} \right)
  2 f_{\mu}\left(p_{\mu}^0\right) \left[ 1 -  f_{e}\left(p_{e}^0\right) \right]
  \left( q_{\bar{\nu}_e} \cdot p_{e} \right)\left( q_{\nu_{\mu}} \cdot p_{\mu} \right) \\
  &= \dfrac{2 \pi E_{\bar{\nu}_e} E_{\nu_{\mu}}}{\Delta^5} \Theta(D) \left( A_1 J_2 + B_1 J_1 + C_1 J_0 \right),
\end{align}
with $\Theta$ being the Heaviside function and
\begin{align}
  A_1 &= - E_{\bar{\nu}_e} E_{\nu_{\mu}} \left(1 - \cos \theta \right)^2
  \left[ E_{\bar{\nu}_e}^2 + E_{\nu_{\mu}}^2 - E_{\bar{\nu}_e} E_{\nu_{\mu}} \left( 3 + \cos \theta \right) \right], \\
  B_1 &= E_{\bar{\nu}_e} E_{\nu_{\mu}}^2 \left(1 - \cos \theta \right)^2
  \left[ E_{\bar{\nu}_e}^2 \left( 1 + 3 \cos \theta \right) + E_{\bar{\nu}_e} E_{\nu_{\mu}} \left( 3 - \cos \theta \right) - 2E_{\nu_{\mu}}^2 \right] \notag \\
  & \ \ \ + Q \left(E_{\nu_{\mu}} + E_{\bar{\nu}_e}\right) \left( 1 - \cos \theta \right)
  \left[ E_{\bar{\nu}_e}^2 - E_{\bar{\nu}_e} E_{\nu_{\mu}} \left( 3 + \cos \theta \right) + E_{\nu_{\mu}}^2 \right], \\
  C_1 &= - E_{\bar{\nu}_e} E_{\nu_{\mu}}^3 \left(1 - \cos \theta \right)^2
  \left[ \dfrac{1}{2} E_{\bar{\nu}_e}^2 \left( 3 \cos^2 \theta -1 \right) + 2 E_{\bar{\nu}_e} E_{\nu_{\mu}} \cos \theta + E_{\nu_{\mu}}^2  \right]
  + \dfrac{m_{e}^2}{2}  E_{\bar{\nu}_e} E_{\nu_{\mu}} \left( 1 - \cos^2 \theta \right) \Delta^2 \notag \\
  & \ \ \ - Q E_{\nu_{\mu}} \left(1 - \cos \theta \right) \left[ E_{\bar{\nu}_e}^3 \cos \theta + E_{\bar{\nu}_e}^2 E_{\nu_{\mu}} \left( 2 - \cos^2 \theta \right) - E_{\bar{\nu}_e} E_{\nu_{\mu}}^2 \cos \theta - E_{\nu_{\mu}}^3\right] \notag \\
  & \ \ \ + Q^2 \left[ E_{\bar{\nu}_e}^2 \cos \theta + \dfrac{1}{2}E_{\bar{\nu}_e} E_{\nu_{\mu}} \left( 3 + \cos^2 \theta \right) + E_{\nu_{\mu}}^2 \cos \theta \right], \\
  \Delta &= \sqrt{E_{\bar{\nu}_e} ^2 + E_{\nu_{\mu}}^2 + 2 E_{\bar{\nu}_e} E_{\nu_{\mu}} \cos \theta}, \\
  Q &= \dfrac{1}{2}\left(m_{\mu}^2 - m_e^2 \right), \\
  D &= \left( 1 - \kappa\right)^2 - \dfrac{2 m_e^2}{E_{\bar{\nu}_e} E_{\nu_{\mu}} \left(1 - \cos \theta\right)}, \\
  \kappa &= \dfrac{Q}{E_{\bar{\nu}_e} E_{\nu_{\mu}} \left(1 - \cos \theta\right)};
\end{align}

$J_i$'s are given by the Fermi-Dirac integrals:
\begin{align}
  J_0 &= \dfrac{T}{1 - e^{\left( E_{\bar{\nu}_e} + E_{\nu_{\mu}} + \mu_e - \mu_{\mu} \right) /T}} \left[ G_0\left( \beta \epsilon_{\text{min}} \right) - G_0\left( \beta \epsilon_{\text{max}} \right) \right], \\
  J_1 &= \dfrac{T^2}{1 - e^{\left( E_{\bar{\nu}_e} + E_{\nu_{\mu}} + \mu_e - \mu_{\mu} \right) /T}} \left\{ \left[ G_1\left(\beta \epsilon_{\text{min}}\right) + \beta \epsilon_{\text{min}} G_0\left( \beta \epsilon_{\text{min}} \right)\right]
  - \left[ G_1\left(\beta \epsilon_{\text{max}}\right) + \beta \epsilon_{\text{max}} G_0\left( \beta \epsilon_{\text{max}} \right)\right]
  \right\}, \\
  J_2 &=\dfrac{T^3}{1 - e^{\left( E_{\bar{\nu}_e} + E_{\nu_{\mu}} + \mu_e - \mu_{\mu} \right) /T}} \left\{ \left[ G_2\left(\beta \epsilon_{\text{min}}\right) + 2 \beta \epsilon_{\text{min}} G_1\left(\beta \epsilon_{\text{min}}\right) + \left( \beta \epsilon_{\text{min}} \right)^2 G_0\left( \beta \epsilon_{\text{min}} \right)\right] \right. \notag \\
  & \ \ \ \ \ \ \ \ \ \ \ \ \ \ \ \ \ \ \ \ \ \ \ \ \ \ \ \ \ \ \ \ \ \ \ \ \ \ \left. - \left[ G_2\left(\beta \epsilon_{\text{max}}\right) + 2 \beta \epsilon_{\text{max}} G_1\left(\beta \epsilon_{\text{max}}\right) + \left( \beta \epsilon_{\text{max}} \right)^2 G_0\left( \beta \epsilon_{\text{max}} \right)\right]
  \right\}, \\
  \epsilon_{\text{min}} &= \max \left\{ m_l, \epsilon_{-}, m_{\mu}-\left( E_{\bar{\nu}_e} + E_{\nu_{\mu}} \right) \right\}, \\
  \epsilon_{\text{max}} &= \epsilon_+ , \\
  \epsilon_{\pm} &= \dfrac{E_{\bar{\nu}_e} + E_{\nu_{\mu}}}{2} \left( \kappa - 1\right) \pm \dfrac{\Delta}{2} \sqrt{\left(1 - \kappa\right)^2 - \dfrac{2 m_e^2}{E_{\bar{\nu}_e} E_{\nu_{\mu}} (1 - \cos \theta)}} = \dfrac{E_{\bar{\nu}_e} + E_{\nu_{\mu}}}{2} \left( \kappa - 1\right) \pm \dfrac{\Delta}{2} \sqrt{D},
\end{align}
where $\mu_{\mu}$ and $\mu_e$ are the chemical potentials of muon and electron, respectively, and $G_i$'s are given as
\begin{equation}
  G_i\left(\beta \epsilon\right) = F_i\left[ \beta \left( \mu_{\mu} - \left( E_{\bar{\nu}_e} + E_{\nu_{\mu}}\right) - \epsilon \right)\right] - F_i\left[ \beta \left( \mu_{e} - \epsilon \right)\right].
\end{equation}
We obtain $J_0$ through $J_2$ at $\mu_{\mu} = \mu_e + E_{\bar{\nu}_e} + E_{\nu_{\mu}} $ by taking appropriate limits:
\begin{align}
  J_0 &= T \left\{ F_{-1}\left[ \beta\left(\mu_{e} - \epsilon_{\text{min}}\right) \right]
  - F_{-1}\left[ \beta\left(\mu_{e} - \epsilon_{\text{max}}\right) \right] \right\}, \\
  J_1 &= T^2 \, \bm{\big(} \left\{ F_0\left[\beta \left(\mu_{e} - \epsilon_{\text{min}}\right) \right] + \beta \epsilon_{\text{min}} F_{-1}\left[ \beta \left(\mu_{e} - \epsilon_{\text{min}}\right) \right] \right\}
  - \left\{ F_0\left[\beta \left(\mu_{e} - \epsilon_{\text{max}}\right) \right] + \beta \epsilon_{\text{max}} F_{-1}\left[ \beta \left(\mu_{e} - \epsilon_{\text{max}}\right) \right] \right\}
  \bm{\big)}, \\
  J_2 &= T^3 \,\bm{\Big(} \left\{ 2 F_1\left[\beta \left(\mu_{e} - \epsilon_{\text{min}}\right) \right]
  + 2 \beta \epsilon_{\text{min}} F_0\left[\beta \left(\mu_{e} - \epsilon_{\text{min}}\right) \right] + \left( \beta \epsilon_{\text{min}} \right)^2 F_{-1}\left[ \beta \left(\mu_{e} - \epsilon_{\text{min}}\right) \right] \right\} \notag \\
  & \ \ \ \ \ \ \ \ \ - \left\{ 2 F_1\left[\beta \left(\mu_{e} - \epsilon_{\text{max}}\right) \right]
  + 2 \beta \epsilon_{\text{max}} F_0\left[\beta \left(\mu_{e} - \epsilon_{\text{max}}\right) \right] + \left( \beta \epsilon_{\text{max}} \right)^2 F_{-1}\left[ \beta \left(\mu_{e} - \epsilon_{\text{max}}\right) \right] \right\}
  \bm{\Big)}.
\end{align}
The other kernel $R^{\text{in}}_{\bar{\nu}_e}$ is obtained from the detailed balance, Eq. (\ref{eq:DetailedBalance_B1}).

\subsection{Pair creations/annihilations \label{appendix:pair}}
In the pair process, $l^- + l^+ \leftrightharpoons \nu + \bar{\nu}$, $\nu$ is any one of $\nu_{e}, \nu_{\mu},  \nu_{\tau}$ and $l$ is either $e$ or $\mu$, and the 4-momenta of neutrino and anti-neutrino are denoted by $q_1^{\alpha}$ and $q_2^{\alpha}$, respectively, and those of lepton and anti-lepton by $p_1^{\alpha}$ and $p_2^{\alpha}$, respectively.
The spin-averaged matrix element squared is given in (\ref{eq:M_pair}) and the coefficients $\alpha$ and $\beta_i$ are listed in Table \ref{tab:pair}.
Here we presents the detailed expression of $R_{\nu_1}^{\text{out}}$ in Eq. (\ref{eq:emi_GroupB}), which is written as
\begin{align}
  R^{\text{out}}_{\nu_1} \left( E_1, E_2, \cos \theta \right) &= \iint
  \dfrac{d^3 \bm{p}_1}{(2 \pi)^3}
  \dfrac{d^3 \bm{p}_2}{(2 \pi)^3} \dfrac{ \langle |\mathcal{M}|^2 \rangle_{\text{pair}}}{16 E_1 E_2 p_1^0 p_2^0} 2 f_{l^-}\left(p_1^0\right) 2 f_{l^+}\left(p_2^0\right)
  \left( 2 \pi \right)^4 \delta^{(4)} \left(q_1^{\alpha} + q_2^{\alpha} - p_1^{\alpha} - p_2^{\alpha}\right) \\
  &= \dfrac{2G_F^2}{(2 \pi)^2 E_1 E_2} \left[ \beta_1 I_1\left( E_1, E_2, \cos \theta \right) + \beta_2 I_2\left( E_1, E_2, \cos \theta \right)  + \beta_3 I_3\left( E_1, E_2, \cos \theta \right) \right], \label{eq:rk_pair}
\end{align}
where $I_1$ can be cast into the following form:
\begin{align}
  I_1\left( E_1, E_2, \cos \theta \right) &= \iint d^3 \bm{p}_1 d^3 \bm{p}_2 \dfrac{1}{p_1^0 p_2^0} \delta^{(4)}\left( q_1^{\alpha} + q_2^{\alpha} - p_1^{\alpha} - p_2^{\alpha} \right)
  f_{l^-}\left(p_1^0\right) f_{l^+}\left(p_2^0\right) \left( q_1 \cdot p_1 \right)\left( q_2 \cdot p_2 \right) \\
  &= \dfrac{2 \pi E_{1}^2 E_{2}^2} {\Delta^5} \left( 1 - \cos \theta \right)^2 \Theta(D) \left( A_1 J_2 + B_1 J_1 + C_1 J_0 \right), \\
  A_1 &= E_1^2 + E_2^2 - E_1 E_2 \left( 3 + \cos \theta \right), \\
  B_1 &= E_1 \left[ -2E_1^2 + E_1 E_2 \left( 3 - \cos \theta \right) + E_2^2 \left( 1 + 3 \cos \theta \right)\right], \\
  C_1 &= E_1^2 \left[ E_1^2 + 2 E_1 E_2 \cos \theta + \dfrac{1}{2} E_2^2 \left( 3 \cos^2 \theta - 1 \right) - \dfrac{m_l^2}{2} \dfrac{1 + \cos \theta}{1 - \cos \theta} \dfrac{\Delta^2}{E_1^2} \right] , \\
  \Delta &= \sqrt{E_1^2 + E_2^2 + 2 E_1 E_2 \cos \theta}, \\
  D &= 1 - \dfrac{2 m_l^2}{E_1 E_2 \left( 1 - \cos \theta \right)}.
\end{align}
Here again $J_i$'s are written in terms of the Fermi-Dirac integrals:
\begin{align}
  J_0 &= \dfrac{T}{e^{\left( E_1 + E_2 \right) /T} - 1} \left[ G_0\left( \beta \epsilon_{\text{min}} \right) - G_0\left( \beta \epsilon_{\text{max}} \right) \right], \\
  J_1 &= \dfrac{T^2}{e^{\left( E_1 + E_2 \right) /T} - 1} \left\{ \left[ G_1\left(\beta \epsilon_{\text{min}}\right) + \beta \epsilon_{\text{min}} G_0\left( \beta \epsilon_{\text{min}} \right)\right]
  - \left[ G_1\left(\beta \epsilon_{\text{max}}\right) + \beta \epsilon_{\text{max}} G_0\left( \beta \epsilon_{\text{max}} \right)\right]
  \right\}, \\
  J_2 &=\dfrac{T^3}{e^{\left( E_1 + E_2 \right) /T} - 1} \left\{ \left[ G_2\left(\beta \epsilon_{\text{min}}\right) + 2 \beta \epsilon_{\text{min}} G_1\left(\beta \epsilon_{\text{min}}\right) + \left( \beta \epsilon_{\text{min}} \right)^2 G_0\left( \beta \epsilon_{\text{min}} \right)\right] \right. \notag \\
  & \ \ \ \ \ \ \ \ \ \ \ \ \ \ \ \ \ \ \ \ \ \ \ \ \ \ \left. - \left[ G_2\left(\beta \epsilon_{\text{max}}\right) + 2 \beta \epsilon_{\text{max}} G_1\left(\beta \epsilon_{\text{max}}\right) + \left( \beta \epsilon_{\text{max}} \right)^2 G_0\left( \beta \epsilon_{\text{max}} \right)\right]
  \right\}, \\
  \epsilon_{\text{min}} &= \max \left\{ m_l, \epsilon_{-} \right\}, \\
  \epsilon_{\text{max}} &= \min \left\{ E_1 + E_2 - m_l, \epsilon_+ \right\}, \\
  \epsilon_{\pm} &= \dfrac{E_{1} + E_{2}}{2} \pm \dfrac{\Delta}{2} \sqrt{1 - \dfrac{2 m_l^2}{E_{1} E_{2} (1 - \cos \theta)}}
  = \dfrac{E_{1} + E_{2}}{2} \pm \dfrac{\Delta}{2} \sqrt{D},
\end{align}
with
\begin{equation}
  G_i\left(\beta \epsilon\right) = F_i\left[ \beta \left( \mu_{l^-} + E_1 + E_2 - \epsilon \right)\right] - F_i\left[ \beta \left( \mu_{l^-} - \epsilon \right)\right] ;
\end{equation}
$I_2$ and $I_3$ are calculated in the similar way as
\begin{align}
  I_2\left( E_1, E_2, \cos \theta \right) &=  I_1\left( E_2, E_1, \cos \theta \right), \\
  I_3\left( E_1, E_2, \cos \theta \right) &= \dfrac{2 \pi m_l^2 E_1 E_2}{\Delta} \left(1 - \cos \theta\right) \Theta \left( D \right)J_0.
\end{align}
The other kernel $R^{\text{in}}_{\nu_1}$ is obtained from the detailed balance, Eq. (\ref{eq:DetailedBalance_B2}).

\subsection{Leptonic annihilations \label{appendix:lep_ani}}
The two reactions, $e^- + \mu^+ \leftrightarrows \nu_{e} + \bar{\nu}_\mu$ and $e^+ + \mu^-  \leftrightarrows \bar{\nu}_e + \nu_{\mu} $, are collectively denoted in this subsection by $e + \mu \leftrightarrows \nu_e + \nu_{\mu}$.
The calculation of $R^{\text{out}}_{{\nu_e}}$ in Eq. (\ref{eq:emi_GroupB}) proceeds similarly to that for the muon decay. It is expressed as
\begin{align}
R^{\text{out}}_{{\nu_e}} \left( E_{{\nu_e}}, E_{\nu_{\mu}}, \cos \theta \right)
&= \iint \dfrac{d^3 \bm{p}_{\mu}}{(2 \pi)^3} \dfrac{d^3 \bm{p}_{e}}{(2 \pi)^3}
\dfrac{ \langle |\mathcal{M}|^2 \rangle_{\text{lep.ann.}}} {16 E_{{\nu_e}} E_{\nu_{\mu}} p_{\mu}^0 p_e^0}
2 f_{e}\left(p_{e}^0\right) 2 f_{\mu}\left(p_{\mu}^0\right)
\left( 2 \pi \right)^4 \delta^{(4)} \left(p_e^{\alpha} + p_{\mu}^{\alpha} - q_{{\nu_e}}^{\alpha} - q_{\nu_{\mu}}^{\alpha}\right) \\
&= \dfrac{ 8 G_F^2}{(2 \pi)^2 E_{{\nu_e}} E_{\nu_{\mu}}} I_1\left( E_{{\nu_e}}, E_{\nu_{\mu}}, \cos \theta \right) , \label{eq:rk_lepani}
\end{align}
where the spin-averaged matrix element squared is given in Eq. (\ref{eq:M_lepani}) and $I_1$ can be cast into the following form:
\begin{align}
  I_1\left( E_{{\nu_e}}, E_{\nu_{\mu}}, \cos \theta \right)
  &= \iint d^3 \bm{p}_{\mu} d^3 \bm{p}_e \dfrac{1}{p_{\mu}^0 p_{e}^0}
  \delta^{(4)}\left( p_e^{\alpha} + p_{\mu}^{\alpha} - q_{{\nu_e}}^{\alpha} - q_{\nu_{\mu}}^{\alpha} \right)
  2 f_{e}\left(p_{e}^0\right) 2 f_{\mu}\left(p_{\mu}^0\right)
  \left( q_{{\nu_e}} \cdot p_{e} \right)\left( q_{\nu_{\mu}} \cdot p_{\mu} \right) \\
  &= \dfrac{2 \pi E_{{\nu_e}} E_{\nu_{\mu}}}{\Delta^5} \Theta(D) \left( A_1 J_2 + B_1 J_1 + C_1 J_0 \right),
\end{align}
where $\Theta$ is again the Heaviside function and other factors are given as follows:
\begin{align}
  A_1 &=  E_{{\nu_e}} E_{\nu_{\mu}} \left(1 - \cos \theta \right)^2
  \left[ E_{{\nu_e}}^2 + E_{\nu_{\mu}}^2 - E_{{\nu_e}} E_{\nu_{\mu}} \left( 3 + \cos \theta \right) \right], \\
  B_1 &= E_{{\nu_e}}^2 E_{\nu_{\mu}} \left(1 - \cos \theta \right)^2
  \left[ E_{\nu_{\mu}}^2 \left( 1 + 3 \cos \theta \right) + E_{\nu_{\mu}} E_{{\nu_{e}}} \left( 3 - \cos \theta \right) - 2E_{{\nu_{e}}}^2 \right] \notag \\
  & \ \ \ - Q \left(E_{{\nu_{e}}} + E_{\nu_{\mu}}\right) \left( 1 - \cos \theta \right)
  \left[ E_{\nu_{\mu}}^2 - E_{\nu_{\mu}} E_{{\nu_{e}}} \left( 3 + \cos \theta \right) + E_{{\nu_{e}}}^2 \right], \\
  C_1 &= - E_{\nu_{\mu}} E_{{\nu_{e}}}^3 \left(1 - \cos \theta \right)^2
  \left[ \dfrac{1}{2} E_{\nu_{\mu}}^2 \left( 3 \cos^2 \theta -1 \right) + 2 E_{\nu_{\mu}} E_{{\nu_{e}}} \cos \theta + E_{{\nu_{e}}}^2  \right]
  - \dfrac{m_{\mu}^2}{2}  E_{\nu_{\mu}} E_{{\nu_{e}}} \left( 1 - \cos^2 \theta \right) \Delta^2 \notag \\
  & \ \ \ + Q E_{{\nu_{e}}} \left(1 - \cos \theta \right) \left[ - E_{\nu_{\mu}}^3 \cos \theta - E_{\nu_{\mu}}^2 E_{{\nu_{e}}} \left( 2 - \cos^2 \theta \right) + E_{\nu_{\mu}} E_{{\nu_{e}}}^2 \cos \theta + E_{{\nu_{e}}}^3\right] \notag \\
  & \ \ \ - Q^2 \left[ E_{\nu_{\mu}}^2 \cos \theta + \dfrac{1}{2}E_{\nu_{\mu}} E_{{\nu_{e}}} \left( 3 + \cos^2 \theta \right) + E_{{\nu_{e}}}^2 \cos \theta \right], \\
  \Delta &= \sqrt{E_{\nu_{\mu}} ^2 + E_{{\nu_{e}}}^2 + 2 E_{\nu_{\mu}} E_{{\nu_{e}}} \cos \theta}, \\
  Q &= \dfrac{1}{2}\left(m_{\mu}^2 - m_e^2 \right), \\
  D &= \left( 1 + \kappa\right)^2 - \dfrac{2 m_{\mu}^2}{E_{\nu_{\mu}} E_{\nu_{e}} \left(1 - \cos \theta\right)}, \\
  \kappa &= \dfrac{Q}{E_{\nu_{\mu}} E_{\nu_{e}} \left(1 - \cos \theta\right)} ;
\end{align}
$J_i$'s are expressed in terms of the Fermi-Dirac integrals:
\begin{align}
  J_0 &= \dfrac{T}{1 - e^{\left( E_{{\nu_e}} + E_{\nu_{\mu}} + \mu_e - \mu_{\mu} \right) /T}} \left[ G_0\left( \beta \epsilon_{\text{min}} \right) - G_0\left( \beta \epsilon_{\text{max}} \right) \right], \\
  J_1 &= \dfrac{T^2}{1 - e^{\left( E_{{\nu_e}} + E_{\nu_{\mu}} + \mu_e - \mu_{\mu} \right) /T}} \left\{ \left[ G_1\left(\beta \epsilon_{\text{min}}\right) + \beta \epsilon_{\text{min}} G_0\left( \beta \epsilon_{\text{min}} \right)\right]
  - \left[ G_1\left(\beta \epsilon_{\text{max}}\right) + \beta \epsilon_{\text{max}} G_0\left( \beta \epsilon_{\text{max}} \right)\right]
  \right\}, \\
  J_2 &=\dfrac{T^3}{1 - e^{\left( E_{{\nu_e}} + E_{\nu_{\mu}} + \mu_e - \mu_{\mu} \right) /T}} \left\{ \left[ G_2\left(\beta \epsilon_{\text{min}}\right) + 2 \beta \epsilon_{\text{min}} G_1\left(\beta \epsilon_{\text{min}}\right) + \left( \beta \epsilon_{\text{min}} \right)^2 G_0\left( \beta \epsilon_{\text{min}} \right)\right] \right. \notag \\
  & \ \ \ \ \ \ \ \ \ \ \ \ \ \ \ \ \ \ \ \ \ \ \ \ \ \ \ \ \ \ \ \ \ \ \ \ \ \ \left. - \left[ G_2\left(\beta \epsilon_{\text{max}}\right) + 2 \beta \epsilon_{\text{max}} G_1\left(\beta \epsilon_{\text{max}}\right) + \left( \beta \epsilon_{\text{max}} \right)^2 G_0\left( \beta \epsilon_{\text{max}} \right)\right]
  \right\}, \\
  \epsilon_{\text{min}} &= \max \left\{ m_{\mu}, \epsilon_{-} \right\}, \\
  \epsilon_{\text{max}} &= \min \left\{ \epsilon_+ ,E_{{\nu_e}} + E_{\nu_{\mu}} - m_e \right\}, \\
  \epsilon_{\pm} &= \dfrac{E_{{\nu_e}} + E_{\nu_{\mu}}}{2} \left( \kappa + 1\right) \pm \dfrac{\Delta}{2} \sqrt{\left(1 + \kappa\right)^2 - \dfrac{2 m_{\mu}^2}{E_{{\nu_e}} E_{\nu_{\mu}} (1 - \cos \theta)}} = \dfrac{E_{{\nu_e}} + E_{\nu_{\mu}}}{2} \left( \kappa + 1\right) \pm \dfrac{\Delta}{2} \sqrt{D},
\end{align}
and
\begin{equation}
  G_i\left(\beta \epsilon\right) = F_i\left[ \beta \left(E_{{\nu_e}} + E_{\nu_{\mu}} - \mu_e - \epsilon \right)\right] - F_i\left[ \beta \left( \mu_{\mu} - \epsilon \right)\right].
\end{equation}
An appropriate limit to $\mu_{\mu} = E_{{\nu_e}} + E_{\nu_{\mu}} - \mu_e $ results in the following expressions of $J_0$ through $J_2$:
\begin{align}
  J_0 &= T \left\{ F_{-1}\left[ \beta\left(\mu_{\mu} - \epsilon_{\text{min}}\right) \right]
  - F_{-1}\left[ \beta\left(\mu_{\mu} - \epsilon_{\text{max}}\right) \right] \right\}, \\
  J_1 &= T^2 \, \bm{\big(} \left\{ F_0\left[\beta \left(\mu_{\mu} - \epsilon_{\text{min}}\right) \right] + \beta \epsilon_{\text{min}} F_{-1}\left[ \beta \left(\mu_{\mu} - \epsilon_{\text{min}}\right) \right] \right\}
  - \left\{ F_0\left[\beta \left(\mu_{\mu} - \epsilon_{\text{max}}\right) \right] + \beta \epsilon_{\text{max}} F_{-1}\left[ \beta \left(\mu_{\mu} - \epsilon_{\text{max}}\right) \right] \right\}
  \bm{\big)} , \\
  J_2 &= T^3 \, \bm{\Big(} \left\{ 2 F_1\left[\beta \left(\mu_{\mu} - \epsilon_{\text{min}}\right) \right]
  + 2 \beta \epsilon_{\text{min}} F_0\left[\beta \left(\mu_{\mu} - \epsilon_{\text{min}}\right) \right] + \left( \beta \epsilon_{\text{min}} \right)^2 F_{-1}\left[ \beta \left(\mu_{\mu} - \epsilon_{\text{min}}\right) \right] \right\} \notag \\
  & \ \ \ \ \ \ \ \ \  - \left\{ 2 F_1\left[\beta \left(\mu_{\mu} - \epsilon_{\text{max}}\right) \right]
  + 2 \beta \epsilon_{\text{max}} F_0\left[\beta \left(\mu_{\mu} - \epsilon_{\text{max}}\right) \right] + \left( \beta \epsilon_{\text{max}} \right)^2 F_{-1}\left[ \beta \left(\mu_{\mu} - \epsilon_{\text{max}}\right) \right] \right\}
  \bm{\Big)}.
\end{align}
Finally, the other kernel $R^{\text{in}}_{\nu_e}$ is obtained from the detailed balance, Eq. (\ref{eq:DetailedBalance_B2}).

\section{Nucleon structure function \label{appendix:CC}}
We give here the detailed calculations of the structure functions of nucleons, which, under the current approximation, are generally written (Eq. (\ref{eq:structure_function})) as
\begin{align}
  \mathcal{S}^{\mu \nu}\left(q^0, q\right) = \iint \dfrac{d^3 \bm{p}_2}{(2 \pi)^3 2E_2^*}
  \dfrac{d^3 \bm{p}_4}{ (2 \pi)^3 2E_4^*} f_2\left( E_2^* \right) \left[ 1 - f_4\left( E_4^* \right) \right]
  \, \Lambda^{\mu \nu} \,
  (2 \pi)^4 \delta^{(4)}(p_1^\mu + p_2^\mu - p_3^\mu - p_4^\mu).
\end{align}
We follow the procedure given in \cite{Roberts2017_muon}.
First, we decompose the hadronic tensor, $\Lambda^{\mu \nu}$, as follows:
\begin{equation}
  \Lambda^{\mu \nu} = A P_1^{\mu \nu} + B P_2^{\mu \nu} + C P_3^{\mu \nu} + D P_4^{\mu \nu} + E P_5^{\mu \nu} + F P_6^{\mu \nu}, \label{eq:Lambda_decomposition}
\end{equation}
where $P_i^{\mu \nu} \ (i = 1, \dots , 6)$ are defined with the transferred 4 momentum $q^{\alpha} = (q^0, 0, 0, q)$ and another vector $n^{\alpha} = (q, 0, 0, q^0)$, orthogonal to $q^{\alpha}$:
\begin{align}
  P_1^{\mu \nu} &= \eta^{\mu \nu} - \dfrac{1}{q_{\alpha}^2} q^{\mu} q^{\nu} - \dfrac{1}{n^2} n^{\mu} n^{\nu}, \\
  P_2^{\mu \nu} &= \dfrac{1}{q_{\alpha}^2} q^{\mu} q^{\nu}, \\
  P_3^{\mu \nu} &= \dfrac{1}{n^2} n^{\mu} n^{\nu}, \\
  P_4^{\mu \nu} &= \dfrac{1}{q_{\alpha}^2} \left( q^{\mu} n^{\nu} + q^{\nu} n^{\mu}\right), \\
  P_5^{\mu \nu} &= \dfrac{1}{q_{\alpha}^2} q_{\alpha} q_{\beta} \epsilon^{\mu \nu \alpha \beta}, \\
  P_6^{\mu \nu} &= \dfrac{1}{q_{\alpha}^2} \left(q^{\mu} n^{\nu} - q^{\nu} n^{\mu}\right),
\end{align}
where $q_{\alpha}^2 = q_{\alpha} q^{\alpha}$, $n^2 = n_{\alpha} n^{\alpha}$ and $\epsilon^{\mu \nu \alpha \beta}$ is the anti-symmetric tensor.
These tensors are orthogonal to one another and satisfy the following relations:
\begin{align}
  P_{1 \, \mu \nu} P_1^{\mu \nu} = 2, \ P_{2 \, \mu \nu} P_2^{\mu \nu} = 1 , \ P_{3 \, \mu \nu} P_3^{\mu \nu} = 1, \ P_{4 \, \mu \nu} P_4^{\mu \nu} = -2 ,\ P_{5 \, \mu \nu} P_5^{\mu \nu} = -2, \ P_{6 \, \mu \nu} P_6^{\mu \nu} = 2.
\end{align}
Then the coefficients in the decomposition (Eq. (\ref{eq:Lambda_decomposition})) are given as follows:
\begin{align}
  A &= 4 m_2^* m_4^* \left(G_V^2 - G_A^2 \right)
  - 4 \left( G_V^2 + G_A^2\right) \dfrac{1}{q_{\alpha}^2} \left[ \left(q \cdot \tilde{p}_2\right) \left(q \cdot \tilde{p}_4\right) - \left(n \cdot \tilde{p}_2\right)\left(n \cdot \tilde{p}_4\right)\right] \notag \\
  &\ \ \ \  + 4 \dfrac{F_2 G_V}{M} \left[ m_2^* \left(q \cdot \tilde{p}_4\right) - m_4^* \left(q \cdot \tilde{p}_2\right)\right]
  + \dfrac{F_2^2}{M^2} \left[ - \left(n \cdot \tilde{p}_2\right)\left(n \cdot \tilde{p}_4\right) + m_2^* m_4^* q_{\alpha}^2 \right] ,\\
  B &= 4 m_2^* m_4^* \left(G_V^2 - G_A^2 \right)
  + 4 \left( G_V^2 + G_A^2\right) \dfrac{1}{q_{\alpha}^2} \left[ 2 \left(q \cdot \tilde{p}_2\right) \left(q \cdot \tilde{p}_4\right) - q_{\alpha}^2 \left( \tilde{p}_2 \cdot \tilde{p}_4\right)\right]
  \notag \\
  &\ \ \ \  -2 \dfrac{F_2^2}{M^2} \left(q \cdot \tilde{p}_2\right) \left(q \cdot \tilde{p}_4\right)
  + 8 \dfrac{G_A G_P}{M} \left[ m_4^* \left(q \cdot \tilde{p}_2\right) - m_2^* \left(q \cdot \tilde{p}_4\right)\right]
  + 4 \dfrac{G_P^2}{M^2} q_{\alpha}^2 \left[ \left(\tilde{p}_2 \cdot \tilde{p}_4\right) - m_2^* m_4^*\right],\\
  C &= 4 m_2^* m_4^* \left(G_V^2 - G_A^2 \right)
  - 4 \left( G_V^2 + G_A^2\right) \dfrac{1}{n^2} \left[ \left(n \cdot \tilde{p}_2\right) \left(n \cdot \tilde{p}_4\right) - n^2 \left( \tilde{p}_2 \cdot \tilde{p}_4\right)\right] \notag \\
 & \ \ \ \  - 4 \dfrac{F_2 G_V}{M} \left[ m_4^* \left(q \cdot \tilde{p}_2\right) - m_2^* \left(q \cdot \tilde{p}_4\right)\right]
  \notag \\
  &\ \ \ \
  + \dfrac{F_2^2}{M^2} \left[ - n^2 \left(\tilde{p}_2 \cdot \tilde{p}_4\right) + 2 \left(n \cdot \tilde{p}_2\right)\left(n \cdot \tilde{p}_4\right) - 2 \left(q \cdot \tilde{p}_2\right)\left(q \cdot \tilde{p}_4\right) + m_2^* m_4^* q_{\alpha}^2 \right], \\
  D &= -4 \left( G_V^2 + G_A^2\right) \dfrac{1}{q_{\alpha}^2} \left[ \left(q \cdot \tilde{p}_2\right) \left(n \cdot \tilde{p}_4\right) + \left(q \cdot \tilde{p}_4\right) \left(n \cdot \tilde{p}_2\right)\right]
  + 2 \dfrac{F_2 G_V}{M} \left[ m_2^* \left(n \cdot \tilde{p}_4\right) - m_4^* \left(n \cdot \tilde{p}_2\right) \right] \notag \\
  & \ \ \ \ + 4 \dfrac{G_A G_P}{M} \left[ m_2^* \left(n \cdot \tilde{p}_4\right) - m_4^* \left(n \cdot \tilde{p}_2\right) \right], \\
  E &= 8 i G_V G_A \dfrac{1}{q_{\alpha}^2} \left[ \left(q \cdot \tilde{p}_4\right) \left(n \cdot \tilde{p}_2\right) - \left(q \cdot \tilde{p}_2\right) \left(n \cdot \tilde{p}_4\right)\right]
  + 4i \dfrac{F_2 G_A}{M} \left[ m_2^* \left(n \cdot \tilde{p}_4\right) + m_4^* \left(n \cdot \tilde{p}_2\right)\right], \\
  F &= -4i G_V G_A \dfrac{1}{q_{\alpha}^2} \left( q_{\mu} n_{\nu} - q_{\nu} n_{\mu} \right) \tilde{p}_{4 \alpha} \tilde{p}_{2 \beta} \epsilon^{\alpha \mu \beta \nu}
  - 2 i \dfrac{F_2 G_P}{M^2} n_{\mu} \tilde{p}_{4 \rho} \tilde{p}_{2 \sigma} q_{\beta} \epsilon^{\rho \sigma \beta \mu}.\label{eq:F}
\end{align}

Now, the calculation of the structure function is reduced to integrations over $\bm{p}_2$ and $\bm{p}_4$ of the following 10 scalar variables:
\begin{align}
  1,
  \left( \tilde{p}_2 \cdot \tilde{p}_4 \right),
  \left( q \cdot \tilde{p}_2 \right),
  \left( q \cdot \tilde{p}_4 \right),
  \left( n \cdot \tilde{p}_2 \right),
  \left( n \cdot \tilde{p}_4 \right),
  \left( q \cdot \tilde{p}_2 \right)\left( q \cdot \tilde{p}_4 \right),
  \left( n \cdot \tilde{p}_2 \right)\left( n \cdot \tilde{p}_4 \right),
  \left( q \cdot \tilde{p}_2 \right)\left( n \cdot \tilde{p}_4 \right),
  \left( n \cdot \tilde{p}_2 \right)\left( q \cdot \tilde{p}_4 \right).
\label{eq:10scalars}
\end{align}
As an example, we show the calculation for 1:
\begin{align}
  \mathcal{I}_1\left(q^0, q\right) = \iint \dfrac{d^3 \bm{p}_2}{(2 \pi)^3 2E_2^*}
  \dfrac{d^3 \bm{p}_4}{ (2 \pi)^3 2E_4^*} f_2\left( E_2^* \right) \left[ 1 - f_4\left( E_4^* \right) \right]
  \cdot 1 \cdot
  (2 \pi)^4 \delta^{(4)}\left(p_1^\mu + p_2^\mu - p_3^\mu - p_4^\mu\right).
\end{align}
The integration over $\bm{p}_4$ is done with the delta function as
\begin{align}
  &\mathcal{I}_1\left(q^0, q\right) \notag \\
  &= \dfrac{1}{16 \pi^2} \iint \dfrac{d^3 \bm{p}_2}{E_2^*}
  d^3 \bm{p}_4
  \int \dfrac{d E_4^*}{E_4^*} \, \delta \left( E_4^* - \sqrt{|\bm{p}_4|^2 + {m_4^*}^2} \right)
  f_2\left( E_2^* \right) \left[ 1 - f_4\left( E_4^* \right) \right]
  \cdot 1 \cdot
  \delta^{(4)}(p_1^\mu + p_2^\mu - p_3^\mu - p_4^\mu) \\
  &= \dfrac{1}{8 \pi^2} \int p_2 \, dE_2^* \, d\cos\alpha \, d\beta
  f_2\left( E_2^* \right) (1 - f_4\left( E_4^* \right))
  \cdot 1 \cdot
  \delta\left( \tilde{p}_4^2 - {m_4^*}^2 \right) \Theta\left(E_4^* - m_4^* \right), \label{eq:I1_after_p4}
\end{align}
where we used the following relations, $\frac{1}{E_4^*} \delta \left( E_4^* - \sqrt{|\bm{p}_4|^2 + {m_4^*}^2} \right) = \delta\left( \tilde{p}_4^2 - {m_4^*}^2 \right)$, $p_2 d p_2 = E_2^* d E_2^*$, and $\alpha$ and $\beta$ are the zenith and azimuth angles of $\bm{p}_2$, respectively, with $\bm{q}$ being the z-axis;
we also note
\begin{align}
  E_4^* &= E_2^* + U_2 - U_4 + q^0 =:E_2^* + \tilde{q}^0, \\
  \bm{p}_4 &= \bm{p}_2 + \bm{q},
\end{align}
and
\begin{align}
  \delta \left( \tilde{p}_4^2 - {m_4^*}^2 \right) &= \dfrac{1}{|2 q p_2|} \delta\left( \cos \alpha - \cos \alpha_0 \right), \\
  \cos \alpha_0 &= \dfrac{ \left( \tilde{q}^0 \right)^2 + 2 \tilde{q}^0 E_2^* - q^2 + {m_2^*}^2 - {m_4^*}^2 }{2 q p_2}.
\end{align}
The integrations over $\beta$ and $\cos \alpha$ can be done to give
\begin{align}
  \mathcal{I}_1\left(q^0, q\right) = \dfrac{1}{8 \pi |q|} \int_{E_{\text{min}}}^{E_{\text{max}}} d E_2^* f_2\left( E_2^* \right) \left[ 1 - f_4\left( E_4^* \right) \right], \label{eq:I1_before_fermi}
\end{align}
where $E_{\text{min}}$ and $E_{\text{max}}$ are given as
\begin{align}
\begin{cases}
  E_{\text{min}} = \max \left\{ m_2^*, m_4^* - \tilde{q}^0, E^*_{2,+} \right\}, \ E_{\text{max}} = \infty & \left( \Delta^2 := \left( \tilde{q}^0 \right)^2 - q^2 < 0 \right) \\
  E_{\text{min}} = \max \left\{ m_2^*, m_4^* - \tilde{q}^0, E^*_{2,-} \right\}, \ E_{\text{max}} = E^*_{2,+} & \left( \Delta^2 > 0 \ \text{and if} \ \Delta^2 < ( {m_2^*}^2 - {m_4^*}^2) \ \text{or} \ \Delta^2 > ( {m_2^*}^2 + {m_4^*}^2) \right)
\end{cases} \label{eq:E2_range}
\end{align}
with
\begin{align}
  E^*_{2, \pm} = - \dfrac{\tilde{q}^0}{2} \kappa \pm \dfrac{|q|}{2} \sqrt{\kappa^2 - \dfrac{4 {m_2^*}^2}{\Delta^2} },
  \ \kappa = 1 + \dfrac{ {m_2^*}^2 - {m_4^*}^2 }{\Delta^2}; \label{eq:E^*_2pm}
\end{align}
otherwise $E_{2, \pm}^*$ has no real solutions for $ ( {m_2^*}^2 - {m_4^*}^2) < \Delta^2 < ( {m_2^*}^2 + {m_4^*}^2)$ and $\mathcal{I}_1 = 0$.

The remaining integration (\ref{eq:I1_before_fermi}) can be written with the Fermi-Dirac integral, which can be easily evaluated numerically in the following way:
\begin{equation}
  \mathcal{I}_1\left(q^0, q\right) = \dfrac{1}{8 \pi |q|} J_0,
\end{equation}
where $J_0$ as well as $J_1$ and $J_2$, which will be needed later, are expressed as
\begin{align}
  J_0 &= \dfrac{T}{1 - \exp \left[ \beta \left( - \tilde{q}^0 - \mu_2 + \mu_4 \right) \right] } \left[ G_0\left( \beta E_{\text{min}}\right) - G_0\left( \beta E_{\text{max}}\right) \right], \\
  J_1 &= \dfrac{T^2}{1 - \exp \left[ \beta \left( - \tilde{q}^0 - \mu_2 + \mu_4 \right) \right]} \left\{ \left[ G_1\left(\beta E_{\text{min}}\right) + \beta E_{\text{min}} G_0\left( \beta E_{\text{min}} \right)\right]
  - \left[ G_1\left(\beta E_{\text{max}}\right) + \beta E_{\text{max}} G_0\left( \beta E_{\text{max}} \right)\right]
  \right\}, \\
  J_2 &=\dfrac{T^3}{1 - \exp \left[ \beta \left( - \tilde{q}^0 - \mu_2 + \mu_4 \right) \right]} \left\{ \left[ G_2\left(\beta E_{\text{min}}\right) + 2 \beta E_{\text{min}} G_1\left(\beta E_{\text{min}}\right) + \left( \beta E_{\text{min}} \right)^2 G_0\left( \beta E_{\text{min}} \right)\right] \right. \notag\\
  & \ \ \ \ \ \ \ \ \ \ \ \ \ \ \ \ \ \ \ \ \ \ \ \ \ \ \ \ \ \ \ \ \ \ \ \ \ \ \ \ \ \ \ \left. - \left[ G_2\left(\beta E_{\text{max}}\right) + 2 \beta E_{\text{max}} G_1\left(\beta E_{\text{max}}\right) + \left( \beta E_{\text{max}} \right)^2 G_0\left( \beta E_{\text{max}} \right)\right]
  \right\},
\end{align}
with
\begin{equation}
  G_i\left(\beta E\right) = F_i\left[ \beta \left(\mu_2 - E \right)\right] - F_i\left[ \beta \left( \mu_{4} - \tilde{q}^0 - E \right)\right].
\end{equation}
Taking appropriate limits, we obtain for $\tilde{q}^0 = \mu_4 - \mu_2$
\begin{align}
  J_0 &= T \left\{ F_{-1}\left[ \beta\left(\mu_2 - E_{\text{min}}\right) \right]
  - F_{-1}\left[ \beta\left(\mu_2 - E_{\text{max}}\right) \right] \right\}, \label{eq:N_J0}\\
  J_1 &= T^2 \, \bm{\big(} \left\{ F_0\left[\beta \left(\mu_2 - E_{\text{min}}\right) \right] + \beta E_{\text{min}} F_{-1}\left[ \beta \left(\mu_2 - E_{\text{min}}\right) \right] \right\} \notag \\
  & \ \ \ \ \ \ \ \ \ - \left\{ F_0\left[\beta \left(\mu_2 - E_{\text{max}}\right) \right] + \beta E_{\text{max}} F_{-1}\left[ \beta \left(\mu_2 - E_{\text{max}}\right) \right] \right\}
  \bm{\big)}, \label{eq:N_J1}\\
  J_2 &= T^3 \, \bm{\Big(} \left\{ 2 F_1\left[\beta \left(\mu_2 - E_{\text{min}}\right) \right]
  + 2 \beta E_{\text{min}} F_0\left[\beta \left(\mu_2 - E_{\text{min}}\right) \right] + \left( \beta E_{\text{min}} \right)^2 F_{-1}\left[ \beta \left(\mu_2 - E_{\text{min}}\right) \right] \right\} \notag \\
  & \ \ \ \ \ \ \ \ \ - \left\{ 2 F_1\left[\beta \left(\mu_2 - E_{\text{max}}\right) \right]
  + 2 \beta E_{\text{max}} F_0\left[\beta \left(\mu_2 - E_{\text{max}}\right) \right] + \left( \beta E_{\text{max}} \right)^2 F_{-1}\left[ \beta \left(\mu_2 - E_{\text{max}}\right) \right] \right\}
  \bm{\Big)}. \label{eq:N_J2}
\end{align}

In a similar way, other integrals for the remaining 9 scalars in Eq. (\ref{eq:10scalars}) can be accomplished to get the followings:
\begin{align}
  \mathcal{I}_{ \left( \tilde{p}_2 \cdot \tilde{p}_4 \right) } \left(q^0, q\right)
  &= \dfrac{1}{8 \pi |q|} \left[ {m_2^*}^2 - \dfrac{1}{2}\left( \Delta^2 + {m_2^*}^2 - {m_4^*}^2\right) \right] J_0,
\end{align}
\begin{align}
  \mathcal{I}_{ \left( q \cdot \tilde{p}_2 \right) } \left(q^0, q\right) &= \dfrac{1}{8 \pi |q|} \left( b_{\left( q \cdot \tilde{p}_2 \right)} J_1 + c_{\left( q \cdot \tilde{p}_2 \right)} J_0 \right), \\
  b_{\left( q \cdot \tilde{p}_2 \right)} &= U_4 - U_2 =: \Delta U ,\\
  c_{\left( q \cdot \tilde{p}_2 \right)} &= -\dfrac{1}{2} \left( \Delta^2 + {m_2^*}^2 - {m_4^*}^2 \right),
\end{align}
\begin{align}
  \mathcal{I}_{ \left( q \cdot \tilde{p}_4 \right) } \left(q^0, q\right) &= \dfrac{1}{8 \pi |q|} \left( b_{\left( q \cdot \tilde{p}_4 \right)} J_1 + c_{\left( q \cdot \tilde{p}_4 \right)} J_0 \right), \\
  b_{\left( q \cdot \tilde{p}_4 \right)} &= \Delta U, \\
  c_{\left( q \cdot \tilde{p}_4 \right)} &= -\dfrac{1}{2} \left( \Delta^2 + {m_2^*}^2 - {m_4^*}^2 \right) + q^0 \tilde{q}^0 - q^2,
\end{align}
\begin{align}
  \mathcal{I}_{ \left( n \cdot \tilde{p}_2 \right) } \left(q^0, q\right) &= \dfrac{1}{8 \pi |q|} \left( b_{\left( n \cdot \tilde{p}_2 \right)} J_1 + c_{\left( n \cdot \tilde{p}_2 \right)} J_0 \right), \\
  b_{\left( n \cdot \tilde{p}_2 \right)} &= q - \dfrac{q^0 \tilde{q}^0}{q}, \\
  c_{\left( n \cdot \tilde{p}_2 \right)} &= -\dfrac{q^0}{2 q} \left( \Delta^2 + {m_2^*}^2 - {m_4^*}^2 \right),
\end{align}
\begin{align}
  \mathcal{I}_{ \left( n \cdot \tilde{p}_4 \right) } \left(q^0, q\right) &= \dfrac{1}{8 \pi |q|} \left( b_{\left( n \cdot \tilde{p}_4 \right)} J_1 + c_{\left( n \cdot \tilde{p}_4 \right)} J_0 \right), \\
  b_{\left( n \cdot \tilde{p}_4 \right)} &= q - \dfrac{q^0 \tilde{q}^0}{q}, \\
  c_{\left( n \cdot \tilde{p}_4 \right)} &= -\dfrac{q^0}{2 q} \left( \Delta^2 + {m_2^*}^2 - {m_4^*}^2 \right) + q \left( \tilde{q}^0 - q^0\right),
\end{align}
\begin{align}
  \mathcal{I}_{ \left( q \cdot \tilde{p}_2 \right)\left( q \cdot \tilde{p}_4 \right) } \left(q^0, q\right)
  &= \dfrac{1}{8 \pi |q|}
  \left( a_{\left( q \cdot \tilde{p}_2 \right)\left( q \cdot \tilde{p}_4 \right)} J_2
  + b_{\left( q \cdot \tilde{p}_2 \right)\left( q \cdot \tilde{p}_4 \right)} J_1
  + c_{\left( q \cdot \tilde{p}_2 \right)\left( q \cdot \tilde{p}_4 \right)} J_0 \right), \\
  a_{\left( q \cdot \tilde{p}_2 \right)\left( q \cdot \tilde{p}_4 \right)} &=  ( \Delta U )^2, \\
  b_{\left( q \cdot \tilde{p}_2 \right)\left( q \cdot \tilde{p}_4 \right)} &= \Delta U \left[ \left( q^0 \tilde{q}^0 - q^2 \right) - \left( \Delta^2 + {m_2^*}^2 - {m_4^*}^2 \right) \right],\\
  c_{\left( q \cdot \tilde{p}_2 \right)\left( q \cdot \tilde{p}_4 \right)} &= \dfrac{1}{4} \left( \Delta^2 + {m_2^*}^2 - {m_4^*}^2\right)^2 - \dfrac{1}{2} \left( \Delta^2 + {m_2^*}^2 - {m_4^*}^2 \right) \left( q^0 \tilde{q}^0 - q^2 \right),
\end{align}
\begin{align}
  \mathcal{I}_{ \left( n \cdot \tilde{p}_2 \right)\left( n \cdot \tilde{p}_4 \right) } \left(q^0, q\right)
  &= \dfrac{1}{8 \pi |q|}
  \left( a_{\left( n \cdot \tilde{p}_2 \right)\left( n \cdot \tilde{p}_4 \right)} J_2
  + b_{\left( n \cdot \tilde{p}_2 \right)\left( n \cdot \tilde{p}_4 \right)} J_1
  + c_{\left( n \cdot \tilde{p}_2 \right)\left( n \cdot \tilde{p}_4 \right)} J_0 \right), \\
  a_{\left( n \cdot \tilde{p}_2 \right)\left( n \cdot \tilde{p}_4 \right)} &=  \left( q - \dfrac{q^0 \tilde{q}^0}{q}\right)^2, \\
  b_{\left( n \cdot \tilde{p}_2 \right)\left( n \cdot \tilde{p}_4 \right)} &= q \left( \tilde{q}^0 - q^0 \right) \left( q - \dfrac{q^0 \tilde{q}^0}{q}\right) - \dfrac{q^0}{q} \left( \Delta^2 + {m_2^*}^2 - {m_4^*}^2 \right) \left( q - \dfrac{q^0 \tilde{q}^0}{q}\right),\\
  c_{\left( n \cdot \tilde{p}_2 \right)\left( n \cdot \tilde{p}_4 \right)} &= \dfrac{1}{4} \left( \dfrac{q^0}{q} \right)^2 \left( \Delta^2 + {m_2^*}^2 - {m_4^*}^2\right)^2 - \dfrac{1}{2} q^0 \left( \tilde{q}^0 - q^0 \right) \left( \Delta^2 + {m_2^*}^2 - {m_4^*}^2 \right) ,
\end{align}
\begin{align}
  \mathcal{I}_{ \left( q \cdot \tilde{p}_2 \right)\left( n \cdot \tilde{p}_4 \right) } \left(q^0, q\right)
  &= \dfrac{1}{8 \pi |q|}
  \left( a_{\left( q \cdot \tilde{p}_2 \right)\left( n \cdot \tilde{p}_4 \right)} J_2
  + b_{\left( q \cdot \tilde{p}_2 \right)\left( n \cdot \tilde{p}_4 \right)} J_1
  + c_{\left( q \cdot \tilde{p}_2 \right)\left( n \cdot \tilde{p}_4 \right)} J_0 \right), \\
  a_{\left( q \cdot \tilde{p}_2 \right)\left( n \cdot \tilde{p}_4 \right)} &= \Delta U \left( q - \dfrac{q^0 \tilde{q}^0}{q}\right), \\
  b_{\left( q \cdot \tilde{p}_2 \right)\left( n \cdot \tilde{p}_4 \right)} &= -\dfrac{1}{2} \left( \Delta^2 + {m_2^*}^2 - {m_4^*}^2\right) \left( q - \dfrac{q^0 \tilde{q}^0}{q}\right)
  + \Delta U \left[ -\dfrac{q^0}{2q} \left( \Delta^2 + {m_2^*}^2 - {m_4^*}^2 \right) + q \left( \tilde{q}^0 - q^0 \right) \right],\\
  c_{\left( q \cdot \tilde{p}_2 \right)\left( n \cdot \tilde{p}_4 \right)} &= -\dfrac{1}{2} \left( \Delta^2 + {m_2^*}^2 - {m_4^*}^2\right) \left[ - \dfrac{q^0}{2q} \left( \Delta^2 + {m_2^*}^2 - {m_4^*}^2\right) + q \left( \tilde{q}^0 - q^0 \right) \right],
\end{align}
\begin{align}
  \mathcal{I}_{ \left( n \cdot \tilde{p}_2 \right)\left( q \cdot \tilde{p}_4 \right) } \left(q^0, q\right)
  &= \dfrac{1}{8 \pi |q|}
  \left( a_{\left( n \cdot \tilde{p}_2 \right)\left( q \cdot \tilde{p}_4 \right)} J_2
  + b_{\left( n \cdot \tilde{p}_2 \right)\left( q \cdot \tilde{p}_4 \right)} J_1
  + c_{\left( n \cdot \tilde{p}_2 \right)\left( q \cdot \tilde{p}_4 \right)} J_0 \right), \\
  a_{\left( n \cdot \tilde{p}_2 \right)\left( q \cdot \tilde{p}_4 \right)} &= \Delta U \left( q - \dfrac{q^0 \tilde{q}^0}{q}\right), \\
  b_{\left( n \cdot \tilde{p}_2 \right)\left( q \cdot \tilde{p}_4 \right)} &= -\dfrac{{q}^0}{2 q} \left( \Delta^2 + {m_2^*}^2 - {m_4^*}^2\right) \Delta U
  + \left( q - \dfrac{q^0 \tilde{q}^0}{q}\right) \left[ -\dfrac{1}{2} \left( \Delta^2 + {m_2^*}^2 - {m_4^*}^2 \right) + q^0 \tilde{q}^0 - q^2 \right],\\
  c_{\left( n \cdot \tilde{p}_2 \right)\left( q \cdot \tilde{p}_4 \right)} &= -\dfrac{q^0}{2q} \left( \Delta^2 + {m_2^*}^2 - {m_4^*}^2\right) \left[ - \dfrac{1}{2} \left( \Delta^2 + {m_2^*}^2 - {m_4^*}^2\right) + q^0 \tilde{q}^0 - q^2 \right].
\end{align}

The structure function is obtained in the following form:
\begin{align}
  \mathcal{S}^{\mu \nu}\left(q^0, q\right) = \bar{A} P_1^{\mu \nu} + \bar{B} P_2^{\mu \nu} + \bar{C} P_3^{\mu \nu} + \bar{D} P_4^{\mu \nu} + \bar{E} P_5^{\mu \nu}, \label{eq:sf_appendix}
\end{align}
where the coefficients are written as
\begin{equation}
  \bar{X} = \iint \dfrac{d^3 \bm{p}_2}{(2 \pi)^3 2E_2^*}
    \dfrac{d^3 \bm{p}_4}{ (2 \pi)^3 2E_4^*} \, f_2\left( E_2^* \right) \left[ 1 - f_4\left( E_4^* \right) \right]
    \, X \,
    (2 \pi)^4 \delta^{(4)}(p_1^\mu + p_2^\mu - p_3^\mu - p_4^\mu) \ \ \ \left( X = A,B,C,D,E\right),
\end{equation}
and are actually some linear combinations of $\mathcal{I}_i$ with $i = 1,
\left( \tilde{p}_2 \cdot \tilde{p}_4 \right), \dots
,\left( n \cdot \tilde{p}_2 \right)\left( q \cdot \tilde{p}_4 \right)$.
Note that the case with $X = F$ (see Eq. (\ref{eq:F})) is omitted because it vanishes as a result of energy momentum conservation,
\begin{align}
  q_{\mu} n_{\nu} \tilde{p}_{2 \alpha} \tilde{p}_{4 \beta} \epsilon^{\mu \nu \alpha \beta}
  &= q_{\mu} n_{\nu} \tilde{p}_{2 \alpha} \left( \tilde{p}_{2 \beta} + q_{\beta} + U_{\beta} \right)\epsilon^{\mu \nu \alpha \beta} \notag \\
  &= (U_2 - U_4) q_{\mu} n_{\nu} \tilde{p}_{2 \alpha} \epsilon^{\mu \nu \alpha 0} \ \ \ \ \left( U^{\beta} = (U_2 - U_4, 0, 0, 0)\right) \notag \\
  &= \Delta U q_{i} n_{j} \tilde{p}_{2 k} \epsilon^{i j k} \notag \\
  &= 0. \ \ \ \ \ \ \ \ \ \ \ \ \ \ \ \ \ \ \left( \because \bm{q} = (0, 0, q), \bm{n} = (0, 0, q^0) \right)
\end{align}

Finally the contraction of the projection tensors with the lepton tensor yields the following results:
\begin{align}
  L_{\mu \nu} P_1^{\mu \nu} &= -16 \left[ \dfrac{1}{q_{\alpha}^2} \left( p_1 \cdot q\right) \left( p_3 \cdot q\right) + \dfrac{1}{n^2} \left( p_1 \cdot n\right) \left( p_3 \cdot n\right) \right], \\
  L_{\mu \nu} P_2^{\mu \nu} &= 8 \left[ \dfrac{2}{q_{\alpha}^2} \left( p_1 \cdot q\right) \left( p_3 \cdot q\right) - \left( p_1 \cdot p_3\right) \right], \\
  L_{\mu \nu} P_3^{\mu \nu} &= 8 \left[ \dfrac{2}{n^2} \left( p_1 \cdot n\right) \left( p_3 \cdot n\right) - \left( p_1 \cdot p_3\right) \right], \\
  L_{\mu \nu} P_4^{\mu \nu} &= \dfrac{16}{q_{\alpha}^2} \left[ \left( p_1 \cdot q\right) \left( p_3 \cdot n\right) + \left( p_1 \cdot n\right) \left( p_3 \cdot q\right) \right], \\
  L_{\mu \nu} P_5^{\mu \nu} &= \pm i \dfrac{16}{q_{\alpha}^2} \left[ \left( p_1 \cdot n\right) \left( p_3 \cdot q\right) - \left( p_1 \cdot q\right) \left( p_3 \cdot n\right) \right] \ \ \ \left( +: \text{neutrino}, -: \text{antineutrino}\right).
\end{align}


\end{document}